\documentclass[physics , phd , 12pt , final]{uleththesis}

\usepackage{amsfonts}
\usepackage{amsmath}
\usepackage{amssymb}
\usepackage{amstext}
\usepackage{amsthm}
\usepackage{booktabs}
\usepackage[tableposition=top]{caption}
\usepackage[usenames,dvipsnames]{color}
\usepackage{colortbl}
\usepackage{enumerate}
\usepackage{float}    \floatstyle{plaintop} \restylefloat{table} 
\usepackage{graphicx}

\usepackage[LGR,T1]{fontenc}
\newcommand{\textgreek}[1]{\begingroup\fontencoding{LGR}\selectfont#1\endgroup}

\usepackage{latexsym}
\usepackage{listings}
\usepackage{marvosym}
\usepackage{multirow}
\usepackage{pslatex}
\usepackage{soul}
\usepackage{url}
\usepackage{xspace}

\usepackage{calc}
\usepackage{subcaption}

\renewcommand{\bar}[1]{\overline{#1}}    

\definecolor{Gray}{gray}{0.9}

\theoremstyle{plain}

\theoremstyle{definition}

\theoremstyle{remark}

\newcommand{\normalarray}{\renewcommand{\arraystretch}{1.1}}

\normalarray

\newcommand{\CE}{\mathcal{C}}

\newcommand{\LL}{\mathcal{L}}

 {\begin{list}{}%
    {\setlength{\leftmargin}{#1}}%
  \item[]%
 }
{\end{list}}

\usepackage{color}

\newcommand{\PM}{P_\mathrm{M}}
\newcommand{\QM}{Q_\mathrm{M}}

\newcommand{\diff}{\mathrm{d}}

\newcommand{\dslash}{\hspace{-0.2em}\not\hspace{-0.1ex}}

\DeclareSymbolFontAlphabet{\mathrm}    {operators}
\DeclareSymbolFontAlphabet{\mathnormal}{letters}
\DeclareSymbolFontAlphabet{\mathcal}   {symbols}
\DeclareMathAlphabet      {\mathbf}{OT1}{cmr}{bx}{n}
\DeclareMathAlphabet      {\mathsf}{OT1}{cmss}{m}{n}
\DeclareMathAlphabet      {\mathit}{OT1}{cmr}{m}{it}
\DeclareMathAlphabet      {\mathtt}{OT1}{cmtt}{m}{n}

\DeclareSymbolFont{operators}   {OT1}{cmr} {m}{n}
\DeclareSymbolFont{letters}     {OML}{cmm} {m}{it}
\DeclareSymbolFont{symbols}     {OMS}{cmsy}{m}{n}

\title{Generalized Uncertainty Principle and Quantum~Gravity~Phenomenology}
\author{Pasquale Bosso}

\degreeyear{2017}

\prevdegrees{Bachelor of Science, Universit\`a degli Studi di Napoli ``Federico II'', 2010\\
			 Master of Science, Universit\`a degli Studi di Roma ``La Sapienza'', 2013}

\defensedate{August 23, 2017}

\addsignatureline{Prof. Saurya Das}{Ph.D.}{Professor}{Supervisor}
\addsignatureline{Prof. Mark Walton}{Ph.D.}{Professor}{Committee Member}
\addsignatureline{Prof. Kent Peacock}{Ph.D.}{Professor}{Committee Member}
\addsignatureline{Prof. Kenneth Vos}{Ph.D.}{Associate Professor}{Internal External Examiner}
\addsignatureline{Prof. David Hobill}{Ph.D.}{Associate Professor}{External Examiner}
\addsignatureline{Prof. Locke Spencer}{Ph.D.}{Assistant Professor}{Chair}

\begin{document}

\frontmatter
 
\ulethtitle
\ulethapproval


\begin{abstract}
	The fundamental physical description of Nature is based on two mutually incompatible theories: Quantum Mechanics and General Relativity.
	Their unification in a theory of Quantum Gravity (QG) remains one of the main challenges of theoretical physics.
	
	Quantum Gravity Phenomenology (QGP) studies QG effects in low-energy systems.
	The basis of one such phenomenological model is the Generalized Uncertainty Principle (GUP), which is a modified Heisenberg uncertainty relation and predicts a deformed canonical commutator.

	In this thesis, we compute Planck-scale corrections to angular momentum eigenvalues, the hydrogen atom spectrum, the Stern--Gerlach experiment, and the Clebsch--Gordan coefficients.
	We then rigorously analyze the GUP-perturbed harmonic oscillator and study new coherent and squeezed states.
	Furthermore, we introduce a scheme for increasing the sensitivity of optomechanical experiments for testing QG effects.
	Finally, we suggest future projects that may potentially test QG effects in the laboratory.
\end{abstract}

\tableofcontents
\listoftables
\listoffigures

\mainmatter

\chapter{Quantum Gravity}

\begin{quote}
	``Or bene,'' gli disse il bravo all'orecchio, ma in tono solenne di comando, ``questo matrimonio non s'ha da fare, n\'e domani, n\'e mai.''
	\begin{flushright}
		I promessi sposi (The Betrothed) -- Alessandro Manzoni
	\end{flushright}
\end{quote}

The current physical description of Nature consists of two theoretical frameworks, General Relativity (GR) and Quantum Mechanics (QM).

GR, proposed by Einstein in 1915 \cite{Einstein1915_1} with the discovery of the field equations \linebreak (\emph{\mbox{Feldgleichungen}} in German) for gravity, is able to predict and describe a large number of physical phenomena in Astrophysics and Cosmology.
Some classic examples include the explanation of the perihelion precession of Mercury \cite{Einstein1915_2}, the deflection of light rays when passing close by massive bodies \cite{Dyson1920_1}, and the gravitational redshift of light \cite{Pound1959_1}.
Since GR describes gravity, it is the theory that is used for studying massive stars and Black Holes \cite{Penrose1969_1}.
For example, the Oppenheimer-Volkoff limit, the upper bound to the mass of a Neutron Star, beyond which a Black Hole can form, was derived while studying the structure of a spherically symmetric body of isotropic material, which is in static gravitational equilibrium in GR \cite{Oppenheimer1939_1}.
Furthermore, since at very large distances gravity is the only relevant interaction (galaxies and galaxy clusters have no net electrical charge), in Cosmology, that is the study of the creation, evolution, and structure of the universe from the Big Bang to its final fate, GR is the relevant theory for a description of the universe.
This description given by the Standard Model of Cosmology can account for several kinds of observations, such as the recession of galaxies \cite{Hubble1929_1,Lemaitre1927_1}, described by a varying scale factor in the metric of the universe, the Cosmic Microwave Background radiation \cite{Penzias1965_1}, the remnant of the thermal radiation from recombination, the Baryonic Acoustic Oscillations, regular fluctuations in the baryonic matter density generated before the epoch of matter-radiation decoupling, and several other phenomena.
All this can be described using a few parameters, like the actual matter density (baryonic and dark matter) and the decoupling time, that can be fixed using cosmological observations obtained by missions like CoBE (Cosmic Background Explorer) \cite{Smoot1999}, WMAP (Wilkinson Microwave Anisotropy Probe) \cite{Bennett2013}, Planck Satellite, as well as sky surveys like SDSS (Sloan Digital Sky Survey) \cite{CollaborationSDSS2016} and 6dFGS (6dF Galaxy Survey) \cite{Jones2009}.
The aim of these observations is to collect data for a better comprehension of the age, the topology, and the composition of the universe using general relativistic models.
Finally, in Cosmology and in Astrophysics, GR is also directly used for observations and to obtain information about the composition of the universe.
An example of these applications is the Gravitational Lensing, which uses the deflection of light to deduce the presence of massive objects and their mass distributions and is heavily used in recent research in Dark Matter \cite{Taylor1998_1}.

QM, on the other hand, has achieved several successes since its foundation.
It can be dated back with Planck hypothesis of quanta of energy of the radiation, proposed in \cite{Planck1900_1} in 1900.
Some examples are the Bohr's model for atoms \cite{Bohr1913_1} in 1913, that allows a theoretical explanation of the line spectra of the hydrogen atom, ``the experimental evidence of the quantization of the direction in magnetic field'' by Gerlach and Stern in 1922 \cite{Gerlach1922_1} and subsequent explanation by Uhlenbeck and Goudsmit with the hypothesis of the electron spin in 1925 \cite{Uhlenbeck1925_1}, the hypothesis of the existence of the positron, the antiparticle of the electron by Dirac \cite{Dirac1928_1}, currently used in PET (Positron Emission Tomography) scans.
Furthermore, the Standard Model of Particle Physics successfully describes three of the four fundamental interactions in Nature.
Indeed, among many other aspects, it includes a unified description of electromagnetism and weak nuclear interaction \cite{Glashow1961_1} as well as a description of the strong interaction.
The Standard Model of Particle Physics, that recently received a new confirmation in the discovery of the Higgs boson \cite{Chatrchyan2012_1}, is ultimately based on QM.
This theory also had deep influences on the technological development of the last century.
Indeed, in QM we can find, for example, the working principles of p-n junctions \cite{Ohl1948}, that is the basis for diodes, transistor and the whole digital electronics, through the quantum-mechanical explanation of conduction and of the properties of semiconductors.
A further example is the microscopic description of superconductors \cite{Bardeen1957_1}, applications of which are, other than in digital circuits, also in high sensitivity detectors, in Magnetic Resonance Imaging, in Nuclear Magnetic Resonance and in creating magnetic confinements in fusion reactors.

The description of Nature by QM and GR is not exempt from problems \cite{Kiefer2006_1,Amelino-Camelia2013}.
GR, for example, is a classical theory, that does not take into consideration any quantum property of matter fields.
Furthermore, it predicts its own demise allowing points in which spacetime is singular.
In fact, the curvature tensor, which determines the strength of gravity, increases without bound at the center of black holes, leading to infinite curvature.
The same problem applies to the universe as a whole:
an initial spacetime singularity is present at Big Bang event, the fundamental hypothesis for the Standard Model of Cosmology and for which there is strong evidence in the cosmic microwave background radiation, in the Hubble's law and in the abundance of primordial elements from nucleosynthesis.
Related to this problem is that of the cosmological initial conditions.
Indeed, because of the breakdown of GR and of the lack of a suitable theory of QM for energies close to the Planck scale, it is not possible to have an insight of how the universe was before the Planck epoch, preventing us from a consistent definition of the initial conditions and their evolution in the present observable universe.
Furthermore, the two theories which constitute the actual framework of theoretical physics, even if very successful in their domains, are not compatible with each other, therefore they cannot be used in contexts that would require both.
For example, while in GR the distribution of massive objects and energy determines the properties of spacetime, which in turn determines the dynamics of massive objects, in QM the spacetime has only the role of a fixed background on which the dynamics of the objects is determined.
Furthermore, being known that black holes, as found in a semi-classical approximation, radiate with thermal spectrum and eventually evaporate (see Section \ref{subsec:quantum_BHT}), for this last stage of black holes evolution a full theory of Quantum Gravity (QG) is needed.
On the other hand, Hawking and Unruh radiations give a hint on the breakdown of the equivalence principle of GR \cite{Singleton2016}.
In fact, straightforward arguments seem to indicate that, using a local device to measure the temperature, so to test the Einstein elevator \emph{Gedankenexperiment}, it is possible to discern between a gravitational field generated by a black hole and an accelerated frame.
Moreover, black holes present an important challenge to QM, because of the ``no-hair'' theorem.
In fact, the information needed to characterize these objects reduces to their mass, electric charge and angular momentum, while all the rest of the information about the matter forming them seems to be lost when the matter crosses the event horizon.
Thus this leads to a non-unitary evolution of the system and, therefore, non-conserved probabilities in QM.
One last motivation comes from Quantum Field Theory (QFT) \cite{Deser1957,Delamotte2004_1}.
Since in this theory divergences arise considering loop diagrams in fundamental interactions, being then ``cured'' through processes of renormalization, a common speculation is that a quantum theory of gravity would naturally let a cutoff at small distances appear, preventing the infinities of the theory.
Each of these aspects serves as a motivation for a quantum theory of gravity, in some cases even identifying possible properties of this new theory.

Being confident that a quantum theory of gravity could solve these problems, physicists tried to develop such a theory following standard procedures in QFT, but it resulted in predictable quantities, such as scattering amplitudes, being infinite.
Furthermore, unlike successful field theories like Quantum Electrodynamics and Quantum Chromodynamics, the QFT for gravity is not renormalizable, that is, its infinities cannot be cured.
The situation was clearly unacceptable from the theoretical point of view. 
New theories have been therefore proposed to solve this and other inconsistencies developing a theory of QG, that is a well-defined theory that would reproduce the effects of GR and QM in their respective domains avoiding their inconsistencies.
After a discussion on the thermodynamic aspects of black hole physics (Section \ref{sec:BHT}), in this Chapter we will review three of the QG approaches.
In Sec.~\ref{sec:string_th}, we will describe String Theory (ST), that relies on the existence of a total of 10 or 11 spacetime dimensions, and supersymmetric partners of observed fundamental particles.
The basic units of the theory, the strings, are one-dimensional objects moving on a fixed classical background.
The excitation of these objects under certain conditions reproduces not only the known elementary particles but also the deviation from a Euclidean spacetime and, therefore, gravity.
In Sec.~\ref{sec:LQG}, we will review Loop Quantum Gravity (LQG).
Keeping the identification between the gravitational field and the metric structure of spacetime, this theory tries to define a theory of QG starting from a canonical quantization of GR and defining a QFT on a manifold {(spin networks), as opposed to a QFT on a metric space}, without any metric being defined \emph{a priori} {\cite{Rovelli2008_1}}.
The fundamental assumption for this theory is that the physics can be described through states with finite norm on one-dimensional structures, like loops and graphs.
These structures give spacetime a sort of granularity that necessarily acts as a cutoff for divergences.
Finally, Sec.~\ref{sec:DSR} is dedicated to Doubly Special Relativity (DSR), that tries to reconcile the kinematic aspects of QM and the Theory of Relativity with the introduction of a second relativistic invariant associated with the Planck scale.
As a result, the momentum space of DSR is defined on a curved manifold and the spacetime is described by a non-commutative manifold.
On the other hand, when this theory is applied to a multi-particle system, some problems arise, like the so-called ``soccer ball problem''.

\section{Black Hole Thermodynamics} \label{sec:BHT}

Black holes are both a blessing and a curse for GR.
On one side they are predicted by this theory, and many observations indicate their existence \cite{Gillessen2009,Munoz2011,Abbott2016}.
On the other hand, they constitute one of the problems of GR, since they are associated with singularities of the metric.
Microscopic black holes could also turn into one of the main key objects for the development of a quantum theory of gravity.
Indeed, on dimensional grounds, the Planck length is of the same order of magnitude of the radius of a black hole when its Schwarzschild radius and its Compton length coincide.
This would, therefore, constitute a hint that QG effects would be crucial for this kind of objects.

Studying classical (non-quantum) black holes in GR, it became clear that many processes are allowed to extract the rest energy of a particle or to extract energy from black holes \cite{Jacobson1996_1}.
These mechanisms are maximally efficient when the horizon area does not change.
Furthermore, it was shown that, under quite general assumptions, the area of an event horizon can never decrease.
Therefore processes that increase the area are irreversible.
All this suggested an analogy between black hole physics and thermodynamics, where one can recognize the area {of the event horizon} playing the role of entropy.
When this analogy was first realized, though, many new questions arose.
In fact, since a classical black hole, by its definition, does not emit {anything}, one cannot associate with it a non-zero temperature.
Furthermore, entropy and horizon area have different dimensions.
Finally, while the area theorem states that the area of \emph{every} black hole does not decrease, from thermodynamics we have that the \emph{total} entropy does not decrease.

The path towards the solution of these problems started from the observation that the surface gravity of the horizon, defined as the acceleration of a stationary zero angular momentum particle just outside the horizon with respect to Killing time, 
plays the role of a temperature for a black hole.
We can also state a zeroth law for black holes in analogy with the zeroth law of thermodynamics, once we observe that the surface gravity is always \emph{constant} over the horizon of a stationary black hole.
The first law can be stated for a charged, rotating black hole as
\begin{equation}
	\mathrm{d}M = \frac{\kappa \hbar}{2 \pi} \frac{\mathrm{d}A}{4 \hbar G} + \Omega \mathrm{d} J + \Phi \mathrm{d}Q~, \label{eqn:1thermo}
\end{equation}
where we considered $c=1$ and where $\kappa$, $\Omega$ and $\Phi$ are the surface gravity, the angular velocity, and the electric potential of the horizon and where $A$, $J$, and $Q$ are the area of the event horizon, the angular momentum of the black hole, and its electric charge.
As we said before, the second law can be stated in terms of the Hawking's area theorem, assuming the Cosmic Censorship.
On the other hand, the third law can be stated noting that surface gravity cannot be reduced to zero in a finite number of steps.

While the interpretation of the second law is obvious, once Hawking's area theorem is taken into account, for the first law, we can consider a quasistatic process in which some small quantity of mass falls in a black hole.
Indeed, in this case, one can prove that an increase in the mass of the black hole, that is an absorption of energy by the system, results in an increasing area, that is an increasing ``entropy'' term.
For the third law, one can check that it is not possible to decrease $\kappa$ to zero with any process in a finite number of steps.
For example, knowing that the surface gravity is inversely related to the mass of the black hole, it would need an infinite amount of mass to make the surface gravity vanishing.

This thermodynamic analogy of black hole physics presents some flaws.
To solve them, Bekenstein proposed that $\eta A/\hbar G$, measured in units of the squared Planck length $\ell_\mathrm{Pl} = \sqrt{\hbar G/c^2}$ and where $\eta$ is a real parameter, is the entropy of the black hole and he conjectured a \emph{generalized second law}:
the sum of the entropy outside the black hole and the entropy of the black hole itself will never decrease \cite{Bekeinstein1973_1}.
This definition of entropy for black holes solves the problem of ``wrong'' dimensionality of the area of the event horizon as an entropy.
We can think, though, of processes that let entropy be absorbed by the black hole without increasing its mass and area.
For example, we can consider lowering a box containing radiation to the horizon of a black hole and finally dropping it, therefore reducing the entropy outside the black hole without changing the black hole entropy.
Thus it seems that the generalized second law can be violated, at least at the classical level.
This level, though, would mean considering the limit for $\hbar \rightarrow 0$, and in this case, an infinitesimal area change would result in a finite change in the Bekenstein entropy.
On the other hand, also the other flaws find a justification in this limit.
Indeed, using the first law, one can define a Bekenstein temperature $T_\mathrm{B} = \hbar \kappa / 8 \pi \eta$, that vanishes in the limit for $\hbar \rightarrow 0$.
The same conjecture, in the classical limit, also solves the third flaw, since a decreasing area would result in an infinite decrease of entropy, making this impossible and supporting the area theorem.

\subsection{Quantum Black Hole Thermodynamics} \label{subsec:quantum_BHT}

It is clear that a classical or a semi-classical approach can only work up to a certain extent.
Indeed the factor $\hbar$ in the definition of entropy and temperature asks for a quantum mechanical or, more appropriately, a quantum gravitational approach.
Unfortunately, as we will see in the rest of this Chapter, a definitive quantum description of gravity is not yet available.
Nonetheless, a semi-classical treatment revealed surprising new phenomena leading towards the idea that the thermodynamic analogy introduced at the beginning of this Section is actually an identity.

A first discovery was made considering one of the energy production processes we touched upon before.
Such process, the so-called ``Penrose process'', consists in using the properties of the ergosphere of rotating black holes in relation to particles, sent from infinity, that would split in two on such a surface.
Considering the same process for waves, one can obtain a phenomenon, called ``super-radiance'', corresponding to stimulated emission in QM.
The attention then moved towards looking for a spontaneous emission for this kind of black holes.
It was then found that even non-rotating black holes would emit, with a thermal spectrum at a temperature $T_\mathrm{H}= \hbar \kappa/2\pi$, corresponding to a Bekenstein temperature with $\eta = 1/4$ \cite{Hawking1974_1}.
From the first law of thermodynamics (\ref{eqn:1thermo}), one can then define the entropy
\begin{equation}
	S_{\mathrm{BH}} = \frac{A}{4 \hbar G}~. \label{eqn:BH_entropy}
\end{equation}
Notice that now we moved from an analogy to an actual identity:
$T_\mathrm{H}$ \emph{is} the temperature of the black hole, understood as the quantity describing the thermal emission of black holes;
$S_{\mathrm{BH}}$ \emph{is} the entropy of the black hole, related to the just mentioned definition of temperature.
Using these definitions, we can then find the temperature and entropy for Schwarzschild and Reissner-Nordstr\"om black holes.
The first is a non-rotating (\mbox{$J=0$}), non-charged ($Q=0$) black hole of mass $M$, with metric in spherical coordinates \cite{Schwarzschild1916}
\begin{equation}
	\diff s^2 = \left(1 - \frac{R_\mathrm{S}}{r} \right) c^2 \diff t^2 - \left(1 - \frac{R_\mathrm{S}}{r} \right)^{-1} \diff r^2 - r^2 (\diff \theta^2 + \sin^2 \theta ~ \diff \phi^2)~,
\end{equation}
where $R_\mathrm{S} = \frac{2 G M}{c^2}$ is the Schwarzschild radius.
The second is a non-rotating, charged black hole of mass $M$ and charge $Q$, with metric \cite{Reissner1916,Nordstrom1918}
\begin{equation}
	\diff s^2 = \left(1 - \frac{R_\mathrm{S}}{r} + \frac{R_\mathrm{Q}^2}{r^2} \right) c^2 \diff t^2 - \left(1 - \frac{R_\mathrm{S}}{r} + \frac{R_\mathrm{Q}^2}{r^2} \right)^{-1} \diff r^2 - r^2 (\diff \theta^2 + \sin^2 \theta ~ \diff \phi^2)~, \label{eqn:RN}
\end{equation}
where $R_\mathrm{Q}^2 = \frac{G Q^2}{4 \pi \epsilon_0 c^4}$ is a characteristic length related to the black hole charge.
Notice that in the limit $Q \rightarrow 0$ we recover the Schwarzschild metric.
We then find the following expressions for the temperature and entropy
\begin{subequations}
\begin{align}
	T_\mathrm{S} & = \frac{\hbar c}{4 \pi R_\mathrm{S} K_\mathrm{B}} & T_\mathrm{RN} & = \frac{\hbar c}{4\pi r_+ K_\mathrm{B}} \left( 1 - \frac{R_\mathrm{Q}^2}{r^2_+} \right) \label{eqn:bh_temperature}\\
	S_\mathrm{S} & = \frac{\pi c^3 K_\mathrm{B}}{\hbar G} R_\mathrm{S}^2 & S_\mathrm{RN} & = \frac{\pi c^3 K_\mathrm{B}}{\hbar G} r_+^2
\end{align}
\end{subequations} 
where
\begin{equation}
	r_+ = \frac{1}{2} \left( R_\mathrm{S} + \sqrt{ R_\mathrm{S}^2 - 4 R_\mathrm{Q}^2} \right)
\end{equation}
is the radius of the event horizon for a charged black hole.

This identification solves all the flaws that we mentioned before.
Now the temperature of a black hole is no longer zero, and the entropy has the correct dimension (using units in which Boltzmann constant $K_\mathrm{B} = 1$, entropy is dimensionless).
Because of Hawking emission, the horizon area, and therefore the black hole entropy, can now locally decrease.
Finally, this treatment also solves the apparent paradox of a decreasing total entropy.
Indeed, in this case, the interaction of the box with the quantum fields in the outer region cannot be neglected.
Basically, when the box is lowered, it needs to radiate positive energy to the surrounding vacuum and needs to be filled with negative energy, until its total energy equals zero at a point where the box would float because of a sort of a buoyancy force.

As we said before, a full description of quantum aspects of black hole thermodynamics, like the interpretation of entropy in terms of the number of microstates, requires a quantum theory of gravity.
Looking at this requirement from a different perspective, because of the robustness of the results in this field, an appropriate theory of QG needs to give a description of black hole thermodynamics.
As we will see in the next sections, the possibility of obtaining the black hole entropy as a result of a particular QG model is often believed to be a discerning element between plausible and implausible theories.

\section{String Theory} \label{sec:string_th}

The aim of ST is two-fold.
On one side it is an attempt to describe the gravitational interaction in quantum terms.
On the other side, a consistent unification of all interactions is sought.
The objects under study are one-dimensional entities, the strings, their dynamics and the evolution of modes on them.
These modes correspond to observable particle fields.
A first step is to define an action on the two-dimensional worldsheet of the string \cite{Schwarz1999_1}.
On this surface, a map from the coordinates of the worldsheet to the embedding target spacetime, the embedding variables $X^\mu$, and an intrinsic worldsheet metric $h_{\alpha\beta}$ are defined.
The target space consists of a fixed pseudo-Riemannian spacetime background.
With respect to these coordinates, the Polyakov action, describing the propagation of a string in a $D$-dimensional target spacetime, is given by \cite{Szabo2002_1}
\begin{equation}
	S \propto \ell^{-2}_{\mathrm{S}} \int \mbox{d}^2 \sigma \sqrt{h} h^{\alpha\beta} \partial_\alpha X^\mu \partial_\beta X^\nu g_{\mu\nu}(X) ~,
\end{equation}
where $g_{\mu\nu}$ is the metric of the $D$-dimensional space and $\ell_\mathrm{S}$ is the string intrinsic length.

To obtain a self-consistent theory, one has to impose some constraints.
For example, the theory has to be able to reproduce Einstein's theory at low-energy.
Furthermore, from some of these self-consistency conditions, one obtains that the dimensions of the target space are restricted to 26 for the bosonic string or 10 or 11 in the case of the superstring.

\subsection{Supersymmetry}

One of the main ingredients of ST is Supersymmetry.
It is a symmetry relating bosons and fermions, in the sense that every boson has a fermionic supersymmetric partner, or superpartner for short, and \emph{vice versa}.
This symmetry was initially thought to be able to solve the so-called ``hierarchy problem'', that is to give an explanation for the different energy scales in particle physics, such as the difference between the standard model ($\sim 100$ GeV) and the Planck scale ($10^{19}$ GeV).
In QFT, even if such a difference is present in some approximation, it is canceled by radiative corrections.
In particular, divergences from loop diagrams move the weak scale to the Planck scale.
It is also true, though, that the bosonic and fermionic divergences appear with a different sign.
Supersymmetry uses this feature to cure divergences of QFT, considering pairs of bosonic and fermionic loops that cancel each other out.
This implies that for every bosonic (fermionic) degree of freedom, we need a corresponding fermionic (bosonic) degree of freedom \cite{Martin1997}.
Furthermore, while a specific supersymmetry transformation can be found for the Lagrangian to let it be invariant \mbox{on-shell}, additional \emph{auxiliary} fields are necessary to have an off-shell invariance.
These fields have no physical interpretation since the corresponding Lagrangian does not contain a kinetic term.
This also implies that, when their equations of motion are considered on-shell, the auxiliary fields vanish, restoring the on-shell invariance of the supersymmetric Lagrangian.

From the Noether theorem, one can define a supercurrent $J^\mu$ and supercharge $Q$
\begin{align}
	\epsilon J^\mu + \epsilon^\dagger J^{\dagger \mu} & \equiv \sum_X \delta X \frac{\delta \mathcal{L}}{\delta (\partial_\mu X)} - K^\mu~, & 	Q_\alpha &= \sqrt{2} \int \mathrm{d}^3 \vec{x} ~ J_\alpha^0~, & Q_{\dot{\alpha}}^\dagger &= \sqrt{2} \int \mathrm{d}^3 \vec{x} ~ J_{\dot{\alpha}}^{\dagger 0},
\end{align}
where $\mathcal{L}$ is a Lagrangian density, $X$ represents the fields involved, $\epsilon^\alpha$ is an infinitesimal, anti-commuting, two-component Weyl fermion object with the role of a parameter for the supersymmetry transformation, and $K^\mu$ is such that $\partial_\mu K^\mu= \delta \mathcal{L}$.
As a quantum mechanical operator, the charge $Q$ must transform bosonic states into fermionic ones and \emph{vice versa}
\begin{align}
	Q |\mathrm{Boson}\rangle & = |\mathrm{Fermion}\rangle~, & Q|\mathrm{Fermion}\rangle & = |\mathrm{Boson}\rangle~,
\end{align}
and must satisfy the following algebra
\begin{align}
	\{Q_\alpha, Q_{\dot{\alpha}}^\dagger \} & = - 2 \sigma^\mu_{\alpha \dot{\alpha}} P_\mu~, & \{Q_\alpha, Q_\beta\} & = \{Q_{\dot{\alpha}}^\dagger , Q_{\dot{\beta}}^\dagger\} = 0~, & [Q_\alpha ,P^\mu] & = [Q_{\dot{\alpha}}^\dagger , P^\mu] = 0~,
\end{align}
where $\sigma^\mu$ are the Pauli matrices and $P^\mu$ the four-momentum operator.

Single-particle states related by this symmetry can be grouped in \emph{supermultiplets}, irreducible representations of the supersymmetry algebra.
Each supermultiplet contains superpartners, that is bosons and fermions related by supersymmetry transformations.
Since the four-momentum, and therefore the squared-mass operator $-P^2$, commutes with the operators $Q$ and $Q^\dagger$, particles in the same supermultiplet must have equal masses.
If this was the case, we would have already found some superpartners of known particles.
Hence supersymmetry has to be a broken symmetry.
Furthermore, the operators $Q$ and $Q^\dagger$ commute with the generators of gauge transformations.
This implies that particles in supermultiplets have also the same electric charge, weak isospin, and color.

Based on this idea, one can extend the standard model of particle physics considering a minimal supersymmetric extension, which would include superpartners for all the particles in the standard model.
The positive aspect of this new model is that it would be manifest in terms of new particles to be discovered.
For example, while the standard model predicts a single doublet of Higgs field to give mass to all quarks and charged leptons, the supersymmetric extension requires two of them, one to give mass to the charged-2/3 quarks, the other to give mass to the charged-1/3 quarks and charged leptons.
Furthermore, supersymmetry predicts the unification of the three non-gravitational interactions when a minimal supersymmetric extension of the standard model is considered \cite{Schwarz1999_1}.
This is not possible in QFT, since extrapolating the high energy behavior of the coupling constants does not allow their values to coincide for any value of energy.
On the other hand, the same extrapolation in the minimal supersymmetric extension for the three coupling constant allows them to meet at a particular energy, where the strength of the three gauge interactions become equal.
Therefore, it is possible to unify these theories into a \emph{grand unified theory}.

Nonetheless, it is important to notice that, although the first estimation for the supersymmetry scale was of the order of TeV, the Large Hadron Collider, operating in that range of energies, did not find any evidence of it.

\subsection{Nonperturbative string theory}

With the so-called \emph{first superstring revolution} of 1984-1985, it became clear that five different perturbative approaches to superstring are present, each with nine spatial and one temporal dimensions.
These five theories are
\begin{itemize}
	\setlength{\itemsep}{-0.5em}
	\item Type I: the string worldsheet is unoriented and it has one spacetime supersymmetry transformation.
	\item Type IIA: the string worldsheet is oriented and it has two spacetime supersymmetry transformations.
		The two Majorana-Weyl spinors defined on the closed string sector are of opposite chirality, therefore the theory is non-chiral.
	\item Type IIB: this theory is similar to type IIA superstring theory. The difference is that the spinors have the same chirality, therefore this theory is chiral.
	\item $E_8 \times E_8$ heterotic: a heterotic string consists of a string the left-moving modes of which are described by $d=26$ bosonic string, while for the right-moving modes we have $d=10$ superstring.
		The remaining 16 degrees of freedom for the right-movers are internal degrees of freedom on a 16-dimensional even self-dual lattice.
		The gauge group associated to the lattice in this theory is $E_8 \times E_8$.
	\item $SO(32)$ heterotic: this theory is similar to the previous one, with the difference of a $SO(32)$ gauge group for the lattice.
\end{itemize}

The advent of the \emph{second superstring revolution} (1994) allowed a nonperturbative treatment of ST.
The main achievement of this revolution is the discovery that the previous perturbative theories are actually equivalent, through a set of dualities, to a single theory.
The former theories could be seen as an expansion of the latter about five different points in the families of consistent vacua contemplated by the perturbative theories.
This nonperturbative model consists of an 11-dimensional Minkowski spacetime.
This theory is denoted M-theory.

This theory, other than containing strings, contemplates also $p$-dimensional objects, called \emph{$p$-branes}.
A special class of $p$-branes consists in objects called Dirichlet $p$-branes (or \emph{D-branes} for short).
They are defined by the boundary conditions at the ends of open strings.
D-branes are therefore $p$-dimensional objects to which string ends are attached.

\subsection{Successes and open problems}

One of the successes of ST consists in the interpretation of the Bekenstein-Hawking entropy in terms of number of quantum microstates for classical black hole configurations \cite{Strominger1994_1}.
Indeed, using D-branes techniques, it is possible to show that the black hole entropy is 1/4 the area of its event horizon.

This theory though is not exempt from problems \cite{Kiefer2006_1}.
For example, since we are able to observe only 4 spacetime dimensions, compactification is required to hide the other dimensions.
But compactification is possible in many ways.
Furthermore, the existence of various brane models would indicate the impossibility of compactifying these extra dimensions.
Also, the metric nature of the target space itself is actually a problem since one needs to fix the background on which strings evolve.
From an observational point of view, ST {has not been able to} recover the Standard Model of Particle Physics.
Moreover, the underlying conceptual principles are not yet understood.
Finally, no experimental result in favor or against the theory is available at the moment.
Even supersymmetry, considered one of the possible first evidence to be discovered and anticipated to be detected in the range of energy probed by the Large Hadron Collider, has not yet been found.


\section{Loop Quantum Gravity} \label{sec:LQG}

In LQG, a background-independent quantization of gravity is attempted.
Therefore we have the first difference to ST:
as we saw in the dedicated section, ST{, as currently formulated,} considers the dynamics of string on a fixed background, where this dynamics and the string evolution generate what can be understood as perturbation of the metric, that is gravity.
On the other hand, in LQG we do not have any separation between background and perturbation metric.
The motivation for this background-independent approach is in the current interpretation of the general theory of relativity:
spacetime metric, as opposed to metric perturbations, and gravitational field are the same physical entity.
Therefore, assuming that such identification still holds for a quantum theory of gravity, it needs to be specified without the definition of a metric, since the theory should be able to define the spacetime metric and the gravitational field at the same time.
Another difference between the two theories is that, while the first attempts to unify the four known interactions, in LQG the effort is focused on a quantum description of gravity and of spacetime.

The birth of this theory can be traced back to the reformulation of GR in 1986 by Abhay Ashtekar.
He introduced a new set of canonical variables that replaced the spacetime metric to obtain a Lagrangian formalism \cite{Rovelli1991_1}.
In this formalism, GR is reformulated in terms of a tetrad field $e^I_\mu$, that is invariant under $SU(2)$ gauge transformations \cite{Ashtekar1986}, and a complex connection $^4A^{IJ}_\mu$.
The index $\mu$ is the usual spacetime index, while the indices $I$, $J$ are internal indices, running from 0 to 3 and raised and lowered with the Minkowski metric $\eta^{IJ}$.
Furthermore, the connections are defined to be self-dual, that is
\begin{equation}
	{}^4 A^{MN}_\mu = - \frac{1}{2} i \epsilon^{MN}_{IJ} ~ {}^4A^{IJ}_\mu~,
\end{equation}
with $\epsilon^{MN}_{IJ}$ the completely antisymmetric tensor.
Introducing the Yang--Mills field strength of ${}^4A^{IJ}_\mu$
\begin{equation}
	{}^4 F^{IJ}_{\mu\nu} = \partial_\mu ~ {}^4A^{IJ}_\nu - \partial_\nu ~ {}^4 A^{IJ}_\mu + {}^4 A^{IM}_\mu ~ {}^4 A^J_{\nu M} - {}^4 A^{IM}_\nu ~ {}^4 A^J_{\mu M}~,
\end{equation}
and considering the action
\begin{equation}
	S[e,{}^4A] = \int \mathrm{d}^4 x ~ e_{\mu I} e_{\nu J} ~ {}^4 F^{IJ}_{\tau \sigma} ~ \epsilon^{\mu\nu\tau\sigma}~, \label{eqn:action}
\end{equation}
one can obtain the following equations of motion
\begin{align}
	\epsilon^{\mu \nu \rho \sigma} ~ e_{\nu J} F^{IJ}_{\rho \sigma} = & 0~, & (\delta^{KI} \delta^{LJ} + \frac{1}{2} i \epsilon^{KLIJ} ) e^{\mu \nu \rho \sigma} \mathcal{D}_\rho (e_{\mu I} e_{\nu J} ) = & 0~,
\end{align}
corresponding to the diffeomorphism and the gauge constraints.
Here, $\mathcal{D}_\rho$ is the covariant derivative defined by ${}^4 A$, $\mathcal{D}_\rho e_{\mu I} = \partial_\rho e_{\mu I} - e_{\mu J} ~ {}^4 A^J_{\rho I}$.
With these new variables satisfying these equations of motion, it is possible to write the metric solution of the vacuum Einstein equations as
\begin{equation}
	g_{\mu \nu}(x) = e^{I}_\mu (x) e^J_\nu (x) \eta_{IJ}~.
\end{equation}
Also, every solution of the vacuum Einstein equations can be written in the same way in terms of solutions of the action (\ref{eqn:action}).

Already from these few preliminary considerations, it is possible to obtain some insights of the advantages and peculiarities of this formalism.
First, we notice that with the use of the connections ${}^4A$, a similarity between GR and Yang--Mills theories is introduced.
Then, since the connections are introduced as complex quantities, the action is also complex.
Though the imaginary part of the action does not affect the dynamics (it results in a topological term), the canonical framework needs to include it and to take in consideration the additional structures implied by the imaginary part.

From this new set of coordinates, using the theory of Wilson loops, it was possible to develop a non-perturbative approach to quantize gravity.
The fundamental element is the Wilson loop of the connection ${}^4A^{IJ}_\mu$, defined as the trace of the holonomy of the same connection on a closed curve $\alpha$
\begin{equation}
	\mathcal{T}[\alpha] = - \mathrm{Tr} \left[U_\alpha\right] = \mathrm{Tr} \left[\mathcal{P} \exp \left(\oint_\alpha A\right)\right]~.
\end{equation}
This loop-space representation is the foundational element of LQG.

\subsection{Loops}

The fundamental entities of LQG are loops or graphs, one-dimensional structures on which states are defined with finite norm.
These loops should not be confused with lines connecting points on spacetime, as for example in the case of Lattice QFT.
Indeed in the latter, a continuous limit of the lattice approach is ill-defined, since the size of the loops shrinks to zero, obtaining infinite-norm states with one-dimensional support.
On the contrary, in LQG, the metric being absent, the operation of ``shrinking'' the loop to zero is not properly defined, while a refinement of the lattice space on which the loops are defined can be simply interpreted as a change of coordinates.
This property is present because the theory is diffeomorphism invariant, a feature inherited by GR since LQG aims towards a quantum version of the metric.
To describe this in a more pictorial way, ``\emph{GR is not physics over a stage, it is the dynamical theory of everything, including the stage itself}'' 
\cite{Rovelli2008_1}.

In practice, defining states corresponding to graph-like excitations of the gravitational field on a differential manifold and factoring away the diffeomorphism invariance, the only relevant information remaining in the framework is the abstract graph structure and its knotting.
An equivalence class of graphs under diffeomorphism, the $s$-knot, represents an elementary quantum excitation of space.

To be more precise, consider a graph $\Gamma$.
A ``coloring'' of this graph is defined by associating an irreducible representation of $SU(2)$ to each link of $\Gamma$ and an invariant tensor $v$ in the tensor product of the representations $j_1 \ldots j_n$ to each node of $\Gamma$ in which links with spin $j_1 \ldots j_n$ meet.
These invariant tensors are called ``intertwiners''.
The set $S = \{\Gamma, \vec{j}, \vec{v} \}$ indicates the colored graph and is denoted as ``spin network''.

Finally, given a loop $\alpha$ on a manifold $M$, one can introduce a state
\begin{equation}
	\psi_\alpha(A) = \langle A | \alpha \rangle = - \mathrm{Tr}[U_\alpha (A)]~.
\end{equation}
Taking the holonomy of the connection along each link of the graph, and contracting their matrix elements in the representation associated with that link with the invariant tensor in each node, one obtains the spin network state
\begin{equation}
	\Psi_S (A) = \langle A | S \rangle~.
\end{equation}
These states are eigenstates of area and volume operators and provide a basis for the quantum theory.

\subsection{Results}

Even though LQG remains a tentative theory, with no experimental or observative confirmation, one can list a number of results from its application \cite{Rovelli2008_1}.
First, a discreteness of space at Planck-scale can be achieved.
This is shown considering that certain operators that describe geometrical quantities like areas and volumes have discrete spectra.
In particular, acting with the area operator $\hat{A}$, defined by writing the metric as a function of the loop variables, on a spin-network state $|S\rangle$, one has for a surface $\Sigma$
\begin{equation}
	\hat{A}[\Sigma] ~ |S\rangle \propto \sum_{i \in \{S \cap \Sigma \} } \sqrt{j_i (j_i + 1)} |S\rangle~, \label{eqn:area_quantization}
\end{equation}
where $i$ labels the intersections between the spin network $S$ and the surface $\Sigma$, and $j_i$ is the spin of the link of $S$ crossing the $i$th intersection.
Similarly, for the volume of a region $R$, one needs to consider the intertwiners of the nodes of the spin network contained in $R$.
Furthermore, singularities, both cosmological and due to black holes, are controlled by the quantum theory of spacetime.
Also, a statistical computation of the Bekenstein entropy can be achieved, leading to the result in (\ref{eqn:BH_entropy}), though not univocally.
This is done counting the microstates corresponding to an ensemble of configurations of the horizon with fixed area, obtaining
\begin{equation}
	S = \frac{c}{\gamma} \frac{A}{4 \hbar G}~,
\end{equation}
where $c$ is a real number of the order unity and $\gamma$ is the Immirzi parameter.
If $c=\gamma$, we obtain the Bekenstein entropy.
Finally, it has been shown that the low-energy limit of LQG coupled with matter in three dimensions is equivalent to a field theory on a noncommutative spacetime.

\subsection{Open Problems and Future Investigations}

LQG is not a definitive theory of QG.
Currently, as for ST, no experimental or observational evidence can support or refute the theory.
Furthermore, LQG is still in a developing phase, with aspects of the theory being investigated and open problems to be faced \cite{Rovelli2008_1}.
One of these problems concerns the low-energy limit of the theory that should resemble the classical theory of gravity.
In fact, so far no convincing proof is available, and the emerging of the low-energy limit from the background-independent theory needs to be understood.
Similarly, $n$-point functions are under study.
These would allow for computing scattering amplitudes and, in turn, for a direct test of the low-energy limit and for computation of the quantum corrections to GR.

Another aspect that future investigations need to take into consideration is what the theory predicts about black holes.
In particular, a derivation of the Bekenstein-Hawking entropy formula from the full quantum theory is not yet available.
It is important to have an understanding of the event horizon in the quantum theory.
This would require an understanding of the quantum dynamics, a description of the classical limit and of the asymptotically-flat limit of the classical theory.
Problems also involve a description in quantum terms of thermodynamics and statistical mechanics.
These issues rely on the absence of a natural local notion of energy in a general covariant context since the number of states with given energy is used to define statistical ensembles.

Also, some conceptual and technical problems are present \cite{Nicolai2005_1}.
For example, it is not clear how two-loop divergences of QG are cured in LQG (notice that one-loop divergences of pure gravity do not pose any problem \cite{tHooft1974}).
On the other hand, the same divergences should be obtained back in the (semi)classical limit, which, as we mentioned above, is not well defined.
Another problem is whether spacetime in this theory possesses full covariance.
This problem is related to the kind of closure of the constraint algebra.

To conclude this non-exhaustive list, notice that the development of a quantum theory of cosmology is one of the main objectives for any quantum theory of gravity.
For LQG, this development is still in progress.
The difficulties on this topic are to provide models fully derived from LQG and to develop these models so that their predictions can be compared with cosmological observations.

\section{Doubly Special Relativity} \label{sec:DSR}

{DSR studies kinematical aspects of QG.
In particular, it focuses on the transformation laws between reference frames, modifying and extending those of Special Relativity.
Therefore, unlike ST and LQG, that are attempt at a full theory of QG, DSR only tries to solve some of the problems of a quantum description of spacetime.}
For this theory the motivation derives from a conceptual question:
``What is the fate of Lorentz symmetry at Planck scale?'' \cite{KowalskiGlikman2005_1}.
Another way to state this question is to address the existence of a mass (or a length, from another point of view) scale $\kappa$ ($\kappa^{-1}$), identified with the Planck mass (length), that can be composed using only constants from GR and QM
\begin{align}
	\mbox{Planck mass} \hspace{0.5em} M_\mathrm{Pl} & \equiv \sqrt{\frac{\hbar c}{G}} \approx 2.2 \times 10^{-8} \mathrm{Kg}, \\
	\mbox{Planck length} \hspace{0.5em} \ell_\mathrm{Pl} & \equiv \sqrt{\frac{\hbar G}{c^3}} \approx 1.6 \times 10^{-35}\mathrm{m},
\end{align}
This scale, as we will see in the next Chapter, is usually referred to as the scale at which QG effects become relevant.
Phenomenologically speaking, many effects are considered in terms of the ratio between a characteristic scale of a system and the Planck scale.
From Special Relativity, though, it is always possible to define an observer for which the characteristic length is of the same order of the Planck length.
Some observers could then observe QG effects while others could not.
Therefore, the existence of a minimal length seems to violate the principle of relativity.

To avoid this inconsistency, a new invariant has been introduced, beside the speed of light: the scale $\kappa$.
One also assumes the standard theory of Special Relativity is restored once the limit $\kappa \propto M_\mathrm{Pl} \rightarrow \infty$ ($\kappa^{-1} \propto \ell_\mathrm{Pl} \rightarrow 0$) is considered.
This class of theory is then called ``Doubly Special Relativity'' since it contains two invariants, or ``Deformed Special Relativity'', since the postulates for this theory impose a modification of the action of symmetry generators.

One of the consequences of these assumptions is that the four-momentum transforms {between two reference frames with relative velocity $v$ along $z$} as \cite{Magueijo2003_1}
\begin{subequations}
\begin{align}
	p'_0 & = \frac{\gamma (p_0 - \beta p_z) }{1 + \ell_\mathrm{Pl} (\gamma - 1) p_0 - \ell_\mathrm{Pl} \gamma \beta p_z}~, & p'_z & = \frac{\gamma (p_z - \beta p_0) }{1 + \ell_\mathrm{Pl} (\gamma - 1) p_0 - \ell_\mathrm{Pl} \gamma \beta p_z}~,\\
	p'_x & = \frac{p_x}{1 + \ell_\mathrm{Pl} (\gamma - 1) p_0 - \ell_\mathrm{Pl} \gamma \beta p_z}~, & p'_y & = \frac{p_y}{1 + \ell_\mathrm{Pl} (\gamma - 1) p_0 - \ell_\mathrm{Pl} \gamma \beta p_z}~,
\end{align}
\end{subequations}
{where $\beta = v/c$ and where $\gamma = 1/\sqrt{1 - \beta^2}$ here is the Lorentz factor.}
Notice that for $\ell_\mathrm{Pl} \rightarrow 0$, that is in the limit of Special Relativity, we obtain back the usual Lorentz transformations.
Furthermore, it is interesting to notice that, unlike Special Relativity, the orthogonal components with respect the direction of the boost (in this case $z$) are changed in DSR.

Furthermore, one also needs to modify the relativistic energy-momentum relation involving non-linear terms, to keep this relation invariant in the new framework.
For example, one of the proposed modifications of the dispersion relation is \cite{AmelinoCamelia2002_1}
\begin{equation}
	2 E_\mathrm{Pl}^2 \left[ \cosh\left(\frac{E}{E_\mathrm{Pl}}\right) - \cosh\left(\frac{m}{M_\mathrm{Pl}}\right) \right] = \vec{p}^2 c^2 e^{E/E_\mathrm{Pl}}
\end{equation}
where $E_\mathrm{Pl}$ is the Planck energy.
{As a result, the speed of massless particles would now be dependent on the energy they carry.}
Notice that in the limit for $\ell_\mathrm{Pl} \rightarrow 0$, this relation reduces to
\begin{equation}
	\vec{p}^2 c^2 = 2 E_\mathrm{Pl}^2 \left[ \frac{E^2}{2 E_\mathrm{Pl}^2} - \frac{m^2 c^4}{2 E_\mathrm{Pl}^2} \right] \rightarrow E^2 = m^2 c^4 + \vec{p}^2 c^2~,
\end{equation}
that is, the usual dispersion relation.
This feature derives from a modified symmetry group for DSR.
In fact, this theory, like standard Special Relativity, possesses a ten-dimensional symmetry group, corresponding to rotations, boosts, and translations.
Whereas this group in the standard theory is the linear Poincar\'e group, in DSR it depends on the parameter $\kappa$.
It is therefore called $\kappa$-Poincar\'e group, governed by relations of a modified Poincar\'e algebra.
In developing this $\kappa$-Poincar\'e algebra, one has to impose that, in the limit $\kappa \rightarrow \infty$, one obtains back the usual Poincar\'e algebra.

In general, the commutators of the $\kappa$-Poincar\'e algebra are expressed as analytic functions of the generators; therefore, the Lorentz and the translation sectors appear deformed.
Since the mathematical framework used in DSR is that of quantum algebras, and since in quantum algebras one can change the basis of generators in an arbitrary, analytic way, it is also possible to change the basis such that the Lorentz part of the algebra becomes classical.
Such a basis is called ``bicrossproduct'' and the model based on such an algebra is called DSR1.
Another realization of DSR, proposed by Magueijo and Smolin, considers a basis in which Lorentz algebra is still not deformed, but there are no deformations also in the brackets of rotations and momenta. Such a realization is called DSR2.
These models were initially presented for a three-dimensional case.
A limit of four-dimensional QG is not known, but some circumstantial evidence is in favor of its existence.

When the kinematics of a particle is considered in DSR, the four-momentum space happens to be a curved manifold of constant curvature $\kappa^{-2}$.
One of the most interesting consequences of this feature is that positions, defined as ``translations'' of momenta, cannot commute and the spacetime of DSR is a non-commutative manifold, called $\kappa$-Minkowski spacetime.

\subsection{Results and Open Problems} \label{ssec:QGP&DSR}

As for the theories of QG that we presented before, so far no experimental result that could be considered in favor or against DSR is available.
Therefore, it is of extreme importance to look for possible experimental signatures of this theory.
One aspect that has been analyzed is the time-of-flight of particles.

Some theories of QG predict a speed of light in vacuum depending on its wavelength.
Also in DSR the velocity of photons, assumed as point-like, massless classical particles, can be calculated, but the results are ambiguous.
Indeed, depending on the approach used, one can obtain a velocity growing with energy or a constant velocity.
Even using an approach based on phase space of DSR, one finds a constant velocity, obtaining a result that seems to contradict the initial statement of an energy-dependent velocity.
However, we need to have in mind that a complete, theoretical solution of the problem of physical particles' velocities is not available.
Therefore, experimental results and future works will be of extreme importance.

Another aspect of this theory that also constitutes one of its biggest problems, concerns the description of multi-particle systems.
The importance for this aspect is obvious:
to compare any theory with the results from scattering processes, we need to understand how to describe the whole system, composed of the scattering particles, what the conserved quantities are, and how these are related to analogous quantities for the individual particles.
With this motivation, one could ask for the description of the addition of momenta for a multi-particle system.
One would then obtain a paradoxical result known as the ``soccer ball'' problem \cite{Hossenfelder2014_1}.
To state this problem, consider a system composed of many particles, like a soccer ball.
If we try to go from an elementary description of the soccer ball to a global description, for example in terms of the \mbox{center-of-mass} degrees of freedom, we do not have any problem in the standard theory, since the momenta sum linearly.
More technically, a description of multi-particle system in QM is implemented considering a tensor product of single-particle states, that is, we are assuming that the particles preserve their identity even when they are considered as part of a larger system.
Suppose we have an operator defined for every single particle of a system.
To act with this operator on the whole system, we need to define a new operator through the co-product, that is a map from the algebra of the original operators to their tensor product.
In the case of the standard theory, the result of this co-product is trivial, acting on each component independently.
In the case of quantum algebras though, the co-product is non-trivial and non-symmetric.
This results in a momentum addition rule that depends not only on the individual momenta of the components but also on ``interference'' terms between pairs of momenta, in number of $N(N-1)/2$.
Therefore, the number of these terms increases with the number of particles and would eventually dominate the total momentum of the system, contradicting the assumption that for classical objects we should recover the standard theory.

\vspace{2em}

As we saw, all the proposals for a theory of QG that we presented here are currently at a development stage.
Nonetheless, they present several features that can be tested and that could help in discerning between good and bad theories, or between features to retain or discard.
The role of testing them is taken by Quantum Gravity Phenomenology, as we will see in the next Chapter.
\chapter{Quantum Gravity Phenomenology}
\label{ch:GUP}

\begin{quote}
	What we observe is not nature in itself but nature exposed to our method of questioning.
	
	\begin{flushright}
		Physics and Philosophy -- Werner Heisenberg
	\end{flushright}
\end{quote}

The etymological origin of the word ``physics'' is the ancient Greek word \textgreek{f'usis}, referring to ``natural things''.
The ultimate aim of Physics is that of finding a suitable description for natural phenomena, one that could both explain and predict physical events.
Therefore, it is clear that the only guidance and judge of physical theories is Nature itself or, on a more practical ground, experimental investigation.
Hence, the only theories that are of interest for physicists are those that can be judged by Nature.
It can though be sneaky, hiding what we look for in places that we cannot probe or in conditions that we cannot test.
The obvious reference, related to what we saw in the previous Chapter, is to the Planck scale physics.
Indeed, we know that using GR and QM at the same time to describe the same phenomenon leads to inconsistencies, and a possible quantum theory of gravity will presumably include new descriptions and phenomena (new with respect to current physical theories) at the Planck scale.
On the other hand, this scale is presently far beyond our reach.
Consider for example Planck energy ($E_{\mathrm{Pl}} \sim 10^{16}~\mathrm{TeV}$) \cite{Planck1899} compared to the energy achieved in the Large Hadron Collider ($E_{\mathrm{LHC}} \sim 10~\mathrm{TeV}$) \cite{OLuanaigh}, or Planck length ($\ell_{\mathrm{Pl}} \sim 10^{-35}~\mathrm{m}$) compared to the uncertainty for the position of the mirrors of LIGO ($h \sim 10^{-18}~\mathrm{m}$) \cite{Abbott2016}, or, finally, Planck time ($t_{\mathrm{Pl}} \sim 10^{-44}~\mathrm{s}$) compared to the shortest light pulse produced in laboratory ($t \sim 10^{-17}~\mathrm{s}$) \cite{Zhao2012}.
It is thus clear that we are at best fifteen orders of magnitude away from achieving the Planck scale.

Is this the end of the story?
Can we only wait for new experiments to fill some of these huge gaps between us and the new Physics?
Or can we peer through these heavy curtains?
In the previous Chapter, we saw what the limitations of the present framework of Physics are.
Reversing this apparently dark side, we can expect that new effects and phenomena will be relevant beyond the current stage of Physics when the characteristic quantities are close to the Planck scale.
Even though these new effects are very small, given the difference of many orders of magnitude between current Physics and the Planck scale, they should be present at every scale.
This last point is the motivation of Quantum Gravity Phenomenology (QGQ) and of this thesis.

Phenomenology is ``the division of any science which is concerned with the description and classification of its phenomena, rather than causal or theoretical explanation'' \cite{Phenomenology}.
In particular, in QGP we are concerned with effects of quantization of gravity and spacetime and the phenomena that these quantizations imply.
From a conceptual point of view, we can distinguish two lines of motivation for a phenomenology of quantum gravity, reflecting the different approaches to the QG problem by the physical community \cite{Amelino-Camelia2013}.
In one case, it can be motivated with the required test on energies achievable with present technology of a full, high energy ``theory-of-everything''.
This conception of QGP requires the existence of such a theory, from which we can extract information on its limit for low energies.
When this is not the case, \emph{i.e.} when a theory-of-everything is yet to come, QGP is motivated by testing known or supposed features of a theory of QG.
In this last case, the phenomenological investigation serves as a guide for the development of the new theory (or series of theories), as studies on the black body radiation and on specific heat for QM, and Maxwell's theory of electromagnetism and Michelson-Morley experiment for Special and General Relativity.
Nonetheless, notice the opposite situation in which the present physical research lies:
while at the end of nineteenth and twentieth centuries, physicists did not imagine the approaching groundbreaking revolution and the apparently contradictory results with respect to the old Physics, in our present case we know we are on the doorstep of a new revolution, but we do not know which door is the correct one (not even how it looks like), and phenomenological investigation directs our steps.

A typical example of QGP investigation is the so-called ``COW experiment'', from the initials of Colella, Overhauser, and Werner.
In this experiment, the authors  investigated the influence of the Earth's gravitational field on a neutron interferometer \cite{Overhauser1974}.
This experiment is particularly relevant since it is the first experiment whose result is based on both $\hbar$ and $G_\mathrm{N}$, Newton's constant.
The main hypothesis for this experiment is that the relevant Hamiltonian is written, following the correspondence principle, as
\begin{equation}
	H = \frac{p^2}{2 M} + M g y~,
\end{equation}
where $M$ is the neutron mass, $g$ the gravitational field, and $y$ the vertical coordinate.
The result of the experiment \cite{Colella1975} is consistent with the theoretical estimations.
Notice, though, that this experiment does not concern the quantization of gravity.
Rather, it investigates the influence of gravity on a quantum system.

Other approaches focus on the expected foaminess of spacetime at the Planck scale.
This aspect of QG is in obvious contrast with the smoothness of spacetime in GR.
Nonetheless, we expect the spacetime to not be smooth at Planck scales.
To realize this, consider what would happen when shorter and shorter distances are probed in QM \cite{Amelino-Camelia2013}.
This can operatively be done considering the scattering of more and more energetic particles.
In QM, the length scale describing the size of the interaction and the length scale we can actually probe is given by the Compton length
\begin{equation}
	R_{\mathrm{C}} = \frac{\hbar}{M c}~,
\end{equation}
where $M$ is the rest mass of the particle that we consider in the scattering process.
The Compton length corresponds to the wavelength of a photon with energy $M c^2$.
On the other hand, GR presents its own length scale, the Schwarzschild radius $R_\mathrm{S}$
\begin{equation}
	\frac{R_{\mathrm{S}}}{2} = \frac{G M}{c^2}~.
\end{equation}
The Schwarzschild radius represents the radius of the event horizon of a Schwarzschild black hole of mass $M$.
We then see that the only parameter in both definitions is the mass, meaning that, while increasing it (or the energy of the scattered particle) will produce a better sensitivity for scattering processes in QM, it will also increase the importance of gravitational interactions.
For a particular value of the mass, these two length scales are the same
\begin{align}
	M_{\mathrm{Pl}} = & \sqrt{\frac{\hbar c}{G}}~, & R_{\mathrm{C}} = \frac{R_{\mathrm{S}}}{2} = \ell_{\mathrm{Pl}} = & \sqrt{\frac{\hbar G}{c^3}}~,
\end{align}
where $M_{\mathrm{Pl}}$ is the Planck mass.
This condition is the base of the argument in \cite{Scardigli1999_1}.

In this paper the author considers the application of the Heisenberg Uncertainty Principle (HUP) to a measuring process.
For relativistic cases, having $E \sim c p$, this principle can be written as
\begin{equation}
	\Delta E \Delta x \gtrsim \frac{\hbar c}{2}~. \label{eqn:relativistic_HUP}
\end{equation}
For this relation, observing a region of size $\Delta x$, fluctuations of the metric with amplitude $\Delta E$ would arise.
As long as the Schwarzschild radius associated with these fluctuations is smaller than the resolution of the instrument, no deviation from the standard theory is required.
On the other hand, when the Schwarzschild radius of the fluctuations becomes of the same order of the resolution of the instrument, considering smaller regions would increase the energy fluctuations and, in turn, the Schwarzschild radius, limiting the precision.
Therefore, we are forced to include a new term in the uncertainty relation \eqref{eqn:relativistic_HUP}, \emph{i.e.} \cite{Scardigli1999_1}
\begin{equation}
	\Delta x \gtrsim \frac{\hbar c}{2 \Delta E} + \frac{2 G \Delta E}{c^4}~.
\end{equation}
We also notice that, writing the metric fluctuation in terms of an effective mass $M = E/c^2$, we can identify in the two terms the Compton length and the Schwarzschild radius, up to factors 2, of the mass $M$.
Summarizing, we cannot arbitrarily increase the sensitivity of quantum-mechanical processes without incurring in no-longer-negligible gravitational interactions.
In particular, we would expect the formation of micro black holes, preventing us from probing distances shorter than the Planck length.
This ultimately suggests the existence of a minimal length scale for the quantum theory of gravity.

\section{Minimal Length}

The existence of a minimal length is present in many other contexts \cite{Garay1995_1}, for example in ST through fixed-angle scattering at high energies \cite{Amati1989_1,Gross1988_1}, in LQG \cite{Rovelli1995}, and in \emph{Gedankenexperimente} involving Heisenberg's microscope-like processes \cite{Mead1964,Maggiore1993_1} or metric fluctuations generating micro-black holes \cite{Scardigli1999_1}, and it is worth to analyze some of them in detail.

\subsection{Revisited Heisenberg Microscope Thought Experiment}

The existence of a minimal measurable length has been established on the base of modification of Heisenberg's microscope thought experiment.
In general, the standard uncertainty relation
\begin{equation}
	\Delta q \gtrsim \frac{\hbar}{\Delta p}~,
\end{equation}
due to the resolution of the microscope, is present in all the proposed \emph{Gedankenexperimente}.
The introduction of gravitational interaction, though, results in the appearance of new terms.

For example, one could repeat Heisenberg's same thought experiment, with a microscope of opening $\varepsilon$, including the gravitational interaction between the photon and the observed particle \cite{Mead1964}.
The Newtonian gravitational field generated by the photon is simply
\begin{equation}
	g \sim \frac{G \hbar}{c^2} \frac{\omega}{r^2}~,
\end{equation}
where $r$ is the radius of the region of strong interaction between the photon and the particle.
Assuming that the particle does not reach relativistic velocities, the photon will escape from the region of radius $r$ in a time $t \sim r/c$.
During this time, the particle acquires a velocity
\begin{equation}
	v \sim \frac{G \hbar}{c^3} \frac{\omega}{r}~,
\end{equation}
traveling in the direction taken by the photon for a distance
\begin{equation}
	d \sim \frac{G \hbar \omega}{c^4}~.
\end{equation}
The unknown direction of the photon is confined inside the microscope opening.
Therefore, the component of the distance traveled by the particle being attracted by the photon along a direction orthogonal to the optical axis of the microscope is approximately $d \sin \varepsilon \lesssim d$.
By Heisenberg's argument, the uncertainty in momentum of the particle is
\begin{equation}
	\Delta p \sim \frac{\hbar \omega}{c} \sin \varepsilon~.
\end{equation}
Therefore, for the uncertainty in position due to the gravitational interaction with the photon we obtain
\begin{equation}
	\Delta q \sim \frac{G}{c^3} \Delta p = \frac{\ell_\mathrm{Pl}^2}{\hbar} \Delta p~.
\end{equation}

A different, but somewhat similar argument consists in replacing the particle of Heisenberg's microscope with an extremal Reissner--Nordstr\"om black hole, where extremal means that the following relation between its mass $M$ and charge $Q$ holds:
\begin{equation}
	\frac{Q^2}{4 \pi \epsilon_0} = G M^2~.
\end{equation}
In this case, the black hole temperature \eqref{eqn:bh_temperature} is zero, which is why it cannot emit via Hawking process and is thus stable.

Let us say, then, that we want to perform the Heisenberg's microscope thought experiment to measure the apparent horizon of this black hole.
Therefore, we send a photon with wavelength $\lambda$ and momentum $p = h/\lambda$.
After the absorption of this photon, the black hole mass is increased by
\begin{equation}
	\Delta M = \frac{h}{\lambda c}~,
\end{equation}
becoming non-extremal.
Hence, it will emit Hawking radiation, which will be observed through the microscope.
In this process, the mass of the black hole reduces discontinuously from $M + \Delta M$ to $M$.
Similarly, its radius will change discontinuously in the process.
Therefore, an intrinsic uncertainty, related to this discontinuity, appears
\begin{multline}
	\Delta r_+ = r_{+,M + \Delta M} - r_{+,M} = \\
	= \frac{G \Delta M}{c^2} + \sqrt{\frac{G}{c^4} \left[ G (M+\Delta M)^2 - \frac{Q^2}{4 \pi \epsilon_0}\right]} - \sqrt{\frac{G}{c^4} \left( G M^2 - \frac{Q^2}{4 \pi \epsilon_0}\right)} \\
	\geq \frac{2 G \Delta M}{c^2} \sim \frac{\ell_\mathrm{Pl}^2}{\lambda} \sim \frac{\ell_\mathrm{Pl}^2}{h} \Delta p~.
\end{multline}

Although these two derivations analyze two different problems (locating a particle and measuring the radius of a black hole), both find a momentum dependent, minimal uncertainty limiting the resolution of the microscope.

\subsection{String Theory}

Historically, the first time that a QG theory predicted a minimal length was studying high-energy, fixed-angle scattering in ST.
Resorting to high-energy scattering is the obvious method when we want to probe smaller and smaller distances.
This is exactly the idea behind particle accelerators and colliders.
Indeed, from Heisenberg principle, higher momentum transfers $p$ in collisions determine shorter resolved distances
\begin{equation}
	\Delta q \sim \frac{\hbar}{p} \sim \frac{c\hbar}{E}~, \label{eqn:Heisenberg_scattering}
\end{equation}
where $E$ is the energy of the colliding particles and where $E \sim p c$ for ultra-relativistic particles.
Therefore, it was clear that, assuming ST is the proper theory of QG, high-energy scattering of strings would give information on the structure of spacetime in this theory.
The simple fact that string interactions are not local, given the extended nature of these objects, gives a first glimpse for the best position resolution since the finite size of the interaction region implies a maximal resolution.
Furthermore, one can also show that the string size increases with its energy \cite{Gross1989}
\begin{equation}
	L_s \sim \frac{\alpha' E}{N}~.
\end{equation}
If we simply add this fundamental uncertainty due to the string size to the other fundamental uncertainty in \eqref{eqn:Heisenberg_scattering}, the spatial resolution in a scattering process will look like
\begin{equation}
	\Delta q \sim \frac{c\hbar}{E} + \frac{\alpha' E}{N} ~,
\end{equation}
implying a minimal uncertainty
\begin{equation}
	\Delta q_{\mathrm{min}} \sim \sqrt{\frac{\alpha' c \hbar}{N}} \qquad \mbox{for} \qquad E \sim  \sqrt{\frac{\alpha'}{N c \hbar}}~.
\end{equation}

In ST, the existence of a minimal length is also embedded in the concept of duality \cite{Witten1996_1}.
In this theory we have a symmetry between the momentum spectra $p_p$ of massless particles on a circle of radius $R$ and of strings wrapped $m$ times around the same circle $p_s$.
From this symmetry, one can find a relation between large and short radii $R$
\begin{equation}
	p_p \sim p_s \Rightarrow R \sim \frac{\alpha'}{R}~.
\end{equation}
Therefore, in the context of compactification of extra dimensions, one finds that a minimal radius for the extra dimensions exists, defined by the string tension
\begin{equation}
	R_{\mathrm{min}} \sim \sqrt{\alpha'}~.
\end{equation}

\subsection{Loop Quantum Gravity}

As we saw in Sec.~\ref{sec:LQG} and in particular in \eqref{eqn:area_quantization}, LQG naturally includes a minimal length in the form of minimal areas and volumes.
Since LQG describes geometrical observables as quantum operators, it predicts a discrete spectrum for areas and volumes, the first related to the intersections between the spin network and a surface, the second defined through the intertwiners of the nodes of the spin network contained in a region of space.
In particular, these quantities are integer multiples of elementary Planck area and volume, $\ell_\mathrm{Pl}^2$ and $\ell_\mathrm{Pl}^3$, respectively \cite{Rovelli1995}.

\subsection{Mass Spectrum of a Kerr--Newman Black Hole}

As a last example of minimal length, we consider the mass spectrum for a Kerr black hole \cite{Bekenstein1974}.
Consider a Kerr--Newman black hole, \emph{i.e.} a charged, rotating black hole generalizing \eqref{eqn:RN} \cite{Newman1965}
\begin{equation}
	\diff s^2 = - \left(\frac{\diff r^2}{\Delta} + \diff \theta^2 \right) \rho^2 + c \diff t - a \sin^2 \theta \diff \phi)^2 \frac{\Delta}{\rho^2} - [(r^2 + a^2) \diff \phi - a c \diff t]^2 \frac{\sin^2 \theta}{\rho^2}~,
\end{equation}
where
\begin{align}
	a = & \frac{J}{M c}~, & \rho^2 = & r^2 + a^2 \cos^2 \theta~, & \Delta = & r^2 - R_\mathrm{S} r + a^2 + R_\mathrm{Q}^2~.
\end{align}
In natural units, the mass $M$, the charge $Q$, and the angular momentum $\vec{J}$ of the black hole are related by
\begin{equation}
	M^2 = M_\mathrm{ir}^2 \left(1 + \frac{Q^2}{4 M_\mathrm{ir}^2}\right) + \frac{J^2}{4 M_\mathrm{ir}^2}~,
\end{equation}
where $M_\mathrm{ir}$ is the irreducible mass, related to the area of the horizon \cite{Christodoulou1971}.
Since a\linebreak\mbox{Kerr--Newman} black hole is a stationary solution of the Einstein--Maxwell equations, in quantum terms it has to correspond to an eigenstate of the energy operator.
Furthermore, a black hole represents a bound state; therefore, its energy spectrum is discrete.
Similarly, also the charge and the angular momentum spectra are discrete.
Therefore, the spectrum of $M_\mathrm{ir}^2$, and consequently the event horizon area, is discrete.
Finally, considering the analogy between the squared irreducible mass and the action integral $\oint p \diff q$ of a periodic quantum system, it is possible to find the eigenvalues of the area
\begin{equation}
	A = 16 \pi M_\mathrm{ir}^2 = 8 \pi \hbar n~, \qquad \qquad n = 1 , 2 , 3 \ldots~,
\end{equation}
with its minimal value for $n=1$.

{\subsection{Lorentz Covariance and Minimal Length}}

{The existence of a minimal length related with QG Physics may suggest violations of Lorentz covariance in the context of QG.
In fact, if one assumes the transformation laws of Special Relativity, one should also assume that, regardless of the actual value of the length scale of a physical system, there exists a reference frame in which such a scale is contracted to the size of the minimal length of QG, generally assumed to be the Planck length.
This would then imply that every physical system should exhibit QG phenomena when observed from particular reference frames, while in other frames such phenomena would not appear, breaking Lorentz covariance.
As we saw in Sec.\,\ref{sec:DSR}, this is the starting point of DSR, in which the Lorentz transformations are modified to preserve the minimal length from contractions and to avoid the paradox described above.}

{This is not case for all theories of QG.
LQG for example, can predict the existence of an invariant minimal length (or more appropriately, a minimal area) scale.
In fact, in this theory, as we saw in Sec.~\ref{sec:LQG}, geometrical properties as areas and volumes appear as eigenvalues of quantum observables.
When two observers in relative motion are considered, the respective area operators do not commute.
Therefore, the spectra of such operators, and in particular their minimal values, result to be uncorrelated, as, for example, in the case of the spectra of two different component of the angular momentum.
On the other hand, a boost continuously changes the probability distribution of observing a particular eigenvalue of the area operator \cite{Rovelli2003}.}

\section{Generalized Uncertainty Principle} \label{sec:GUP}

As we have seen, QG theories predict a minimal uncertainty in position at the Planck scale.
On the other hand, the Heisenberg principle, one of the pillars of QM stating that it is not possible to simultaneously measure with arbitrary precision both position and momentum of a particle, allows for any arbitrary value of $\Delta q$, as long as the product of the position and momentum uncertainties is larger than a given quantity
\begin{equation}
	\Delta q \Delta p \geq \frac{\hbar}{2}~.
\end{equation}
Many authors have tried to account for this difference between QM and QG.
In particular, we can identify three different approaches to the Generalized Uncertainty Principle (GUP) \cite{Scardigli2016a}.

The first of them, proposed for the first time by Kempf, Mangano, and Mann in 1995 \cite{Kempf1995_1}, consists in a modified commutation relation between position and momentum, that for one-dimensional systems reads
\begin{equation}
	[q,p] = i \hbar \left[1 + \beta p^2\right]~, \label{eqn:commutator_kmm}
\end{equation}
where $\beta$ is a parameter that describes the scale at which quantum-gravitational effects become relevant.
Using the Schr\"odinger--Robertson uncertainty relation for two quantum operators $A$ and $B$
\begin{equation}
	(\Delta A)^2 (\Delta B)^2 \geq \left|\frac{\left<AB + BA\right>}{2} - \langle A \rangle \langle B \rangle\right|^2 + \left|\frac{\left<[A,B]\right>}{2}\right|^2~, \label{eqn:Sch-Ro}
\end{equation}
where
\begin{align}
	(\Delta A)^2 = & \langle A^2 \rangle - \langle A \rangle^2~, & (\Delta B)^2 = & \langle B^2 \rangle - \langle B \rangle^2~,
\end{align}
it is possible to show that the relation in (\ref{eqn:commutator_kmm}) implies a minimal uncertainty in position.
{In fact, for the case of the one-dimensional harmonic oscillator, as we will see in \mbox{Chapter \ref{ch:HO}}, we can find the following uncertainty relation}
\begin{figure}
	\centering
	\includegraphics[width=0.5\textwidth]{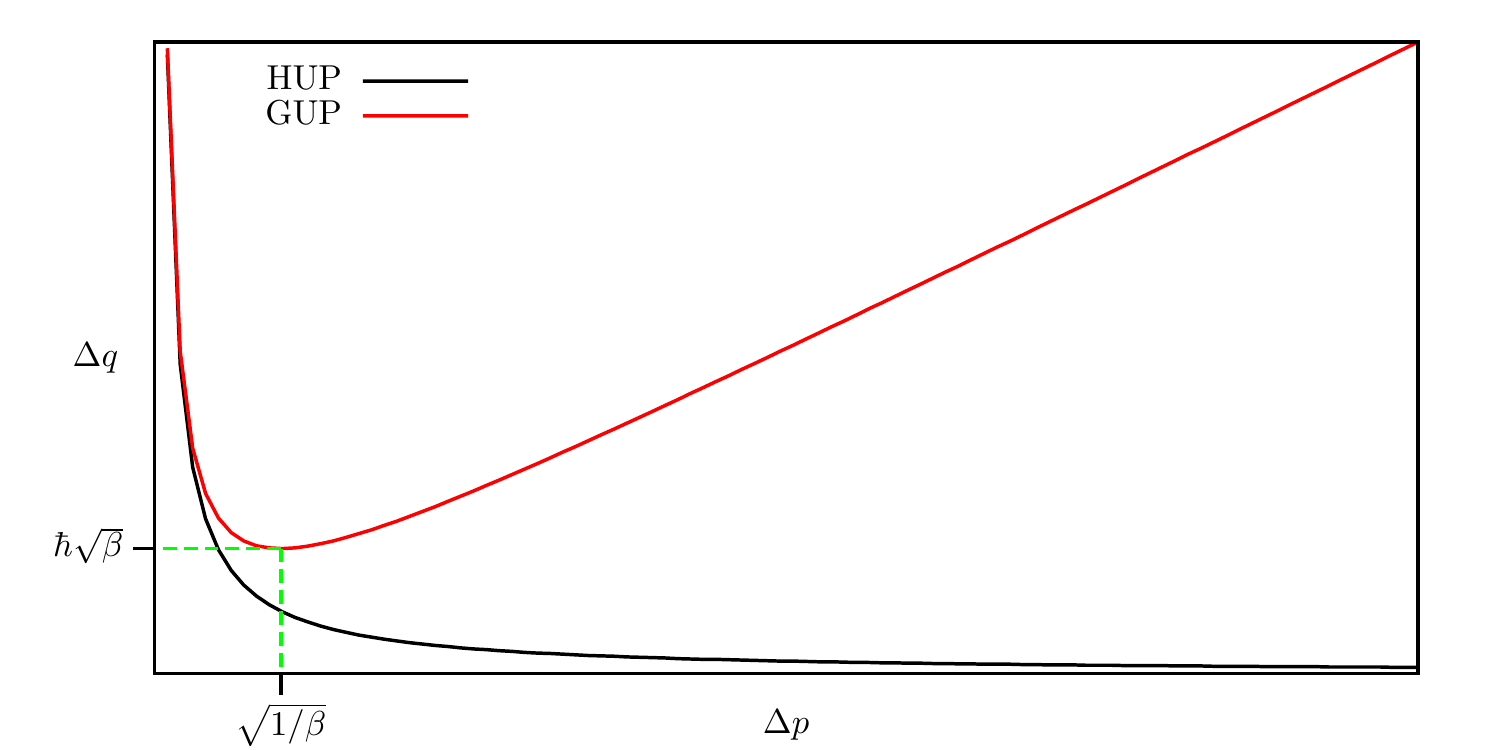}
	\caption[Comparison between HUP and GUP]{Comparison between Heisenberg's Uncertainty Principle (black line) and GUP (red line).
		Values for the minimal uncertainty in position and for the corresponding uncertainty in momentum are shown.} \label{fig:GUP}
\end{figure}
\begin{equation}
	\Delta q \geq \frac{\hbar}{2} \left[ \frac{1}{\Delta p} + \beta \Delta p \right]~, \label{eqn:unc_rel_quad}
\end{equation}
that makes evident the existence of a minimal uncertainty in position
\begin{equation}
	\Delta q_{\mathrm{min}} = \hbar \sqrt{\beta} \qquad \mbox{for} \qquad \Delta p = \sqrt{\frac{1}{\beta}}~,
\end{equation}
as shown in Fig. \ref{fig:GUP}.
Notice that this model does not include a minimal or a maximal uncertainty in momentum.

The second group of works considers a deformation of classical Newtonian mechanics, introducing a modified version of the standard Poisson brackets resembling the quantum commutator in \eqref{eqn:commutator_kmm}
\begin{equation}
	\{q,p\} = (1 + \beta p^2)~,
\end{equation} 
where in this case $q$ and $p$ are the classical position and momentum, respectively.
It is furthermore possible to identify two subgroups.
In one of them \cite{Benczik2002}, the deformed Newtonian mechanics is studied and the results are interpreted as extra contributions to the known results of Newtonian mechanics and GR.
This poses the conceptual problem of why Planck scale effects should be simply superposed to GR.
A slightly different approach considers a similar modification of the Poisson bracket in covariant form.
First developed in Minkowski spacetime then generalized to curved ones, the authors found that imposing such modification implies a deformed geodesic equation, with a consequent violation of the Equivalence Principle.
Other authors \cite{Scardigli2015} found that this violation is not due to the covariant formalism, but it is characteristic of the deformed Poisson brackets.

Finally, the third group of works \cite{Scardigli2016a,Scardigli2015} approached the problem converting the uncertainty relation in other observable features of quantum mechanical effects, as \emph{e.g.}, the Hawking evaporation.
Then, connecting these effects to the classical description of spacetime, it is possible to obtain information on observable quantities, as the perihelion precession in GR, and on the GUP parameter itself without requiring any representation of commutators or modifying the classical equations of motion.
Furthermore, in contrast to the second approach, they obtain that the Equivalence Principle is preserved and that the description of the motion of a test particle is still given by the standard geodesic equation.

In this thesis, we will consider the first approach.
In particular, we will consider a model including a linear and a quadratic dependence on the momentum.
While the presence of the quadratic term is dictated by ST \cite{Gross1988_1,Amati1989_1} and \emph{Gedankenexperimente} in black hole physics \cite{Maggiore1993_1,Scardigli1999_1}, a linear momentum-dependent term is motivated by DSR \cite{AmelinoCamelia2002_1} and as a generalization of the quadratic model.
To include both linear and quadratic terms, Ali, Das, and Vagenas in 2011 proposed the following model \cite{Ali2011_1}
\begin{equation}
	[x_i,p_j] = i \hbar \left\{ \delta_{ij} - \gamma \delta \left( p \delta_{ij} + \frac{p_i p_j}{p}\right) + \gamma^2 \left[ \epsilon p^2 \delta_{ij} + (2 \epsilon + \delta^2) p_i p_j \right] \right\}~, \label{eqn:GUP}
\end{equation}
where $\gamma$ is related to the scale at which we expect quantum-gravitational effects to become relevant, usually defined as a quantity proportional to the inverse Planck momentum
\begin{equation}
	\gamma = \frac{\gamma_0}{M_{\mathrm{Pl}} c} \qquad \mbox{with} \qquad \gamma_0 \sim 1~. \label{Mpscale}
\end{equation}
In (\ref{eqn:GUP}) the dimensionless parameters $\delta$ and $\epsilon$ are included to highlight the terms originating from the linear and quadratic contribution to GUP.
For the one-dimensional case with $\delta=0$ and $\epsilon = 1/3$ we obtain the same model as in (\ref{eqn:commutator_kmm}).

In \cite{Ali2011_1}, it has also been {proven} that GUP in (\ref{eqn:GUP}) is equivalent to {modifying} the standard position and momentum operators.
In particular, one can consider the following definitions
\begin{align}
	q_i = & q_{0,i} & p_i = & p_{0,i} \left[1 - \delta p_0 + (\epsilon + \delta^2) p_0^2 \right]~, \label{eqn:expansion_momentum}
\end{align}
where $\vec{q}_0$ and $\vec{p}_0$ are the usual, ``low-energy'' position and momentum operators, satisfying the standard commutation relation
\begin{equation}
	[q_{0,i},p_{0,j}] = i \hbar \delta_{i,j}~.
\end{equation}
One can indeed show that the commutation relation between $q_i$ and $p_i$ given by \eqref{eqn:expansion_momentum} is \eqref{eqn:GUP}.
On the other hand, notice that $p_{0,i} = - i \hbar \partial / \partial q_i $ is the generator of translations.
Therefore, we see that the physical momentum $\vec{p}$ in (\ref{eqn:expansion_momentum}) is no longer a generator of translations, but rather it can be interpreted as a composition of translations depending on all the components of the momentum through the module $p_0$.

Following these and other GUP models, in the last few years, several schemes to probe deviations from the Heisenberg principle were proposed, for example analyzing the normal modes of gravitational bar detectors \cite{Bawaj2014_1,Marin2013_1,Marin2014_1}, studying the Wheeler's conception of ``quantum foam'' \cite{Bekenstein2012_1}, or considering the optomechanical interaction in an optical cavity \cite{Pikovski2012_1}.
Nonetheless, some of these works, as well as estimations from known physical phenomena \cite{Das2008_1,Ali2011_1}, found no direct observations of these modifications.
Upper bounds for the parameters for the modification, though, have been deduced.

Considering what we have seen in the previous Chapter and in the present one, this thesis and the research work from which it originates has a twofold importance:
following the QGP program, we look for evidence for GUP that would not only imply the correctness of the existence of a minimal length{, realized through the existance of a minimal spatial uncertainty or through the discrete nature of spacetime}, but also support or refute various theories of QG, leading to a new and better description of Nature itself.
\chapter{Uncertainty Relation for Angular Momentum}
\label{ch:angular-momentum}

\begin{quote}
	 Nec adeo mirum fuerit, si quis pr\ae ter illam cotidianam revolutionem, alium quendam terr\ae ~motum opinaretur, nempe terram volvi, atque etiam pluribus motibus vagantem, \& unam esse ex astris Philolaus Pythagoricus sensisse fertur, Mathematicus non vulgaris, utpote cuius visendi gratia Plato non distulit Italiam petere, quemadmodum qui vitam Platonis scripsere, tradunt.	 
	 \begin{flushright}
	 	De revolutionibus orbium c\oe lestium -- Nicolaus Copernicus
	 \end{flushright}
\end{quote}

Angular momentum is of fundamental importance in Physics, both in the classical and the quantum theory.
For example, the second Kepler's law, empirically derived by Johannes Kepler from data previously collected by Tycho Brahe, can be obtained from the conservation of angular momentum.
This law of conservation, though, is not restricted to planetary astronomy only.
Rather, together with the conservation of linear momentum, it constitutes one of the cornerstones of Classical Physics.
On the other hand, conservation of angular momentum can be thought of as a consequence of a symmetry of Nature through Noether's theorem, \emph{i.e.} symmetry under rotations \cite{Goldstein_H}.
In fact, assume that a system described by a Lagrangian $\LL(q,p)$ admits a symmetry transformation involving a generalized coordinate $q_i$, that is, a transformation that changes $q_i$ leaving the Lagrangian invariant.
This means that the Lagrangian does not depend on the particular coordinate $q_i$ ($q_i$ is then said \emph{ignorable}).
Therefore, the Lagrangian equation for this particular coordinate reads
\begin{equation}
	{\frac{\diff}{\diff~t} \frac{\partial \LL(q,p)}{\partial \dot{q}_i} = \frac{\partial \LL(q,p)}{\partial q_i} = 0 \qquad \Rightarrow \qquad \frac{\diff}{\diff~t} \frac{\partial \LL(q,p)}{\partial \dot{q}_i} \equiv \dot{p}_i = 0~.}
\end{equation}
Therefore, the momentum conjugate to $q_i$ is a constant of motion.
For the particular case of rotational symmetry, describing the system in spherical coordinates and since the kinetic energy does not depend on any angle, the requirement that the potential does not depend on the azimuthal angle $\theta$ suffices to say that the component of the angular momentum orthogonal to the plane of constant polar angle $\phi=\mathrm{const}$ is a constant of motion.
In case of a system with a central force, like a planetary system, since the kinetic energy is the only term containing any dependence on $\dot{\theta}$ we have
\begin{equation}
	{\frac{\diff}{\diff~t} \frac{\partial \LL(\vec{r},\vec{v})}{\partial \dot{\theta}}} = \frac{\diff}{\diff~t} r m \dot{\theta} = \frac{\diff L_z}{\diff~t} = 0~,
\end{equation}
where $r$ is the radial coordinate and $m$ the mass of the object under consideration.

In QM, angular momentum acquires an even more important role.
Since the first quantum mechanical description of atomic systems by Bohr, Sommerfeld, and Wilson \cite{Wilson1915}, the quantization of angular momentum becomes a fundamental ingredient in any theory aimed to a description of electrons around a charged nucleus.
Furthermore, not only the \mbox{Bohr-Sommerfeld} model could explain why atoms do not decay, avoiding the paradox of Rutherford's model of electrons collapsing on the nucleus, but it could also give a detailed explanation of the spectroscopic series.

With the advent of the New Quantum Mechanics, the role of angular momentum became essential: angular momentum, through its magnitude $L^2$ and one of its components $L_z$, allows for a fundamental description of the elements of the periodic table \cite{Born}.
Therefore, the atom predicted by the Schr\"odinger equation, not only explains why each element occupies that particular position in the periodic table, but also gives a fundamental description of atomic interactions, becoming \emph{ipso facto} the fundamental ground of Chemistry.
As we will see in Sec.~\ref{sec:h-atom}, the dynamics of an electron in a hydrogen atom is characterized in QM by three quantum numbers: the principal quantum number $n$, describing the energy level of the electron and responsible for the periods of the periodic table; the azimuthal quantum number $l$, describing the total angular momentum; the magnetic quantum number $m$, describing one component of the angular momentum and responsible together with $l$ of the groups of the periodic table.
These three numbers, with the addition of the electronic spin $s$ and the atomic number $Z$, ultimately characterize an atom in QM in absence of external fields, from its chemical properties to the probability distribution of the electrons \footnote{Notice that for atoms containing more than one electron, one has to consider also electron-electron interactions, requiring a perturbative description of the atom.
Nonetheless, most of the properties can be obtained from the independent electron approximation and through variational computations.}.

All this said, the questions concerning modifications to the theory governing the angular momentum and their observations become relevant and require to be asked.
Therefore, carrying forward the phenomenological program of this thesis, in this Chapter we compute QG corrections to an important theoretical as well as experimental areas of QM, namely the angular momentum of elementary quantum systems and their experimental applications.
We indeed show that there are potentially measurable Planck scale effects to well-understood phenomena such as line spectra from the hydrogen atom, the Stern--Gerlach experiment, Larmor frequency and Clebsh-Gordan coefficients \cite{Bosso2017}.

As for the uncertainty principle, the motivation for applying GUP to the theory of angular momentum comes directly from the existence of a minimal measurable length and an angle-angular momentum uncertainty relation.
Consider, \emph{e.g.}, the observation of an extended object of mass $m$ at a distance $R$ moving with average angular velocity $\omega$.
For simplicity, let us neglect the uncertainty on $R$.
The angular size $\theta$ of the object can then be derived from its transiting a reference point, occurring in a time interval $T = x / v$, where $x$ is the proper size of the object and $v = \omega R$ is the linear velocity of the object.
These last two quantities are along an orthogonal direction to the line of sight.
We therefore have
\begin{equation}
	\Delta \theta = \Delta (\omega T) = \Delta \left( \frac{x}{R} \right)~.
\end{equation}
In standard QM, the HUP then implies for the angular resolution
\begin{align}
	\Delta \theta = & \frac{\hbar}{2 \Delta L} & \Rightarrow & & \Delta \theta \Delta L = & \frac{\hbar}{2}~,
\end{align}
where $L=Rmv$ is the angular momentum of the object with respect the observer.
On the other hand, a minimal uncertainty in the position of the object derived from GUP will imply a minimal angular resolution.
We are then forced to consider a GUP also for angle variables and their conjugate momenta, \emph{i.e.}, angular momenta.
Therefore, we expect a modification of the angular momentum algebra.
In the present Chapter, we will obtain this modification as a consequence of the GUP in position and momentum.

\section{Modified Angular Momentum Algebra} \label{sec:algebra}

In this Section, we start with the standard definition of angular momentum in Classical Mechanics,
\begin{equation}
	\vec{L} = \vec{q} \times \vec{p} = \left(
	\begin{array}{c}
		yp_z - zp_y\\
		zp_x - xp_z\\
		xp_y - yp_x
	\end{array}
	\right)~, \label{def:L}
\end{equation}
since it will allow us to directly apply a modified commutator between $q$ and $p$.
Furthermore, we will show how this quantity is modified when GUP is considered.
In standard QM, one often defines the angular momentum as the generator of rotations and through its algebraic properties.
One then usually shows that, when it comes to orbital angular momentum, the definitions in \eqref{def:L} and as the generator of rotations are equivalent.
However, as well as the momentum defined as the generator of translations cannot give rise to any minimal uncertainty relation (see below for more details), the angular momentum as the generator of rotations does not allow for any minimal angular resolution, required by GUP as explained above.
For this reason, for the moment we prefer to focus on the orbital angular momentum, defined by \eqref{def:L}.
We will then see that the two definitions for the angular momentum are no longer equivalent when GUP is included.
Rather, the orbital angular momentum acts like a composition of rotations and translations.

Using the definition in \eqref{def:L} and the commutation relation between the position and momentum operators, one can obtain the commutator of angular momentum components.
For example, using Heisenberg algebra $[q_i, p_j] = i \hbar \delta_{ij}$, we find the usual algebra for the angular momentum operators
\begin{equation}
	[L_i,L_j] = i\hbar \epsilon_{ijk}L_k,
\end{equation}
where $\epsilon_{ijk}$ are the usual Levi-Civita symbols.
However, if we consider a generic commutation relation between the position and momentum operators, this is not true anymore, \emph{i.e.}, GUP now implies the existence of minimal measurable angles.

Consider the commutator between two components of the angular momentum
\begin{equation}
	[L_i,L_j] = \epsilon_{imn} \epsilon_{irs} [q_m p_n, q_r p_s] = \epsilon_{imn} \epsilon_{irs} \{ q_m [ p_n, q_r ] p_s + q_r [q_m , p_s ] p_n \}~, \label{eqn:commutator_ang_mom}
\end{equation}
Using the GUP commutator in (\ref{eqn:GUP}), we obtain
\begin{multline*}
	[L_i,L_j] = i \hbar \epsilon_{imn} \epsilon_{irs} \times \\
		\qquad \times \left \{ q_r p_n \left[ \delta_{ms} - \delta \gamma \left(\delta_{ms} p + \frac{p_m p_s}{p} \right) + \gamma^2 (\epsilon \delta_{ms} p^2 + (2 \epsilon + \delta^2) p_m p_s ) \right] + \right. \\
			\qquad \qquad \left. - q_m p_s \left[ \delta_{nr} - \delta \gamma \left( \delta_{nr} p + \frac{p_n p_r}{p} \right) + \gamma^2 ( \epsilon \delta_{nr} p^2 + (2 \epsilon + \delta^2) p_n p_r ) \right] \right\} = \\
	= i \hbar ( \epsilon_{mni} \epsilon_{mjr} q_r p_n - \epsilon_{nim} \epsilon_{nsj} q_m p_s ) ( 1 - \delta \gamma p + \epsilon \gamma^2 p^2) = \\
	 = i \hbar [ ( \delta_{ir} \delta_{nj} - \delta_{nr} \delta_{ij} ) q_r p_n - ( \delta_{is} \delta_{mj} - \delta_{ij} \delta_{ms} ) q_m p_s ] ( 1 - \delta \gamma p + \epsilon \gamma^2 p^2)~,
\end{multline*}
{\begin{equation}
	[L_i,L_j] = i \hbar \epsilon_{ijk} L_k ( 1 - \delta \gamma p + \epsilon \gamma^2 p^2)~, \label{eqn:generalized_commutator_ang_mom}
\end{equation}}
where we used the following properties of Levi-Civita symbols
\begin{equation}
	\epsilon_{mni} \epsilon_{mjr} = \delta_{ir} \delta_{nj} - \delta_{nr} \delta_{ij}~.
\end{equation}
It is also possible to show that Jacobi identity
\begin{equation}
	[L_i,[L_j,L_k]] + [L_j,[L_k,L_i]] + [L_k,[L_i,L_j]] = 0
\end{equation}
is trivially satisfied by this modified commutation relation.
Note that we recover the standard commutation relation $[L_i,L_j] = i \hbar \epsilon_{ijk} L_k$ for $\gamma=0$.

We can also show that, including GUP, the following relation still holds
\begin{equation}
	[L^2,L_j] = 0~. \label{eqn:commutator_L2_Lj}
\end{equation}
In fact, consider that
\begin{multline}
	[L_i,p_l] = \epsilon_{ijk}(q_j p_k p_l - p_l q_j p_k) = \\
	= i \hbar \epsilon_{ijk} \left\{ \delta_{jl} p_k - \delta \gamma \left( \delta_{jl} p p_k + \frac{p_j p_l p_k}{p} \right) + \gamma^2 \left[ \epsilon \delta_{jl} p^2 p_k + (2 \epsilon + \delta^2) p_j p_l p_k \right] \right\} = \\
	= i \hbar \epsilon_{ilk} p_k \left( 1 - \delta \gamma p + \epsilon \gamma^2 p^2 \right)~, \label{eqn:commutator_Li_pl}
\end{multline}
where we used the model in (\ref{eqn:GUP}).
We then have
\begin{equation}
	[L_i,p^2] = p_l[L_i,p_l] + [L_i,p_l]p_l = 2 i \hbar \epsilon_{ilk} p_k p_l \left( 1 - \delta \gamma p + \epsilon \gamma^2 p^2 \right) = 0~. \label{eqn:commutator_Li_p2}
\end{equation}

{Given \eqref{eqn:commutator_Li_pl} and observing that in this relation the commutator between the vectors $\vec{L}$ and $\vec{p}$ depends only on $\vec{p}$, t}o find the commutation relation between a component of the angular momentum and the magnitude of the linear momentum, we will first suppose that such a commutator depends only on the vector $\vec{p}$
\begin{equation}
	[L_i,p] = f(\vec{p})~.
\end{equation}
In this way, using the result just found we have
\begin{equation}
	0 = [L_i,p^2] = p[L_i,p] + [L_i,p]p = 2p[L_i,p]~,
\end{equation}
that means, for $p\not=0$,
\begin{equation}
	[L_i,p] = 0~. \label{eqn:commutator_Li_p}
\end{equation}
Finally, we find
\begin{multline}
	[L^2,L_j] = L_i[L_i,L_j] + [L_i,L_j]L_i = \\
	= i \hbar \epsilon_{ijk} [L_i L_k (1 - \delta \gamma p + \epsilon \gamma^2 p^2) + L_k (1 - \delta \gamma p + \epsilon \gamma^2 p^2) L_i] = \\
	= i \hbar \epsilon_{ijk} (L_i L_k + L_k L_i) (1 - \delta \gamma p + \epsilon \gamma^2 p^2) = 0~.
\end{multline}
Therefore, even with a modified angular momentum algebra, we can define simultaneous eigenstates of $L^2$ and $L_z$.
Furthermore we notice that for (\ref{eqn:commutator_Li_p2}) and (\ref{eqn:commutator_Li_p}) also $p$ and $p^2$ commute with $L_z$ and thus they also commute with $L^2$
\begin{subequations}
	\begin{align}
		[L^2,p] & = L_i[L_i,p] + [L_i,p]L_i = 0~, \\
		[L^2,p^2] & = L_i[L_i,p^2] + [L_i,p^2]L_i = 0~,
	\end{align} \label{eqn:commutator_L2_p_p2}
\end{subequations}
that is, we obtain the same results of the standard theory.
Hence we can define simultaneous eigenstates of the operators $L^2$, $L_z$, and $p$.
Moreover, such a state is also an energy eigenstate for the free Hamiltonian.
From a mathematical point of view, we then see that the angular momentum algebra does not close anymore.
However, its closure can be achieved including the linear momentum.
More on this in the next Subsection.

For a better understanding of the meaning of this modified algebra, following \cite{Ali2011_1} we would like to describe the physical angular momentum $\vec{L}$ in terms of low-energy quantities $\vec{L}_0$ and $\vec{p}_0$.
Using \eqref{eqn:expansion_momentum}, we can write the momentum $\vec{p}$ in terms of the generator of translations $\vec{p}_0$.
On the other hand, introducing the generators of rotations we necessarily obtain
\begin{equation}
	[L_{0,i},L_{0,j}] = i \hbar \epsilon_{ijk} L_{0,k}~.
\end{equation}
Since from standard QM we know that $L_{0,k}$, representing the orbital angular momentum, fulfill similar relations to (\ref{def:L}),
we are allowed to expand the physical angular momentum in terms of generators of translations and rotations,  $\vec{p}_0$ and $\vec{L}_0$, respectively,
\begin{equation}
	L_i = L_{0,i} \left[ 1 - \delta \gamma p_0 + (\epsilon + \delta^2) \gamma^2 p^2_0 \right] ~. \label{eqn:ang_mom_henergy_lenergy}
\end{equation}
This relation is consistent with the previous result on the physical angular momentum algebra
\begin{multline}
	[L_i,L_j] 
	= [L_{0,i},L_{0,j}] \left[1 - \delta \gamma p_0 + (\epsilon + \delta^2) \gamma^2 p^2_0\right]^2 = \\
	= i \hbar \epsilon_{ijk} L_{0,k} \left[1 - 2 \delta \gamma p_0 + (2 \epsilon + 3 \delta^2) \gamma^2 p^2_0 \right]
	= i \hbar \epsilon_{ijk} L_k (1 - \delta \gamma p + \epsilon \gamma^2 p^2)~. \label{eqn:com_rel_ang_mom_lenergy}
\end{multline}
We note that now the angular momentum does not coincide with the generator of rotations, and therefore it is not conserved in general even for rotationally invariant Hamiltonians, as in presence of a central potential (see Sec.~\ref{sec:h-atom}).
However, $\vec{L}$ can be expanded in terms of generators of rotations and translations.
On the other hand, notice that $\vec{L}_0$ is still conserved for rotationally invariant systems.

To conclude this section, we show that linear and angular momenta are orthogonal to each other, \emph{i.e.} $\vec{p}\cdot\vec{L} = 0$, in agreement with Classical and standard Quantum Mechanics.
In fact, consider the scalar product
\begin{equation}
	\vec{p}\cdot\vec{L} = p_i \epsilon_{ijk} x_j p_k = p_i \epsilon_{ijk} [x_j, p_k] + p_i \epsilon_{ijk} p_k x_j=0~.
\end{equation}
Both terms in the last expression vanish, since momentum operators commute and\linebreak\mbox{$[x_i,p_j] = [x_j,p_i]$}, as one can easily prove from (\ref{eqn:GUP}).

\subsection{QG-Modified Angular Momentum} \label{ssec:QG-mod_ang_mom}

As we saw in the introduction to the present Chapter, phenomenological arguments suggest a modification of the physical interpretation of the angular momentum and of its mathematical structure.
Moreover, using dimensional analysis we can obtain hints on how the angular momentum theory should be modified.
These arguments will not be sufficient to characterize possible modifications, but they will motivate what we found in this Section.

In QGP one usually expects to observe quantum gravitational effects at Planck scales, and describes these effects in terms of ratios between characteristic quantities and homologous Planck scale ones defined through appropriate products of fundamental constants (see Chapter \ref{ch:GUP}), \emph{e.g.} in terms of ratios of linear momenta $p/P_\mathrm{Pl}$, energies $E/E_\mathrm{Pl}$, lengths $l/\ell_\mathrm{Pl}$, etc.
However, the only quantity with units of angular momentum composed using fundamental constants is $\hbar$.
No terms involving special ($c$) or general relativity ($G$) are present, making of the ``angular momentum scale'' $\hbar$ a purely quantum mechanical scale.
If this was the end of the story, possible modifications involving angular momenta only, in the form of ratios such as $L^2/\hbar^2$, would be observable also in standard QM.
From experiments in QM and QFT, no terms that would justify such a modification are found.
Nonetheless, as explained at the beginning of the present Chapter, we do expect quantum gravitational effects on the angular momentum.
To describe these effects we are then forced to define the angular momentum as function of observables whose relative Planck scale is defined not only by $\hbar$.
Following these arguments, one may even argue that QG deeply affects the same definition and concept of angular momentum, since it is required to be accompanied by other physical quantities.

Furthermore, we can also have some insights of whether the modified angular momentum algebra still closes in QG.
Let us assume the following form for an effective modification of angular momentum 
\begin{equation}
	L_i = L_{0,i} [ 1 + f_i(\zeta)]~, \label{eqn:ang_mom_mod}
\end{equation}
where $\zeta$ represents the set of observables that can be used to modify the angular momentum, and $L_{0,i}$ is the standard angular momentum.
We also assumed that each angular momentum component depends on the homologous component of the vector function $\vec{f}(\zeta)$.
We can then prove that the algebra of $\vec{L}$ does not close.
In fact, we find
\begin{equation}
	[L_i, L_j] = [L_{0,i}, L_{0,j}] [1 + f_i(\zeta) + f_j(\zeta)] + L_{0,i} [ f_i(\zeta) , L_{0,j} ] + L_{0,j}[L_{0,i} , f_j(\zeta)] ~, \label{eqn:ang_mom_mod_com}
\end{equation}
where we neglected products of $\vec{f}$, since we are assuming small modifications.
For a vector operator the following relation holds \cite{Messiah}
\begin{equation}
	[L_{0,i} , f_j] = i \hbar \epsilon_{ijk} f_k~.
\end{equation}
Thus we obtain
\begin{equation}
	[L_i, L_j] = i \hbar \epsilon_{ijk} L_{0,k} [1 + f_i(\zeta) + f_j(\zeta)] + i \hbar \epsilon_{ijk} ( L_{0,i} + L_{0,j} ) f_k(\zeta) \not =i \hbar \epsilon_{ijk} L_{0,k} [ 1 + f_k(\zeta)] ~.
\end{equation}
Therefore, the standard angular momentum algebra does not close with the modification \eqref{eqn:ang_mom_mod}.
On the other hand, considering a scalar modification
\begin{equation}
	L_i = L_{0,i} [ 1 + g (\zeta)]~,
\end{equation}
again from \eqref{eqn:ang_mom_mod_com} we have
\begin{equation}
	[L_i, L_j] = [L_{0,i}, L_{0,j}] [1 + 2 g(\zeta)] \not = i \hbar \epsilon_{ijk} L_{0,k} [ 1 + g(\zeta)] ~,
\end{equation}
where we used the following property for scalars
\begin{equation}
	[L_{0,i} , g] = 0~.
\end{equation}
Notice that \eqref{eqn:ang_mom_henergy_lenergy} corresponds to this last case.

\section{Modified Angular Momentum Spectrum}\label{sec:spectrum} \label{sec:angular_momentum}

We want now to find the spectrum of the angular momentum with GUP.
Using a standard procedure to find it, we will see that it will have a similar structure with respect to the standard theory, but the quantum numbers $l$ and $m$ are in principle not integer, being modified by GUP.

As mentioned above, we are able to consider simultaneous eigenstates of $p$, $L_z$, and $L^2$
\begin{align}
	L^2|p\lambda m\rangle = & \hbar^2 \lambda |p\lambda m\rangle & \lambda &\geq 0~, & \qquad && L_z|p\lambda m\rangle = & \hbar m |p\lambda m\rangle & m^2 &\leq \lambda~. \label{eqn:eigenvalues}
\end{align}
Defining the usual ladder operators
	\begin{align}
		L_+ &= L_x + i L_y~, & L_- &= L_x - i L_y~, \label{def:L_pm}
	\end{align}
and using (\ref{eqn:generalized_commutator_ang_mom}), we get
\begin{subequations}
	\begin{align}
		[L_z,L_\pm] &= [L_z,L_x] \pm i [L_z,L_y] = \hbar (iL_y \pm L_x)(1 - \delta \gamma p + \epsilon \gamma^2 p^2) = \pm \hbar L_\pm (1 - \mathcal{C}) ~,\label{eqn:commutator_Lz_Lpm} \\
		[L^2,L_\pm] &= [L^2,L_x] \pm i [L^2,L_y] = 0~, \label{eqn:commutator_L2_Lpm} \displaybreak\\
		[L_+,L_-] &= [L_x,L_x] - i [L_x,L_y] + i[L_y,L_x] + [L_y,L_y] = -2i[L_x,L_y] = 2 \hbar L_z (1 - \mathcal{C} )~, \label{eqn:commutator_L+_L-}
	\end{align}
\end{subequations}
where $\mathcal{C} \equiv \delta \gamma p - \epsilon \gamma^2 p^2$ represents the modification due to the GUP.
From (\ref{eqn:commutator_L2_Lpm}) we obtain
\begin{equation}
	L^2 L_\pm |p \lambda m\rangle = L_\pm L^2 | p \lambda m \rangle = \hbar^2 \lambda L_\pm |p \lambda m\rangle~.
\end{equation}
Thus, even for GUP, the ladder operators do not change the eigenvalues of the magnitude of the angular momentum.
On the other hand, using (\ref{eqn:commutator_Lz_Lpm}) we get
\begin{equation}
	L_z L_\pm |p \lambda m\rangle = L_\pm [ L_z \pm(1- \mathcal{C} )]|p \lambda m\rangle = \hbar [m \pm (1- \mathcal{C} )] L_\pm|p \lambda m\rangle~.
\end{equation}
That is, while $L_\pm$ still act as ladder operators, the spacing between two consecutive $L_z$ eigenstates undergoes modifications due to GUP.
Furthermore, from (\ref{def:L_pm}) and (\ref{eqn:generalized_commutator_ang_mom}) we get
\begin{equation}
	L_\mp L_\pm = L^2 - L_z[L_z \pm \hbar(1 - \mathcal{C} )]~,
\end{equation}
from which we obtain the norm of the states $L_\pm |p \lambda m\rangle$ as
\begin{equation}
	||L_\pm | p \lambda m\rangle ||^2 = \langle p \lambda m | L_\mp L_\pm | p \lambda m \rangle = \hbar^2 \{ \lambda - m[m \pm (1 - \delta \gamma p + \epsilon \gamma^2 p^2)]\} \geq 0~. \label{eqn:norm_L+-}
\end{equation}
We note that if we considered eigenstates of $L^2$ and $L_z$ only, we would have obtained
\begin{equation}
	||L_\pm |\lambda m\rangle ||^2 =
	\langle \lambda m | L_\mp L_\pm | \lambda m \rangle =
	\hbar^2 \{ \lambda - m[m \pm (1-\langle \mathcal{C} \rangle)]\} \geq 0~, \label{eqn:module_exp_val}
\end{equation}
that is, the modification due to the GUP would appear as $\langle \cal{C} \rangle$.
This observation will come in handy when we will consider GUP effects for the hydrogen atom model in Sec.~\ref{sec:h-atom}.
We will postpone further discussion on this feature to the next Subsection.

Furthermore, notice that the point $\CE = 0$, for $\gamma \not = 0$, and for $p \not = 0$ and $p^2 \not = 0$, corresponds to the Planck scale, for which $p = \frac{\delta}{\epsilon \gamma_0} M_\mathrm{Pl} c$.
We then expect that the present GUP model will not be reliable at this scale since higher order terms of a full theory of QG will become relevant.

As for the eigenvalues of $L^2$ and $L_z$, since we expect $\langle L^2 \rangle \geq \langle L_z^2 \rangle$ we require $\lambda \geq m^2$.
This condition implies the existence of upper and lower bounds for $m$, which we denote $m_+$ and $m_-$, respectively.
From (\ref{eqn:norm_L+-}), it follows
\begin{equation}
	L_\pm |p \lambda m_\pm\rangle = 0 \qquad \Rightarrow \qquad \lambda = m_\pm[m_\pm \pm (1- \mathcal{C} )]~. \label{eqn:condition_lambda_m}
\end{equation}
We can reach these values starting from any value $m$ and applying $s$ times $L_+$ or $t$ times $L_-$, with $s, t \in \mathbb{N}$
\begin{align}
		m_+ = & m + s(1- \mathcal{C} )~, & m_- = & m - t(1- \mathcal{C} )~.
\end{align}
Combining these two relations we find
\begin{equation}
	m_+ = m_- + (s+t)(1 - \mathcal{C} ) = m_- + n(1- \mathcal{C} )~,
\end{equation}
with $n \in \mathbb{N}$ giving the distance between the two bounds.
Inserting this last relation in one of the rightmost equations of (\ref{eqn:condition_lambda_m}) and solving for $m_+$ or $m_-$, we find the two relations
\begin{equation}
	m_+ = \frac{n}{2}(1- \mathcal{C} ) = l (1- \mathcal{C} ), \qquad m_- = -\frac{n}{2}(1- \mathcal{C} ) = -l (1- \mathcal{C} )~, 
\end{equation}
where we defined $l \equiv n/2$.
\begin{figure}
	\centering
	\begin{subfigure}[t]{0.44\linewidth}
		\includegraphics[width=\textwidth]{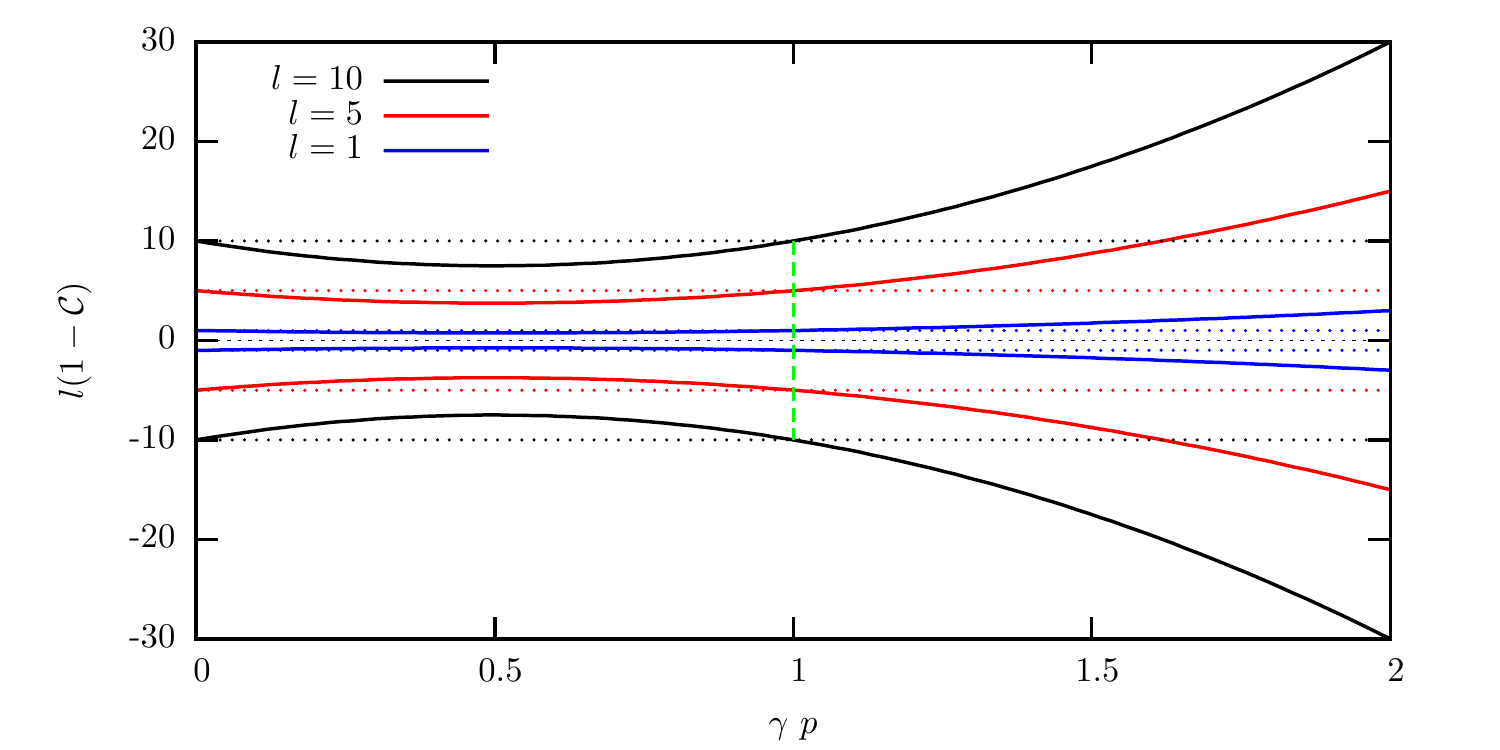}
    		\caption{Plot of the constraints with respect to $\gamma p$ for three different values of the azimuthal quantum number. ${l=1}$, $l=5$ and $l=10$.} \label{fig:linear+quadratic}
    \end{subfigure}
    \qquad
    \begin{subfigure}[t]{0.44\linewidth}
    	\includegraphics[width=\textwidth]{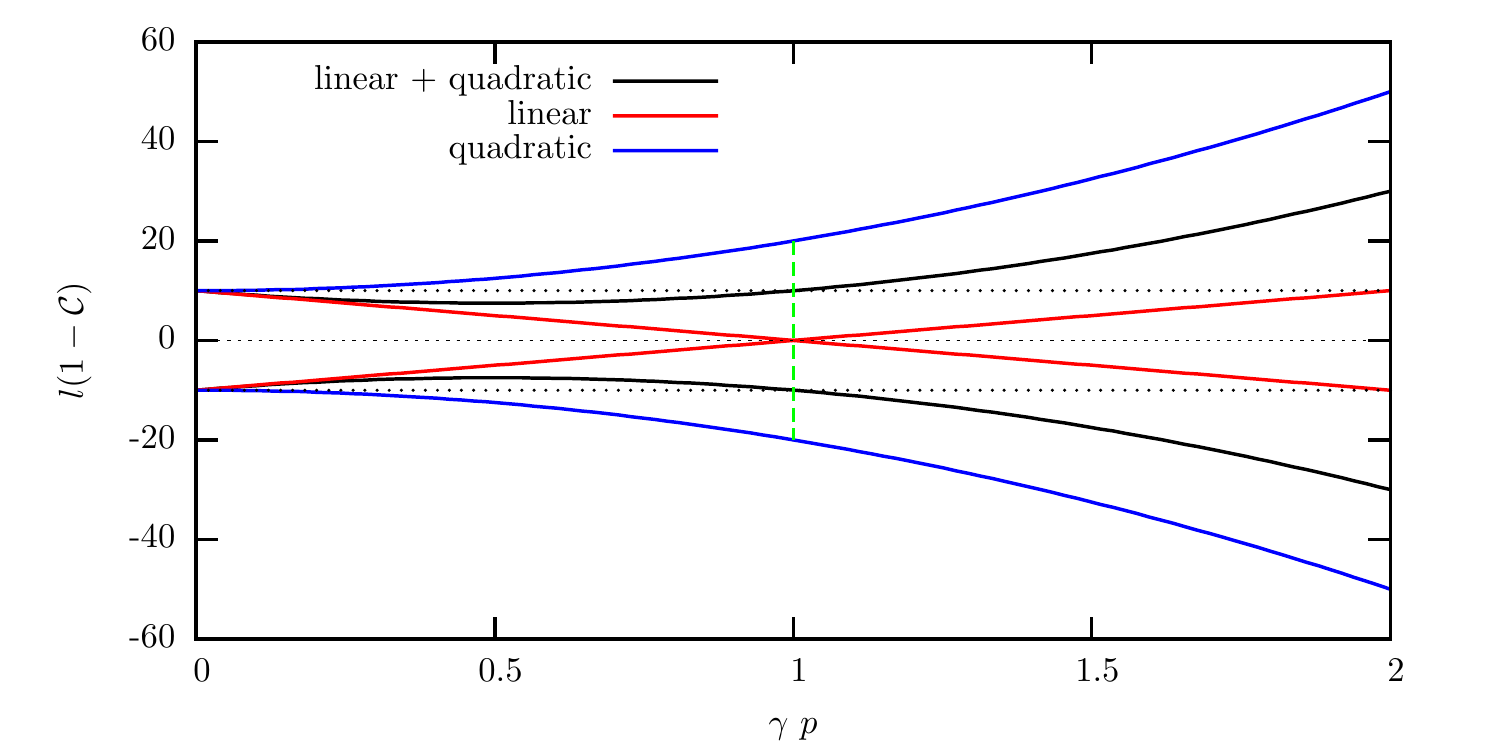}
    	\caption{Plot of the constraints with respect to $\gamma p$ for $l=10$ considering three GUP models.}\label{fig:models}
	\end{subfigure} \caption[(a) Constraints on $m$ as a function of $\gamma p$ for different value of $l$.\protect\\ (b) Constraints on $m$ as a function of $\gamma p$ for different GUP models.]{}
\end{figure}
In Figure \ref{fig:linear+quadratic}, the bounds of $m$ are represented for the model in \cite{Ali2011_1} with $\delta = \epsilon = 1$ for three values of $l$. 
Solid lines represent the constraints found in GUP, dotted lines the constraints in the standard theory.
In Figure \ref{fig:models}, the constraints for $l=10$ for three different parametrizations are plotted: a linear model ($\delta = 1, ~ \epsilon = 0$, red line), a quadratic model ($\delta = 0, ~ \epsilon = 1$, black line) and a linear and quadratic model ($\delta = \epsilon = 1$, blue line).
In both the figures, the vertical dashed line corresponds to $\gamma p = 1$.

Using (\ref{eqn:condition_lambda_m}) we find
\begin{equation}
	\lambda = l(1- \mathcal{C} )[l(1- \mathcal{C} ) + (1- \mathcal{C} )] = l(l+1)(1- \mathcal{C} )^2~. 
\end{equation}
Notice that, in general, $m$ is not an integer and that two consecutive eigenstates of $L_z$, with eigenvalues $m_1$ and $m_2$, have $m_2 - m_1 = 1- \mathcal{C} $.
We can thus redefine the magnetic quantum number as
\begin{equation}
	m \rightarrow m(1- \mathcal{C} )~.
\end{equation}
This ``new'' $m$ is an integer bounded as follows
\begin{equation}
	-l \leq m \leq l~. \label{eqn:constraint_m}
\end{equation}
Next, for the eigenvalues equations (\ref{eqn:eigenvalues}),
\begin{align}
	L^2|p l m\rangle = & \hbar^2 l(l+1) (1- \mathcal{C} )^2 |p l m\rangle~, & L_z|p l m\rangle = & \hbar m (1- \mathcal{C} ) |p l m\rangle~. \label{eqn:eigenvalues_L2_Lz_final}
\end{align}
Furthermore, for the $L_{\pm}$ we find
\begin{equation}
	L_{\pm}|plm\rangle = \hbar(1- \mathcal{C} )\sqrt{l(l+1) - m(m \pm 1)}|pl(m\pm1)\rangle~. \label{eqn:L+-}
\end{equation}

Finally, we can show that the eigenvalues of $L_z$ and $L^2$ summarized in (\ref{eqn:eigenvalues_L2_Lz_final}) are compatible with the uncertainty relation for angular momentum implied by (\ref{eqn:generalized_commutator_ang_mom}).
First, considering the definition of the operators $L_\pm$ in (\ref{def:L_pm}), we can show that the expectation values for $L_x$ and $L_y$ in an $L^2$ and $L_z$ eigenstate are zero
\begin{equation}
	\langle L_x \rangle  = \langle L_y \rangle = 0~.
\end{equation}
Indeed, since we can reverse (\ref{def:L_pm}) as
\begin{align}
	L_x = & \frac{L_+ + L_-}{2} & L_y = & i \frac{L_- - L_+}{2}~,
\end{align}
then we have
\begin{equation}
	\langle L_x \rangle = \frac{\langle L_+ \rangle + \langle L_- \rangle}{2} = 0~.
\end{equation}
Similarly for $\langle L_y \rangle$.
This implies that the variances of these quantities for the same eigenstates are
\begin{equation}
	(\Delta L_x)^2 = \langle L_x^2 \rangle, \qquad (\Delta L_y)^2 = \langle L_y^2 \rangle~.
\end{equation}
Furthermore, for the second of (\ref{eqn:eigenvalues_L2_Lz_final}), the uncertainty relation for $L_x$ and $L_y$ on a $L_z$ eigenstate with eigenvalue $m$ is
\begin{equation}
	\Delta L_x \Delta L_y = (\Delta L_x)^2 = (\Delta L_y)^2 \geq \frac{|\langle[L_x,L_y]\rangle|}{2} = \frac{\hbar}{2}|\langle L_z \rangle (1 - \langle \mathcal{C} \rangle) |= \frac{\hbar^2}{2} |m| (1 - \langle \mathcal{C} \rangle)^2~,
\end{equation}
where we used the equivalence between $L_x$ and $L_y$ in an $L_z$ eigenstate.
Inserting these last results in the first of (\ref{eqn:eigenvalues_L2_Lz_final}), we obtain
\begin{equation}
	\langle L^2 \rangle = \hbar^2 l(l+1) (1 - \langle \mathcal{C} \rangle)^2 = \langle L_x^2 \rangle + \langle L_y^2 \rangle + \langle L_z^2 \rangle \geq \hbar^2 (m^2 +|m|) (1 - \langle \mathcal{C} \rangle)^2~,
\end{equation}
where the equality holds for $m=\pm l$.
We see that $\langle L_z^2 \rangle$ is bound by $\langle L^2 \rangle$, because of the uncertainty on the other components, as in standard QM.

\subsection{Angular Momentum Spectrum for $|l,m\rangle$} \label{ssec:spectrum_w/p}

So far we considered simultaneous eigenstates of $L^2$, $L_z$, and $p$.
In the standard theory, though, one usually refers to eigenstates of $L^2$ and $L_z$ only.
We will then try to analyze the modified angular momentum spectrum for a state $|l,m\rangle$.

Since we defined these states as eigenstates of $L^2$ and $L_z$, we expect the matrix elements of these operators on $|l,m\rangle$ to be diagonal.
In particular we expect
\begin{equation}
	\langle l', m'| L_z | l , m \rangle = \hbar m \delta_{l',l} \delta_{m',m} (1 + \mathcal{G})~, \label{eqn:L_z_diagonal}
\end{equation}
where $\mathcal{G}$ is a real number containing GUP modifications up to second order in $\gamma$.
Expanding $L_z$ as in (\ref{eqn:ang_mom_henergy_lenergy}), we find
\begin{equation}
	\langle l', m'| L_{0,z} \left[ 1 - \delta \gamma p_0 + (\epsilon + \delta^2) \gamma^2 p_0^2\right] | l , m \rangle = \hbar m' \left[1 - \langle l', m'| \delta \gamma p_0 - (\epsilon + \delta^2) \gamma^2 p_0^2 | l , m \rangle\right]~. \label{eqn:exp_Lz}
\end{equation}
Recalling that in spherical coordinates we have
\begin{align}
	p_0^2 = & - \hbar^2 \nabla^2 = - \frac{\hbar^2}{r^2} \left[ \frac{\partial}{\partial r} \left(r^2 \frac{\partial}{\partial r} \right) + \frac{1}{\sin \theta} \frac{\partial}{\partial \theta} \left(\sin \theta \frac{\partial}{\partial \theta} \right) + \frac{1}{\sin^2 \theta} \frac{\partial^2 }{\partial \phi^2} \right] = \nonumber \\
	= & - \frac{\hbar^2}{r^2} \frac{\partial}{\partial r} \left(r^2 \frac{\partial}{\partial r} \right) + \frac{L^2}{r^2}~, \label{eqn:momentum_spherical}
\end{align}
and considering that a state $|l,m\rangle$ depends only on the angles $\theta$ and $\phi$ when it is described by spherical harmonics, we have
\begin{equation}
	\langle l', m'|p_0^2 | l , m \rangle = \frac{\hbar^2 l (l+1)}{r^2} \delta_{l',l} \delta_{m',m}~.
\end{equation}
In position representation, this is a diagonal matrix.
It is worth noting the position dependence in the previous relation.
This is something that we could have already expected from dimensional analysis, as we pointed out in Sec.~\ref{ssec:QG-mod_ang_mom}.

Going back to our original aim, the only unknown term in \eqref{eqn:exp_Lz} is $\langle l', m'|p_0 | l , m \rangle$, that we will assume to be diagonal as well, \emph{i.e.}, it can be written as a composition of $L^2$, $L_z$, and $r$ in position representation, other than derivatives with respect $r$.
Therefore, we have that the real number $\mathcal{G}$ in (\ref{eqn:L_z_diagonal}) is given by
\begin{multline}
	\mathcal{G} = - \left( \delta \gamma \langle p_0 \rangle - (\epsilon + \delta^2) \gamma^2 \langle p_0^2 \rangle \right) = - \left\{ \delta \gamma \langle p \rangle - \gamma^2 \left[ \epsilon \langle p^2 \rangle + \delta^2 (\Delta p)^2 \right] \right\} = \\
	= - \langle \mathcal{C} \rangle + \gamma^2 \delta^2 (\Delta p)^2 ~.
\end{multline}
Furthermore, notice that our assumption of $p_0$ being diagonal implies that 
\begin{equation}
	\gamma^2 (\Delta p)^2 = \gamma^2 ( \langle p_0^2 \rangle - \langle p_0 \rangle^2) = 0 \Rightarrow \mathcal{G} = - \langle \CE \rangle ~. \label{eqn:mod_is_diag}
\end{equation}

We are then led to considering that the GUP modification appears as an expectation value
\begin{align}
	\langle l', m'| L_z | l , m \rangle = & \hbar m \delta_{l',l} \delta_{m',m} (1 - \langle\mathcal{C}\rangle) & \Leftarrow & & L_z | l , m \rangle = & \hbar m (1 - \langle\mathcal{C}\rangle) | l , m \rangle~. \label{eqn:C_exp_val}
\end{align}
This relation also gives further hint on the diagonality of the matrix element $\langle l', m'|p_0 | l , m \rangle$.
In fact, consider the expectation value of $L_z^2$ on a generic state $|l,m\rangle$ (its matrix element has to be diagonal by definition).
We can compute it following two different strategies.
On one hand, we can expand $L_z$ in low energy operators, obtaining
\begin{equation}
	\langle L_z^2 \rangle = \hbar^2 m^2 \langle 1 - 2 \delta \gamma p_0 + (2 \epsilon + 3 \delta^2) \gamma^2 p_0^2 \rangle 
	= \hbar^2 m^2 \left[1 - 2 \delta \gamma \langle p_0 \rangle + (2 \epsilon + 3 \delta^2) \gamma^2 \langle p_0^2 \rangle \right]~.
\end{equation}
On the other hand, we can use (\ref{eqn:C_exp_val}) and write
\begin{multline}
	\langle L_z^2 \rangle 
	= \langle l,m | L_z \left( \sum_{l',m'} |l',m'\rangle \langle l',m'|\right) L_z | l,m \rangle 
	= \langle L_z \rangle^2 = \\
	= \hbar^2 m^2 (1 - \delta \gamma \langle p_0 \rangle + \gamma^2 (\epsilon + \delta^2) \langle p_0^2 \rangle )^2 = \\
	= \hbar^2 m^2 \left[1 - 2 \delta \gamma \langle p_0 \rangle + \gamma^2 ( \delta^2 \langle p_0 \rangle^2 + 2 (\epsilon + \delta^2) \langle p_0^2 \rangle ) \right]~.
\end{multline}
We then see that, in both cases, if $p_0$ was not diagonal on angular momentum eigenstates, we would have $\langle L_z^2 \rangle \geq \langle L_z \rangle^2$ and in general a nonvanishing uncertainty for $L_z$, contradicting the condition that $|l,m\rangle$ is a $L_z$ eigenstate.
Therefore, while this shows that the matrix elements of $p_0$ are diagonal, at the same time this confirms the results \eqref{eqn:mod_is_diag} and \eqref{eqn:C_exp_val}, that is, GUP modifications are effectively described by expectation values of $\mathcal{C}$.

Furthermore, notice that, when the linear ($\epsilon=0$) or the quadratic ($\delta=0$) models are considered singularly, to first and second order in $\gamma$, respectively, the argument above is unnecessary, since terms like $\gamma^2 \langle p_0 \rangle^2$ would not be present.
In particular, for the quadratic model (as in \cite{Kempf1995_1}), using (\ref{eqn:momentum_spherical}) we have
\begin{align}
	L_z | l , m \rangle = & \hbar m \left[1 + \epsilon \gamma^2 \frac{l (l+1)}{ r^2 } \right] | l , m \rangle~, & 
	L^2 | l , m \rangle = & \hbar^2 l(l+1) \left[1 + 2 \epsilon \gamma^2 \frac{l (l+1)}{ r^2 } \right] | l , m \rangle~.
\end{align}

Also, notice that the sign of $1 - \langle \mathcal{C} \rangle$ depends on the particular GUP model in use.
Indeed, since
\begin{equation}
	(\Delta p)^2 = \langle p^2 \rangle - \langle p \rangle^2 = 0 \qquad \Rightarrow \qquad \langle p^2 \rangle = \langle p \rangle^2 ~ , \label{eqn:p2_p}
\end{equation}
one can write
\begin{equation}
	1 - \langle \mathcal{C} \rangle = 1 - \delta \gamma \langle p \rangle + \epsilon \gamma^2 \langle p \rangle^2 ~ .
\end{equation}
Therefore, $1 - \langle \mathcal{C} \rangle > 0$ for $\delta^2 < 4 \epsilon$.
Models as in \cite{Kempf1995_1} ($\delta=0, ~ \epsilon=1/3$) and in \cite{Ali2011_1} ($\delta = \epsilon = 1$), for example, fulfill this condition.
On the other hand, for models with $\delta^2 > 4 \epsilon$, there may exist intervals of values of $\langle p \rangle$ 
such that $1 - \langle \mathcal{C} \rangle < 0$, signifying that in these intervals the role of $L_+$ and $L_-$ are exchanged, \emph{i.e.} repeated use of $L_+$ ($L_-$) will reduce (increase) the magnetic quantum number $m$.
This is a very interesting feature, since it may imply a threshold for $m$ different from the standard one.

Moreover, notice that in general we have
\begin{equation}
	(L_\pm)^2 |l,m\rangle \not = \hbar^2 (1 - \langle \CE \rangle)^2 \sqrt{l(l+1) - m(m \pm 1)} \sqrt{l(l+1) - (m \pm 1)( m \pm 2)} |l,m \pm 2\rangle~.
\end{equation}
Indeed, considering the assumption stated above, \emph{i.e.}
\begin{equation}
	\CE |l,m\rangle \rightarrow \langle l,m | \CE | l,m \rangle | l,m \rangle \equiv \langle \CE \rangle_{l,m} |l,m\rangle~,
\end{equation}
each application of $L_\pm$ will introduce a different expectation value $\langle \CE \rangle$, obtaining
\begin{multline}
	(L_\pm)^2 |l,m\rangle = \hbar^2 (1 - \langle \CE \rangle_{l,m}) (1 - \langle \CE \rangle_{l,m-1}) \sqrt{l(l+1) - m(m \pm 1)} \times \\
	\times \sqrt{l(l+1) - (m \pm 1)( m \pm 2)} |l,m \pm 2\rangle~.
\end{multline}
In general, we would then have
\begin{equation}
	(L_\pm)^n |l,m\rangle = \\
	= \hbar^n \prod_{i=0}^{n-1} (1 - \langle \CE \rangle_{l,m \pm i}) \sqrt{l(l+1) - (m \pm i)[ m \pm (i + 1)]} |l,m \pm n\rangle~,
\end{equation}
for $n < l \mp m$.
This observation will be crucial in computing Clebsch--Gordan Coefficients in Appendix \ref{apx:CG}.

To conclude this Section, we investigate whether the spherical harmonics are eigenstates, other than of $L^2_0$ and $L_{0,z}$, also of $L^2$ and $L_z$, as we have implicitly assumed so far.
In spherical coordinates, spherical harmonics are the solutions of the following equations
\begin{subequations}
\begin{align}
	L^2_0 Y_{lm} (\theta,\phi) & = - \hbar^2 \left[ \frac{1}{\sin \theta} \frac{\partial}{\partial \theta} \left(\sin \theta \frac{\partial}{\partial \theta} \right) + \frac{1}{\sin^2 \theta} \frac{\partial^2 }{\partial \phi^2} \right] Y_{lm}(\theta,\phi) = \hbar^2 l (l+1) Y_{lm} (\theta,\phi) \\
	L_{0,z} Y_{lm} (\theta,\phi) & = - \hbar \frac{\partial}{\partial \phi} Y_{lm} (\theta,\phi) = \hbar m Y_{lm} (\theta,\phi)~.
\end{align}
\end{subequations}
Considering GUP, though, and expanding the angular momentum using \eqref{eqn:ang_mom_henergy_lenergy}, additional terms proportional to $p_0$ and $p_0^2$ have to be considered.
For simplicity, we will consider the case of a quadratic model.
In this case, the eigenvalue equations read
\begin{subequations} \label{eqns:ang_mom+GUP_eigeneq}
\begin{align}
	L^2 \mathcal{Y}(\vec{r}) & = L^2_0 \left[1 + 2 \gamma^2 \epsilon \left(p_{0,r}^2 + \frac{L_0^2}{r^2} \right) \right] \mathcal{Y}(\vec{r}) = \hbar^2 \lambda \mathcal{Y}(\vec{r}) \\
	L_z \mathcal{Y}(\vec{r}) & = L_{0,z} \left[1 + \gamma^2 \epsilon \left(p_{0,r}^2 + \frac{L_0^2}{r^2} \right) \right] \mathcal{Y}(\vec{r}) = \hbar m \mathcal{Y}(\vec{r})~.
\end{align}
\end{subequations}
Since
\begin{equation}
	p_{0,r}^2 Y_{lm} (\theta,\phi) = 0~,
\end{equation}
we notice that spherical harmonics are solutions of \eqref{eqns:ang_mom+GUP_eigeneq}.
Therefore, they are also eigenfunctions of $L^2$ and $L_z$.

\section{Modified Energy Levels of the Hydrogen Atom} \label{sec:h-atom}

As we pointed out at the beginning of the present Chapter, angular momentum has a fundamental role in the description of atomic systems.
It is then important to apply what we have found in the previous Sections to the case of the hydrogen atom.
In particular, in this Section, we study GUP corrections to the hydrogen atom energy levels and spectra.
We will follow the standard steps to study a system of reduced mass $m$ and subject to a central force with potential energy $U(r) = - e^2/r$, where $e$ is the charge of the electron and where we used electrostatic units.
Let us consider the usual Hamiltonian for a central potential \cite{Messiah}
\begin{equation}
	H \psi(\vec{r}) = \left[ \frac{p_r^2}{2 m} + \frac{L^2}{2 m  r^2} - \frac{e^2}{r} \right] \psi(\vec{r}) ~, \label{eqn:hamiltonian}
\end{equation}
where $p_r$ is understood to be the radial momentum.
Unlike the standard theory, in principle, this problem is not separable since the second term will depend on both angular and radial part of the state.
However, as already mentioned in Sec.~\ref{sec:angular_momentum}, considering an eigenstate of $L^2$ and $L_z$ we are tempted to include GUP effects in terms of expectation values, \emph{i.e.} $\langle \mathcal{C} \rangle = \delta \gamma \langle p \rangle  - \epsilon \gamma^2 \langle p^2 \rangle$.
In this Section, we will extend this observation, assuming that in general GUP modifications appear only through the expectation value $\langle \mathcal{C} \rangle$, \emph{i.e.}
\begin{subequations} \label{eqn:assumption_exp_value}
	\begin{align}
		L^2|lm\rangle &= \hbar^2 l(l+1) (1 - \langle\mathcal{C}\rangle )^2 |lm\rangle~, & L_z|lm\rangle &= \hbar m (1 - \langle\mathcal{C}\rangle ) |lm\rangle~,
	\end{align}
	\begin{equation}
		\vec{p} |lm\rangle = \vec{p}_{0} (1 - \langle\mathcal{C}\rangle ) |lm\rangle~.
	\end{equation}
\end{subequations}
In this way, the correction term $\langle \mathcal{C} \rangle$ represents an average value for the GUP corrections.
This assumption can be further motivated noticing that, for experimental purposes, expectation values are the relevant quantities.
This suffices to estimate Planck scale corrections to observable quantities.
From (\ref{eqn:assumption_exp_value}) we can now see that $\vec{L}$ is simply proportional to $\vec{L}_0$ and $L_i$ and $L^2$ commute with $H$.
For the case of a quadratic model, as we found in the previous Section, this assumption reduces to
\begin{equation}
	L^2 |nlm\rangle = L_0^2 \left[ 1 - 2 \epsilon \langle p_r^2 \rangle - 2 \epsilon \hbar^2 \frac{l (l+1)}{\langle r^2 \rangle} \right] |nlm\rangle ~,
\end{equation}
and the problem of the hydrogen atom is still separable and exactly solvable.

With the assumption (\ref{eqn:assumption_exp_value}), the GUP modified radial part of the Schr\"odinger equation is 
\begin{multline}
	\left[\frac{p_r^2}{2m} + \frac{\hbar^2 l(l+1)}{2m r^2}(1- \langle \mathcal{C} \rangle)^2 - \frac{e^2}{r}\right] y_l(r) = \\
	= \left[\frac{p_{0,r}^2}{2m}(1 - \langle \mathcal{C} \rangle )^2 + \frac{\hbar^2 l(l+1)}{2m r^2}(1- \langle \mathcal{C} \rangle )^2 - \frac{e^2}{r}\right] y_l(r) = Ey_l(r)~, \label{eqn:radial_equation_1}
\end{multline}
where $E$ is the energy eigenvalue.
It is interesting noticing that we found this last equation considering the expansion in low-energy terms of the Hamiltonian \eqref{eqn:hamiltonian} following the prescriptions in \eqref{eqn:expansion_momentum} and \eqref{eqn:ang_mom_henergy_lenergy}.
On the other hand, though, we could have obtained a similar equation using only \eqref{eqn:expansion_momentum} and avoiding separating the radial and the angular coordinates, \emph{i.e.}
\begin{equation}
	H = \frac{p_0^2}{2m}(1 - \langle \CE \rangle)^2 - \frac{e^2}{r}~.
\end{equation}
where ${p_0^2} = p_{0,r}^2 + L^2_{0}/r^2$ is the total linear momentum squared.

In what follows, we will use the following definitions
\begin{align}
	\chi = & \frac{\sqrt{-2mE}}{\hbar}~, &
	a = & \frac{\hbar^2}{m e^2}~, &
	\nu = & \frac{1}{a \chi}~, &
	z = & \frac{2 \chi r}{1 - \langle \mathcal{C} \rangle} ~, & 
	y_l (r) = & z^{(l+1)} e^{-\frac{z}{2}} v(z)~.\label{def:chi-a-nu}
\end{align}
In this way, the radial equation (\ref{eqn:radial_equation_1}) becomes
\begin{equation}
	\left[z\frac{\diff^2}{\diff \, z^2} + (2l + 2 - z)\frac{\diff}{\diff~z} + \left(\frac{\nu}{1 - \langle \mathcal{C} \rangle }- l - 1\right)\right]v(z) = 0~, \label{eqn:laplace}
\end{equation}
which reduces to the well-known Laguerre equation for $\gamma=0$
\begin{equation}
	\left[x\frac{\diff^2}{\diff~x^2} + \left(2l + 2 - x\right)\frac{\diff}{\diff~x} + n'\right]v(x) = 0~.
\end{equation}
With the condition
\begin{equation}
	n' \equiv \frac{\nu}{1 - \langle \mathcal{C} \rangle } - l - 1 \in \mathbb{N}~, \label{eqn:def_and_condition_n'}
\end{equation}
the solutions of (\ref{eqn:laplace}) are the {associated} Laguerre polynomials
\begin{equation}
	L^{(2l + 1)}_{n'} (z) = \sum_{i=0}^{n'} (-1)^i \frac{[(n' + 2l + 1)!]^2}{(n' - i)! (2l + 1 + i)!}\frac{z^i}{i!}~.
\end{equation}
Therefore, the solution for the radial Schr\"odinger's equation (\ref{eqn:radial_equation_1}) is
\begin{equation}
	y_l(r) = z^{(l+1)} e^{-\frac{z}{2}}\frac{n'! (2l + 1)!}{[(n' + 2l + 1)!]^2}L_{n'}^{(2l + 1)}(z) = 
	z^{(l+1)} e^{-\frac{z}{2}}\sum_{i=0}^{n'} (-1)^i \frac{n'! (2l + 1)!}{(n'-i)! (2l + 1 + i)}\frac{z^i}{i!}~.
\end{equation}
and the \emph{generalized principal quantum number} is
\begin{equation}
	n = \nu = \frac{e^2}{\hbar}\sqrt{\frac{m}{-2E}} = (n' + l + 1)(1 - \langle \mathcal{C} \rangle) = n_0 (1 - \langle \mathcal{C} \rangle)~,
\end{equation}
where $n_0$ is the principal quantum number of the standard theory.
In conclusion, the GUP modified energy levels of the hydrogen atom are given by
\begin{align}
	E_n = & - \frac{e^4}{\hbar^2}\frac{m}{2[(n' + l + 1)(1 - \langle \mathcal{C} \rangle)]^2} = -\left(\frac{e^2}{\hbar c}\right)^2 \frac{mc^2}{2n^2} \nonumber \\
	\simeq & {E^{(0)}_n [1 + 2 \delta \gamma \langle p_0 \rangle + \gamma^2 (3 \delta^2 \langle p_0 \rangle^2 - 2 \epsilon \langle p_0^2 \rangle) ]~,} \label{eqn:energy_levels_GUP}
\end{align}
where $E_{n(0)}$ is the corresponding energy level of the standard theory and where $\langle p_0 \rangle$ and $\langle p_0^2 \rangle$ are the expectation values of $p_0$ and $p_0^2$ in a $|n_0 l m\rangle$ eigenstate, respectively.
As before, we recover the results of the standard theory of the hydrogen atom for $\gamma=0$.

From QM, we also know that the energy emitted or absorbed in a transition between two atomic energy levels is simply given by the difference in energy.
Therefore, the frequency and the wavelength of the corresponding absorption or emission line is given by
\begin{equation}
	\frac{1}{\lambda} = \frac{\nu}{c} = \frac{|E_i - E_f|}{h c}~,
\end{equation}
where $E_i$ and $E_f$ are the energy of the initial and final levels, respectively.
In the standard theory, this quantity is proportional to the difference of the inverse squared principal quantum numbers describing the initial and final energy levels.
This was one of the great successes of QM since it exactly reproduced the wavelengths of the already known spectroscopic series.
On the other hand, from (\ref{eqn:energy_levels_GUP}) we have that the energies do not depend simply on the inverse of the principal quantum number.
Therefore, the wavelength of photons emitted when the atom transits from an energy level $E_i$ to $E_f$ changes as follows
\begin{equation}
	\frac{1}{\lambda} = \frac{|E_i - E_f|}{h c} = R_\infty \left|\frac{1}{n_{0,f}^2(1 - \langle \mathcal{C}_f \rangle)^2} - \frac{1}{n_{0,i}^2(1 - \langle \mathcal{C}_i \rangle)^2}\right|~, \label{eqn:h-spectrum_GUP}
\end{equation}
where $R_\infty$ is the Rydberg constant.
We see now that the GUP-corrected spectrum depends not only on the principal quantum number, but also on the expectation values of the electron's momentum, as well as its angular momentum quantum numbers (the latter as we know also happens for the relativistic hydrogen atom).

\subsection{Corrections for a Quadratic Model}

The angular momentum dependence of the energy levels and of the wavelengths of the emitted photons becomes particularly evident in the quadratic model.
In this case we find
\begin{equation}
	E_n \simeq E^{(0)}_n \left\{ 1 - 2 \epsilon \gamma^2 \left[ \langle p_r^2 \rangle + \hbar^2 \left< \frac{1}{r^2} \right> l (l+1) \right] \right\}~.
\end{equation}
Therefore, we will compute here expectation values of the square of the radial momentum and the inverse of the square radius on a hydrogen atom eigenstate.

Recalling that such a state is given by
\begin{equation}
	\psi (x) =  \sqrt{\left(\frac{2}{n a_0}\right)^3 \frac{(n-l-1)!}{2n (n l)!}} e^{- x/2} x^l L^{2l+1}_{n-l-1} (x)~,
\end{equation}
where $x = 2r/n a_0$ and $a_0 = 4 \pi \epsilon_0 \hbar^2 / m e^2$ is the Bohr radius, the expectation value of the squared radial momentum is given by
\begin{multline}
	{\langle p_r^2\rangle =} \frac{\hbar^2 (n-l-1)!}{a_0^2 n^3 (n+l)!} \int_0^\infty \diff x e^{-x/2} x^{2l} \left\{- 2 x^2 L^{2l+1}_{n-l-1} (x) L^{2l+3}_{n-l-3} (x) + \right. \\
	+ 2 (2l+1) x L^{2l+1}_{n-l-1} (x) L^{2l+2}_{n-l-2} (x) 
	- 2 x^2 L^{2l+1}_{n-l-1} (x) L^{2l+2}_{n-l-2} (x) + (2l+1) x \left[L^{2l+1}_{n-l-1} (x)\right]^2 + \\
	\left. - 2 \left[L^{2l+1}_{n-l-1} (x)\right]^2 - \frac{1}{2} x^2 \left[L^{2l+1}_{n-l-1} (x)\right]^2 \right\}~,
\end{multline}
where we used the following property of {the associated} Laguerre polynomials
\begin{equation}
	\frac{\diff^k}{\diff x^k} L_n^\alpha (x) = {\left\{
		\begin{array}{ccc}
			(-1)^k L_{n-k}^{\alpha+k} (x) & \mbox{if} & k \leq n~, \\
			0 & \mbox{if} & k > n~.
		\end{array}
	\right.}
\end{equation}
Considering the additional properties
\begin{subequations}
\begin{align}
	\int_0^\infty \diff x e^{-x} x^\alpha L^\alpha_n(x) L^\alpha_m (x) = & \frac{(n+\alpha)!}{n!} \delta_{nm}~, \\
	L_n^\alpha (x) = & \sum_{i=0}^k (-1)^i \binom{k}{i} L_{n-i}^{\alpha+k} (x)~, \\
	L_{n}^\alpha (x) = & \sum_{i=0}^n \binom{k+n-i-1}{n-i} L_i^{\alpha-k} (x)~,
\end{align}
\end{subequations}
it is easy to prove that the integral of the first two terms vanishes, while for the rest we have
\begin{equation}
	\langle p_r^2 \rangle = \frac{\hbar^2}{a_0^2 n^3} \left[ n - 1 - \frac{2 l^2 (n-l-1)!}{(n+l)!} \sum_{i=0}^{n-l-1} \frac{(2l+i)!}{i!} \right]~. \label{eqn:exp_p_r2}
\end{equation}
As for the last term in parenthesis, notice that we can write
\begin{equation}
	\sum_{i=0}^{n-l-1} \frac{(2l+1)!}{i!} = (2l)! \sum_{i=0}^{n-l-1} \binom{2l + i}{2l} = (2l)! \sum_{i=2l}^{n+l-1} \binom{i}{2l}~.
\end{equation}
Then, using the Hockey-stick identity
\begin{equation}
	\sum_{i=k}^p \binom{i}{k} = \binom{p+1}{k+1}~,
\end{equation}
we finally find
\begin{multline}
	\langle p_r^2 \rangle = \frac{\hbar^2}{a_0^2 n^3} \left[ n - 1 - \frac{2 l^2 (n-l-1)!}{(n+l)!} (2l)! \binom{n+l}{2l+1} \right] = \frac{\hbar^2}{a_0^2 n^3} \left[ n - 1 - \frac{2 l^2 }{2l+1} \right] = \\
	= \frac{\hbar^2}{a_0^2 n^3} \left[ n - \frac{2 l(l+1) + 1 }{2l+1} \right]~.
\end{multline}
Furthermore, one also has
\begin{equation}
	\left<\frac{1}{r^2}\right> = \left( \frac{2}{n a_0} \right)^2 \frac{1}{2n (2l+1)}~.
\end{equation}
Finally, for the corrections to the energy level of the hydrogen atom up to second order in $\gamma$ we find
\begin{equation}
	E_n \simeq E^{(0)}_n \left\{ 1 - 2 \epsilon \gamma^2 \frac{\hbar^2}{a_0^2 n^3} \left[ n - \frac{ 1 }{2l+1} \right] \right\}~.
\end{equation}
It is worth noticing that the degeneracy on the total angular momentum is removed, even though for high {energy levels} the difference in energy between levels with different $l$ is negligible, \emph{i.e.}
\begin{equation}
	E_n \simeq E^{(0)}_n \left[ 1 - 4 \epsilon \gamma^2 \frac{\hbar^2}{a_0^2 n^2} \right]~. \label{eqn:h-atom_GUP_correction}
\end{equation}
%
%

\section{Inclusion of a Magnetic Field} \label{sec:magnetic_field}
Since a magnetic field interacts with the angular momentum of an atom via its magnetic moment, we expect that GUP modifications of the angular momentum theory will result in a modification of this interaction and will have observable consequences.
In what follows we will consider atoms with only one electron in an S shell ($l=0$) and all other levels being filled.
Examples of this kind of atoms are those of the first group of the periodic table and the elements in the group of copper.
This may allow us to test the direct consequences of GUP on angular momenta studying its effects on the electronic spin.

Spin: Our motivation to apply the same modification to spin is the same as in standard QM, wherein spin and orbital angular momentum are treated in the same way, including their algebras. Furthermore, as far as we know, experiments cannot distinguish between spin and orbital angular momentum (for example, a Stern-Gerlach device could not see the difference between orbital and spin angular momenta).

We assume the spin operators to satisfy the same modified algebra for the angular momentum (\ref{eqn:generalized_commutator_ang_mom}) and the same low-energy expansion (\ref{eqn:ang_mom_henergy_lenergy}).
{This is motivated by the fact that in standard QM spin and angular momentum have the same mathematical structure and are treated in similar ways.
Furthermore, experiments are not able to distinguish between spin and angular momentum.
With this assumption in mind, t}he magnetic moment of an electron is
\begin{equation}
	\vec{M} = - \frac{g_S \mu_B}{\hbar}\vec{S}~,
\end{equation}
where 
\begin{equation}
	\mu_B = \frac{e \hbar}{2 m}~,
\end{equation}
is the Bohr magneton, $g_S$ is the electron g-factor and $\vec{S}$ is the spin operator.
Therefore, we set 
\begin{equation}
	\vec{M} = \vec{M}_0[1 - \delta \gamma p_0 + (\epsilon + \delta^2) \gamma^2 p_0^2]~, \label{eqn:mag_mom_henergy_lenergy}
\end{equation}
where
\begin{equation}
	\vec{M}_0 = - \frac{g_S \mu_B}{\hbar}\vec{S}_0~,
\end{equation}
satisfying the standard algebra, is interpreted as the magnetic moment at low energies.

For magnetic fields less than $\sim 10^6$ T \cite{Goswami}, the quadrupole term appearing in the Hamiltonian for the magnetic interaction on an atomic system is negligible with respect the dipole term.
Therefore, we can write the Hamiltonian considering only a term involving the scalar product between the magnetic moment and the magnetic field itself
\begin{equation}
	H = \frac{p^2}{2m} - \vec{M} \cdot \vec{B}~.
\end{equation}
Using (\ref{eqn:expansion_momentum}) and (\ref{eqn:mag_mom_henergy_lenergy}), we can rewrite this relation in terms of low-energy quantities
\begin{equation}
	H = \frac{p^2_0}{2m} \left[1 - 2 \delta \gamma p_0 + (2 \epsilon + 3 \delta^2) \gamma^2 p_0^2\right] - [1 - \delta \gamma p_0 + (\epsilon + \delta^2) \gamma^2 p_0^2] \vec{M}_0\cdot\vec{B}~.
\end{equation}
Looking at the second term, \emph{i.e.} $[1 - \delta \gamma p_0 + (\epsilon + \delta^2) \gamma^2 p_0^2] \vec{M}_0\cdot\vec{B}$, we notice that it acts on both the space and the spin variables through the operators $p_0$ and $\vec{M}_0$ respectively.
As in the previous Section, we replace $p_0$ and $p_0^2$ with $\langle p_0 \rangle$ and $\langle p_0^2 \rangle$, respectively.
In this way, the wavefunction can be factorized in its space and spin parts as follows
\begin{equation}
	\Psi(\vec{r},t) = \psi(\vec{r},t)[\alpha(t) | + \rangle + \beta(t) | - \rangle]~,
\end{equation}
with $\psi$ being the spatial wave function, $\alpha$ and $\beta$ being functions of time such that $|\alpha|^2 + |\beta|^2 = 1$ and $|+\rangle$ and $|-\rangle$ eigenstates of the $z$-component of the magnetic moment operator
\begin{align}
	M_{0,z} |+ \rangle &= \mu_0 |+\rangle~, & M_{0,z} |-\rangle &= - \mu_0 |-\rangle~.
\end{align}
Therefore, the Schr\"odinger equation also splits into
\begin{subequations} \label{eqn:schrödinger_spin}
	\begin{align}
		i\hbar \frac{\partial}{\partial t}\psi(\vec{r},t) &= \frac{p_0^2}{2m}( 1 - \langle \mathcal{C} \rangle )^2\psi(\vec{r},t)~,\\
		i\hbar \frac{\diff}{\diff \, t}(\alpha(t)|+\rangle + \beta(t)|-\rangle) &= - \vec{M}_0\cdot \vec{B}(1-\langle \mathcal{C} \rangle)(\alpha(t) | + \rangle + \beta(t) | - \rangle)~. \label{eqn:se_spin}
	\end{align}
\end{subequations}

We now show how this modification affects the magnetic interaction and how it could in principle be tested.

\subsection{Uniform Magnetic Field}
We have seen that a magnetic field will impose a motion of a system characterized by a magnetic moment $\vec{\mu}$.
In classical mechanics, and for a magnetic field slowly varying on length scales larger than the physical size of the system, this motion corresponds to a precession of the magnetic moment around the magnetic field lines.
Indeed, the magnetic field will generate a torque $\vec{\tau} = \vec{\mu} \times \vec{B}$.
In QM, we have to study the evolution of the state of the system following the Schr\"odinger equation (\ref{eqn:schrödinger_spin}).

Let us consider a uniform magnetic field along the $z$-axis \cite{bransden2000quantum}.
Therefore, for each component of the spinor \eqref{eqn:se_spin} can be written as
\begin{align}
	i \hbar \frac{\diff}{\diff\,t}\alpha(t) & = - \mu_0 B(1- \langle \mathcal{C} \rangle)\alpha(t)~, & i \hbar \frac{\diff}{\diff\,t}\beta(t) &= \mu_0 B(1- \langle \mathcal{C} \rangle)\beta(t)~,
\end{align}
that is, the coefficients will evolve as
\begin{align}
	\alpha(t) = & \alpha(0) e^{ 2 i g_s \mu_B s_+ B (1- \langle \mathcal{C} \rangle) t / \hbar}~, & \beta(t) = & \beta(0) e^{ - 2 i g_s \mu_B s_- B (1- \langle \mathcal{C} \rangle) t / \hbar}~,
\end{align}
where $s_\pm = \pm 1/2$.
From these last relations, we find the modified Larmor frequency of the system
\begin{equation}
	\omega_L = - \frac{2\mu_0B}{\hbar}(1- \langle \mathcal{C} \rangle)~. \label{eqn:larmor_frequency_GUP}
\end{equation}
Furthermore, since the magnitude of the linear momentum is not changed by the magnetic field, we can obtain the same form for the equations of motion as found in the standard theory
\begin{equation}
	\frac{\diff}{\diff\,t}\langle \vec{M} \rangle = \vec{\Omega} \times \langle \vec{M} \rangle~,
\end{equation}
where $\vec{\Omega} = \omega_L \hat{u}_z$.
We, therefore, see that the Larmor frequency is modified by the GUP, although the form of the precession equation remains unchanged.

\subsection{Non-Uniform Magnetic Field: Stern--Gerlach Experiment} \label{subsec:Stern--Gerlach}

The Stern--Gerlach experiment is one of the experiments that laid the foundations for the modern quantum theory \cite{Gerlach1922_1,Bernstein2010_1}.
In this subsection we will study the effects of GUP on this kind of experiment.

Consider a magnetic field with a gradient along the $z$-direction \cite{Basdevant}
\begin{equation}
	\vec{B}(\vec{r}) = B_z(\vec{r}) \hat{u}_z - b'x \hat{u}_x, \qquad \mbox{where} \qquad B_z(\vec{r}) = B_0 + b'z~,
\end{equation}
with $b'\Delta z \ll B_0$, where $\Delta z$ is the width along the $z$-axis of the beam used for the experiment.
The term $-b'x\hat{u}_x$ is necessary to ensure $\vec{\nabla}\cdot\vec{B}=0$.
If the dominant part of the field along $z$ is much more intense that the transverse component over the transverse extension of the wave packet, that is
\begin{equation}
	\langle M_z \rangle B_z \simeq \langle M_z \rangle B_0 \gg \langle M_x \rangle b' \Delta x~,
\end{equation}
then the eigenstates of $-\vec{M}\cdot\vec{B}$ remain practically equal to $|\pm\rangle_z$, since we can use the approximation $\vec{M} \cdot \vec{B} \simeq M_z B_z$, and neglect the transverse component.
For the original Stern--Gerlach experiment \cite{Bretislav2003_1} the following values were used
\begin{align}
	B_0 &\simeq 0.1\mbox{ T}~, & b' &\simeq 1\mbox{ T/mm}~, & \Delta z &\simeq \Delta x \simeq 0.03 \mbox{ mm}~,
\end{align}
that is
\begin{equation}
	\frac{B_0}{b' \Delta x} \simeq 3.3~.
\end{equation}
With this assumption, the Schr\"odinger's equation for the two spin components are
\begin{equation}
	i \hbar \frac{\partial}{\partial t} \psi_\pm = \left[\frac{p_0^2}{2m} ( 1 - \langle \mathcal{C} \rangle )^2 \mp \mu_0(B_0 + b' z)(1- \langle \mathcal{C} \rangle )\right] \psi_\pm~.
\end{equation}

The expectation values of position and momentum are given by
\begin{subequations}
	\begin{align}
		\langle \vec{r}_\pm\rangle &= \frac{\int \vec{r} |\psi_\pm(\vec{r},t)|^2 \mbox{d}^3r}{\int |\psi_\pm(\vec{r},t)|^2 \mbox{d}^3r}~,\\
		\langle \vec{p}_\pm\rangle &= \frac{\int \psi_\pm^*(\vec{r},t) \vec{p}\psi_\pm(\vec{r},t) \mbox{d}^3r}{\int |\psi_\pm(\vec{r},t)|^2 \mbox{d}^3r} = \langle \vec{p}_{0\pm}\rangle (1 - \langle \mathcal{C} \rangle)~.
	\end{align}
\end{subequations}
From the Ehrenfest Theorem, we can obtain the time evolution of these quantities
\begin{subequations}
	\begin{align}
		\frac{\mbox{d}}{\mbox{d} t}\langle \vec{r}_\pm\rangle &= \frac{\langle \vec{p}_\pm\rangle}{m} = \frac{\langle \vec{p}_{0\pm}\rangle}{m}(1 - \langle \mathcal{C} \rangle)~,\\
		\frac{\mbox{d}}{\mbox{d} t} \langle p_{x\pm}\rangle &= - \left\langle \mp \frac{\partial}{\partial x}\mu_0 B (1- \langle \mathcal{C} \rangle)\right\rangle = 0~,\\
		\frac{\mbox{d}}{\mbox{d} t} \langle p_{y\pm}\rangle &= - \left\langle \mp \frac{\partial}{\partial y}\mu_0 B (1-\langle \mathcal{C} \rangle)\right\rangle = 0~,\\
		\frac{\mbox{d}}{\mbox{d} t} \langle p_{z\pm}\rangle &= - \left\langle \mp \frac{\partial}{\partial z}\mu_0 B (1- \langle \mathcal{C} \rangle)\right\rangle = \pm \mu_0 b' (1- \langle \mathcal{C} \rangle)~.
	\end{align}
\end{subequations}
We assume that at $t=0$ a beam of atoms enters the apparatus, which we assume to be at the origin of the coordinate system.
We also assume that the initial momentum is directed along the $y$-axis, without any component along the $x$ or $z$-axis, that is
\begin{subequations}
	\begin{align}
		\langle x_\pm \rangle(0) = \langle y_\pm \rangle(0) = \langle z_\pm \rangle(0) &= 0~,\\
		\langle p_{x\pm}\rangle(0) = \langle p_{z\pm}\rangle(0) &= 0~,\\
		\langle p_{y\pm}\rangle(0) &= mv~,
	\end{align}
\end{subequations}
We then obtain the following equations for its motion
\begin{align}
	\langle x_\pm \rangle(t) &= 0~, & \langle y_\pm \rangle(t) &= vt~, & \langle z_\pm \rangle(t) &= \pm \frac{\mu_0 b' t^2}{2m}(1- \langle \mathcal{C} \rangle)~.
\end{align}
As for the standard theory, these equations represent a beam splitting in two along the $z$-axis.
In this case though, the term $\langle \mathcal{C} \rangle$ will impose a dependence of the splitting on the expectation values of the momentum and the momentum squared of the electron in the outer S shell, the splitting being
\begin{equation}
	\delta z = \frac{\mu_0 b'}{m}\frac{L^2}{v^2}(1- \langle \mathcal{C} \rangle)~, \label{eqn:separation}
\end{equation}
where $L$ is the length of the apparatus.

Again, for the Stern--Gerlach apparatus described in \cite{Bretislav2003_1}, one has for an electron in the 5S state of silver atoms
\begin{align}
		\langle p^2_0 \rangle &= 2.83\times10^{-26}\mbox{ N}^2\mbox{s}^2~, & \langle p_0 \rangle &= 0\mbox{ Ns}~.
	\end{align}
We then have that the ratio between the expected splitting in GUP and the splitting in the standard theory is
\begin{equation}
	\frac{\delta z_{\mathrm{GUP}}}{\delta z_0} - 1 = - \langle \mathcal{C} \rangle = \frac{\gamma_0^2 \langle p^2_0 \rangle}{(M_{\mathrm{Pl}} c)^2} \simeq \gamma_0^2 6.66 \times 10^{-28}~. \label{eqn:cor_Stern--Gerlach}
\end{equation}
More generally, considering a model with $\delta^2 = 0$, we have
\begin{equation}
	\langle \CE \rangle = \gamma_0^2 \epsilon \frac{G \hbar}{a_0^2 c^3 n^3} (n - 1)~.
\end{equation}
For splittings of the order of that achieved in the original experiment ($\sim 0.2$ mm), the difference between the GUP and the standard cases would not be observable.
But if larger splittings could be produced, for example with a longer apparatus or lower velocities of the atoms in the beam, a better resolution could be achieved.
On the other hand, one could also use atoms other than silver.
In the model used to describe the Stern--Gerlach experiment, though, a variation of the mass leads to two contrasting effects.
Consider, for example, the case of a higher mass than that of a silver atom.
Since the inverse of the mass appears in (\ref{eqn:separation}), this will reduce the separation of the two spots.
On the other hand, higher atomic numbers lead to higher momenta for the external electrons, and hence to higher $|\langle \mathcal{C} \rangle|$.
The two effects will thus compete with each other.

\section{Multi-Particles Systems}\label{sec:multi-particles}

In many theories of QG, some problems arise from the modification of the momentum composition law.
As we saw for example in Sec.~\ref{sec:DSR}, the soccer ball problem is the result of such modification.
When the composition law presents terms that are not linear in the constituent momenta, depending on its form, we would face enhancement or suppression of these terms with the number of constituents.
In this Section, we will examine how GUP affects multiparticle angular momentum algebra.

\subsection{Dependence of $[L_i,L_j]$ on the number of particles}

Consider a system of $N$ particles with angular momentum $\vec{l}_n$, with $n = 1, \ldots,N$, and total angular momentum
\begin{equation}
	\vec{L} = \sum_{n=1}^N \vec{l}_n~.
\end{equation}
The commutator between components of the total angular momentum is
\begin{multline}
	[L_i,L_j] = \sum_{n=1}^N \sum_{m=1}^N[l_{i,n},l_{j,m}] = \sum_{n=1}^N[l_{i,n},l_{j,n}] = \sum_{n=1}^N i \hbar \epsilon_{ijk} l_{k,n} (1 - \delta \gamma p_n + \epsilon \gamma^2 p_n^2) \\
	= i \hbar \epsilon_{ijk} [L_k - \sum_{n=1}^N l_{k,n} ( \delta \gamma p_n - \epsilon \gamma^2 p_n^2)] =\\
	= i \hbar \epsilon_{ijk} \left[L_k (1 - \delta \gamma P + \epsilon \gamma^2 P^2) + \qquad \qquad \qquad \qquad \qquad \right. \\
	\left. - \delta \gamma \sum_{n=1}^N \left(l_{k,n} p_n - \frac{L_k P}{N}\right) + \epsilon \gamma^2 \sum_{n=1}^N\left(l_{k,n} p_n^2 - \frac{L_k P^2}{N}\right)\right]~, \label{eqn:commutator_multi_particles}
\end{multline}
where
\begin{equation}
	P^2 = \sum_{n=1}^N \left[ p_n^2 + 2 \sum_{m>n}^{N} \vec{p}_n \cdot \vec{p}_m \right]~,
\end{equation}
and where we assumed
\begin{equation}
	[l_{i,m},l_{j,n}] = 0~, \qquad m \not = n~, \label{eqn:different_part_commutes}
\end{equation}
\emph{i.e.} the angular momentum components of different particles commute.

For a better interpretation of \eqref{eqn:commutator_multi_particles}, consider for example the case in which all the particles in the system have the same angular momentum, \emph{e.g.} particle in a rotating ring,
\begin{equation}
	l_{n,k} = \frac{L_k}{N}~, \qquad n = 1, \ldots, N~,
\end{equation}
in which case from (\ref{eqn:commutator_multi_particles}), one obtains
\begin{equation}
	[L_i,L_j] = i \hbar \epsilon_{ijk} L_k \left[1 - \delta \gamma P + \epsilon \gamma^2 P^2 - \delta \frac{\gamma}{N} \sum_{n=1}^N \left(p_n - P\right) + \epsilon \frac{\gamma^2}{N} \sum_{n=1}^N\left(p_n^2 - P^2\right)\right]~. \label{eqn:ang_mom_com_rel_same_ang_mom}
\end{equation}
Furthermore, when a quadratic model is considered, we find
\begin{multline}
	[L_i,L_j] = i \hbar \epsilon_{ijk} L_k \left\{ 1 + \epsilon \frac{\gamma^2}{N^2} P^2 + \epsilon \gamma^2 \sum_{m=1}^3  \left[ \frac{1}{N} \sum_{n=1}^N p_{m,n}^2 - \left( \frac{1}{N} \sum_{n=1}^N p_{m,n} \right)^2\right] \right\} = \\
	= i \hbar \epsilon_{ijk} L_k \left\{ 1 + \epsilon \frac{\gamma^2}{N^2} P^2 + \epsilon \gamma^2 \sum_{m=1}^3  \left[ \langle p_{m}^2 \rangle - \langle p_{m} \rangle^2\right] \right\} = \\
	= i \hbar \epsilon_{ijk} L_k \left\{ 1 + \epsilon \frac{\gamma^2}{N^2} P^2 + \epsilon \gamma^2 \sum_{m=1}^3  \left( \Delta p_m \right)^2 \right\}~,\label{eqn:ang_mom_com_rel_same_ang_mom_q}
\end{multline}
where $\Delta p_m$ is the standard deviation of the $m$-th component of the constituents' momentum.

As a second example, we consider particles with the same linear momentum, \emph{e.g.}, a rigid body in pure translation, for which
\begin{equation}
	p_n = \frac{P}{N}~, \qquad n = 1,\ldots,N~,
\end{equation}
in which case we find
\begin{multline}
	[L_i,L_j] = i \hbar \epsilon_{ijk} \left[L_k (1 - \delta \gamma P + \epsilon \gamma^2 P^2) - \delta \frac{\gamma}{N} P\sum_{n=1}^N \left(l_{k,n} - L_k\right) + \right. \\
	\left. + \epsilon \frac{\gamma^2}{N^2} P^2 \sum_{n=1}^N\left(l_{k,n} - N L_k\right)\right] = \\
	= i \hbar \epsilon_{ijk} L_k \left(1 - \delta \frac{\gamma}{N} P + \epsilon \frac{\gamma^2}{N^2} P^2\right)~. \label{eqn:ang_mom_com_rel_same_lin_mom}
\end{multline}
Note that the RHS of both \eqref{eqn:ang_mom_com_rel_same_ang_mom} and \eqref{eqn:ang_mom_com_rel_same_lin_mom} scales as powers of $N$, and therefore, for large $N$ GUP effects become negligible.
Nonetheless, \eqref{eqn:ang_mom_com_rel_same_ang_mom_q} gives the possibility of testing GUP for a composite system, because of the presence of the term $\gamma^2 \sum_{m=1}^3  \left( \Delta p_m \right)^2$ related to the dispersion of the system in momentum space.

Notice that in deriving these results we assumed that momenta and angular momenta add following the usual rules of linear algebra, that is we did not consider here any possible modification of the addition laws for linear and angular momentum vectors.
However, it is worth noticing that the enhancing effect of the soccer ball problem and the scaling that we found here could in principle oppose and cancel each other out.

\subsection{Addition of Angular Momentum}

In this Subsection we examine the problem of addition of angular momentum including GUP.

Consider a system composed of $N$ particles, with $l_n$ and $m_n$ the azimuthal and magnetic quantum numbers of the particles, with $n = 1, \ldots, N$.
From (\ref{eqn:eigenvalues_L2_Lz_final}) we have
\begin{subequations}
	\begin{align}
	l_n^2 |l_n,m_n\rangle &= \hbar^2 (1- \langle \mathcal{C}_n \rangle)^2l_n(l_n+1) |l_n,m_n\rangle~, \\
	l_{n,z} |l_n,m_n\rangle &= \hbar (1- \langle \mathcal{C}_n \rangle )m_n|l_n,m_n\rangle~.
	\end{align}
\end{subequations}
The $z$-component of the angular momentum for the composite system is
\begin{equation}
	L_z = \sum_{n=1}^N l_{n,z}~.
\end{equation}
Operating on the combined state
\begin{equation}
	|\{l_n,m_n\}\rangle  = \bigotimes_{n=1}^N |l_n,m_n\rangle~,
\end{equation}
we obtain
\begin{equation}
	L_z|\{l_n,m_n\}\rangle = \sum_{n=1}^N l_{n,z} |\{l_n,m_n\}\rangle = \sum_{n=1}^N\hbar(1- \langle \mathcal{C}_n \rangle)m_n|\{l_n,m_n\}\rangle~. \label{eqn:lz_composite_system}
\end{equation}
For the eigenvalue of $L_z$ in (\ref{eqn:lz_composite_system}), we thus have
\begin{multline}
	\sum_{n=1}^N\hbar(1- \langle \mathcal{C}_n \rangle )m_n = \sum_{n=1}^N\hbar m_n (1 - \delta \gamma \langle p_n \rangle + \epsilon \gamma^2 \langle p_n^2 \rangle) = \\
	= \hbar \left[ M (1 - \delta \gamma \langle P \rangle + \epsilon \gamma^2 \langle P^2 \rangle) - \delta \gamma \sum_{n=1}^N \left(m_n \langle p_n \rangle - \frac{M \langle P \rangle }{N}\right) + \right. \\
	\left. + \epsilon \gamma^2 \sum_{n=1}^N \left(m_n \langle p_n^2 \rangle - \frac{M \langle P^2 \rangle}{N}\right)\right]~,
\end{multline}
where $M = \sum m_n$.
Furthermore, for (\ref{eqn:constraint_m}) we have
\begin{equation}
	M_{\mathrm{min}} \equiv - \sum_{n=1}^N l_n \leq M \leq \sum_{n=1}^N l_n \equiv M_{\mathrm{max}}~,
\end{equation}
while, since the following inequalities hold
\begin{equation}
	-\sum_{n=1}^N l_n (1- \langle \mathcal{C}_n \rangle ) \leq \sum_{n=1}^N m_n (1- \langle \mathcal{C}_n \rangle ) \leq \sum_{n=1}^N l_n (1- \langle \mathcal{C}_n \rangle)~,
\end{equation}
we obtain for the eigenvalue of $L_z$
\begin{multline}
	M (1 - \langle \bar{\mathcal{C}} \rangle ) - \delta \gamma \sum_{n=1}^N \left(m_n \langle p_n \rangle - \frac{M \langle P \rangle }{N}\right) + \epsilon \gamma^2 \sum_{n=1}^N \left(m_n \langle p_n^2 \rangle - \frac{M \langle P^2\rangle }{N}\right) \geq \\
	\geq M_{\mathrm{min}} (1 - \langle \bar{\mathcal{C}}\rangle ) + \delta \gamma \sum_{n=1}^N \left(l_n \langle p_n \rangle + \frac{M_{\mathrm{min}} \langle P \rangle }{N}\right) - \epsilon \gamma^2 \sum_{n=1}^N \left(l_n \langle p_n^2 \rangle + \frac{M_\mathrm{min} \langle P^2 \rangle}{N}\right)
\end{multline}
and
\begin{multline}
	M (1 - \langle \bar{\mathcal{C}} \rangle ) - \delta \gamma \sum_{n=1}^N \left(m_n \langle p_n \rangle - \frac{M \langle P \rangle }{N}\right) + \epsilon \gamma^2 \sum_{n=1}^N \left(m_n \langle p_n^2 \rangle - \frac{M \langle P^2 \rangle }{N}\right) \leq \\
	\leq M_{\mathrm{max}} (1 - \langle \bar{\mathcal{C}} \rangle ) - \delta \gamma \sum_{n=1}^N \left(l_n \langle p_n \rangle- \frac{M_{\mathrm{max}} \langle P \rangle}{N}\right) + \epsilon \gamma^2 \sum_{n=1}^N \left(l_n \langle p_n^2 \rangle - \frac{M_\mathrm{max} \langle P^2 \rangle}{N}\right)~,
\end{multline}
where $\langle \bar{\mathcal{C}} \rangle = \delta \gamma \langle P \rangle - \epsilon \gamma^2 \langle P^2 \rangle $.

If $L$ is the azimuthal quantum number for the complete system, since $|M|\leq L$, we find that $L$ has the value
\begin{equation}
	L = \sum_{n=1}^N l_n~. \label{eqn:total_azimuthal_qn}
\end{equation}
Higher values of $L$ are not allowed since they would imply $M > M_\mathrm{max}$.
Thus
\begin{equation}
	L_\mathrm{max} = \sum_{n=1}^N l_n
\end{equation}
or, for the case of two particles, useful for the next Section,
\begin{equation}
	L_\mathrm{max} = l_1 + l_2~.\label{eqn:max_azimuthal_qn2}
\end{equation}
Note that the results (\ref{eqn:total_azimuthal_qn} - \ref{eqn:max_azimuthal_qn2}) correspond to those in standard QM.
Therefore, following similar reasoning one gets
\begin{equation}
	|l_1 - l_2| \leq L \leq l_1 + l_2~.
\end{equation}

Next, we define the ladder operators for the combined system
\begin{equation}
	L_{\pm} = \sum_{n=1}^N l_{n,\pm}~.
\end{equation}
Then it follows from (\ref{eqn:commutator_Lz_Lpm}) that 
\begin{equation}
	[l_{n,z},l_{m,\pm}] = \pm \delta_{nm} \hbar l_{n,\pm} ( 1 - \langle \mathcal{C}_n \rangle)~,
\end{equation}
from which one gets
\begin{multline}
	L_z L_\pm |l_1,m_1;\ldots;l_N,m_N\rangle = [L_\pm L_z \pm \hbar \sum_{n=1}^N l_{n,\pm} (1- \langle \mathcal{C}_n \rangle )] |l_1,m_1;\ldots;l_N,m_N\rangle = \\
	= \hbar \sum_{n=1}^N [M \pm (1- \langle \mathcal{C}_n \rangle)] l_{n,\pm} |l_1,m_1;\ldots;l_N,m_N\rangle~.
\end{multline}
Note that the RHS is no longer an eigenstate of $L_z$, unlike the $\gamma \rightarrow 0$ case.
Also, we get
\begin{equation}
	[L_+,L_-] = -i[L_x,L_y] +i[L_y,L_x] = -2i[L_x,L_y] = 2\hbar\sum_{n=1}^N l_{n,z}(1- \langle \mathcal{C}_n \rangle)~,
\end{equation}
where we have used (\ref{eqn:different_part_commutes}) and (\ref{eqn:generalized_commutator_ang_mom}).
Notice that this last commutator cannot be written in terms of the total angular momentum operator $L_z$ alone.

Next, we specialize to the case of two angular momenta (\emph{i.e.} $N=2$)
\begin{equation}
	\begin{split}
		L_z L_\pm |l_1,m_1;l_2,m_2\rangle &= \hbar \{ [M \pm (1- \langle \mathcal{C}_1 \rangle)]l_{1,\pm} + [M \pm (1- \langle \mathcal{C}_2 \rangle)]l_{2,\pm} \}|l_1,m_1;l_2,m_2\rangle = \\
		&= \hbar \{ [M \pm (1- \langle \mathcal{C}_1 \rangle)]L_{\pm} \pm (\langle \mathcal{C}_1 \rangle - \langle \mathcal{C}_2 \rangle)l_{2,\pm} \}|l_1,m_1;l_2,m_2\rangle = \\
		&= \hbar \{ [M \pm (1- \langle \mathcal{C}_2 \rangle)]L_{\pm} \pm (\langle \mathcal{C}_2 \rangle -  \langle \mathcal{C}_1 \rangle)l_{1,\pm} \}|l_1,m_1;l_2,m_2\rangle~.
	\end{split}
\end{equation}
It is worth noticing that these equivalent results show that not only we can obtain the results of standard QM by taking $\gamma=0$, but also when $\langle \mathcal{C}_1 \rangle = \langle \mathcal{C}_2 \rangle$.
We will find that this feature persists for the remainder of the Section.

\subsection{Clebsch--Gordan Coefficients}\label{subsec:CG-coefficients}

As in standard QM, the following commutation relations still hold ($n=1,2$)
\begin{align}
	[L_i,l_n^2] & = 0 = [L^2,l_n^2] & [L_z,l_{n,z}] & = 0~,
\end{align}
but in general
\begin{equation}
	[L^2,l_{n,z}]\not=0~.
\end{equation}
This means that $\{l_1^2,l_2^2,l_{1,z},l_{2,z}\}$ and $\{l_1^2,l_2^2,L^2,L_z\}$ form complete sets of observables also considering GUP.
Since both the systems $\{|l_1,m_1;l_2,m_2 \rangle\}$ and $\{|l_1,l_2,L,M\rangle\}$ form complete sets of eigenstates, we can use the completeness relations
\begin{align}
	\sum_{m_1,m_2}|l_1,m_1;l_2,m_2 \rangle \langle l_1,m_1;l_2,m_2| =& 1~, & \sum_{M,L} |l_1,l_2,L,M\rangle \langle l_1,l_2,L,M| =& 1~,
\end{align}
and write
\begin{subequations}
	\begin{align}
		|l_1,m_1;l_2,m_2 \rangle &= \sum_{M,L} |l_1,l_2,L,M\rangle \langle l_1,l_2,L,M|l_1,m_1;l_2,m_2 \rangle~,\label{eqn:definition_CG}\\
		|l_1,l_2,L,M\rangle &= \sum_{m_1,m_2}|l_1,m_1;l_2,m_2 \rangle \langle l_1,m_1;l_2,m_2|l_1,l_2,L,M\rangle~, \label{eqn:definition_CG_conjugate}
	\end{align}
\end{subequations}
where
\begin{equation}
	\langle l_1,m_1;l_2,m_2|l_1,l_2,L,M\rangle = \langle l_1,l_2,L,M|l_1,m_1;l_2,m_2 \rangle^*
\end{equation}
are the Clebsch--Gordan (CG) coefficients.

From here on, when the azimuthal quantum numbers of the single constituents $l_1$ and $l_2$ are obvious, we will use a simpler notation
\begin{align}
	|l_1,m_1;l_2,m_2 \rangle & \equiv |m_1;m_2\rangle~, & |l_1,l_2,L,M\rangle & \equiv |L,M\rangle~.
\end{align}

\subsubsection{Orthogonality relations}
Using the definition of CG coefficients in (\ref{eqn:definition_CG}) and (\ref{eqn:definition_CG_conjugate}) we can find the following orthogonality relations
\begin{subequations}
	\begin{align}
		\sum_{m1,m2} \langle l_1,l_2,L',M'|l_1,m_1;l_2,m_2 \rangle \langle l_1,m_1;l_2,m_2|l_1,l_2,L,M\rangle &= \delta_{M',M}\delta_{L',L}~, \label{eqn:orthogonality_relation_total}\\
		\sum_{M,L} \langle l_1,m_1';l_2,m_2'|l_1,l_2,L,M\rangle \langle l_1,l_2,L,M|l_1,m_1;l_2,m_2 \rangle &=\delta_{m_1',m_1}\delta_{m_2',m_2}~,\label{eqn:orthogonality_relation_parts}
	\end{align}
\end{subequations}
identical to those in standard QM.

\subsubsection{Clebsch--Gordan Recursion Relation}
From (\ref{eqn:L+-}) we have,
\begin{multline}
	L_\pm|L,M\rangle = \hbar(1- \langle \bar{\mathcal{C}} \rangle )\sqrt{L(L+1) - M(M\pm1)}|L,M\pm1\rangle = \\
	= \hbar(1- \langle \bar{\mathcal{C}} \rangle )\sqrt{L(L+1) - M(M\pm1)} \sum_{m_1,m_2}|m_1;m_2\rangle\langle m_1;m_2|L,M\pm1\rangle~,\label{eqn:rec_rel_step1}
\end{multline}
where $\langle P \rangle $ and $\langle P^2 \rangle$ are expectation values of the momentum and the momentum squared of the combined system.
On the other hand, since $L_\pm = l_{1,\pm} + l_{2,\pm}$, we have
\begin{multline}
	L_\pm|L,M\rangle = L_\pm\sum_{m_1,m_2}|m_1;m_2\rangle\langle m_1;m_2|L,M\rangle = \\
	= \hbar \sum_{m_1,m_2} \left[(1- \langle \mathcal{C}_1 \rangle )\sqrt{l_1(l_1+1)-m_1(m_1\pm1)}|m_1 \pm 1;m_2\rangle + \right. \\
	\left. + (1- \langle \mathcal{C}_2 \rangle )\sqrt{l_2(l_2+1)-m_2(m_2\pm1)}|m_1;m_2\pm1\rangle \right] \langle m_1;m_2|L,M\rangle = \\
	= \hbar \sum_{m_1,m_2}|m_1,m_2\rangle[(1- \langle \mathcal{C}_1 \rangle )\sqrt{l_1(l_1+1)-m_1(m_1\mp1)} \langle m_1\mp1;m_2|L,M\rangle + \\
	+ (1- \langle \mathcal{C}_2 \rangle )\sqrt{l_2(l_2+1)-m_2(m_2\mp1)}\langle m_1;m_2\mp1|L,M\rangle]~. \label{eqn:rec_rel_step2}
\end{multline}
Equating the RHS of (\ref{eqn:rec_rel_step1}) and (\ref{eqn:rec_rel_step2}) we get
\begin{multline}
	(1- \langle \bar{\mathcal{C}} \rangle )\sqrt{L(L+1) - M(M\pm1)}\langle m_1;m_2|L M\pm1\rangle = \\
	= (1- \langle \mathcal{C}_1 \rangle )\sqrt{l_1(l_1+1)-m_1(m_1\mp1)} \langle m_1\mp1;m_2|L,M\rangle + \\
	+ (1- \langle \mathcal{C}_2 \rangle )\sqrt{l_2(l_2+1)-m_2(m_2\mp1)}\langle m_1;m_2\mp1|L,M\rangle~, \label{eqn:CG_recursive_relation}
\end{multline}
which reduces to the standard result if $\gamma = 0$, \emph{i.e.}, $\langle \bar{\mathcal{C}} \rangle=\langle \mathcal{C}_1 \rangle=\langle \mathcal{C}_2 \rangle=0$.

\subsubsection{Clebsch--Gordan Coefficient Tables} \label{sssec:CG_tables}

In this Section, using the results in Appendix \ref{apx:CG} we provide explicit expressions of  CG coefficients for some simple cases including GUP.
In the following tables, we apply the \mbox{Condon--Shortley} phase convention.

\vspace{2em}

{\large $\displaystyle{\mathbf{l_1 = \frac{1}{2},~ l_2=\frac{1}{2}}}$}

\vspace{1em}

$M=1$

\begin{table}[h]
\begin{tabular}{rc|c|}
	\cline{3-3}
	$L=$ && 1 \\
	\cline{2-3}
	\multicolumn{1}{c|}{$m_1,~m_2=$}&$\frac{1}{2},~\frac{1}{2}$ & 1 \\
	\cline{2-3}
\end{tabular}
\caption{Clebsch--Gordan coefficients for $M=1, ~ l_1 = \frac{1}{2}, ~ l_2=\frac{1}{2}$}
\end{table}

\vspace{1em}

$M=0$

\begin{table}[h]
\begin{tabular}{rc|c|c|}
	\cline{3-4}
	$L=$ && 1 & 0\\
	\cline{2-4}
	\multicolumn{1}{c|}{\multirow{3}{*}{$m_1,~m_2=$}}
	&$\frac{1}{2},~ - \frac{1}{2}$
	& $\displaystyle{\frac{1}{\sqrt{2}}}$
	& $ \displaystyle{\frac{1}{\sqrt{2}}}$ \\
	\cline{2-4}
	&\multicolumn{1}{|c|}{$-\frac{1}{2},~\frac{1}{2}$}
	&$\displaystyle{\frac{1}{\sqrt{2}}}$
	&$\displaystyle{- \frac{1}{\sqrt{2}}}$ \\
	\cline{2-4}
\end{tabular}
\caption{Clebsch--Gordan coefficients for $M = 0, ~ l_1 = \frac{1}{2}, ~ l_2=\frac{1}{2}$}
\end{table}

\vspace{2em}

{\large $\displaystyle{\mathbf{l_1 = 1, ~ l_2=\frac{1}{2}}}$}

\vspace{1em}

$M=\frac{3}{2}$

\begin{table}[h]
\begin{tabular}{rc|c|}
	\cline{3-3}
	$L=$ && $\frac{3}{2}$\\
	\cline{2-3}
	\multicolumn{1}{c|}{\multirow{1}{*}{$m_1,~ m_2=$}}&$1,~ \frac{1}{2}$ & 1\\
	\cline{2-3}
\end{tabular}
\caption{Clebsch--Gordan coefficients for $M = \frac{3}{2}, ~ l_1 = 1, ~ l_2=\frac{1}{2}$}
\end{table}

\clearpage

$M=\frac{1}{2}$

\begin{table}[h]
\begin{tabular}{rc|c|c|}
	\cline{3-4}
	$L=$ && $\frac{3}{2}$ & $\frac{1}{2}$\\
	\cline{2-4}
	\multicolumn{1}{c|}{\multirow{3}{*}{$m_1,~m_2=$}}
	& $1,~ -\frac{1}{2}$ 
	& $\displaystyle{\frac{1- \langle \mathcal{C} \rangle_{\frac{1}{2},\frac{1}{2}}}{\sqrt{ \mathcal{S}_1}}}$ 
	& $ \displaystyle{\frac{\sqrt{2}(1- \langle \mathcal{C} \rangle_{1,1} )}{\sqrt{ \mathcal{S}_1}}}$ \\
	\cline{2-4}
	&\multicolumn{1}{|c|}{$0,~\frac{1}{2}$}
	& $\displaystyle{\frac{\sqrt{2}(1- \langle \mathcal{C} \rangle_{1,1} )}{\sqrt{ \mathcal{S}_1}}}$
	&$\displaystyle{- \frac{1- \langle \mathcal{C} \rangle_{\frac{1}{2},\frac{1}{2}}}{\sqrt{ \mathcal{S}_1}}}$ \\
	\cline{2-4}
\end{tabular}
\caption{Clebsch--Gordan coefficients for $M = \frac{1}{2}, ~ l_1 = 1, ~ l_2=\frac{1}{2}$}
\end{table}
with $$\mathcal{S}_1 = 2 (1- \langle \mathcal{C} \rangle_{1,1} )^2 + (1- \langle \mathcal{C} \rangle_{\frac{1}{2},\frac{1}{2}} )^2~.$$

\vspace{1em}

$M=-\frac{1}{2}$

\begin{table}[h]
\begin{tabular}{rc|c|c|}
	\cline{3-4}
	$L=$ && $\frac{3}{2}$ & $\frac{1}{2}$\\
	\cline{2-4}
	\multicolumn{1}{c|}{\multirow{3}{*}{$m_1,~ m_2=$}}
	& $0,~ - \frac{1}{2}$ 
	& $\displaystyle{\frac{\sqrt{2}(1- \langle \mathcal{C} \rangle_{\frac{1}{2},\frac{1}{2}} )}{\sqrt{\mathcal{S}_2}}}$ 
	& $ \displaystyle{\frac{(1- \langle \mathcal{C} \rangle_{1,1} )^2 2 - (1- \langle \mathcal{C} \rangle_{\frac{1}{2},\frac{1}{2}} )^2 }{\sqrt{ \mathcal{S}_3 }}}$ \\
	\cline{2-4}
	&\multicolumn{1}{|c|}{$-1, ~ \frac{1}{2}$}
	& $\displaystyle{\frac{(1- \langle \mathcal{C} \rangle_{1,0} ) }{\sqrt{\mathcal{S}_2}}}$
	& $\displaystyle{- \frac{\sqrt{2} (1- \langle \mathcal{C} \rangle_{1,0} ) (1- \langle \mathcal{C} \rangle_{\frac{1}{2},\frac{1}{2}} )}{\sqrt{ \mathcal{S}_3 }}}$ \\
	\cline{2-4}
\end{tabular}
\caption{Clebsch--Gordan coefficients for $M = -\frac{1}{2}, ~ l_1 = 1, ~ l_2=\frac{1}{2}$}
\end{table}
with $$\mathcal{S}_2 = (1 - \langle \CE \rangle_{1,0})^2 + 2 (1 - \langle \CE \rangle_{\frac{1}{2},\frac{1}{2}})^2~,$$
	$$\mathcal{S}_3 = 2 (1- \langle \mathcal{C} \rangle_{1,0} )^2 (1 - \langle \mathcal{C} \rangle_{\frac{1}{2},\frac{1}{2}} )^2 + \left[ 2 (1- \langle \mathcal{C} \rangle_{1,1} )^2 - (1- \langle \mathcal{C} \rangle_{\frac{1}{2},\frac{1}{2}} )^2 \right]^2~.$$

\vspace{2em}


{\large $\mathbf{l_1=1, ~ l_2=1}$}

\vspace{1em}

$M=2$

\begin{table}[h]
\begin{tabular}{rc|c|}
	\cline{3-3}
	$L=$ && 2\\
	\cline{2-3}
	\multicolumn{1}{c|}{\multirow{1}{*}{$m_1,~ m_2=$}}&$1,~ 1$ & 1 \\
	\cline{2-3}
\end{tabular}
\caption{Clebsch--Gordan coefficients for $M = 2, ~ l_1 = 1, ~ l_2=1$}
\end{table}

\clearpage

$M=1$

\begin{table}[h]
\begin{tabular}{rc|c|c|}
	\cline{3-4}
	$L=$ && 2 & 1\\
	\cline{2-4}
	\multicolumn{1}{c|}{\multirow{3}{*}{$m_1,~ m_2=$}}
	& $1,~ 0$ 
	& $\displaystyle{\frac{1}{\sqrt{2}}}$ 
	& $ \displaystyle{\frac{1}{\sqrt{2}}}$ \\
	\cline{2-4}
	& \multicolumn{1}{|c|}{$0, ~ 1$}
	& $\displaystyle{\frac{1}{\sqrt{2}}}$
	& $\displaystyle{- \frac{1}{\sqrt{2}}}$ \\
	\cline{2-4}
\end{tabular}
\caption{Clebsch--Gordan coefficients for $M = 1, ~ l_1 = 1, ~ l_2=1$}
\end{table}

\vspace{1em}

$M=0$

\begin{table}[h]
\begin{tabular}{rc|c|c|c|}
	\cline{3-5}
	$L=$ && 2 & 1 & 0\\
	\cline{2-5}
	\multicolumn{1}{c|}{\multirow{5}{*}{$m_1, ~ m_2=$}}
	& $1, ~ - 1$ 
	& $\displaystyle{\frac{(1- \langle \mathcal{C} \rangle_{1, 0} )}{\sqrt{2\mathcal{S}_4}}}$
	& $\displaystyle{\frac{1}{\sqrt{2}}}$ 
	& $\displaystyle{\frac{(1- \langle \mathcal{C} \rangle_{1, 1})}{\sqrt{\mathcal{S}_4}} } $ \\
	\cline{2-5}
	&\multicolumn{1}{|c|}{$0, ~ 0$} 
	& $\displaystyle{\frac{\sqrt{2} (1- \langle \mathcal{C} \rangle_{1, 1} )}{\sqrt{\mathcal{S}_4}}}$
	& $ 0 $
	& $\displaystyle{- \frac{ (1- \langle \mathcal{C} \rangle_{1, 0})}{\sqrt{\mathcal{S}_4}}}$ \\
	\cline{2-5}
	&\multicolumn{1}{|c|}{$-1, ~ 1$} 
	& $\displaystyle{\frac{(1- \langle \mathcal{C} \rangle_{1, 0} )}{\sqrt{2 \mathcal{S}_4}}}$
	& $\displaystyle{ - \frac{1}{\sqrt{2}}}$ 
	& $\displaystyle{\frac{(1- \langle \mathcal{C} \rangle_{1, 1})}{\sqrt{\mathcal{S}_4}}}$ \\
	\cline{2-5}
\end{tabular}
\caption{Clebsch--Gordan coefficients for $M = 0, ~ l_1 = 1, ~ l_2=1$}
\end{table}
with $$\mathcal{S}_4 =  (1- \langle \mathcal{C} \rangle_{1, 0} )^2 + 2 (1- \langle \mathcal{C} \rangle_{1, 1} )^2~.$$

\vspace{2em}

{\large $\mathbf{l_1=\frac{3}{2}, ~ l_2=\frac{1}{2}}$}

\vspace{1em}

$M=2$

\begin{table}[h]
\begin{tabular}{rc|c|}
	\cline{3-3}
	$L=$ && 2\\
	\cline{2-3}
	\multicolumn{1}{c|}{\multirow{1}{*}{$m_1,~ m_2=$}}&$\frac{3}{2},~ \frac{1}{2}$ & 1 \\
	\cline{2-3}
\end{tabular}
\caption{Clebsch--Gordan coefficients for $M = 2, ~ l_1 = \frac{3}{2}, ~ l_2=\frac{1}{2}$}
\end{table}

\clearpage

$M=1$

\begin{table}[h]
\begin{tabular}{rc|c|c|}
	\cline{3-4}
	$L=$ && 2 & 1\\
	\cline{2-4}
	\multicolumn{1}{c|}{\multirow{3}{*}{$m_1,~ m_2=$}}
	& $\frac{3}{2},~ - \frac{1}{2}$ 
	& $\displaystyle{\frac{(1-\langle \CE \rangle_{\frac{1}{2},\frac{1}{2}})}{\sqrt{\mathcal{S}_5}}}$ 
	& $ \displaystyle{\frac{\sqrt{3}(1-\langle \CE \rangle_{\frac{3}{2},\frac{3}{2}})}{\sqrt{\mathcal{S}_5}}}$ \\
	\cline{2-4}
	& \multicolumn{1}{|c|}{$0, ~ 1$}
	& $\displaystyle{\frac{\sqrt{3}(1-\langle \CE \rangle_{\frac{3}{2},\frac{3}{2}})}{\sqrt{\mathcal{S}_5}}}$
	& $\displaystyle{\frac{(1-\langle \CE \rangle_{\frac{1}{2},\frac{1}{2}})}{\sqrt{\mathcal{S}_5}}}$ \\
	\cline{2-4}
\end{tabular}
\caption{Clebsch--Gordan coefficients for $M = 1, ~ l_1 = \frac{3}{2}, ~ l_2=\frac{1}{2}$}
\end{table}
with $$ \mathcal{S}_5 = 3 (1-\langle \CE \rangle_{\frac{3}{2},\frac{3}{2}})^2 + (1-\langle \CE \rangle_{\frac{1}{2},\frac{1}{2}})^2$$

\vspace{1em}

$M=0$

\begin{table}[h]
\begin{tabular}{rc|c|c|}
	\cline{3-4}
	$L=$ && 2 & 1\\
	\cline{2-4}
	\multicolumn{1}{c|}{\multirow{5}{*}{$m_1, ~ m_2=$}}
	& $\frac{1}{2}, ~ - \frac{1}{2}$ 
	& $\displaystyle{\frac{(1 - \langle \CE \rangle_{\frac{1}{2},\frac{1}{2}})}{\sqrt{\mathcal{S}_6}}}$
	& $\displaystyle{- \frac{3 (1 - \langle \CE \rangle_{\frac{3}{2},\frac{3}{2}})^2 - (1 - \langle \CE \rangle_{\frac{1}{2},\frac{1}{2}})^2}{\sqrt{\mathcal{S}_7}}}$\\
	\cline{2-4}
	&\multicolumn{1}{|c|}{$-\frac{1}{2}, ~ \frac{1}{2}$} 
	& $\displaystyle{\frac{(1 - \langle \CE \rangle_{\frac{3}{2},\frac{1}{2}})}{\sqrt{\mathcal{S}_6}}}$
	& $\displaystyle{\frac{2 (1 - \langle \CE \rangle_{\frac{3}{2},\frac{1}{2}}) (1 - \langle \CE \rangle_{\frac{1}{2},\frac{1}{2}})}{\sqrt{\mathcal{S}_7}}}$\\
	\cline{2-4}
\end{tabular}
\caption{Clebsch--Gordan coefficients for $M = 0, ~ l_1 = \frac{3}{2}, ~ l_2=\frac{1}{2}$}
\end{table}
with
$$ \mathcal{S}_6 = (1 - \langle \CE \rangle_{\frac{3}{2},\frac{1}{2}})^2 + (1 - \langle \CE \rangle_{\frac{1}{2},\frac{1}{2}})^2$$
$$ \mathcal{S}_7 = 4 (1 - \langle \CE \rangle_{\frac{3}{2},\frac{1}{2}})^2 (1 - \langle \CE \rangle_{\frac{1}{2},\frac{1}{2}})^2 + \left[ 3 (1 - \langle \CE \rangle_{\frac{3}{2},\frac{3}{2}})^2 - (1 - \langle \CE \rangle_{\frac{1}{2},\frac{1}{2}})^2 \right]^2$$

Notice that these tables reduce to the corresponding CG tables of standard QM, as given \emph{e.g.} in \cite{Goswami}, when $\gamma=0$.

Note that the CG that we collected in these tables are found starting from a state of maximum total angular momentum and applying $L_-$ or applying orthonormality conditions with states previously analyzed.
Inverting the direction of our derivation, though, leads to slightly different coefficients for systems with components characterized by different angular momenta.
Consider, for example, a system composed of two particles with $l_1=1$ and $l_2=1/2$ and total angular momentum $L=3/2$ and $M=-1/2$.
Using the corresponding table we find
\begin{equation}
	\left\langle 0 ; - \frac{1}{2} \middle| \frac{3}{2}, - \frac{1}{2} \right\rangle = \frac{\sqrt{2}(1- \langle \mathcal{C} \rangle_{\frac{1}{2},\frac{1}{2}} )}{\sqrt{(1 - \langle \CE \rangle_{1,0})^2 + 2 (1 - \langle \CE \rangle_{\frac{1}{2},\frac{1}{2}})^2}}~.
\end{equation}
On the other hand, starting from the state $|3/2,-3/2\rangle$ and applying the operator $L_+$, we find for the same CG
\begin{equation}
	\left\langle 0 ; - \frac{1}{2} \middle| \frac{3}{2} , - \frac{1}{2} \right\rangle = \frac{\sqrt{2}(1- \langle \mathcal{C} \rangle_{1,-1} )}{\sqrt{(1 - \langle \CE \rangle_{\frac{1}{2},-\frac{1}{2}})^2 + 2 (1 - \langle \CE \rangle_{1,-1})^2}}~.
\end{equation}
The same result can be obtained for every linear and quadratic GUP model, \emph{i.e.} for every $\delta$ and $\epsilon$.
It is particularly evident for a quadratic model.
Indeed, in this case, the expectation value of $\CE$ depends only on the azimuthal quantum number.
Ultimately this ambiguity is generated by the commutation relations (\ref{eqn:commutator_Lz_Lpm}).
These relations are in turn directly derived from (\ref{eqn:generalized_commutator_ang_mom}).
They are therefore implications of the modification of Heisenberg algebra and of the classical definition of angular momentum.
This difference in CG coefficients is related to the fact that the present modification of the angular momentum algebra is sensitive to the angular momentum quantum numbers; therefore, it will distinguish which particle's angular momentum is changed.
It is also interesting to notice that this ambiguity is not present in some CG coefficients in systems composed of identical particles, reinforcing our observation of a distinction based on the angular momenta of the constituents.
{Notice that in this thesis we considered GUP up to second order in $\gamma$.
In case of an uncertainty principle derived from a full theory of QG, this issue may not be present as a consequence of higher order terms.}
Further work is {therefore} required for a better understanding of this problem.
\chapter{Harmonic Oscillator with Modified Uncertainty Relation} \label{ch:HO}

\begin{quote}
The career of a young theoretical physicist consists of treating the harmonic oscillator in ever-increasing levels of abstraction.

- Sidney Coleman
\end{quote}

A Harmonic Oscillator (HO) is a physical oscillating system described by a mass $m$ and an angular frequency $\omega$.
Given its simplicity and versatility, it constitutes one of the best and most studied systems.
Furthermore, in QM the HO is one of the few problems that is exactly solvable.
This makes of the HO a priceless tool for studying other systems, like atoms bounded in molecules, electrons in conductors or in crystals, and, given the affinity of the Hamiltonian for the electromagnetic radiation with the HO Hamiltonian, also light (see Chapter \ref{ch:QO} and Sec.~\ref{ssec:fund_QO}).

In QM, following the correspondence principle, the Hamiltonian for a one-dimensional HO is simply given by
\begin{equation}
	H = \frac{p^2}{2m} + \frac{m \omega^2}{2} q^2~.
\end{equation}
As in many cases, two different approaches to solve this system are possible \cite{Messiah}.
From the calculus point of view, one can solve the corresponding differential equation in position representation, looking for eigenvalues and eigenstates of the Hamiltonian
\begin{equation}
	\left[- \frac{\hbar^2}{2m}\frac{\mathrm{d}^2}{\mathrm{d} q^2} + \frac{m \omega^2}{2} q^2 \right] \psi(q) = E \psi(q)~.
\end{equation}
One would then find that the spectrum of this eigenvalue equation is discrete, and the eigensolutions are given by
\begin{equation}
	u_n (Q) = \frac{e^{-Q^2/2}}{\sqrt{2^n n! \sqrt{\pi}}} H_n (Q)~, \mbox{ with eigenvalue } E_n = \hbar \omega \left(n + \frac{1}{2} \right)~,
\end{equation}
where $H_n(Q)$ is the $n$-th Hermite polynomial and $Q = q \sqrt{\frac{m \omega}{\hbar}}$.
A different but equivalent approach, due to Dirac, consists of defining a set of operators that allow us to move from one eigenstate to the other.
In fact, one can define the following operators
\begin{align}
	a = & \sqrt{\frac{m \omega}{2\hbar}} \left( q + \frac{i}{m \omega} p \right)~, & a^\dagger = & \sqrt{\frac{m \omega}{2\hbar}} \left( q - \frac{i}{m \omega} p \right)~.
\end{align}
Notice that these two operators do not commute.
Indeed, we can find
\begin{equation}
	[a, a^\dagger] = -\frac{i}{\hbar} [q,p] = 1~, \label{eqn:std_a-ad_com_rel}
\end{equation}
where for the moment we considered a standard commutation relation $[q,p] = i \hbar$.
With these definitions, the Hamiltonian reads
\begin{equation}
	H = \hbar \omega \left( N + \frac{1}{2} \right)~, \label{eqn:HO_Hamiltonian}
\end{equation}
where $N = a^\dagger a$ and where we used the commutation relation \eqref{eqn:std_a-ad_com_rel}.
Therefore, the eigenvalue problem for the HO reduces to the eigenvalue problem for the operator $N$.
One can then prove the following properties
\begin{align}
	a |n\rangle = & \sqrt{n} |n-1\rangle~, & {a^\dagger |n\rangle} = & \sqrt{n+1} |n+1\rangle~, & N |n\rangle = & n |n\rangle~, \label{eqn:prop_standard_operators}
\end{align}
where $n \in \mathbb{N}$.
We then see that the eigenvalue spectra of the two approaches coincide, while we also have
\begin{equation}
	\langle q | n \rangle = u_n (q)~.
\end{equation}

This last method is extremely convenient and widely used also in other areas of quantum physics.
For example, QFT follows very closely this formalism.
Indeed, since the values in the energy spectrum are equally spaced, one can interpret a system described by the Hamiltonian \eqref{eqn:HO_Hamiltonian} as a system of $n$ particles, each with energy $\hbar \omega$ and total energy \mbox{$E_n = \hbar \omega (n + 1/2)$}.
Then acting with the operators $a$ or $a^\dagger$ on a given state $|n\rangle$, the energy will decrease or increase by the amount $\hbar \omega$, respectively, as particles with that particular value of energy were destroyed or created.
This feature suggested the names of ``annihilation'' and ``creation'' operators for $a$ or $a^\dagger$, respectively.

Given its simplicity, the HO is usually one of the first ground models where one can test effects of a new quantum theory.
This was also the case for GUP.
Indeed, in the seminal paper \cite{Kempf1995_1}, the energy spectrum of the HO with a quadratic GUP is studied, finding deviations resulting in a no longer equally spaced energy spectrum.
Notice that in that paper an analytical approach is pursued.
Furthermore, potential experimental tests have been proposed considering microscopic \cite{Bawaj2014_1} or macroscopic \cite{Marin2013_1} HO's or using quantum optomechanical systems \cite{Pikovski2012_1,Bosso2016_2} (see also Chapter \ref{ch:QO}).

In this Chapter, we would like to propose a rigorous study of the HO including both linear and quadratic GUP, focusing on an algebraic approach that will show several analogies with the theory of HO in QM.
Furthermore, a study of the coherent and squeezed states of the HO is carried out, showing potentially observable features \cite{Bosso2017a}.


\section{Harmonic Oscillator in GUP} \label{sec:HO_GUP}

The commutator between position and momentum has a fundamental role in determining the properties of the HO.
In fact, consider the Hamiltonian for the one-dimensional HO
\begin{equation}
 H = \frac{p^2}{2m} + \frac{1}{2}m\omega^2 q^2 =
 -\frac{\hbar \omega}{4} (a^\dagger - a)^2 + \frac{\hbar\omega}{4}(a + a^\dagger)^2 = 
 \hbar\omega\left(N + \frac{1}{2}[a,a^\dagger]\right)~, \label{eqn:Hamiltonian}
\end{equation}
Though in the standard theory $[a,a^\dagger]=1$, since this commutator explicitly appears in the Hamiltonian, we see that a modification of this commutator will influence the properties of the system.

Using the model in (\ref{eqn:GUP}) and the expansion \eqref{eqn:expansion_momentum}, we can write the Hamiltonian for the HO as
\begin{equation}
	H = \frac{p_0^2}{2m} + \frac{1}{2}m\omega^2 q_0^2 - \delta \gamma \frac{p_0^3}{m} + (3 \delta^2 + 2 \epsilon) \gamma^2 \frac{p_0^4}{2 m}~, \label{eqn:perturbed_Hamiltonian}
\end{equation}
to second order in the parameter $\gamma$.
The same Hamiltonian can be described in terms of a modified $[a,a^\dagger]$ commutator
\begin{equation}
	[a,a^\dagger] = - \frac{i}{\hbar}[q,p] = 1 - 2 \delta \gamma p + (3 \epsilon + \delta^2) \gamma^2 p^2 = 1 - 2 \delta \gamma p_0 + 3(\epsilon + \delta^2) \gamma^2 p_0^2~.
\end{equation}
Using these results, in this Section we will find the perturbed eigenstates and eigenvalues for the HO Hamiltonian with GUP.
Furthermore, since $\langle 0 | p_0 | 0 \rangle = 0$ but $\langle 0 | p_0^2 | 0 \rangle \not = 0$, we also expect that the ground state energy is changed by GUP.
In the next Section, we will then see that this is the case.

The previous Hamiltonian can be analyzed considering the terms arising from GUP as a perturbation.
It is therefore worth at this point to review time-independent perturbation theory in QM.

\subsection{Perturbation Theory in Quantum Mechanics}

Consider a stationary perturbation of a Hamiltonian for which a solution to the eigenvalue problem in the unperturbed case is known.
Let us assume that we can describe the perturbed Hamiltonian, $H$, as the unperturbed one, $H_0$, and an additional term written in the form $\lambda V$, where $\lambda$ is a small, real parameter
\begin{equation}
	H = H_0 + \lambda V. \label{eqn:pert_unpert}
\end{equation}
Notice that $\lambda$ and $V$ have dimensions such that the product $\lambda V$ has dimensions of an energy.
The assumption that a solution to the eigenvalue problem for the unperturbed Hamiltonian is known means that the energy spectrum of the system under exam is known with the corresponding energy eigenstates
\begin{equation}
	H_0 | E_i^{(0)} \rangle = E_i^{(0)} | E_i^{(0)} \rangle, \qquad \mbox{with } i=1,2,\ldots ~ ,
\end{equation}
Notice that we are considering the perturbation theory for non-degenerate states.
The purpose of the perturbation theory is to find a solution to the eigenvalue problem for the perturbed Hamiltonian, with the unperturbed case as a limit for $\lambda \rightarrow 0$.

Let $E_i$ be an eigenvalue of $H$ for an eigenstate $|E_i\rangle$
\begin{equation}
	H |E_i\rangle = E_i |E_i\rangle~. \label{eqn:pert_eigen}
\end{equation}
Let $E_i$ and $|E_i\rangle$ tend to $E_i^{(0)}$ and $|E_i^{(0)}\rangle$, respectively, in the limit $\lambda \rightarrow0$.
If the perturbation $\lambda V$ is sufficiently small, we can expand $E_i$ and $|E_i\rangle$ in terms of a power series of $\lambda$ centered on $\lambda=0$
\begin{subequations} \label{eqns:expansions_eigenvalues-eigenstates}
\begin{align}
	E_i & = E_i^{(0)} + \lambda \epsilon_{i,(1)} + \lambda^2 \epsilon_{i,(2)} + \ldots = E_i^{(0)} + \sum_{n=1}^\infty \lambda^n \epsilon_{i,(n)}~, \\
	|E_i\rangle & = |E_i^{(0)}\rangle + \lambda |\eta_{i,(1)}\rangle + \lambda^2 |\eta_{i,(2)}\rangle + \ldots = |E_i^{(0)}\rangle + \sum_{n=1}^\infty \lambda^n |\eta_{i,(n)}\rangle~,
\end{align}
\end{subequations}
where $\epsilon_{i,(n)}$ and $|\eta_{i,(n)}\rangle$ are the $n$-th order correction to the energy eigenvalue and eigenstate, respectively.
In what follows, a subscript in parenthesis indicates a correction term, while a {superscript} in parenthesis indicates a corrected quantity up to a given order, \emph{e.g.}
\begin{subequations}
\begin{align}
	E_i^{(2)} & = E_i^{(0)} + \lambda \epsilon_{i,(1)} + \lambda^2 \epsilon_{i,(2)}~, & |E_i^{(2)} \rangle & = |E_i^{(0)}\rangle + \lambda |\eta_{i,(1)}\rangle + \lambda^2 |\eta_{i,(2)}\rangle~.
\end{align}
\end{subequations}
We will furthermore impose that the corrections to the eigenstate are orthogonal to the unperturbed eigenstate \cite{esposito2004classical}
\begin{equation}
	\langle E_i^{(0)} | E_i \rangle = \langle E_i^{(0)} | E_i^{(0)} \rangle = 1 \Rightarrow \langle E_i^{(0)} | \eta_{i,(n)} \rangle = 0 \qquad \forall n
\end{equation}
We can then prove the following result

\newtheorem{Energy_eigenvalue_eigenstate_pert}{Result}

\begin{Energy_eigenvalue_eigenstate_pert} \label{the:nperturbation}
The $n$-th order corrections to the energy eigenvalue and eigenstate are
\begin{subequations}
\begin{align}
	\epsilon_{i,(n)} & = \langle E_i^{(0)} | V | \eta_{i,(n-1)} \rangle~, \label{eqn:nperturbation_value}\\
	|\eta_{i,(n)} \rangle & = \sum_{j \not= i} \frac{|E_j^{(0)} \rangle \langle E_j^{(0)} |}{E_i^{(0)} - E_j^{(0)}}[(V - \epsilon_{i,(1)}) | \eta_{i,(n-1)} \rangle - \epsilon_{i,(2)} | \eta_{i,(n-2)} \rangle - \ldots - \epsilon_{i,(n-1)} | \eta_{i,(1)} \rangle]~. \label{eqn:nperturbation_state}
\end{align}
\end{subequations}
\end{Energy_eigenvalue_eigenstate_pert}

\begin{proof}
	Substituting (\ref{eqn:pert_unpert}) and (\ref{eqns:expansions_eigenvalues-eigenstates}) in (\ref{eqn:pert_eigen}) we find
\begin{equation}
	(H_0 + \lambda V) \left( |E_i^{(0)}\rangle + \sum_{n=1}^\infty \lambda^n |\eta_{i,(n)}\rangle \right) = \left( E_i^{(0)} + \sum_{n=1}^\infty \lambda^n \epsilon_{i,(n)} \right) \left( |E_i^{(0)}\rangle + \sum_{n=1}^\infty \lambda^n |\eta_{i,(n)}\rangle \right)~.
\end{equation}	
We can analyze this equation for every power of $\lambda$ separately, obtaining
\refstepcounter{equation}
\begin{subequations}
\begin{align}
	( H_0 - E_i^{(0)} ) |E_i^{(0)}\rangle & = 0~, \tag{$\theequation^{(0)}$} \\
	( H_0 - E_i^{(0)} ) |\eta_{i,(1)}\rangle + ( V - \epsilon_{i,(1)} ) |E_i^{(0)}\rangle & = 0~, \tag{$\theequation^{(1)}$} \label{eqn:1-order} \\
	\vdots & \nonumber\\
	( H_0 - E_i^{(0)} ) |\eta_{i,(n)}\rangle + ( V - \epsilon_{i,(1)} ) |\eta_{i,(n-1)}\rangle - \epsilon_{i,(2)} |\eta_{i,(n-2)}\rangle - \ldots - \epsilon_{i,(n)} |E_i^{(0)}\rangle & = 0~, \tag{$\theequation^{(n)}$} \label{eqn:n-order}\\
	\vdots & \nonumber
\end{align}
\end{subequations}
Projecting onto $|E_i^{(0)}\rangle$, we have the corrections to the energy eigenvalue
\begin{equation}
	\epsilon_{i,(n)} = \langle E_i^{(0)} | V |\eta_{i,(n-1)}\rangle~.
\end{equation}
As for the eigenstate corrections, projecting \eqref{eqn:n-order} on the subspace orthogonal to $|E_i^{(0)}\rangle$ and noticing that
\begin{equation}
	\langle E_i^{(0)} | \eta_{i,(n)} \rangle = 0 \Rightarrow \left( \sum_{j \not = i} |E_j^{(0)} \rangle \langle E_j^{(0)} | \right) |\eta_{i,(n)}\rangle = |\eta_{i,(n)}\rangle ~,
\end{equation}
we have
\begin{equation}
	|\eta_{i,(n)}\rangle = \sum_{j \not = i} \frac{ |E_j^{(0)} \rangle \langle E_j^{(0)} | }{ E_i^{(0)} - E_j^{(0)} } [ ( V - \epsilon_{i,(1)} ) |\eta_{i,(n-1)}\rangle - \epsilon_{i,(2)} |\eta_{i,(n-2)}\rangle - \ldots - \epsilon_{i,(n-1)} |\eta_{i,(1)}\rangle ] ~.
\end{equation}
\end{proof}

We can apply this result to the explicit case of first and second order perturbations

\newtheorem{1perturbation}{Corollary}

\begin{1perturbation} \label{cor:1perturbation}
	The corrections to the energy eigenvalue and eigenstate to the first order in $\lambda$ are
	\begin{align}
		\epsilon_{i,(1)} & = \langle E_i^{(0)} | V | E_i^{(0)} \rangle & | \eta_{i,(1)} \rangle & = \sum_{j \not = i} \frac{ \langle E_j^{(0)} | V | E_i^{(0)} \rangle }{E_i^{(0)} - E_j^{(0)}} |E_j^{(0)} \rangle
	\end{align}
\end{1perturbation}

\begin{proof}
	As for the correction to the energy eigenvalue, it can be derived from a direct application of (\ref{eqn:nperturbation_value}), with $|\eta_{i,(0)}\rangle \equiv |E^{(0)}_i\rangle$.
	For the correction to the energy eigenstate, it is easier to start from (\ref{eqn:1-order}), projecting this relation on the subspace orthogonal to $|E^{(0)}_i\rangle$, obtaining
	\begin{equation}
		| \eta_{i,(1)} \rangle = \sum_{j \not = i} \frac{ \langle E_j^{(0)} | V | E_i^{(0)} \rangle }{E_i^{(0)} - E_j^{(0)}} |E_j^{(0)} \rangle
	\end{equation}
\end{proof}

The corrected energy up to first order in $\lambda$ therefore is
\begin{equation}
	E_i^{(1)} = E_i^{(0)} + \langle E_i^{(0)} | \lambda V | E_i^{(0)} \rangle + \mathcal{O}(\lambda^2) = \langle E_i^{(0)} | H | E_i^{(0)} \rangle + \mathcal{O}(\lambda^2)~,
\end{equation}
while for the state, we have
\begin{equation}
	| E_i^{(1)} \rangle = |E_i^{(0)} \rangle + \sum_{j \not = i} \frac{ \langle E_j^{(0)} | \lambda V | E_i^{(0)} \rangle }{E_i^{(0)} - E_j^{(0)}} |E_j^{(0)} \rangle + \mathcal{O}(\lambda^2)~. \label{eqn:norm_1order}
\end{equation}
Notice that the perturbed state is normalized up to order $\lambda$, since $ \langle \eta_i^{(1)} | E_i^{(0)} \rangle = 0$ by construction and since
\begin{equation}
	\langle E_i^{(1)} | E_i^{(1)} \rangle = 1 + \mathcal{O}(\lambda^2)
\end{equation}

\newtheorem{2perturbation}[1perturbation]{Corollary}

We can also find the corrections to second order in $\lambda$
\begin{2perturbation} \label{cor:2perturbation}
	The corrections to the energy eigenvalue and eigenstate to the second order in $\lambda$ are
	\begin{align}
		\epsilon_{i,(2)} & = \sum_{j \not = i} \frac{|\langle E_j^{(0)} | V | E_i^{(0)} \rangle |^2 }{E_i^{(0)} - E_j^{(0)}} \\
		| \eta_{i,(2)} \rangle & = \left( \sum_{j \not = i} \sum_{k \not= i} \frac{ \langle E_j^{(0)} | V | E_k^{(0)} \rangle \langle E_k^{(0)} | V | E_i^{(0)} \rangle }{(E_i^{(0)} - E_j^{(0)}) (E_i^{(0)} - E_k^{(0)})} - \sum_{j \not = i}\frac{ \langle E_j^{(0)} | V | E_i^{(0)} \rangle \langle E_i^{(0)} | V | E_i^{(0)} \rangle }{(E_i^{(0)} - E_j^{(0)})^2} \right) \times \nonumber \\
			& \qquad \times |E_j^{(0)} \rangle~.
	\end{align}
\end{2perturbation}
These results are obtained simply applying the results of Result \ref{the:nperturbation} and of Corollary \ref{cor:1perturbation}.
Higher order corrections are obtained iterating the same methods.
We can also show that the eigenstate corrected to the second order is not normalized.
To see this, we first notice that the terms $\langle \eta_{i,(1)} | \eta_{i,(2)} \rangle$ and $\langle \eta_{i,(2)} | \eta_{i,(2)} \rangle$ are terms of order higher than second.
We also see that, similarly to the previous case, $\langle \eta_{i,(2)} | E_i^{(0)} \rangle = 0$ by construction.
Therefore, the only non-zero terms are $\langle E_i^{(0)} | E_i^{(0)} \rangle = 1$ and $\lambda^2 \langle \eta_{i,(1)} | \eta_{i,(1)} \rangle$.
In particular, this last term is of order $\lambda^2$ and needs to be retained in the second order expansion.
{One can also show that the eigenstate corrected up to higher orders is not normalized as well \cite{esposito2004classical}.}

{Similarly, we can find expressions for the corrections to the expectation values of the Hamiltonian on its eigenstates, $\langle E_i | H | E_i \rangle \equiv \langle H \rangle_i$.
Defining the perturbed series for the expectation values
\begin{equation}
	\langle E_i | H | E_i \rangle = \sum_{n=0}^\infty \lambda^n \langle H \rangle_{i,(n)}~, \qquad \mbox{with} \qquad \langle H \rangle_{i,(0)} = \langle H \rangle_i^{(0)}
\end{equation}
and using the perturbation series for the terms in the LHS we find the following equations
\refstepcounter{equation}
\begin{subequations}
\begin{align}
	\langle E_i^{(0)} | H_0 | E_i^{(0)} \rangle = & \langle H \rangle_i^{(0)}~,\tag{$\theequation_{(0)}$}\\
	\langle E_i^{(0)} | V | E_i^{(0)} \rangle = & \langle H \rangle_{i,(1)}~,\tag{$\theequation_{(1)}$}\\
	\langle \eta_{i,(1)} | H_0 | \eta_{i,(1)} \rangle + \langle E_i^{(0)} | V | \eta_{i,(1)} \rangle + \langle \eta_{i,(1)} | V | E_i^{(0)} \rangle = & \langle H \rangle_{i,(2)}~,\tag{$\theequation_{(2)}$}\\
	\vdots \qquad  = & \quad \vdots \nonumber \\
	\sum_{m=1}^{n-1} \left[\langle \eta_{i,(m)} | H_0 | \eta_{i,(n-m)} \rangle\right] + \sum_{m=0}^{n-1} \left[\langle \eta_{i,(m)} | V | \eta_{i,(n-m-1)} \rangle\right] = & \langle H \rangle_{i,(n)}~,\tag{$\theequation_{(n)}$}\\
	\vdots \qquad  = & \quad \vdots \nonumber~.
\end{align}
\end{subequations}
It is then evident that $\langle H \rangle_{i,(n)} \not = \epsilon_{i,(n)}$ and therefore, in general to a given order $n$
\begin{equation}
	\langle E_i^{(n)} | H | E_i^{(n)} \rangle = \langle H \rangle_i^{(0)} + \lambda \langle H \rangle_{i,(1)} + \ldots + \lambda^n \langle H \rangle_{i,(n)} \not = E_i^{(n)}~,
\end{equation}
or, in other words, the Hamiltonian expectation value to a given order is not equal to the Hamiltonian eigenvalue to the same order.}

{We can then evaluate the following difference to a given order $n\geq 1$
\begin{equation}
	\langle H \rangle_{i,(n)} - \epsilon_{i,(n)} = \sum_{m=1}^{n-1} \left[\langle \eta_{i,(m)} | H_0 | \eta_{i,(n-m)} \rangle + \langle \eta_{i,(m)} | V | \eta_{i,(n-m-1)} \rangle\right]~,
\end{equation}
or, in terms of corrected quantities up to order $n$
\begin{multline}
	\langle H \rangle_i^{(n)} - E_i^{(n)} 
	= \sum_{\nu=2}^n \lambda^\nu \sum_{m=1}^{\nu-1} \left[\langle \eta_{i,(m)} | H_0 | \eta_{i,(\nu-m)} \rangle + \langle \eta_{i,(m)} | V | \eta_{i,(\nu-m-1)} \rangle\right] = \\
	= \sum_{\nu=2}^{n-1} \lambda^\nu \sum_{m=1}^{\nu-1} \left[\langle \eta_{i,(m)} | H | \eta_{i,(\nu-m)} \rangle\right] + \lambda^n \left\{\sum_{m=1}^{n-1} \left[\langle \eta_{i,(m)} | H_0 | \eta_{i,(n-m)} \rangle\right] + \epsilon_{i,(n)}^\star\right\}~.
\end{multline}}

\subsection{Perturbed Harmonic Oscillator Eigenstates and Eigenvalues}

Applying the results of the previous Subsection to the Hamiltonian \eqref{eqn:Hamiltonian}, with $\gamma$ as perturbation parameter, after tedious but straightforward computations, we obtain the following expression for the normalized perturbed Hamiltonian eigenstates up to second order in $\gamma$
\begin{multline}
	|n^{(2)} \rangle = - \frac{\delta^2}{72} \gamma^2 \frac{\hbar m \omega}{2} \sqrt{n^{\underline{6}}} |n-6\rangle
	+ \frac{\gamma^2}{8} \frac{\hbar m \omega}{2} \left( 2 \delta^2 n + \epsilon \right) \sqrt{n^{\underline{4}}} |n-4\rangle + \\
	- i \delta \frac{\gamma}{6} \sqrt{\frac{\hbar m \omega}{2}} \sqrt{n^{\underline{3}}}|n-3\rangle
	- \frac{\gamma^2}{8} \frac{\hbar m \omega}{2} \left[ \delta^2 \left( 7n^2 - 7 n - 5 \right) + 4 \epsilon (2n - 1) \right] \sqrt{n^{\underline{2}}} |n-2\rangle  + \\
	+ i \frac{3}{2} \delta \gamma \sqrt{\frac{\hbar m \omega}{2}} n \sqrt{n^{\underline{1}}} |n-1\rangle
	+ \left[1 - \delta^2 \frac{\gamma^2}{72} \frac{\hbar m \omega}{2} (164 n^3 + 246 n^2 + 256 n + 87)\right]|n\rangle + \\
	+ i \frac{3}{2} \delta \gamma \sqrt{\frac{\hbar m \omega}{2}} (n+1) \sqrt{(n+1)^{\overline{1}}} |n+1\rangle + \\
	- \frac{\gamma^2}{8} \frac{\hbar m \omega}{2} \left[ \delta^2 \left( 7n^2 + 21 n + 9 \right) - 4 \epsilon (2n + 3) \right] \sqrt{(n+1)^{\overline{2}}} |n+2\rangle + \\
	- i \delta \frac{\gamma}{6} \sqrt{\frac{\hbar m \omega}{2}} \sqrt{(n+1)^{\overline{3}}} |n+3\rangle
	+ \frac{\gamma^2}{8} \frac{\hbar m \omega}{2} \left[ 2 \delta^2 ( n + 1) - \epsilon \right] \sqrt{(n+4)^{\overline{4}}} |n+4\rangle + \\
	- \frac{\delta^2}{72} \gamma^2\frac{\hbar m \omega}{2} \sqrt{(n+1)^{\overline{6}}} |n+6\rangle~, \label{eqn:perturbed_state}
\end{multline}
where we used the notation
\begin{align}
	(n+1)^{\overline{k}} & = \frac{(n+k)!}{n!} ~, & n^{\underline{k}} & = \frac{n!}{(n-k)!} \quad \mbox{for }k \leq n~, & n^{\underline{k}} & = 0 \quad \mbox{for } k > n ~,
\end{align}
and where $|n\rangle = |n^{(0)}\rangle$.
The perturbed energy eigenvalue is
\begin{equation}
	E^{(2)} = \hbar \omega \left\{ \left( n + \frac{1}{2} \right) - \frac{\hbar m \omega}{2} \frac{\gamma^2}{2} [ (6 n^2 + 6 n + 1) \delta^2 - (2 n^2 + 2 n + 1) 3 \epsilon ] \right\} = E^{(0)} + \Delta E ~, \label{eqn:energ_eigenstate}
\end{equation}
where
\begin{align}
	E^{(0)} = & \hbar \omega \left(n + \frac{1}{2} \right)~, & \Delta E = - \hbar \omega \frac{\hbar m \omega}{2} \frac{\gamma^2}{2} [ (6 n^2 + 6 n + 1) \delta^2 - (2 n^2 + 2 n + 1) 3 \epsilon ]~.
\end{align}
Furthermore, the spacing of the energy levels is
\begin{equation}
	E^{(2)}(n+1) - E^{(2)}(n) = \hbar \omega - 6 \hbar \omega (n + 1) \frac{\hbar m \omega}{2} \gamma^2 [ \delta^2 - \epsilon ]~.
\end{equation}
Note that the linear and the quadratic contributions to \eqref{eqn:GUP}, identifiable through the parameters $\delta$ and $\epsilon$, compete against each other in the determination of the energy.
It is indeed worth noting that, for the class of models with $\delta^2 = \epsilon$, the correction $\Delta E$ is independent of $n$, obtaining an equally spaced energy spectrum.
For $\delta^2 < \epsilon$, the correction is a positive function of the number $n$.
In particular, for $\delta = 0$ and $\epsilon = 1/3$, we obtain the results presented in \cite{Kempf1995_1} and \cite{Hossenfelder2003}.
Also, notice that the particular energy spectrum presented there can be obtained only for a quadratic model.
Indeed, to obtain the same spectrum for a linear and quadratic GUP one would require an imaginary coefficient for the linear term, making the momentum operator not Hermitian.
Finally, we see that for $\delta^2 > \epsilon$ the spacing between energy levels decreases and one can find the value of $n$ corresponding to the maximal attainable energy
\begin{equation}
	n_\mathrm{max} = \left\lfloor \frac{1}{6 \frac{\hbar m \omega}{2} \gamma^2 (\delta^2 - \epsilon)} - 1 \right\rfloor~.
\end{equation}
Notice that the existence of a maximal energy or a maximal number is connected with Planck scale effects and with the break down of the present GUP model since higher order effects become relevant.

To estimate the magnitude of these corrections, we compute
\begin{multline}
	\frac{|\Delta E|}{E^{(0)}} 
	= \frac{\hbar m \omega}{2} \frac{\gamma^2}{2} \frac{\left| (6 n^2 + 6 n + 1) \delta^2 - (2 n^2 + 2 n + 1) 3 \epsilon \right|}{n + \frac{1}{2}} 
	{\simeq} 3 n \frac{\hbar m \omega}{2} \gamma^2 \left|\delta^2 - \epsilon\right| \\
	\simeq \frac{3}{2} m E^{(0)} \gamma^2 \left|\delta^2 - \epsilon\right|~,
\end{multline}
where we considered the limit for large numbers $n$.
Conversely, if we performed an experiment with a sensitivity $\Delta = |\Delta E|/ E^{(0)}$, we could detect a deviation from the unperturbed energies at
\begin{align}
	E^{(0)} 
	= & \frac{2 \Delta}{3 m \gamma^2 \left|\delta^2 - \epsilon\right|} = \frac{28.372\Delta}{m \gamma_0^2 \left|\delta^2 - \epsilon\right|} \mathrm{J} ~, & n = & \frac{\Delta}{3 \frac{\hbar m \omega}{2} \gamma^2 \left|\delta^2 - \epsilon\right|} = \frac{28.892 \Delta}{ m \omega \gamma_0^2 \left|\delta^2 - \epsilon\right|} \times 10^{34} \label{eqn:detection}
\end{align}
for $m$ and $\omega$ measured in Kg and in Hz, respectively.
Although these values appear huge, as we show below and later in the Chapter, these Planck scale effects are potentially accessible to a number of current and future experiments.

As particular examples, we considered the cases of the mechanical oscillator considered in \cite{Pikovski2012_1}, the oscillators considered in \cite{Bawaj2014_1}, the resonant-mass bar AURIGA in its first longitudinal mode \cite{Marin2013_1}, the mirrors in LIGO oscillating at a frequency in the middle of its detection band, and a system with a mass equivalent to the reduced mass of the system that generated the gravitational wave GW150914, oscillating at a frequency of 100 Hz \cite{Abbott2016}.
\begin{table}
\begin{center}
\begin{tabular}{lcccc}
	\toprule
	Type & Ref. & $m$ (Kg) & $\omega/2\pi$ (Hz) & $n / \displaystyle{\frac{\Delta}{\gamma_0^2 \left|\delta^2 - \epsilon\right|}}$ \\
	\midrule
	Optomechanical system & \cite{Pikovski2012_1} & $10^{-11}$ & $10^5$ & $ 3 \times 10^{41}$ \\
	Bar detector AURIGA & \cite{Marin2013_1} & $1.1 \times 10^{3}$ & $900$ & $3 \times 10^{29}$ \\
	\multirow{4}{*}{Mechanical oscillators} & \multirow{4}{*}{\cite{Bawaj2014_1}} & $3.3 \times 10^{-5}$ & $5.64 \times 10^3$ & $1.5 \times 10^{36}$ \\
	&& $7.7 \times 10^{-8}$ & $1.29 \times 10^5$ & $3 \times 10^{37}$ \\
	&& $2 \times 10^{-8}$ & $1.42 \times 10^5$ & $1.0 \times 10^{38}$ \\
	&& $2 \times 10^{-11}$ & $7.47 \times 10^5$ & $2 \times 10^{40}$ \\
	LIGO detector & \cite{Abbott2016} & 40 & 200 & $4 \times 10^{31}$\\
	GW150914 & \cite{Abbott2016} & $3.2 \times 10^{31}$ & 100 & 90\\
	\bottomrule
\end{tabular}
\caption[Harmonic Oscillator energy threshold]{Some relevant examples of HO are considered, including their mass, frequency and levels at which GUP effects become dominant, as given by \eqref{eqn:detection}.} \label{tab:HO_energy}
\end{center}
\end{table}
From Table \ref{tab:HO_energy}, we then see that massive and/or rapidly oscillating systems most significantly enhance GUP effects.

\section{New GUP-modified ladder operators} \label{sec:new_operators}

Given the form of the perturbed eigenstates in terms of the standard number states (\ref{eqn:perturbed_state}), it is straightforward to show that the standard annihilation and creation operators do not act anymore as ladder operators.
In this Section, therefore, we define and study a new set of operators that will serve as new ladder operators, useful for constructing coherent and squeezed states.
We start from the following definitions
\begin{align}
	\tilde{a} | n^{(2)} \rangle = & \sqrt{n} ~ | (n-1)^{(2)} \rangle~, & \widetilde{a^\dagger} | n^{(2)} \rangle = & \sqrt{n+1} ~ | (n+1)^{(2)} \rangle~, & \tilde{N} | n^{(2)} \rangle = & n ~ | n^{(2)} \rangle~. \label{def:new_ops}
\end{align}
Notice that these properties are identical to those for the standard operators \eqref{eqn:prop_standard_operators}.
We would like then to write these new operators in terms of the old ones.
Let us consider the case of $\tilde{a}$, and let us assume that it can be written in the following form
\begin{equation}
	\tilde{a} = A + \delta \gamma B + \gamma^2 ( \delta^2 C + \epsilon D)~, \label{eqn:tilde_a_expansion}
\end{equation}
where $A$, $B$, $C$, and $D$ are functions of the usual operators $a$, $a^\dagger$, and $N$.
Using the definition in \eqref{def:new_ops}, we will require that, for vanishing $\gamma$, the old and new annihilation operators act in the same way when applied on a number state
\begin{equation}
	\tilde{a} |n^{(2)}\rangle |_{\gamma=0} = A | n \rangle = \sqrt{n} | n - 1\rangle = a | n \rangle \qquad \Rightarrow \qquad A \equiv a~.
\end{equation}
Considering the expansions of $\tilde{a}$ and $|n^{(2)}\rangle$ up to order $\gamma$, \eqref{eqn:tilde_a_expansion} and \eqref{eqn:perturbed_state}, respectively, we find
\begin{equation*}
	\tilde{a} |n^{(1)}\rangle = (a + \delta \gamma B) | n^{(1)} \rangle = \left\{a  - i \delta \frac{\gamma}{2}\sqrt{\frac{\hbar m \omega}{2}} \left[3 a^2 + 3 (2 N + 1) - a^\dagger \right] \right\} | n^{(1)} \rangle \qquad \Rightarrow
\end{equation*}
\begin{equation}
	\Rightarrow \qquad B \equiv - \frac{i}{2}\sqrt{\frac{\hbar m \omega}{2}} \left[3 a^2 + 3 (2 N + 1) - a^\dagger {}^2 \right]
\end{equation}
In the same way, one finds
\begin{align}
	C \equiv & - \frac{\gamma^2}{2} \frac{\hbar m \omega}{2} \left[ 6 a^3 + 4 a N + 6 N a^\dagger -  3 {a^\dagger}^3 \right]~,\\
	D \equiv & - \frac{\gamma^2}{2} \frac{\hbar m \omega}{2} \left[ - 2 a^3 + 6 N a^\dagger - {a^\dagger}^3 \right]~.
\end{align}
A similar procedure can be used for $\tilde{a^\dagger}$ and $\tilde{N}$, finding the following expressions
\begin{subequations} \label{eqns:new_ops}
\begin{align}
	\tilde{a} = & a
	- \delta i \frac{\gamma}{2} \sqrt{\frac{\hbar m \omega}{2}} \left[ 3 a^2 + 3 (2N+1) - {a^\dagger}^2 \right] + \nonumber \\
	& - \frac{\gamma^2}{2} \frac{\hbar m \omega}{2} \left[ 2 \left( 3 \delta^2 - \epsilon \right) a^3 + 4 \delta^2 a N + 6 \left(  \delta^2 + \epsilon \right) N a^\dagger - \left( 3 \delta^2 + \epsilon \right) {a^\dagger}^3 \right]~, \\
	\widetilde{a^\dagger} = & a^\dagger 
	+ \delta i \frac{\gamma}{2} \sqrt{\frac{\hbar m \omega}{2}} \left[ 3{a^\dagger}^2 + 3(2N+1) - a^2 \right] + \nonumber \\
	& - \frac{\gamma^2}{2} \frac{\hbar m \omega}{2} \left[ 2 \left( 3 \delta^2 - \epsilon \right) {a^\dagger}^3 + 4 \delta^2 N a^\dagger + 6 ( \delta^2 + \epsilon ) a N - \left(3 \delta^2 + \epsilon \right) a^3 \right]~, \\
	\tilde{N} = & N 
	- \delta i \frac{\gamma}{2} \sqrt{\frac{\hbar m \omega}{2}} \left[ a^3 - 3 ( a N - N a^\dagger ) - {a^\dagger}^3 \right] + \nonumber \\
	& + \frac{\gamma^2}{4} \frac{\hbar m \omega}{2} \left\{ (3 \delta^2 + 2 \epsilon) [ a^4 - 2 (2 a N a + a^2) - 2 (2 a^\dagger N a^\dagger + {a^\dagger}^2) + {a^\dagger}^4 ] + \right. \nonumber \\
	& \left. + \delta^2 (30 N^2 + 30 N + 11) \right\}~,
\end{align}
\end{subequations}
Notice that, iterating this procedure one can extend the relations in (\ref{eqns:new_ops}) to any arbitrary order.
Furthermore, note that these operators obey the following relations
\begin{align}
	\widetilde{a^\dagger} = & \tilde{a}^\dagger~, & \tilde{N} = & \tilde{a}^\dagger \tilde{a} ~, & [\tilde{a},\tilde{a}^\dagger] = & 1~.
\end{align}

Relations (\ref{eqns:new_ops}) can be inverted, obtaining
\begin{subequations} \label{ref}
\begin{align}
	a = & \tilde{a} + i \frac{\delta}{2} \sqrt{\frac{\hbar m \omega}{2}} \gamma \left[3 \tilde{a}^2 + 3 \left(2 \tilde{N} + 1\right) - \tilde{a}^\dagger{}^2\right] + \nonumber \\
	& - \frac{\hbar m \omega}{2} \frac{\gamma^2}{2} \left[ 2 (3 \delta^2 + \epsilon) \tilde{a}^3 + 4 \delta^2 \tilde{a} \tilde{N} + 6 (\delta^2 - \epsilon) \tilde{N} \tilde{a}^\dagger + ( 3 \delta^2 + \epsilon) \tilde{a}^\dagger{}^3 \right] ~, \\
	a^\dagger = & \tilde{a}^\dagger - i \frac{\delta}{2} \sqrt{\frac{\hbar m \omega}{2}} \gamma \left[3 \tilde{a}^\dagger {}^2 + 3 \left(2 \tilde{N} + 1\right) - \tilde{a}^2\right] + \nonumber \\
	& - \frac{\hbar m \omega}{2} \frac{\gamma^2}{2} \left[ 2 (3 \delta^2 + \epsilon) \tilde{a}^\dagger {}^3 + 4 \delta^2 \tilde{N} \tilde{a}^\dagger + 6 (\delta^2 - \epsilon) \tilde{a} \tilde{N} + ( 3 \delta^2 + \epsilon) \tilde{a}^3 \right]~,
	\\
%
%
%
	N = & \tilde{N}
	- i \frac{\delta}{2} \sqrt{\frac{\hbar m \omega}{2}} \gamma \left(\tilde{a}^\dagger {}^3 - \tilde{a}^3 + 3 \tilde{a} \tilde{N} - 3 \tilde{N} \tilde{a}^\dagger \right) + \nonumber \\
	& - \frac{\hbar m \omega}{2} \frac{\gamma^2}{4} \left[ \left( 9 \delta^2 + 2\epsilon \right) \tilde{a}^\dagger {}^4 + 2 \left( 3 \delta^2 - 2 \epsilon \right) (2 \tilde{a} \tilde{N} \tilde{a} + \tilde{a}^2 ) - \delta^2 ( 30 \tilde{N}^2 + 30 \tilde{N} + 11) + \right. \nonumber \\
	& \phantom{- \frac{\hbar m \omega}{2} \frac{\gamma^2}{4} \left[ \right]} \left. + 2 \left( 3 \delta^2 - 2 \epsilon \right) (2 \tilde{a}^\dagger \tilde{N} \tilde{a}^\dagger + \tilde{a}^\dagger {}^2 ) + \left( 9 \delta^2 + 2 \epsilon \right) \tilde{a}^4\right] ~.
\end{align}
\end{subequations}
Using these expressions, we can write the Hamiltonian of the HO, the physical position and momentum operators in a simple form in terms of the $\sim$-operators
\begin{equation}
	H = \hbar \omega \left\{ \left(\tilde{N} + \frac{1}{2} \right) - \frac{\hbar m \omega}{2} \frac{\gamma^2}{2} [ 6 (\tilde{N}^2 + \tilde{N}) (\delta^2 - \epsilon) + \delta^2 - 3 \epsilon]\right\}~,
\end{equation}
\vspace{-3em}
\begin{subequations} \label{eqns:expansions_q&p}
\begin{align}
	q = & (\tilde{a}^\dagger + \tilde{a}) \sqrt{\frac{\hbar}{2 m \omega}} 
		- 2 i (\tilde{a}^\dagger {}^2 - \tilde{a}^2 ) \delta \frac{\hbar}{2} \gamma + \nonumber \\
		& - [ 3 ( \tilde{a}^3 + \tilde{a}^\dagger {}^3 ) ( 3 \delta^2 + \epsilon) + 2 (\tilde{a} \tilde{N} + \tilde{N} \tilde{a}^\dagger) (5 \delta^2 - 3 \epsilon) ] \frac{\hbar}{2} \sqrt{\frac{\hbar m \omega}{2}} \frac{\gamma^2}{2}~, \\
 	p = & i (\tilde{a}^\dagger - \tilde{a}) \sqrt{\frac{\hbar m \omega}{2}} 
 		+ 2 (\tilde{a}^\dagger {}^2 + 2 \tilde{N} + 1 + \tilde{a}^2) \delta \frac{\hbar m \omega}{2} \gamma + \nonumber \\
 		& + 3 i [( \tilde{a}^3 - \tilde{a}^\dagger {}^3 ) ( 3 \delta^2 + \epsilon) + 4 (\tilde{a} \tilde{N} - \tilde{N} \tilde{a}^\dagger ) \delta^2 ] \left(\frac{\hbar m \omega}{2}\right)^{3/2} \frac{\gamma^2}{2} ~.
\end{align}
\end{subequations}
As we anticipated in the previous Section, the ground state energy is changed by GUP since now we have
\begin{equation}
	\langle 0^{(2)} | H | 0^{(2)} \rangle = \hbar \omega \left[ \frac{1}{2} - \frac{\hbar m \omega}{2} \gamma^2 (\delta^2 - 3 \epsilon) \right]~.
\end{equation}
One can also write these operators in terms of position-like and momentum-like $\sim$-op\-er\-a\-tors
\begin{subequations} \label{eqns:qp_qp-tilde}
\begin{align}
	\tilde{q} & = \sqrt{\frac{\hbar}{2 m \omega}} (\tilde{a}^\dagger + \tilde{a})~,\\
	\tilde{p} & = i \sqrt{\frac{\hbar m \omega}{2}} (\tilde{a}^\dagger - \tilde{a})~,\\
	[\tilde{q},\tilde{p}] & = i \hbar~, \\
	H & = \frac{\tilde{p}^2}{2 m} + \frac{m \omega^2}{2} \tilde{q}^2 - \frac{3}{8 m} (\delta^2 - \epsilon) \gamma^2 (m^4 \omega^4 \tilde{q}^4 + 2 m^2 \omega^2 \tilde{q} \tilde{p}^2 \tilde{q} + \tilde{p}^4) ~,\\
	q & = \tilde{q} - \delta \gamma(\tilde{q} \tilde{p} + \tilde{p} \tilde{q}) + \frac{\gamma^2}{8} \left[ (-19 \delta^2 + 3 \epsilon) m^2 \omega^2 \tilde{q}^3 + (17\delta^2 + 15 \epsilon) \tilde{p} \tilde{q} \tilde{p} \right]~,\\
	p & = \tilde{p} - 2 \delta \gamma m^2 \omega^2 \tilde{q}^2 - \frac{3 \gamma^2}{8} \left[ (\delta^2 - \epsilon) \tilde{p}^3 + (13 \delta^2 + 3 \epsilon) m^2 \omega^2 \tilde{q} \tilde{p} \tilde{q} \right]~.
\end{align}
\end{subequations}
Notice that $\tilde{q}$ and $\tilde{p}$ do not have a direct physical interpretation, but they allow for a simple treatment of the theory.
It is then easy to find the expectation values and the uncertainties in position and momentum for the perturbed number states.
Defining $\langle A \rangle = \langle n^{(2)} | A | n^{(2)}\rangle$ for any operator $A$, we find
\begin{subequations}
\begin{align}
	\langle q \rangle = & 
	0 ~, \\
	\langle q^2 \rangle 
	= & \frac{\hbar}{2 m \omega} (2 n + 1) - 2 \gamma^2 \frac{\hbar^2}{4} \left[ \delta^2 (6n^2 + 6n + 1) - 3 \epsilon (2 n^2 + 2 n + 1) \right]~ ,\\
	(\Delta q)^2 = & \langle q^2 \rangle ~, \\
	\langle p \rangle = & 
	2 \delta \gamma \frac{\hbar m \omega}{2} (2n + 1) ~, \\
	\langle p^2 \rangle 
	= & \frac{\hbar m \omega}{2} (2n + 1)~,\\
	(\Delta p)^2 = & \frac{\hbar m \omega}{2} (2n + 1) \left[ 1 - 4 \delta^2 \gamma^2 \frac{\hbar m \omega}{2} (2n + 1) \right]~, \\
	(\Delta q)^2 (\Delta p)^2 = & \frac{\hbar^2}{4} (2n + 1)^2 + \nonumber \\	
	& \qquad - 2 \gamma^2 \frac{\hbar^2}{4} \frac{\hbar m \omega}{2} (2n + 1) [ \delta^2 (14n^2 + 14n + 3) - 3\epsilon (2n^2 + 2n + 1)] \label{eqn:unc_rel_HO}
\end{align}
\end{subequations}
The same results can be obtained starting from the expansion in (\ref{eqn:perturbed_state}).
Notice that the linear term leads to some interesting features.
First, the expectation value of the momentum does not vanish while the expectation value of the position does.
This is a consequence of the linear part of the GUP model in \eqref{eqn:GUP}, which singles out a preferred direction and so breaks translation invariance.
Furthermore, this same term implies that there exists a value for $n$ such that $(\Delta p)^2 < 0$, showing that a linear model leads to critical results for high energy systems, \emph{i.e.} when the energy is close to the Planck scale.
We expect that at this energy scale a full theory of QG has to be considered, with non-negligible higher order corrections.

{We can now prove the result that we anticipated in \eqref{eqn:unc_rel_quad} for a quadratic model ($\delta = 0$, $\epsilon=1/3$).
To find the minimal uncertainty product, we have to use \eqref{eqn:Sch-Ro}.
Since we have
\begin{equation}
	\langle p q \rangle = - i \frac{\hbar}{2} \left[1 + \gamma^2 (\Delta p)^2\right]~,
\end{equation}
we can easily see that the first term in \eqref{eqn:Sch-Ro} vanishes.
Furthermore, since for a quadratic GUP we can write $\langle p^2 \rangle = (\Delta p)^2$, we finally have
\begin{equation}
	\Delta q \Delta p \geq \left|\frac{\langle[q,p]\rangle}{2}\right| \Rightarrow \Delta q \geq \frac{\hbar}{2} \left[\frac{1}{\Delta p} + \gamma^2 \Delta p\right], 
\end{equation}
as in \eqref{eqn:unc_rel_quad}.}

To conclude this Section, we emphasize that $|n^{(2)}\rangle$ are the eigenstates of the full Hamiltonian with GUP (\ref{eqn:Hamiltonian}) up to $\mathcal{O}(\gamma^2)$, while the standard states $|n\rangle$ are not.
Therefore, we will consider the former as the physical number states.
Furthermore, the $\sim$--operators that we have defined correctly act as ladder operators.
For this reason, we apply the above new operators to define coherent and squeezed states.

\section{Coherent States} \label{sec:coherent_states}

\begin{figure}
\begin{center}
	\includegraphics[width=0.5\textwidth]{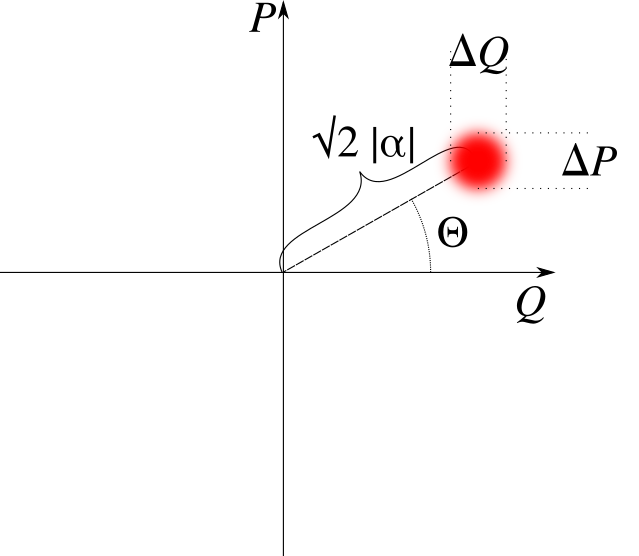}
\end{center}
\caption{Quadrature-space representation of a coherent state.} \label{fig:coherent_state}
\end{figure}

We now study the properties of coherent states with GUP.
In QM, coherent states are the ``most classical'' states, in the sense that the expectation values of position and momentum for a coherent state reproduce the dynamics of a classical HO, like a body attached to a spring.
This is evident, for example, in the Ehrenfest theorem applied to a coherent state of the HO
\begin{align}
	\frac{\diff~\langle q \rangle}{\diff~t} = & \frac{\langle p \rangle}{m}~, & \frac{\diff~\langle p \rangle}{\diff~t} = & - m \omega^2 \langle q \rangle~, \label{eqn:std_ehrenfest}
\end{align}
whose solutions are
\begin{align}
	\langle q \rangle = & A \sin (\omega t) + B \cos (\omega t)~, & \langle p \rangle = & m \omega \left[ A \cos (\omega t) - B \sin (\omega t)\right]~,
\end{align}
with $A$ and $B$ two constants with dimensions of position.
Other important characteristics of coherent states in QM are their position and momentum uncertainties.
In fact, these states are minimal uncertainty states, in the sense that the Heisenberg principle is saturated for these states (this was actually the original motivation for their definition and study), while, when expressed in terms of the dimensionless quadratures
\begin{align}
	Q = & \sqrt{\frac{m \omega}{\hbar}} q~, & P = \frac{1}{\sqrt{\hbar m \omega}} p ~, \label{def:quadratures}
\end{align}
the uncertainty relation is
\begin{equation}
	\Delta Q \Delta P = \frac{1}{2} ~, \qquad \mbox{with} \qquad \Delta Q = \Delta P~.
\end{equation}
In the standard theory, one of the possible ways of defining coherent states is as eigenstates of the annihilation operator.
In QM, using this definition we can also find the following relations
\begin{subequations} \label{eqns:exp_unc_coherent_noGUP}
\begin{align}
	\langle q \rangle = & \Re(\alpha) \sqrt{\frac{2 \hbar}{m \omega}} ~, & \langle Q \rangle = & \sqrt{2} \Re (\alpha)~,\\
	(\Delta q)^2 = & \frac{\hbar}{2 m \omega} ~, & (\Delta Q)^2 = & \frac{1}{2} \\
	\langle p \rangle = & \Im \left(\alpha\right) \sqrt{2 \hbar m\omega} ~, & \langle P \rangle = & \sqrt{2} \Im (\alpha)~,\\
	(\Delta p)^2 = & \frac{\hbar m \omega}{2} ~, & (\Delta P)^2 = & \frac{1}{2} \\
	(\Delta q)^2 (\Delta p)^2 = & \frac{\hbar^2}{4} ~, & (\Delta Q)^2 (\Delta P)^2 = & \frac{1}{4}~,
\end{align}
\end{subequations}
These relations let us represent coherent states in phase-space, or more precisely in the quadrature-space, in a very efficient way.
Indeed, we observe that the expectation value of position is proportional to the real part of $\alpha$, while the expectation value of momentum to its imaginary part.
Therefore, as done in Fig. \ref{fig:coherent_state}, we can locate a coherent state on a complex plane through the quantity $\alpha=|\alpha| e^{i \Theta}$, where $|\alpha|$ represents the distance from the origin of the complex plane and $\Theta$ the angle with respect the real axis.
Identifying the real axis with the $Q$ axis and the imaginary axis with the $P$ axis, we can also describe the uncertainties of coherent states as a disk around the position $(Q,P) = \sqrt{2} |\alpha| e^{i \Theta}$, the size of which graphically describes the uncertainties.
These and many other features (coherence, statistical distribution, etc...), make of coherent states valuable tools for research.
Given their interesting properties and their wide use in many areas of physics, they offer a possible testing ground for possible deviations due to QG.
We will indeed see that GUP affects several aspects of coherent states, with potentially observable modifications.

As in the standard theory, we define a coherent state as the eigenstate of the annihilation operator $\tilde{a}$
\begin{equation}
	\tilde{a} | \alpha^{(2)} \rangle = \alpha |\alpha^{(2)} \rangle \label{def:coherent_states}
\end{equation}
to order $\gamma^2$.
The choice for this definition is motivated by the interpretation of $|n^{(2)}\rangle$ as the physical number state and by $\tilde{a}$ being the appropriate annihilation operator.
Also, since the theory constructed from the $\sim$-operators is similar to the standard theory of the HO, \emph{i.e.} they share the same commutation relation, one obtains similar equations for the coherent states.
In particular, following the results of the previous Section, we can find that $\sim$-coherent states can be expanded in terms of the number states in the standard way.
In fact, consider a generic coherent state as a linear combination of number states
\begin{equation}
	|\alpha^{(2)}\rangle = \sum_{n=0}^\infty a_n |n^{(2)}\rangle~,
\end{equation}
where $a_n$ are the coefficients of the combination.
Applying the definition \eqref{def:coherent_states} we find
\begin{equation}
	\tilde{a} |\alpha^{(2)}\rangle = \sum_{n=1}^\infty a_n \sqrt{n} |(n-1)^{(2)}\rangle = \alpha |\alpha^{(2)}\rangle~,
\end{equation}
from which we find the following recursion relation
\begin{equation}
	a_n = \frac{\alpha }{\sqrt{n}} a_{n-1} = \frac{\alpha^n}{\sqrt{n!}} a_0~.
\end{equation}
Furthermore, assuming $a_0 \in \mathbb{R}$ and imposing the normalization for coherent states, we find
\begin{equation}
	\langle \alpha^{(2)} | \alpha^{(2)} \rangle = 1 = \sum_{n=0}^\infty |a_n|^2 = \sum_{n=0}^\infty \frac{|\alpha|^{2n}}{n!} a_0^2 \Rightarrow a_0 = e^{ - \frac{|\alpha|^2}{2}}~.
\end{equation}
Therefore, we can write a $\sim$-coherent state in the following form
\begin{equation}
	|\alpha^{(2)} \rangle = e^{- \frac{|\alpha|^2}{2}} \sum_{n=0}^\infty \frac{\alpha^n}{\sqrt{n!}} |n^{(2)}\rangle~, \label{eqn:expansion_coherent_states}
\end{equation}
hence showing a Poisson distribution of number states, and with the chance of defining a displacement operator of the form
\begin{align}
	\tilde{\mathcal{D}}(\alpha) = & e^{\alpha \tilde{a}^\dagger - \alpha^* \tilde{a}} & \mbox{such that} & & |\alpha^{(2)}\rangle = \tilde{\mathcal{D}} (\alpha) |0^{(2)}\rangle~.
\end{align}
Furthermore, given the expansions in (\ref{eqns:expansions_q&p}), one can easily obtain the following results for the expectation values and uncertainties of position and momentum of coherent states
\begin{subequations} \label{eqns:exp_unc_coherent}
\begin{align}
	\langle q \rangle = & (\alpha^\star + \alpha) \sqrt{\frac{\hbar}{2 m \omega}} 
	- 2 i (\alpha^\star - \alpha) (\alpha + \alpha^\star) \delta \frac{\hbar}{2} \gamma + \nonumber \\
	& - \left[ \delta^2 (10 + 9 \alpha^2 + |\alpha|^2 + 9 \alpha^\star {}^2) + 3 \epsilon (-2 + \alpha^2 - 3 |\alpha|^2 + \alpha^\star {}^2) \right] \times \nonumber \\
	& \qquad \times (\alpha + \alpha^\star) \sqrt{\frac{\hbar^3 m \omega}{8}} \frac{\gamma^2}{2} ~, \label{eqn:exp_q_coherent} \\
	\langle q^2 \rangle = & ( \alpha^\star {}^2 + 2 |\alpha|^2 + \alpha^2 + 1) \frac{\hbar}{2 m \omega} + \nonumber \\
	& - 4 i (\alpha^\star - \alpha) (1 + \alpha^2 + 2 |\alpha|^2 + \alpha^\star {}^2) \delta \sqrt{\frac{\hbar^3}{8  m \omega}} \gamma
	+ \nonumber \\
	& - \left\{ \delta^2 [4 + 57 (\alpha^2 + \alpha^\star {}^2) + 26 (\alpha^4 + \alpha^\star {}^4) + 48 |\alpha|^2 + 38 |\alpha|^2 (\alpha^2 + \alpha^\star {}^2) \right. + \nonumber \\ 
	& \qquad \qquad + 24 |\alpha|^4] + \nonumber \\
	& + 3 \epsilon [- 4 - 3 (\alpha^2 + \alpha^\star {}^2) + 2 (\alpha^4 + \alpha^\star {}^4)- 16 |\alpha|^2 - 2 |\alpha|^2 (\alpha^2 + \alpha^\star {}^2) + \nonumber \\
	& \left. \qquad \qquad - 8 |\alpha|^4] \right\} \frac{\hbar^2}{4} \frac{3\gamma^2}{2} ~, \displaybreak \\
	(\Delta q)^2 = & \frac{\hbar}{2 m \omega} - 4 i \left( \alpha^\star - \alpha\right) \sqrt{\frac{\hbar^3}{8 m \omega}} \delta \gamma + \nonumber \\
	& - \left[ \delta^2 ( 4 + 37 \alpha^2 + 8 |\alpha|^2 + 37 {\alpha^\star}^2) + 3 \epsilon (- 4 + \alpha^2 - 8 |\alpha|^2 + {\alpha^\star}^2) \right] \frac{\hbar^2}{4} \frac{\gamma^2}{2} ~, \\ 
	\langle p \rangle = & i \left(\alpha^\star-\alpha\right) \sqrt{\frac{\hbar m\omega}{2}} + 2 \delta \left( {\alpha^\star}^2 + 2 |\alpha|^2 + \alpha^2 + 1 \right) \frac{\hbar m \omega}{2} \gamma + \nonumber \\
	& - 3 i \left\{ \delta^2 [ 3 ( {\alpha^\star}^3 - \alpha^3) + 4 (1 + |\alpha|^2) (\alpha^\star - \alpha) ] + \right. \nonumber \\
	& \left. \qquad \qquad + \epsilon ( {\alpha^\star}^3 - \alpha^3) \right\} \left( \frac{\hbar m \omega}{2} \right)^{3/2} \frac{\gamma^2}{2} ~, \\
	\langle p^2 \rangle = & \left(- {\alpha^\star}^2 + 2 |\alpha|^2 - \alpha^2 + 1 \right) \frac{\hbar m \omega}{2} + \nonumber \\
	& + 4 i \left[ {\alpha^\star}^3 - \alpha^3 + (\alpha^\star - \alpha)(1 + |\alpha|^2) \right] \delta \left(\frac{\hbar m \omega}{2} \right)^{3/2} \gamma + \nonumber \\
	& + \left\{ \delta^2 [ 26 ( {\alpha^\star}^4 + \alpha^4) + 19 ( {\alpha^\star}^2 + \alpha^2 ) (3 + 2 |\alpha|^2)] + \right. \nonumber \\
	& \left. + 3 \epsilon [2 ( {\alpha^\star}^4 + \alpha^4) - ( {\alpha^\star}^2 + \alpha^2 ) (3 + 2 |\alpha|^2) ] \right\} \left( \frac{\hbar m \omega}{2} \right)^2 \frac{\gamma^2}{2} ~, \\ 
	(\Delta p)^2 = & \frac{\hbar m \omega}{2} + \left\{ \delta^2 [ 17 ( {\alpha^\star}^2 + \alpha^2) + 4 (4 |\alpha|^2 - 2) ] - 9 \epsilon ( {\alpha^\star}^2 + \alpha^2) \right\} \left(\frac{\hbar m \omega}{2}\right)^2 \frac{\gamma^2}{2} ~,\\
	(\Delta q)^2 (\Delta p)^2 = & \frac{\hbar^2}{4} \left\{ 1
		- 4 i \delta \left( \alpha^\star - \alpha \right) \sqrt{\frac{\hbar m \omega}{2}} \gamma + \right. \nonumber \\
		& \left. - 2 \left[ \delta^2 ( 3 + 5 \alpha^2 + 5 {\alpha^\star}^2 - 2 |\alpha|^2 ) + 3 \epsilon \left( - 1 + \alpha^2 + \alpha^\star {}^2 - 2 |\alpha|^2\right) \right] \frac{\hbar m \omega}{2} \gamma^2 \right\}~, \label{eqn:prod_unc}
\end{align}
\end{subequations}
where we see that the last term of the last relation exhibits interesting properties.
If $\epsilon > 0$ and $\delta=0$ then the uncertainty $(\Delta q)^2 (\Delta p)^2 $ is greater than $\frac{\hbar^2}{4}$ whereas if $\epsilon = 0$ and $\delta>0$ then this quantity is smaller than $\frac{\hbar^2}{4}$.
Notice that these results reduce to \eqref{eqns:exp_unc_coherent_noGUP} in the limit $\gamma\rightarrow 0$.
Furthermore, note that with GUP a coherent state is no longer a smooth disk in \mbox{quadrature-space}, but rather its shape depends on the particular position in \mbox{quadrature-space}, \emph{i.e.} on the phase of $\alpha$.

Notice that the same results can be obtained starting from the expansion in \eqref{eqns:qp_qp-tilde}.
Indeed, since
\begin{align}
	\tilde{\mathcal{D}}^\dagger (\alpha) \tilde{q} \tilde{\mathcal{D}} (\alpha) = & \tilde{q} + \sqrt{\frac{\hbar}{2 m \omega}} (\alpha^\star + \alpha) ~, 
	& \tilde{\mathcal{D}}^\dagger (\alpha) \tilde{p} \tilde{\mathcal{D}} (\alpha) = & \tilde{p} + i \sqrt{\frac{\hbar m \omega}{2}} (\alpha^\star - \alpha)~,
\end{align}
we have, for example
\begin{multline}
	\langle \alpha^{(2)} | q | \alpha^{(2)} \rangle = \sqrt{\frac{\hbar}{2 m \omega}} (\alpha^\star + \alpha) - \delta \gamma \left\{ 2 i (\alpha^\star {}^2 - \alpha^2) \frac{\hbar}{2} + \langle 0^{(2)} | \tilde{q} \tilde{p} + \tilde{p} \tilde{q} | 0^{(2)} \rangle \right\} + \\
	+ \frac{\gamma^2}{8} \left[ (-19 \delta^2 + 3 \epsilon) m^2 \omega^2 \left\{ \left(\frac{\hbar}{2 m \omega}\right)^{3/2} (\alpha^\star + \alpha)^3 + 3 \sqrt{\frac{\hbar}{2 m \omega}} (\alpha^\star + \alpha) \langle 0^{(2)} | \tilde{q}^2 | 0^{(2)} \rangle \right\} \right. + \\
	+ (17\delta^2 + 15 \epsilon) \left\{ - \frac{\hbar}{2} \sqrt{\frac{\hbar m \omega}{2}} (\alpha^\star {}^2 - \alpha^2) (\alpha^\star - \alpha) + \sqrt{\frac{\hbar m \omega}{2}} (\alpha^\star - \alpha) \langle 0^{(2)} | \tilde{q} \tilde{p} + \tilde{p} \tilde{q} | 0^{(2)} \rangle + \right. \\
	\left. \left. + \sqrt{\frac{\hbar}{2 m \omega}} (\alpha^\star + \alpha) \langle 0^{(2)} | \tilde{p}^2 | 0^{(2)} \rangle \right\} \right] = \\
	= \sqrt{\frac{\hbar}{2 m \omega}} (\alpha^\star + \alpha) 
	- 2 i \delta \gamma (\alpha^\star {}^2 - \alpha^2) \frac{\hbar}{2} 
	+ \left\{ (-19 \delta^2 + 3 \epsilon) (\alpha^\star + \alpha) [ (\alpha^\star + \alpha)^2 + 3 ]  \right. + \\
	+ \left. (17\delta^2 + 15 \epsilon) (\alpha^\star + \alpha) [ - (\alpha^\star - \alpha)^2 + 1 ] \right\} \sqrt{\frac{\hbar^3}{8 m \omega}} \frac{\gamma^2}{8} ~.
\end{multline}
This expression is equivalent to \eqref{eqn:exp_q_coherent}.
This second way of computing expectation values can be useful when we notice that
\begin{equation}
	\langle \alpha^{(2)} | f(\tilde{q},\tilde{p}) | \alpha^{(2)} \rangle = \langle \alpha | f(q_0,p_0) | \alpha \rangle~,
\end{equation}
where $f(\tilde{q},\tilde{p})$ is an arbitrary function of $\tilde{q}$ and $\tilde{p}$.
It, therefore, allows us to use the results of standard QM for expressions involving GUP.

At this point, we wonder if coherent states with GUP are still minimal uncertainty states.
Hence, we look for the minimal uncertainty product, which is the smallest uncertainty product predicted by QM as found using the Schr\"odinger--Robertson relation \eqref{eqn:Sch-Ro}.
Furthermore, notice that the minimal uncertainty product depends on the particular state we are considering through the expectation value $\langle[A,B]\rangle$.
In fact, from the model in (\ref{eqn:GUP}), we obtain the following minimal value for the uncertainty product for coherent states
{
\begin{multline}
	[(\Delta q)^2 (\Delta p)^2]_\mathrm{min} = \frac{\hbar^2}{4} \left\{1 - 4 i \delta \left( \alpha^\star - \alpha \right) \sqrt{\frac{\hbar m \omega}{2}} \gamma + \right. \\
	\left. - 2 \left[ \delta^2 \left( 3 + 5 \alpha^2 + 5 {\alpha^\star}^2 - 2 |\alpha|^2 \right) + 3 \epsilon \left( - 1 + \alpha^2 + \alpha^\star {}^2 - 2 |\alpha|^2 \right)\right] \frac{\hbar m \omega}{2} \gamma^2 \right\}~,
\end{multline}
}
and again nonzero $\epsilon$ makes this quantity larger than the standard QM value whereas non-zero $\delta$ makes it smaller.
Furthermore, we see that coherent states are also minimal uncertainty states when the GUP is included
\begin{equation}
	(\Delta q)^2 (\Delta p)^2 - [(\Delta q)^2 (\Delta p)^2]_\mathrm{min} = 0 \qquad \forall \alpha, \gamma, \delta, \epsilon~. \label{eqn:diff_coherent}
\end{equation}
The quantity $(\Delta q)^2 (\Delta p)^2$ is the actual product of the uncertainties in position and momentum.
Therefore, we see that coherent states as defined in \eqref{def:coherent_states} satisfy all the known properties of the standard theory; \emph{i.e.}, they follow a Poisson distribution, are eigenstates of the annihilation operator, and are minimal uncertainty states.

Furthermore, notice that the uncertainty product $(\Delta q)^2 (\Delta p)^2$ in \eqref{eqn:prod_unc} is negative for some values of $\alpha$ and of $\sqrt{\frac{\hbar m \omega}{2}} \gamma$. This is due to the linear term in GUP and will go negative for
\begin{equation}
	\delta^2 > 3 \frac{1 + 4 \Im^2(\alpha)}{3 + 8 \Re^2 (\alpha) - 4 \Im^2 (\alpha)} \epsilon~.
\end{equation}
In this case, the model will present anomalies when the following relation is fulfilled
\begin{equation}
	\sqrt{\frac{\hbar m \omega}{2}} \gamma = {\frac{ - \Im(\alpha) \delta \pm \sqrt{2} \sqrt{ [3 + 8 \Re^2 (\alpha) - 4 \Im^2 (\alpha)] \delta^2 + 3 [1 + 4 \Im^2(\alpha)] \epsilon}}{ 2 \{ 3[1 + 4 \Im^2(\alpha)] \epsilon + [3 + 8 \Re^2 (\alpha) - 12 \Im^2 (\alpha)] \delta^2\} }} \label{eqn:negative_unc_pro}
\end{equation}
Notice that this quantity depends on the phase of $\alpha$ and that one cannot \emph{a priori} choose the sign in the expression.

It is finally interesting to note that many of the other known properties of coherent states remain when a minimal length is included.
It is indeed straightforward to prove that $\sim$-coherent states form a complete set for the states of the HO.
In fact, we have
\begin{align}
	\int |\alpha^{(2)}\rangle \langle \alpha^{(2)} | \diff^2\alpha = & \int e^{-|\alpha|^2} \sum_{n=0}^\infty \sum_{m=0}^\infty \frac{\alpha^n (\alpha^*)^m}{\sqrt{n!}\sqrt{m!}}|n^{(2)}\rangle\langle m^{(2)}| \diff^2 \alpha = \nonumber \\
	= & \sum_{n=0}^\infty \sum_{m=0}^\infty \frac{|n^{(2)}\rangle\langle m^{(2)}|}{\sqrt{n!}\sqrt{m!}}\int_0^\infty |\alpha|^{n+m+1} e^{-|\alpha|^2} \diff |\alpha| \int_0^{2\pi}e^{i(n-m)\Theta} \diff \Theta = \nonumber \\
	= & \sum_{n=0}^\infty \sum_{m=0}^\infty \frac{|n^{(2)}\rangle\langle m^{(2)}|}{\sqrt{n!}\sqrt{m!}}2\pi \delta_{nm}\int_0^\infty |\alpha|^{n+m+1} e^{-|\alpha|^2} \diff |\alpha| = \nonumber \\
	= & \sum_{n=0}^\infty \frac{|n^{(2)}\rangle\langle n^{(2)}|}{n!}2\pi \int_0^\infty |\alpha|^{2n+1} e^{-|\alpha|^2}\diff |\alpha| = \pi \sum_{n=0}^\infty |n^{(2)}\rangle\langle n^{(2)}|~.
\end{align}
Furthermore, coherent states with different eigenvalues of the annihilation operator $\alpha$ and $\alpha'$ are not orthogonal, and their scalar product is the same as in the standard theory
\begin{equation}
	\langle \alpha^{(2)} | \alpha'^{(2)} \rangle = e^{-\frac{|\alpha|^2 + |\alpha'|^2}{2}}\sum_{n=0}^\infty \frac{(\alpha^*)^n \alpha'}{n!} = e^{-\frac{|\alpha - \alpha'|^2}{2}}~.
\end{equation}
Therefore, $\sim$-coherent states form an overcomplete set, and we can expand a coherent state in the following way
\begin{equation}
	|\alpha^{(2)}\rangle = \frac{1}{\pi} \int |\alpha'^{(2)} \rangle \langle \alpha'^{(2)} | \alpha^{(2)} \rangle \diff^2 \alpha' =  \frac{1}{\pi}\int |\alpha'^{(2)} \rangle e^{-\frac{|\alpha' - \alpha|^2}{2}} \diff^2 \alpha'~.
\end{equation}

\subsection{Time Evolution} \label{subsec:time_coherent}

We would like now to study the time evolution of coherent states.
We will consider this problem using both Schr\"odinger and Heisenberg pictures since each of them will give interesting insights.
We define a simpler notation
\begin{equation}
	\chi = \frac{\hbar m \omega}{2} \gamma^2 (\delta^2 - \epsilon)~.
\end{equation}
Notice that for $\delta^2 = \epsilon$ we have $\chi = 0$.

Let us start from the Schr\"odinger picture.
Using the expansion of coherent states in terms of number states (\ref{eqn:expansion_coherent_states}), one can obtain the time evolved coherent states
\begin{equation}
	|\alpha_0 ,t\rangle= e^{- \frac{|\alpha|^2}{2}} e^{-i \frac{\omega t}{2} [1 - \frac{\hbar m \omega}{2} \gamma^2 (\delta^2 - 3\epsilon)]} \sum_{n=0}^\infty \frac{\alpha^n(t)}{\sqrt{n!}} e^{3 i \chi \omega t n^2} |n^{(2)}\rangle ~,
\end{equation}
where
\begin{equation}
	\alpha(t) = \alpha_0 e^{-i \omega t [1 - 3 \chi]}
\end{equation}
Because of the presence of the $n^2$ term on the rightmost exponential, inherited from the $n^2$ term in the energy eigenvalue (\ref{eqn:energ_eigenstate}), this object is no more an eigenstate of $\tilde{a}$ if $\delta^2 \not = \epsilon$.
Indeed we find
\begin{multline}
	\tilde{a} |\alpha,t\rangle = \alpha(t) e^{- \frac{|\alpha|^2}{2}} e^{-i \frac{\omega t}{2} [1 + \frac{\hbar m \omega}{2} \gamma^2 (5\delta^2 - 3\epsilon)]} \sum_{n=1}^\infty \left[\frac{\alpha^{n-1}(t)}{\sqrt{(n-1)!}} e^{3 i \chi \omega t (n-1)^2} \times \right. \\
	\left. \phantom{\frac{\alpha^{n-1}(t)}{\sqrt{(n-1)!}}} \times |(n-1)^{(2)}\rangle \right] e^{6 i \chi \omega t n }~.
\end{multline}
Nonetheless, the statistical properties of the time-evolved coherent states do not change.
This can be seen noting that all the moments of the number states distribution are not affected by the time evolution.
In fact, consider the expectation value of the generic power $\tilde{N}^k$ as a sum of expectation values of $\tilde{a}^\dagger {}^j \tilde{a}^j$, with $j=1, \ldots, k$.
Calculating these terms, the phases cancel out, leaving time-independent quantities.
Notice that we can write the time evolved coherent state as
\begin{equation}
	|\alpha_0 ,t\rangle = e^{- \frac{|\alpha|^2}{2}} e^{i \Theta/2} e^{-i \frac{\omega t}{2} [1 + \hbar m \omega \gamma^2 \delta^2]} \sum_{n=0}^\infty \frac{\left(\alpha_0 e^{i\Theta}\right)^n(t)}{\sqrt{n!}} e^{i \theta_n} |n^{(2)}\rangle ~, \label{eqn:time_coherent_phases}
\end{equation}
where we defined an overall phase and a peculiar phase for each number component, given respectively by
\begin{align}
	\Theta = & - \omega t [1 - 3 \chi]~, 
	& \theta_n = & 3 \chi \omega t n^2~. \label{eqn:coherent_phases}
\end{align}
Notice that, if $\theta_n$ vanished in (\ref{eqn:time_coherent_phases}), the state would still be an eigenstate of the annihilation operator with eigenvalue $\alpha = \alpha_0 e^{i\Theta}$, as in the standard theory.
However, the condition $\theta_n = 0$ can be satisfied only for $\delta^2 = \epsilon$ or in the limit $\gamma \rightarrow 0$.
In both cases, also the GUP term in $\Theta$ would vanish, obtaining a standard coherent state.

To better understand the evolution in phase-space of coherent states, we compute the position and momentum expectation values:
\begin{subequations}
\begin{align}
	\langle q \rangle = & 2 e^{-\alpha_0^2} \sqrt{\frac{\hbar}{2 m \omega}} \sum_{n=0}^\infty \frac{\alpha_0^{2n}}{n!} \left\{ 
	\alpha_0 \cos (- \Theta + \theta_n - \theta_{n+1}) + \nonumber \right.\\
	& + 2 \alpha_0^2 \delta \sqrt{\frac{\hbar m \omega}{2}} \gamma \sin (- 2 \Theta + \theta_n - \theta_{n+2}) + \nonumber \\
	& - \alpha_0 \frac{\hbar m \omega}{2} \frac{\gamma^2}{2} \left[ 3 \alpha_0^2 (3 \delta^2 + \epsilon) \cos (- 3 \Theta + \theta_n - \theta_{n+3}) + \right. \nonumber \\
	& \left. \left. + 2 (n+1) (5 \delta^2 - 3 \epsilon ) \cos (- \Theta + \theta_n - \theta_{n+1}) \right] \right\} = \\ 
%
%
%
%
%
%
	\simeq & 2 \sqrt{\frac{\hbar}{2 m \omega}} \left\{ 
	 \alpha_0 e^{-\alpha_0^2} \sum_{n=0}^\infty \frac{\alpha_0^{2n}}{n!}\cos \left\{ \omega t \left[ 1 - 6 \chi (n+1) \right] \right\} + \right. \nonumber \\
	& + 2 \alpha_0^2 \delta \sqrt{\frac{\hbar m \omega}{2}} \gamma \sin (2 \omega t) \nonumber \\
	& \left. - \alpha_0 \frac{\hbar m \omega}{2} \frac{\gamma^2}{2} \left[ 3 \alpha_0^2 (3 \delta^2 + \epsilon) \cos (3 \omega t) + 2 (\alpha_0^2 + 1) (5 \delta^2 - 3 \epsilon ) \cos (\omega t) \right] \right\} \displaybreak \\
	\langle p \rangle = & - 2 e^{-\alpha_0^2} \sqrt{\frac{\hbar m \omega}{2}} \sum_{n=0}^\infty \frac{\alpha_0^{2n}}{n!} \left\{ 
	\alpha_0 \sin (- \Theta + \theta_n - \theta_{n+1}) + \nonumber \right.\\
	& + 2 \delta \sqrt{\frac{\hbar m \omega}{2}} \gamma \left\{ \alpha_0^2 \left[ \cos (- 2 \Theta + \theta_n - \theta_{n+2}) + 1 \right] + \frac{1}{2} \right\}+ \nonumber \\
	& \left. + 3 \alpha_0 \frac{\hbar m \omega}{2} \frac{\gamma^2}{2} \left[ \alpha_0^2 (3 \delta^2 + \epsilon) \sin (- 3 \Theta + \theta_n - \theta_{n+3}) + 4 \delta^2 \sin (- \Theta + \theta_n - \theta_{n+1}) \right] \right\} = \\ 
%
%
%
%
%
%
	\simeq & 2 \sqrt{\frac{\hbar m \omega}{2}} \left\{ 
	- \alpha_0 e^{-\alpha_0^2} \sum_{n=0}^\infty \frac{\alpha_0^{2n}}{n!} \sin \left\{ \omega t \left[ 1 - 6 \chi (n+1) \right] \right\} \right. \nonumber \\
	& + 2 \delta \sqrt{\frac{\hbar m \omega}{2}} \gamma \left[ \alpha_0^2 \left[ \cos (2 \omega t) + 1 \right] + \frac{1}{2} \right] + \nonumber \\
	& + \left. 3 \alpha_0^3 \frac{\hbar m \omega}{2} \frac{\gamma^2}{2} \left[ (3 \delta^2 + \epsilon) \sin (3 \omega t) + 4 \delta^2 \sin (\omega t) \right] \right\} ~,
\end{align}
\end{subequations}
where we considered $\alpha_0 \in \mathbb{R}$.
In obtaining these expressions, we expanded in Taylor series about $\gamma=0$ the terms explicitly multiplied by $\gamma$, retaining only the terms up to second order, while we did not expand the leading term.
Notice that in the limit $\gamma\rightarrow0$ one obtains the standard behavior for a HO coherent state.
Secondly, notice that focusing on the leading term, one can define a frequency for the number state components
\begin{equation}
	\omega_n = \omega \left[ 1 - 6 \chi (n+1) \right]~,
\end{equation}
with second order corrections in $\gamma$.
We can distinguish three cases.
If $\delta^2 = \epsilon$, the frequency does not depend on $n$.
An initially coherent state remains coherent and all the number state components move in phase-space with the same phase.
Remember that, as we saw in Sec.~\ref{sec:HO_GUP}, this condition also reproduces an equally spaced HO energy spectrum.
If $\delta^2 < \epsilon$, the frequency increases with $n$.
Therefore, larger number states move in phase-space faster than smaller number states, with a consequent spreading of the packet.
Finally, for $\delta^2 > \epsilon$, the frequency decreases with $n$.
In particular, there will be a value of $n$ after which the frequency will become negative.
Also in this case, each number state will move differently in \mbox{phase-space}, and the states will spread.

Notice that the ``relative velocities'' of each number state with respect the standard frequency $\omega$  are harmonics of $\Delta \omega = - 6 \chi \omega$.
Thus, although the states spread in \mbox{phase-space}, they evolve back together after a period of the fundamental harmonic
\begin{equation}
	T = \frac{\pi}{3 |\chi| \omega} \label{eqn:period_ev_coherent_states}~.
\end{equation}

From a different perspective, notice that the leading term can be written in a simpler way.
In fact, consider, \emph{e.g.}, the leading term for $\langle q \rangle$.
We can write
\begin{multline}
	2 \sqrt{\frac{\hbar}{2 m \omega}} \alpha_0 e^{-\alpha_0^2} \sum_{n=0}^\infty \frac{\alpha_0^{2n}}{n!}\cos \left\{ \omega t \left[ 1 - 6 \chi (n+1) \right] \right\} = \\
	= \sqrt{\frac{\hbar}{2 m \omega}} \alpha_0 e^{-\alpha_0^2} \left[ e^{ i \omega t (1 - 6 \chi) } \sum_{n=0}^\infty \frac{\alpha_0^{2n}}{n!} e^{ - 6 i \chi \omega t n } + e^{ - i \omega t (1 - 6 \chi) } \sum_{n=0}^\infty \frac{\alpha_0^{2n}}{n!} e^{ 6 i \chi \omega t n } \right] = \\
	\simeq \sqrt{\frac{\hbar}{2 m \omega}} \alpha_0 \left\{ e^{ i \omega t [1 - 6 \chi (\bar{n} + 1)] } + e^{ - i \omega t [1 - 6 \chi (\bar{n} + 1)] } \right\} = \\
	= 2 \sqrt{\frac{\hbar}{2 m \omega}} \alpha_0 \cos \left\{ \omega t [1 - 6 \chi (\bar{n} + 1)] \right\}~,
\end{multline}
where we considered that $\bar{n} = \langle \alpha^{(2)}| \tilde{N} | \alpha^{(2)} \rangle$ is the average number of phonons in a coherent state.
Similarly, for the leading term of the momentum, we can find
\begin{multline}
	- 2 \sqrt{\frac{\hbar m \omega}{2}} \alpha_0 e^{-\alpha_0^2} \sum_{n=0}^\infty \frac{\alpha_0^{2n}}{n!} \sin \left\{ \omega t \left[ 1 - 6 \chi (n+1) \right] \right\} = \\
	= - 2 \sqrt{\frac{\hbar m \omega}{2}} \alpha_0 \sin \left\{ \omega t \left[ 1 - 6 \chi (\bar{n}+1) \right] \right\}~.
\end{multline}
We then see that the phase $\Theta$ in Figure \ref{fig:coherent_state}, including terms depending on $\gamma$, varies differently than the standard value $\Theta_0 = \omega t$.
Using these expressions, it becomes easy to see that coherent states, as described by these terms, still follow very closely the classical behavior.
In particular, we obtain the following deviations from the classical kinematics of a HO
\begin{subequations} \label{eqns:ehrenfest}
\begin{align}
	\frac{\diff~\langle q \rangle}{\diff~t} = & \frac{\langle p \rangle}{m} \left[ 1 - 6 \chi (\bar{n} + 1)\right] ~, \\
	\frac{\diff~\langle p \rangle}{\diff~t} = & - m \omega^2 \langle q \rangle \left[ 1 - 6 \chi (\bar{n}+1) \right]~,
\end{align}
\end{subequations}
whose solutions are similar to \eqref{eqn:std_ehrenfest}
\begin{align}
	\langle q \rangle = & A \sin (\tilde{\omega} t) + B \cos (\tilde{\omega} t)~, & \langle p \rangle = & m \tilde{\omega} \left[ A \cos (\tilde{\omega} t) - B \sin (\tilde{\omega} t)\right]~,
\end{align}
with $\tilde{\omega} = \omega \left[ 1 - 6 \chi (\bar{n} + 1)\right]$.
We, therefore, have an effective change in the oscillation frequency, depending on the average number $\bar{n}$ and, therefore, on the amplitude $\alpha$.
We also observe that for $\delta^2 = \epsilon$, that is, for $\chi=0$, we obtain the standard Ehrenfest theorem \eqref{eqn:std_ehrenfest}.

To compute the time-dependent uncertainties in position and momentum for a coherent state, it is convenient to switch to Heisenberg picture.
We then find the time-evolved annihilation operator
\begin{equation}
	\tilde{a} (t) = e^{i H t/\hbar} \tilde{a} e^{-i H t / \hbar} = \sum_{k=0}^\infty \frac{1}{k!} \left( \frac{i t}{\hbar} \right)^k [^k H , \tilde{a}]~, \label{eqn:ev_a_1}
\end{equation}
where we used the notation of Appendix \ref{apx:exps}.
We can then prove that
\begin{equation}
	\left( \frac{i t}{\hbar} \right)^k [^k H , \tilde{a}] = (-i \omega t)^k \tilde{a} \sum_{j=0}^k \binom{k}{j} (- 6 \chi \tilde{N} )^j~. \label{eqn:commutators}
\end{equation}
It is easy to see that this is the case for $k=0$ and $k=1$.
Proceeding by induction, assuming the relation in \eqref{eqn:commutators} for a generic $k$, we want to prove it for $k+1$.
We then have
\begin{multline}
	\left( \frac{i t}{\hbar} \right)^{k+1} [^k H , \tilde{a}] 
	= (-1)^k (i \omega t)^{k+1} \left\{ \sum_{j=0}^k \binom{k}{j} (- 6 \chi)^j \left[\tilde{N} - 3 \chi (\tilde{N}^2 + \tilde{N}), \tilde{a} \tilde{N}^j \right] \right\} = \\
	= (-1)^k (i \omega t)^{k+1} \left\{ \sum_{j=0}^k \binom{k}{j} (- 6 \chi)^j \left( - \tilde{a} + 6 \chi \tilde{a} \tilde{N} \right) \tilde{N}^j \right\} = \\ 
	= (- i \omega t)^{k+1} \left\{ \sum_{j=0}^k \binom{k}{j} (- 6 \chi)^j \tilde{a} \tilde{N}^j + \sum_{j=1}^{k+1} \binom{k}{j-1} (- 6 \chi)^j \tilde{a} \tilde{N}^j \right\} = \\
	= (- i \omega t)^{k+1} \left\{ \tilde{a} + \sum_{j=1}^{k} \left[ \binom{k}{j} + \binom{k}{j-1} \right] (- 6 \chi)^j \tilde{a} \tilde{N}^j + (- 6 \chi)^{k+1} \tilde{a} \tilde{N}^{k+1} \right\} = \\
	= (- i \omega t)^{k+1} \left\{ \tilde{a} + \sum_{j=1}^{k} \binom{k+1}{j} (- 6 \chi)^j \tilde{a} \tilde{N}^j + (- 6 \chi)^{k+1} \tilde{a} \tilde{N}^{k+1} \right\} = \\
	= (- i \omega t)^{k+1} \sum_{j=0}^{k+1} \binom{k+1}{j} (- 6 \chi)^j \tilde{a} \tilde{N}^j ~.
\end{multline}
Using \eqref{eqn:ev_a_1}, we finally find
\begin{multline}
	\tilde{a} (t) = \sum_{k=0}^{\infty} \frac{1}{k!} (- i \omega t)^{k} \sum_{j=0}^{k} \binom{k}{j} (- 6 \chi)^j \tilde{a} \tilde{N}^j  = \tilde{a} \sum_{k=0}^{\infty} \frac{1}{k!} (- i \omega t)^{k} \left(1 - 6 \chi \tilde{N}\right)^k = \\
	= \tilde{a} e^{- i \omega t \left(1 - 6 \chi \tilde{N}\right)}~. \label{chain}
\end{multline}
Similarly, we can also find
\begin{equation}
	\tilde{a}^\dagger (t) = e^{i \omega t \left(1 - 6 \chi \tilde{N}\right)} \tilde{a}^\dagger ~.
\end{equation}
For computational purposes, it is useful to write these operators in the following form
\begin{align}
	\tilde{a} (t) = & e^{- i \omega t \left[1 - 6 \chi (\tilde{N} + 1)\right]} \tilde{a} ~, & \tilde{a}^\dagger (t) = & \tilde{a}^\dagger e^{ i \omega t \left[1 - 6 \chi (\tilde{N} + 1)\right]}~, \label{eqn:ev_annihil_creat_ops}
\end{align}
where to find the first of these relations we used the chain of equalities \eqref{chain}, observing that
\begin{equation}
	\tilde{a} \tilde{N}^j = (\tilde{N} + 1)^j \tilde{a}~,
\end{equation}
and similarly for $\tilde{a}^\dagger$.
Inserting the time-evolved operators \eqref{eqn:ev_annihil_creat_ops} in \eqref{eqns:expansions_q&p} and noticing that
\begin{multline}
	\langle \alpha^{(2)} | e^{i \beta \tilde{N}} | \alpha^{(2)} \rangle 
	= e^{-|\alpha|^2} \sum_{n=0}^\infty \frac{\alpha^\star {}^n}{\sqrt{n!}} \left[ \sum_{m=0}^{\infty} \frac{\alpha^m}{\sqrt{m!}} \langle n^{(2)} | e^{i \beta \tilde{N}} |m^{(2)}\rangle \right] = \\
	= e^{-|\alpha|^2} \sum_{n=0}^\infty \frac{|\alpha|^{2n}}{n!} e^{i \beta n} = e^{- |\alpha|^2 \left(1 - e^{i \beta} \right)}~,
\end{multline}
where we used the expansion \eqref{eqn:expansion_coherent_states}, we find the time-evolved position and momentum operators
\begin{subequations} \label{eqns:expansions_q&p_time-ev}
\begin{align}
	q(t) = & \sqrt{\frac{\hbar}{2 m \omega}} \left\{ e^ {- i \omega t [1 - 6 \chi (1 + \tilde{N})] } \tilde{a} + \tilde{a}^\dagger e^{i \omega t [1 - 6 \chi (1 + \tilde{N})]} \right\} + \nonumber \\
	& + 2 i \delta \frac{\hbar}{2} \gamma \left\{ e^{- i \omega t \left[2 - 6 \chi (3 + 2 \tilde{N})\right] } \tilde{a}^2 - \tilde{a}^\dagger {}^2 e^{i \omega t \left[2 - 6 \chi (3 + 2 \tilde{N})
 \right]}\right\} + \nonumber \\
	& - \epsilon \sqrt{\frac{\hbar^3 m \omega}{8}} \frac{3 \gamma^2}{2} \left\{ \tilde{a}^\dagger {}^3 e^{3 i \omega t \left[1 - 6 \chi (2 + \tilde{N})\right]} 
	- 2 \tilde{N} \tilde{a}^\dagger e^{i \omega t \left[1 - 6 \chi (1 + \tilde{N})\right]} + \right. \nonumber \\
	& \qquad \left.
	- 2 e^{- i \omega t \left[1 - 6 \chi (1 + \tilde{N})\right] } \tilde{a} \tilde{N} 
	+ e^{- 3 i \omega t \left[1 - 6 \chi (2 + \tilde{N})\right] } \tilde{a}^3 \right\} + \nonumber \\
	& - \delta^2 \sqrt{\frac{\hbar^3 m \omega}{8}} \frac{\gamma^2}{2} \left\{9 \tilde{a}^\dagger {}^3 e^{3  i \omega t \left[1 - 6 \chi (2 + \tilde{N})\right]} 
	+ 10 \tilde{N} \tilde{a}^\dagger e^{i \omega t \left[1 - 6 \chi (1 + \tilde{N})\right]} \right. \nonumber \\
	& \qquad \left. + 10 e^{- i \omega t \left[1 - 6 \chi (1 + \tilde{N})\right] } \tilde{a} \tilde{N} 
	+ 9 e^{- 3 i \omega t \left[1 - 6 \chi (2 + \tilde{N}) \right] } \tilde{a}^3 \right\} \label{eqn:expansion_q_time-ev} \\
	p(t) = & i \sqrt{\frac{\hbar m \omega}{2}} \left\{\tilde{a}^\dagger e^{i \omega t \left[1 - 6 \chi (1 + \tilde{N})\right]} - e^{- i \omega t \left[1 - 6 \chi (1 + \tilde{N})\right] } \tilde{a}\right\} + \nonumber \\
	& + 2 \delta \frac{\hbar m \omega}{2} \gamma \left\{e^{- i \omega t \left[2 - 6 \chi (3 + 2 \tilde{N})\right]} \tilde{a}^2 + \tilde{a}^\dagger {}^2 e^{i \omega t \left[2 - 6 \chi (3 + 2 \tilde{N})\right]} + 2 \tilde{N} + 1\right\} + \nonumber \\
	& + i \epsilon \left(\frac{\hbar m \omega}{2}\right)^{3/2} \frac{3\gamma^2}{2} \left\{e^{- 3 i \omega t \left[1 - 6 \chi (2 + \tilde{N})\right]} \tilde{a}^3 
	- \tilde{a}^\dagger {}^3 e^{3 i \omega t \left[1 - 6 \chi (2 + \tilde{N})\right]}\right\} + \nonumber \\
	& + i \delta^2 \left(\frac{\hbar m \omega}{2}\right)^{3/2} \frac{3\gamma^2}{2} \left\{3 e^{- 3 i \omega t \left[1 - 6 \chi (2 + \tilde{N})\right]} \tilde{a}^3 
	+ 4 e^{- i \omega t \left[1 - 6 \chi (1 + \tilde{N})\right] } \tilde{a} \tilde{N} + \right. \nonumber \\
	& \qquad \left. - 4 \tilde{N} \tilde{a}^\dagger e^{i \omega t \left[1 - 6 \chi (1 + \tilde{N})\right]} 
	- 3 \tilde{a}^\dagger {}^3 e^{3 i \omega t \left[1 - 6 \chi (2 + \tilde{N})\right]}\right\} ~.
\end{align}
\end{subequations}
Notice that for $t=0$ we obtain \eqref{eqns:expansions_q&p}.
It is also interesting to notice that, contrary to the standard theory, $q(T/4) \not = p$, where $T = 2 \pi/\omega$ is the period of one oscillation.
Using these relations, we find the following uncertainties
\begin{subequations}
\begin{align}
	(\Delta q)^2 \simeq & \frac{\hbar}{2 m \omega} \left\{1 + 2 \alpha_0^2 \left\{- e^{2 \alpha_0^2 \left[\cos (6 \omega t \chi) - 1\right]} + \right. \right. \nonumber \\
	& \left. \left. \qquad \qquad - 2 \sin \left\{\omega t \left[2 - 3 \chi (5 + 4 \alpha_0^2) \right] \right\} \sin ( 3 \omega t \chi) + 1\right\} \right\} + \nonumber \\
	& + \sqrt{\frac{\hbar^3}{8 m \omega}} 8 \alpha_0 \delta \sin (\omega t) \gamma
	+ \frac{\hbar^2}{4} \left[2 (2 \alpha_0^2 + 1) (3 \epsilon - \delta^2) - \alpha_0^2 (37 \delta^2 + 3 \epsilon) \cos (2 \omega t) \right] \gamma^2 
	 \nonumber \\
	\simeq & \frac{\hbar}{2 m \omega} \left[1 - 4 \alpha_0^2 \sin (2 \omega t) \sin ( 3 \omega t \chi)\right] + \nonumber \\
	& + \sqrt{\frac{\hbar^3}{8 m \omega}} 8 \alpha_0 \delta \sin (\omega t) \gamma
	+ \frac{\hbar^2}{4} (2 (2 \alpha_0^2 + 1) (3 \epsilon - \delta^2) - \alpha_0^2 (37 \delta^2 + 3 \epsilon) \cos (2 \omega t)) \gamma^2 \\
	(\Delta p)^2 \simeq & \frac{\hbar m \omega}{2} \left\{1 + 2 \alpha_0^2 \left\{1 + 2 \sin \left\{\omega t \left[ 2 - 3 \chi ( 5 + 4 \alpha_0^2)\right]\right\} \sin( 3 \omega t \chi) + \right.\right. \nonumber \\
	& \qquad \qquad \left.\left. + e^ {- 2 \alpha^2 \left[1-\cos (6 \omega t \chi)\right] }\right\}\right\} + \nonumber \\
	& + \left( \frac{\hbar m \omega}{2} \right)^2 \left\{ \alpha_0^2 (17 \delta^2-9 \epsilon) \left[\cos^2 (\omega t) - \sin^2 (\omega t)\right] - 4 (1 - 2 \alpha_0^2) \delta^2 \right\} \gamma^2 \nonumber \\
	\simeq & \frac{\hbar m \omega}{2} \left[1 + 4 \alpha_0^2 \sin (2 \omega t) \sin(3 \omega t \chi)\right] + \nonumber \\
	& + \left( \frac{\hbar m \omega}{2} \right)^2 \left( \alpha_0^2 (17 \delta^2-9 \epsilon) (\cos^2 (\omega t) - \sin^2 (\omega t)) - 4 (1 - 2 \alpha_0^2) \delta^2 \right) \gamma^2
\end{align}
\end{subequations}
Notice that, in the limit of standard theory $\gamma \rightarrow 0$, we obtain the usual results
\begin{equation}
	(\Delta q)^2 = \frac{\hbar}{2 m \omega} \qquad (\Delta p)^2 = \frac{\hbar m \omega}{2}~,
\end{equation}
while for $t=0$, we obtain, as expected, the results in \eqref{eqns:exp_unc_coherent}.

These two expressions indicate that the GUP imputes important effects on the uncertainties of position and momentum.
Consider the main term of each expression.
We notice an oscillation in the uncertainties, governed by the term $\sin (2 \omega t) \sin(- 3 \omega t \chi)$, with amplitude proportional to the standard uncertainties multiplied by $|\alpha|^2$.
This term, initially 0 for $t=0$, varies with time with an overall period
\begin{equation}
	T = \frac{\pi}{3 |\chi| \omega}~. \label{eqn:unc_period}
\end{equation}
The subsequent two terms in each expression present an oscillation with a similar period.
This oscillation is present only for models with $\delta^2 \not = \epsilon$.
Notice that this period is consistent with the period found for the spreading of coherent states in the Schr\"odinger picture.
We also notice that the behavior imposed by the term $\sin(2\omega t)$ is what we would expect if the uncertainty of a coherent state was not represented by a disk in quadrature-space, but rather by an ellipse, with its major axis parallel to the main diagonal, and rotating around its center with the same angular velocity $\omega$, similar to what we would expect for a squeezed state (see next Section).
Furthermore, the size of this ellipse, as we saw before, oscillates with a frequency $3 \omega \chi$.

For the systems considered at the end of Sec.~\ref{sec:HO_GUP}, we have the Table \ref{tbl:unc_osc_period}.
\begin{table}
\begin{center}
\begin{tabular}{lcccc}
	\toprule
	Type & Ref. & $m$ (Kg) & $\omega/2\pi$ (Hz) & $T \gamma_0^2 \left|\delta^2 - \epsilon\right| (s)$ \\
	\midrule
	Optomechanical system & \cite{Pikovski2012_1} & $10^{-11}$ & $10^5$ & $ 2.14 \times 10^{35}$ \\
	Bar detector AURIGA & \cite{Marin2013_1} & $1.1 \times 10^{3}$ & $900$ & $5.40 \times 10^{25}$ \\
	\multirow{4}{*}{Mechanical oscillators} & \multirow{4}{*}{\cite{Bawaj2014_1}} & $3.3 \times 10^{-5}$ & $5.64 \times 10^3$ & $2.04 \times 10^{31}$ \\
	&& $7.7 \times 10^{-8}$ & $1.29 \times 10^5$ & $1.67 \times 10^{31}$ \\
	&& $2 \times 10^{-8}$ & $1.42 \times 10^5$ & $5.31 \times 10^{31}$ \\
	&& $2 \times 10^{-11}$ & $7.47 \times 10^5$ & $1.91 \times 10^{33}$ \\
	LIGO detector & \cite{Abbott2016} & 40 & 200 & $1.34 \times 10^{28}$\\
	GW150914 & \cite{Abbott2016} & $3.2 \times 10^{31}$ & 100 & $0.0669$\\
	\bottomrule
\end{tabular}
\caption[Period of oscillation of $(\Delta q)^2$ and $(\Delta p)^2$ for several systems]{Period of oscillation of $(\Delta q)^2$ and $(\Delta p)^2$ for several systems, as given by \eqref{eqn:unc_period}. \label{tbl:unc_osc_period}}
\end{center}
\end{table}
We see that in most of the cases we have a period of the uncertainties oscillation several orders of magnitude larger than the age of the universe.
However, the last case is associated with a period of the order of tenth of seconds.
This is due to the very large characteristic mass for the system that generated GW150914 as compared to the other systems.

\begin{figure}
\center
\begin{subfigure}[t]{0.45\textwidth}
\center
\includegraphics[width=\textwidth]{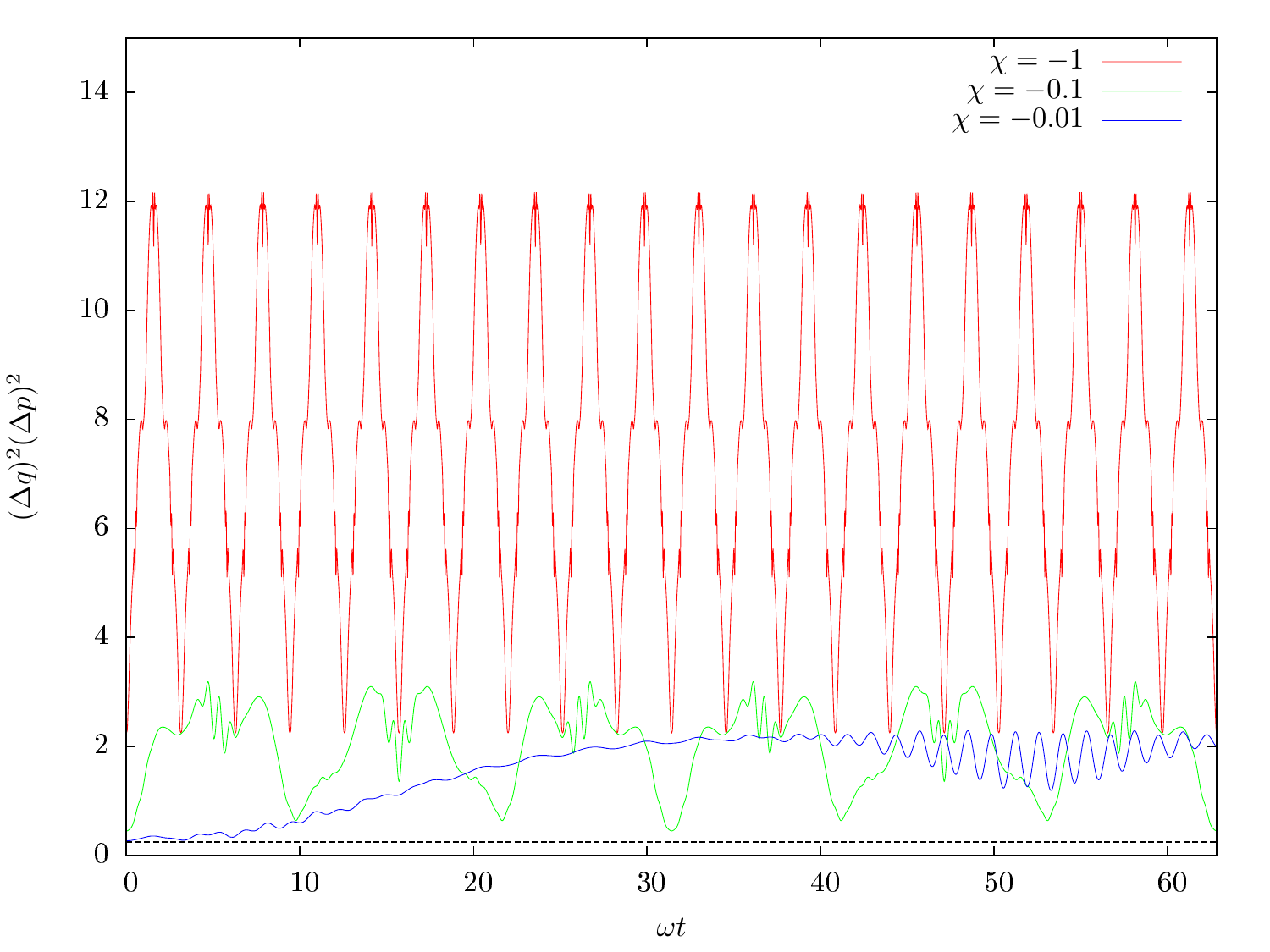}
\caption{The time evolution of the uncertainty product for coherent states is plotted for three values of $\chi$ in a model with $\delta=0$ and $\epsilon=1$ (quadratic GUP).
The black dashed line corresponds to the standard case.
The results are given in natural units.
Notice that each case presents small oscillations superposed to the dominant $\sin^2$ term.
Also, notice that the minimal value of the uncertainty product increases with $|\chi|$.} \label{fig:unc_pro_t_q}
\end{subfigure}
\qquad
\begin{subfigure}[t]{0.45\textwidth}
\includegraphics[width=\textwidth]{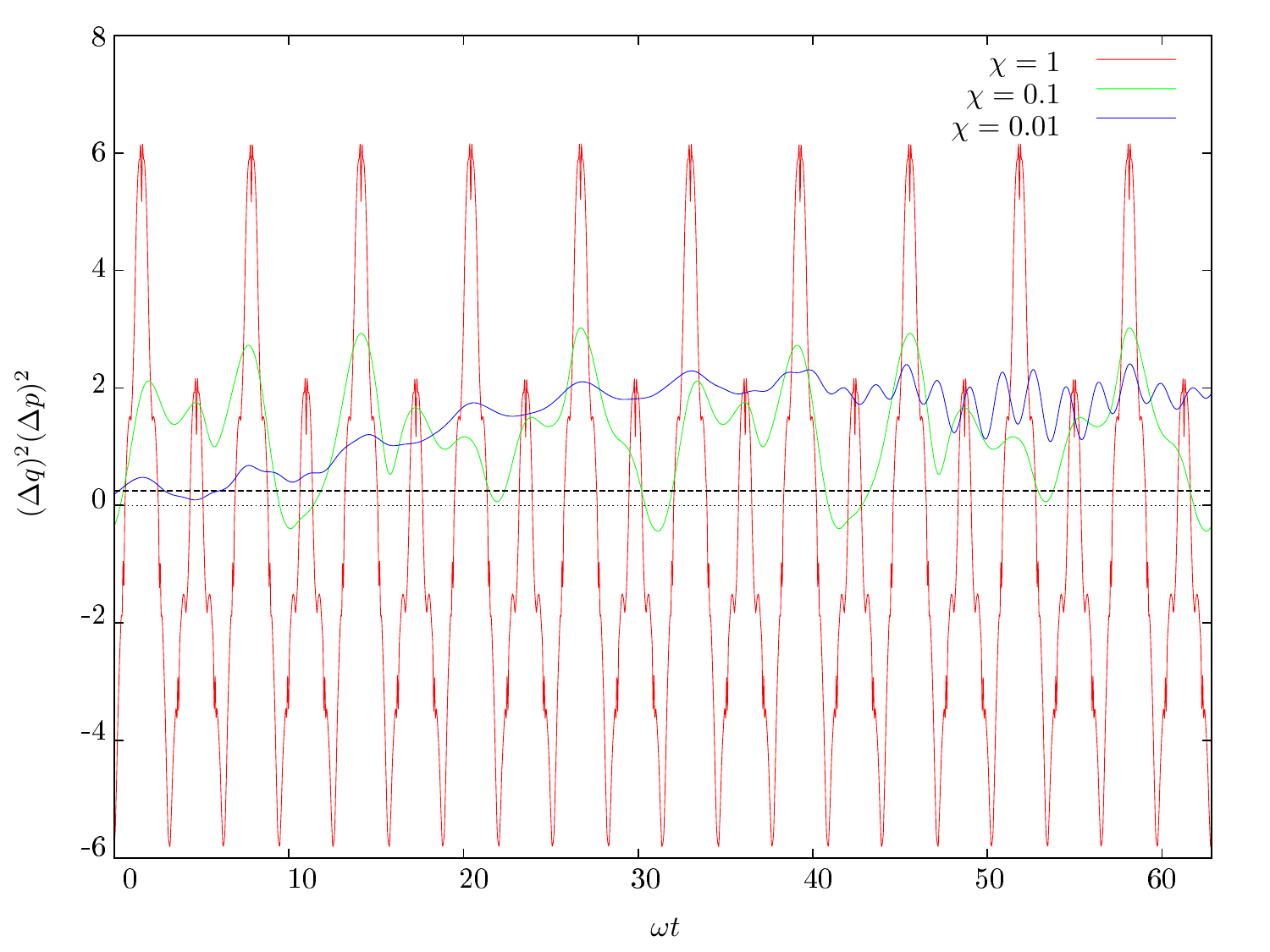}
\caption{Time evolution of the uncertainty product for coherent states for three values of $\chi$ in a model with $\delta=1$ and $\epsilon=0$ (linear GUP).
Similar observations as in the previous figure are valid here.
Here it is evident that a linear model can imply negative uncertainties and uncertainty product.} \label{fig:unc_pro_t_l}
\end{subfigure}
\begin{subfigure}{\textwidth}
\center
\includegraphics[scale=0.54]{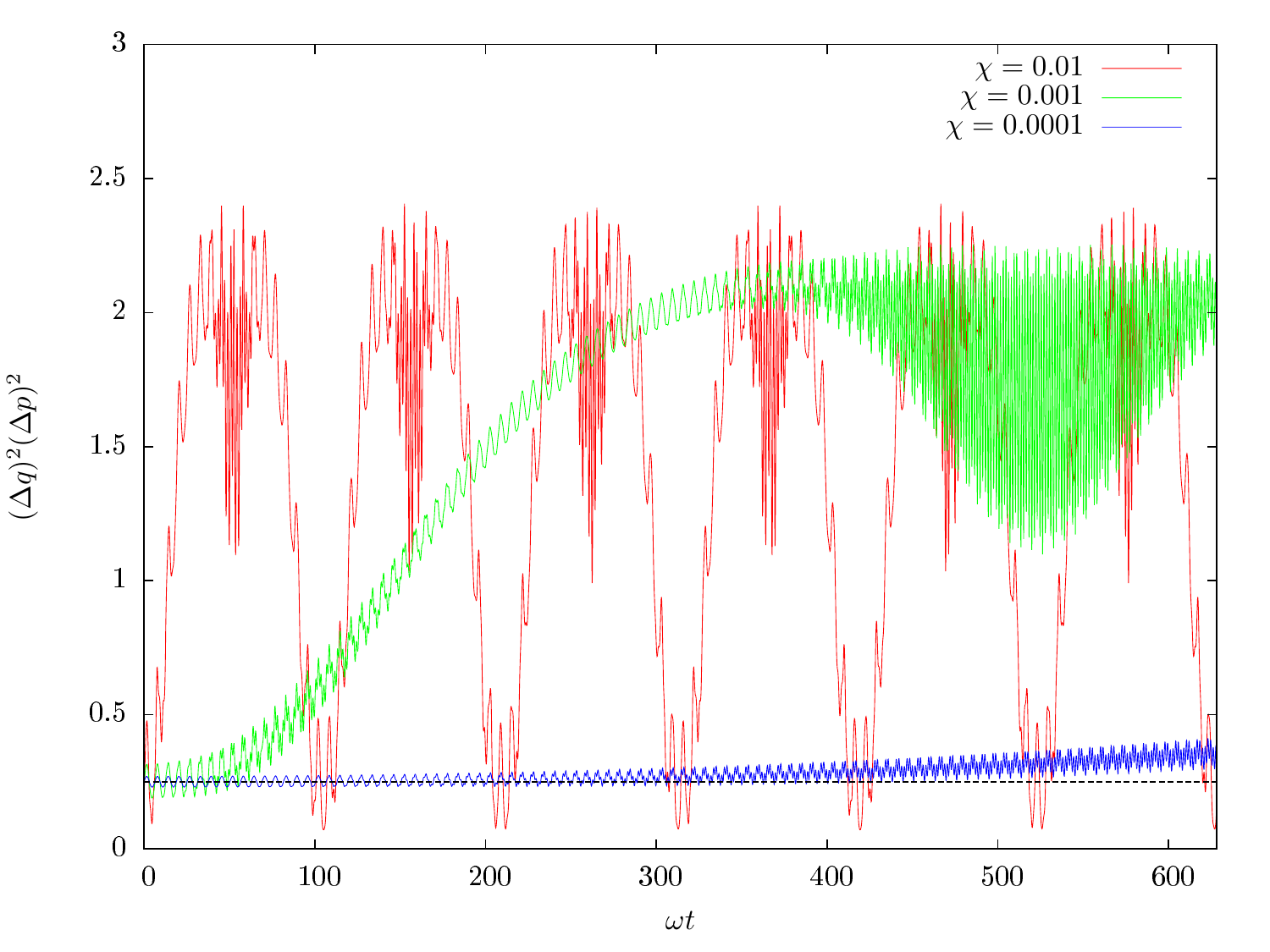}
\caption{Similar case as Figure \ref{fig:unc_pro_t_l}, but with the condition \eqref{eqn:negative_unc_pro} not satisfied. In this case we see that the uncertainty products are positive.} \label{fig:unc_pro_t_l1}
\end{subfigure} \caption[(a) Time-evolved coherent states for a quadratic GUP model. \protect\\
	(b) Time-evolved coherent states for a linear GUP model. \protect\\
	(c) Time-evolved coherent states for a linear GUP model with positive uncertainties.]{}\label{fig:time-ev-unc-prod}
\end{figure}

As for the uncertainty product, we have
\begin{multline}
	\left(\Delta q\right)^2 \left(\Delta p\right)^2 \simeq 
	\frac{\hbar^2}{4} \left\{1 - 16 \alpha_0^2 \sin^2(2 \omega t) \sin^2 (3 \omega t) + 8 \sqrt{\frac{\hbar m \omega}{2}} \delta \alpha_0 \sin ( \omega t) \gamma + \right. \\
	\left. - 2 \frac{\hbar m \omega}{2} \left[3 (\delta^2 - \epsilon) - 2 \alpha_0^2 (\delta^2 + 3 \epsilon) + 2 \alpha_0^2 ( 5 \delta^2 + 3 \epsilon ) \cos (2 \omega t) \right] \gamma^2 \right\}
\end{multline}
and it shows similar variations to those previously analyzed {(see also Fig. \ref{fig:time-ev-unc-prod})}.

\section{Squeezed States} \label{sec:squeezed_states}

The Heisenberg principle, as we saw in Chapter \ref{ch:GUP}, establishes that the product of the uncertainties of simultaneous position and momentum measurements cannot be smaller than $\hbar/2$.
On the other hand, though, it does not limit the individual uncertainties.
However, at the beginning of the previous Section we saw that coherent states in the standard theory are minimal uncertainty states, in the sense that they saturate the uncertainty inequality, but in terms of dimensionless quadratures the uncertainties have the same value.
It was therefore immediate to think of states that, remaining minimal uncertainty states, allowed different values of uncertainties in position and momentum.
In particular, one could ask how to define states whose uncertainties in position or momentum are ``squeezed'', while the other is increased.
The result is a class of states with no classical analogue and several different features with respect coherent states.
{However, the possibility of infinitely squeeze the uncertainty in position is in contrast with GUP.
We then expect that, when a minimal uncertainty in position is implemented for squeezed states, this will result in observable features.
Here we will see that this is the case.
Furthermore, as we will see in Sec.~\ref{ssec:LIGO}, the properties of squeezed states with GUP have direct implications on Michelson-Morley interferometers as LIGO.}

In this Section, complying with \cite{Lu1972}, we follow similar steps in GUP with respect to the standard theory, showing a number of new effects.
In particular, we define a new set of operators
\begin{align}
	\tilde{a}_r = & \tilde{a} \cosh r - e^{i\zeta} \tilde{a}^\dagger \sinh r~, & \tilde{a}^\dagger_r = \tilde{a}^\dagger \cosh r - e^{- i\zeta} \tilde{a} \sinh r~, \label{def:squeezed_ac_operators}
\end{align}
where $r$ is the squeeze parameter and $\zeta$ a phase angle.
Notice that in principles one could consider operators with different coefficients, but the choice of hyperbolic functions ensures the usual commutation relation $[\tilde{a}_r , \tilde{a}^\dagger_r] = 1$.
We can proceed as in the standard theory, defining a squeezed state such that
\begin{align}
	\tilde{a}_r |\alpha^{(2)}\rangle_r = & \alpha |\alpha^{(2)}\rangle_r ~, & \tilde{a}_r |0^{(2)}\rangle_r = & 0~. \label{eqn:annihilation_squeezed_states}
\end{align}
We can then find an expansion of this state in terms of number states.
Consider therefore the following expansion
\begin{equation}
	|0^{(2)}\rangle_r = \sum_{n=0}^\infty b_{n} |n^{(2)}\rangle~,
\end{equation}
where $b_n$ are in principle complex coefficients.
Using \eqref{eqn:annihilation_squeezed_states} we then have
\begin{multline}
	\tilde{a}_r |\alpha^{(2)}\rangle_r 
	= \sum_{n=0}^\infty b_{n} (\tilde{a} \cosh r - e^{i\zeta} \tilde{a}^\dagger \sinh r) |n^{(2)}\rangle = \\
	= \sum_{n=0}^\infty b_{n} (\sqrt{n} \cosh r | (n-1)^{(2)} \rangle - e^{i\zeta} \sqrt{n+1} \sinh r |(n+1)^{(2)}\rangle)
	= \alpha \sum_{n=0}^\infty b_{n} |n^{(2)}\rangle ~.
\end{multline}
Equating the coefficients for the same number states, we have the following recursion relation
\begin{align}
	\sqrt{n} b_n \cosh r - e^{i\zeta} \sqrt{n-1} b_{n-2} \sinh{r} = & \alpha b_{n-1} & \Rightarrow & & b_n = & \frac{\alpha b_{n-1} + e^{i\zeta} \sqrt{n-1} b_{n-2} \sinh{r}}{\sqrt{n} \cosh r} ~.
\end{align}
When a squeezed vacuum state $|0^{(2)}\rangle_r$ is considered, we find
\begin{subequations}
\begin{align}
	b_{2n} = & b_{2(n-1)} e^{i\zeta} \tanh r \sqrt{\frac{2n-1}{2n}} = b_0 \left( e^{i \zeta} \tanh r \right)^n \sqrt{\frac{(2n-1)!!}{(2n)!!}}~,\\
	b_1 = & 0 \Rightarrow b_{2n+1} = 0~.
\end{align}
\end{subequations}
Finally, we can determine $b_0$ assuming it real and using the normalization condition
\begin{align*}
	{}_r\langle 0^{(2)} | 0^{(2)} \rangle_r = & 1 & \Rightarrow & & \sum_{n=0}^\infty |b_{2n}|^2 = b_0^2 \left[ 1 + \sum_{n=1}^\infty (\tanh r)^{2n} \frac{(2n-1)!!}{2n!!} \right] = 1 \Rightarrow
\end{align*}
\begin{equation}
	\Rightarrow b_0 = \left[ 1 + \sum_{n=1}^\infty (\tanh r)^{2n} \frac{(2n-1)!!}{2n!!} \right]^{-1/2}
\end{equation}

Another equivalent way to define squeezed states is through the squeeze operator \linebreak \mbox{$\tilde{\mathcal{S}}(z) = \exp[-\frac{1}{2} ( z \tilde{a}^2 - z^\star \tilde{a}^\dagger {}^2) ]$}, where $z=r e^{i\zeta}$.
In this case we have
\begin{align}
	\tilde{a}_z = & \tilde{\mathcal{S}}^\dagger (z) ~ \tilde{a} ~ \tilde{\mathcal{S}}(z)~, & |0^{(2)}\rangle_r = \tilde{\mathcal{S}}(z) |0^{(2)}\rangle~.
\end{align}
We can then find the expectation values of position and momentum for a squeezed state
\begin{subequations} \label{eqns:exp_q&p_squeezed}
\begin{align}
	\langle q \rangle_r = & \sqrt{\frac{\hbar}{2 m \omega}} \left\{ (\alpha + \alpha^\star) 
	- 2 \delta \sqrt{\frac{\hbar m \omega}{2}} \gamma \left[ i (\alpha^\star + \alpha) (\alpha^\star - \alpha) - \sinh (2 r) \sin \zeta\right] \right. + \nonumber\\
	& + \frac{1}{2} \frac{\hbar m \omega}{2} \gamma^2 \left\{\delta^2 \left\{ - (\alpha^\star+\alpha) \left[10 \cosh (2 r) + 9 \alpha^\star{}^2 + |\alpha|^2 + 9 \alpha^2\right] + \right. \right. \nonumber \\
	& \left. + \frac{27}{2} \sinh (2 r) (\alpha^\star e^{- i \zeta} + \alpha e^{i \zeta}) + 5 \sinh (2 r) (\alpha^\star e^{i \zeta} + \alpha e^{- i \zeta}) \right\} + \nonumber \\
	& - 3 \epsilon \left\{ - (\alpha^\star + \alpha) \left[2 \cosh (2r) - \alpha^\star {}^2 + 3 |\alpha|^2 - \alpha^2\right] + \right. \nonumber \\
	& \left. \left. \left. - 3 \sinh (2 r) (\alpha^\star e^ {- i \zeta} + \alpha e^{i \zeta}) + \sinh (2 r) (\alpha^\star e^{i \zeta} +  \alpha e^ {- i \zeta}) \right\} \right\} \right\}~, \displaybreak \\
	\langle p \rangle_r = & i \sqrt{\frac{\hbar m \omega}{2}} \left\{ (\alpha^\star - \alpha)
	- 2 i \delta \sqrt{\frac{\hbar m \omega}{2}} \left[ (\alpha + \alpha^\star)^2 + \cosh (2 r) - \sinh (2 r) \cos \zeta \right] \gamma + \nonumber \right. \\
	& - \frac{3}{4} \frac{\hbar m \omega}{2} \gamma^2 \left\{\epsilon \left[ (\alpha^\star - \alpha) ( \alpha^2 + |\alpha|^2 + \alpha^\star {}^2 ) - \frac{3}{2} \sinh (2 r) (\alpha^\star e^ {- i \zeta} - \alpha e^{i \zeta}) \right] \right. \nonumber \\
	& + \delta^2 \left\{ 2 \sinh (2 r) (\alpha^\star e^{i \zeta} - \alpha e^{- i \zeta}) - \frac{9}{2} \sinh (2 r) (\alpha^\star e^ {- i \zeta} - \alpha e^{i \zeta}) + \right. \nonumber \\
	& \left. \left. \left. + (\alpha^\star - \alpha) ( 4 \cosh (2 r) + 3 \alpha^2 + 7 |\alpha|^2 + 3 \alpha^\star {}^2 )  \right\} \right\} \right\}~.
\end{align}
\end{subequations}
Considering the case $\zeta = 0$, we can invert the relations (\ref{def:squeezed_ac_operators}) and study the properties of squeezed states using (\ref{eqns:expansions_q&p}), obtaining the following expressions for the uncertainties
\begin{subequations}
\begin{align}
	(\Delta q)^2 = & \frac{\hbar}{2 m \omega} \left\{ e^{-2r} - 4 i (\alpha^\star - \alpha) \delta \sqrt{\frac{\hbar m \omega}{2}} e^{- 2 r} \gamma + \right. \nonumber \\
	& + \frac{\hbar m \omega}{2} \frac{\gamma^2}{4} \left\{ 3 \epsilon (5 - 2 (\alpha^\star - \alpha)^2 e^{- 2 r} + 12 |\alpha|^2 e^{- 2 r} + 3  e^{- 4 r} ) + \right. \nonumber \\
	& \left. \left. + \delta^2 \left[ 16 (\alpha^\star + \alpha)^2 e^{2 r} + 49 - 90 (\alpha^\star + \alpha)^2 e^{- 2 r} + 132 |\alpha|^2 e^{- 2 r} - 57  e^{- 4 r} \right] \right\} \right\} \\
	(\Delta p)^2 = & \frac{\hbar m \omega}{2} \left\{ e^{2r} + \frac{\hbar m \omega}{2} \frac{\gamma^2}{4} \left\{ 9 \epsilon (e^{4 r} - 2 \alpha^\star {}^2 e^{2 r} - 2 \alpha^2 e^{2 r} - 1) + \right. \right. \nonumber \\
	& \left. \left. - \delta^2 \left[ - 32 e^{- 4 r} - 64 (\alpha^\star + \alpha)^2 e^{- 2 r} + 39 + 30 (\alpha^\star + \alpha)^2 e^{2 r} + \right. \right. \right. \nonumber \\
	& \left. \left. \left. \qquad + 36 |\alpha|^2 e^{2 r} + 9 e^{4 r} \right] \right\} \right\} ~, \\
	(\Delta q)^2 (\Delta p)^2 = & \frac{\hbar^2}{4} \left\{ 1 - 4 i \delta \sqrt{\frac{\hbar m \omega}{2}} \gamma (\alpha^\star - \alpha) + 2 \left\{ 3 \epsilon  (e^{2r} - (\alpha^\star - \alpha)^2) + \right. \right. \nonumber \\
	& + \delta^2 \left[ 2 (\alpha^\star + \alpha)^2 (e^{2r} - 2 e^{-2r})^2 + e^{2r} - 7 \alpha^\star {}^2 - 2 |\alpha|^2 - 7 \alpha^2 - 4 e^{-2r} + \right. \nonumber \\
	& \left. \left. \left. + 16 e^{- 2 r } \sinh^2 (2r) \right] \right\} \frac{\hbar m \omega}{2} \gamma^2 \right\}~,
\end{align}
\end{subequations}
to order $\gamma^2$. 
As expected, these reduce to quantities in the standard theory for a vanishing GUP parameter $\gamma \rightarrow 0$.
In this case too, we can compute the theoretical minimal uncertainty through the \mbox{Schr\"odinger--Robertson} relation \eqref{eqn:Sch-Ro}
{
\begin{multline}
	[(\Delta q)^2 (\Delta p)^2]_{\mathrm{min}} = \frac{\hbar^2}{4} \left\{ 1 - 4 i (\alpha^\star - \alpha) \delta \sqrt{\frac{\hbar m \omega}{2}} \gamma + 2 \left\{ 3 \epsilon (e^{2r} - (\alpha^\star - \alpha)^2) + \right. \right. \\
	+ \left. \left. \delta^2 \left[ 2 (\alpha^\star + \alpha)^2 (e^{2r} - 2 e^{-2r})^2 + e^{2r} - 7 \alpha^\star {}^2 - 2 |\alpha|^2 - 7 \alpha^2 - 4 e^{-2r}\right] \right\} \frac{\hbar m \omega}{2} \gamma^2 \right\}
\end{multline}
and thus find 
\begin{equation}
	(\Delta q)^2 (\Delta p)^2 - [(\Delta q)^2 (\Delta p)^2]_{\mathrm{min}} = 32 \delta^2 \frac{\hbar^2}{4} \frac{\hbar m \omega}{2} e^{-2r} \sinh^2 (2 r) \gamma^2 \label{eqn:diff_unc}
\end{equation}
for the difference with the minimal uncertainty product.
Several features are noteworthy.
First, the above difference is positive for all values of $r$ and does not depend on $\alpha$.
Second, the difference with the minimal product is of the second order in the GUP parameter.
Finally, it is present only for models with a nonzero linear term; it is not present in the model introduced in \cite{Kempf1995_1}.
Finding evidence for the relation \eqref{eqn:diff_unc} could therefore help distinguish the models.}

{Furthermore, w}e can observe a number of features that discriminate the standard results from those presented above.
\begin{figure}
\begin{center}
	\includegraphics[width=0.5\textwidth]{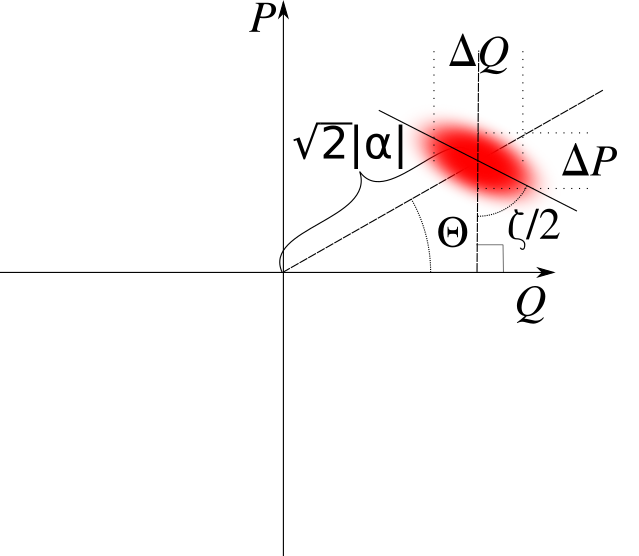}
	\caption{Quadrature-space representation of a squeezed state.}\label{fig:squeezed_state}
\end{center}
\end{figure}
First, we observe that the expectation values of position and momentum in the standard case, obtained from \eqref{eqns:exp_q&p_squeezed} in the limit $\gamma \rightarrow 0$, do not depend on $r$ or $\zeta$.
As for GUP, for simplicity, consider the case of $\alpha=\alpha_0 \in \mathbb{R}$.
The expressions in \eqref{eqns:exp_q&p_squeezed} now read
\begin{subequations}
\begin{align}
	\langle q \rangle_r = & \sqrt{\frac{\hbar}{2 m \omega}} \left\{ 2 \alpha_0 
	+ 2 \delta \sqrt{\frac{\hbar m \omega}{2}} \gamma \sinh (2 r) \sin \zeta \right. + \nonumber\\
	& + \frac{1}{2} \frac{\hbar m \omega}{2} \gamma^2 \left\{\delta^2 \alpha_0 \left[ - 20 \cosh (2 r) + 37 \sinh (2 r) \cos(\zeta) - 38 \alpha_0^2 \right] + \right. \nonumber \\
	& \left. + 6 \epsilon \alpha_0 \left[2 \cosh (2r) + 2 \sinh (2 r) \cos (\zeta) + \alpha_0^2 \right] \right\}~, \\
	\langle p \rangle_r = & - i \sqrt{\frac{\hbar m \omega}{2}} \left\{ 2 i \delta \sqrt{\frac{\hbar m \omega}{2}} \left[ 4 \alpha_0^2 + \cosh (2 r) - \sinh (2 r) \cos \zeta \right] \gamma + \nonumber \right. \\
	& \left. + \frac{3}{4} i \frac{\hbar m \omega}{2} \gamma^2 \alpha_0 \sinh (2 r) \sin (\zeta) ( 3 \epsilon + 13 \delta^2) \right\}
\end{align}
\end{subequations}
We then see that in this case, the expectation values of position and momentum, defining the center of the uncertainty ellipse in the quadrature-space, do depend on $r$ and $\zeta$.

Furthermore, in the standard case, the uncertainties are given by ($\zeta=0$)
\begin{align}
	(\Delta q)^2 = & \frac{\hbar}{2 m \omega} e^{-2r}~, & (\Delta p)^2 = & \frac{\hbar m \omega}{2} e^{2r}~.
\end{align}
Therefore, we see that the uncertainties depend only on the parameter $r$, while in GUP they also depend on the phase of $\alpha$, that in turn, to 0-th order in $\gamma$, depends on time.

\subsection{Minimal Uncertainties for Squeezed States}

As we said at the beginning of this Section, squeezed states saturate the uncertainty relation allowing for reduced uncertainties, with respect to coherent states, in either position or momentum.
In principle, the uncertainty in position (momentum) can be made arbitrarily small for HUP-saturating states.
However, the  GUP  introduces  a minimal uncertainty in position.
Here we study the implications of this fact.

To better observe the features of optimal squeezing in GUP, we consider the particular model examined in \cite{Kempf1995_1}.
This model corresponds to $\delta = 0$ and $\epsilon = 1/3$.
The position uncertainty for this model is
\begin{equation}
	(\Delta q)^2 = \frac{\hbar}{2 m \omega} \left\{ e^{-2r} + \frac{\hbar m \omega}{2} \frac{\gamma^2}{4} \left[5 - 2 (\alpha^\star - \alpha)^2 e^{- 2 r} + 12 |\alpha|^2 e^{- 2 r} + 3  e^{- 4 r} \right] \right\}~.
\end{equation}
We obtain a minimal position uncertainty for $e^{-r} = 0 \Rightarrow r = + \infty$.
It is
\begin{equation}
	(\Delta q)^2 = \frac{\hbar^2}{4} \frac{5}{4} \gamma^2  > (\Delta q)^2_{\mathrm{min}} = \hbar^2 \frac{\gamma^2}{4}
\end{equation}
whereas the latter quantity has been previously computed \cite{Kempf1995_1}.
Therefore, the maximally squeezed uncertainty in position is always larger than the minimal uncertainty predicted by the model.

The momentum uncertainty for the same model is
\begin{equation}
	(\Delta p)^2 = \frac{\hbar m \omega}{2} \left[ e^{2r} + \frac{3}{4} \frac{\hbar m \omega}{2} \gamma^2 \epsilon (e^{4 r} - 2 \alpha^\star {}^2 e^{2 r} - 2 \alpha^2 e^{2 r} - 1) \right] ~,
\end{equation}
and is minimized for  $e^{2r} = 0 \Rightarrow r = - \infty$, the minimal uncertainty being
\begin{equation}
	(\Delta p)^2 = - \frac{3}{4} \left(\frac{\hbar m \omega}{2} \right)^2 \gamma^2  
\end{equation}
which is negative. This means that this model cannot describe  infinite squeezing.
Rather, by inverting the reasoning, there must exist a lower value of $r$ for the interval of squeezing in which GUP can be used.
In fact, when we have
\begin{multline}
	e^{2r} = \frac{ - 4 + 3 (\alpha^\star {}^2 + \alpha^2) \hbar m \omega \gamma^2 + \sqrt{ 9\hbar^2 m^2 \omega^2 \gamma^4 + [ 4 - 3 (\alpha^2 + \alpha^\star {}^2) \hbar m \omega \gamma^2 ]^2 }}{3 \hbar m \omega \gamma^2} \\
	\simeq \frac{3 \hbar m \omega \gamma^2}{8}~, \label{eqn:max_squeezing_mom_kmm}
\end{multline}
the squeezed uncertainty in momentum vanishes.

We next turn to the consequences for the model in \cite{Ali2011_1}, for which $\delta=1$ and $\epsilon=1$.
For simplicity, we consider a squeezed vacuum state, \emph{i.e.} $\alpha=0$.
The position uncertainty in this case is
\begin{equation}
	\left(\Delta q\right)^2 = \frac{\hbar}{2 m \omega}  e^{-2r} + 4 \frac{\hbar^2}{4} (4 - 3 e^{-4r}) \gamma^2~.
\end{equation}
It has a minimum at $e^{-r} = 0 \Rightarrow r = + \infty$, corresponding to
\begin{equation}
	\left(\Delta q\right)^2 = 16 \frac{\hbar^2}{4} \gamma^2 >  \left(\Delta q\right)^2_{\mathrm{min}} = 8 \frac{\hbar^2}{4} \gamma^2
\end{equation}
where the latter quantity is  the minimal length allowed by the model in \cite{Ali2011_1}. 

As for the uncertainty in momentum,  we find
\begin{equation}
	(\Delta p)^2 = \frac{\hbar m \omega}{2} e^{2 r} - 4 \left(\frac{\hbar m \omega}{2}\right)^2 (2 e^{-4r} - 3) \gamma^2 
	\end{equation}
for this model.
Given the presence of both decreasing and increasing exponentials in $r$, there is no minimal value for the momentum uncertainty. Instead
we see that $\Delta p$
  vanishes for
\begin{equation}
	e^{2r_\mathrm{min}} = 4 \frac{\hbar m \omega}{2} \gamma^2 + \frac{32}{3} \left(\frac{\hbar m \omega}{2} \right)^{2/3} \gamma^{4/3} - 2 \left( \frac{\hbar m \omega}{2} \right)^{1/3} \gamma^{2/3} + \mathcal{O}(\gamma^{8/3})~.
\end{equation}
As with the previous model, the value $r_\mathrm{min}$ for the squeeze parameter can be interpreted as the smallest physical value of $r$ 
permitted in this model, since smaller values yield \mbox{$(\Delta p)^2 < 0$}.

Finally, in both models we observe similar features: the uncertainties in position have a non-vanishing minimal value larger than the minimal length allowed by the models.
These uncertainties are achieved for infinite squeezing.
On the other hand, minimal uncertainties in momentum, as computed by the GUP models, can be negative, signaling the break down of GUP for high energies.
Therefore, GUP can describe squeezing in momentum uncertainties only up to particular limits for the squeeze parameter.

\chapter{Amplification of Planck scale effects in quantum optical system} \label{ch:QO}

\begin{quote}
	\begin{verse}
		Ognuno sta solo sul cuor della terra\\
		trafitto da un raggio di sole:\\
		ed \`e subito sera.
	\end{verse}
	
	\begin{flushright}
		Salvatore Quasimodo
	\end{flushright}
\end{quote}

So far, we have focused our attention on the theoretical aspects of incorporating a minimal length in the uncertainty principle.
As we stated in Chapter \ref{ch:GUP}, experiments are of vital importance in Physics.
We will, therefore, dedicate this Chapter to an experimental proposal to test GUP effects in a quantum optomechanical system.

\section{Quantum Cavity Optomechanics}

One of the most active research areas in Physics is Quantum Optics, in particular Quantum Optomechanics.
It is ultimately motivated by investigations on the quantum properties of light and its interactions with matter, resulting in mechanical effects.
Cavity Optomechanics, for example, which will be the main topic of this Chapter, concerns with the interaction of a light field in a cavity.
Applications of this branch of Physics span from sensitivity enhancement in gravitational wave detection, where the needed precision requires taking into account for quantum effects of light as well as reducing noises \cite{Caves1981_1}, to fundamental research in Quantum Optics and QG, as we will see below.

In this Section, we will briefly review the Hamiltonian formulation of Cavity Optomechanics, following \cite{Aspelmeyer2013_1}.
To derive the Hamiltonian describing the optomechanical interaction in a cavity, we consider only the closest optical mode to resonance with the driving laser and one mechanical normal mode.
This choice will provide a valid approximation as long as the dynamics is linear with independently evolving normal modes.

Consider uncoupled cavity (optical) and mechanical modes, with frequency $\omega_\mathrm{C}$ and $\omega_\mathrm{M}$ respectively, represented by two harmonic oscillators.
The Hamiltonian for this system is thus
\begin{equation}
	H_0 = \hbar \omega_\mathrm{C} a^\dagger a + \hbar \omega_\mathrm{M} b^\dagger b~, \label{eqn:hamiltonian_cavity}
\end{equation} 
where $a$ ($a^\dagger$) and $b$ ($b^\dagger$) are the annihilation (creation) operators for the cavity and the mechanical modes, respectively.
Consider now a cavity with a movable end mirror and a parametric coupling between the optical and mechanical modes.
This means that the cavity resonance frequency is modulated by the {displacement of the movable mirror} \cite{Aspelmeyer2013_1}
{\begin{equation}
	\omega_\mathrm{C}(q) \approx \omega_{\mathrm{C},0} + q \frac{\partial \omega_\mathrm{C}}{\partial q} + \ldots~,
\end{equation}
where $\omega_{\mathrm{C},0} \equiv \omega_\mathrm{C}(0)$ is the cavity frequency when the mirror is at its equilibrium position.}
For a simple cavity of length $L$, considering a mechanical amplitude $|q| \ll L$, we can define the optical frequency shift per displacement
\begin{equation}
	{G \equiv - \frac{\partial \omega_\mathrm{C}}{\partial q} \simeq \frac{\omega_{\mathrm{C},0}}{L}}~.
\end{equation}
{For the last approximate equality, consider that, f}or a fixed cavity, the resonance frequency is related to the length $L$ by means of the relation
\begin{equation}
	\omega_\mathrm{C} = \frac{c}{\lambda}2\pi = \frac{2nc}{L}2\pi~,
\end{equation}
where the integer $n$ selects the mode of interest.
Considering a cavity with a movable end mirror, the length $L$ varies with the mechanical amplitude $q$, \emph{i.e.}
\begin{equation}
	\omega_\mathrm{C} = \frac{4\pi nc}{L+q}~.
\end{equation}
Hence, the variation of the cavity frequency is given by
\begin{equation}
	\frac{\partial \omega_\mathrm{C}}{\partial q} = - \frac{4\pi nc}{(L+q)^2} = - \frac{\omega_\mathrm{C}}{L+q}\simeq - \frac{\omega_{\mathrm{C},0}}{L}~,
\end{equation}
where we considered $|q| \ll L$.
Therefore, the optical part of the Hamiltonian in (\ref{eqn:hamiltonian_cavity}) in the case of an optomechanical coupling is
\begin{equation}
	H^\mathrm{opt} = \hbar \omega_\mathrm{C} a^\dagger a \approx \hbar (\omega_\mathrm{C} - G q)a^\dagger a = \hbar \omega_\mathrm{C} a^\dagger a - \hbar g_0 a^\dagger a Q_\mathrm{M}~, \label{eqn:hamiltonian_optics}
\end{equation}
where $Q_\mathrm{M}$ is the position quadrature for the mechanical oscillator, as defined by \eqref{def:quadratures}, and where we defined the vacuum optomechanical coupling strength expressed as a frequency
\begin{equation}
	g_0 = G \sqrt{\frac{\hbar}{m \omega_\mathrm{M}}}~.
\end{equation}
It describes the interaction between a single phonon and a single photon.

Using Hamilton equations \cite{Goldstein_H} the radiation pressure force for this interaction is
\begin{equation}
	F = - \frac{\diff~H_\mathrm{int}^\mathrm{opt}}{\diff~q} = \hbar G a^\dagger a = g_0 \sqrt{\hbar m \omega_\mathrm{M}} a^\dagger a.
\end{equation}
It is convenient to change the description of the optical mode by switching to a frame rotating at the laser frequency $\omega_\mathrm{L}$.
Given a unitary operator $U$ representing a transformation, the Hamiltonian changes as
\begin{equation}
	H' = U H U^\dagger + i \hbar \frac{\partial U}{\partial t} U^\dagger~. \label{eqn:rotated_hamiltonian}
\end{equation}
In our particular case, rotating the system with a unitary transformation $U = \exp(i \omega_\mathrm{L} a^\dagger a t)$ makes the driving terms time-independent.
In fact, such terms, for a coherent input field, are of the form \cite{PikovskiPhD}
\begin{equation}
	H_\mathrm{d} = i \hbar (E e^{-i \omega_\mathrm{L} t} a^\dagger - E^\star e^{i \omega_\mathrm{L} t} a) ~,
\end{equation}
where $E$ is related to the input power $P$ by
\begin{equation}
	E = \sqrt{\frac{2 \kappa P}{\hbar \omega_\mathrm{L}}}~,
\end{equation}
with $\kappa$ the loss rate at the input cavity mirror \cite{Aspelmeyer2013_1}.
Applying the transformation $U$ to these terms we have
\begin{equation}
	U (E e^{-i \omega_\mathrm{L} t} a^\dagger - E^\star e^{i \omega_\mathrm{L} t} a) U^\dagger = E U a U^\dagger e^{-i\omega_\mathrm{L} t} + E^\star U a^\dagger U^\dagger e^{i\omega_\mathrm{L} t}~.
\end{equation}
Considering the following commutation relations
\begin{align}
	[a^\dagger a, a] = &  - a & [a^\dagger a, a^\dagger] = & a^\dagger~,
\end{align}
and using the identity \eqref{eqn:hadamard}
\begin{equation}
	e^{A} B e^{-A} = B + [A,B] + \ldots + \frac{1}{n!} \underbrace{[A,[A\ldots,[A,[A,}_{\mbox{n times}}B]]\ldots]] + \ldots~,
\end{equation}
we find
\begin{multline}
	\left. \begin{array}{cc}
		U a U^\dagger = & e^{i\omega_L a^\dagger a t} a e^{- i\omega_\mathrm{L} a^\dagger a t} = a e^{- i\omega_\mathrm{L} t} \\
		U a^\dagger U^\dagger = & e^{i\omega_\mathrm{L} a^\dagger a t} a^\dagger e^{- i\omega_\mathrm{L} a^\dagger a t} = a^\dagger e^{i\omega_\mathrm{L} t} 
	\end{array} \right\}
	\Rightarrow U (E e^{-i \omega_\mathrm{L} t} a^\dagger - E^\star e^{i \omega_\mathrm{L} t} a) U^\dagger = \\
	= E a^\dagger - E^\star a~.
\end{multline}
Using \eqref{eqn:rotated_hamiltonian}, this unitary operator generates a new Hamiltonian
\begin{equation}
	H' = -\hbar \Delta a^\dagger a + \hbar \omega_\mathrm{M} b^\dagger b - \hbar g_0 a^\dagger a Q_\mathrm{M} + \ldots~, \label{eqn:hamiltonian_rotated}
\end{equation}
where
\begin{equation}
	\Delta = \omega_L - \omega_\mathrm{cav}
\end{equation}
is the laser detuning.
The ellipsis is for additional terms describing driving, decay and fluctuations that are not of interest in what follows.

To interpret the Hamiltonian \eqref{eqn:hamiltonian_rotated}, in particular the interaction term, consider a strong coherent optical field $|\alpha\rangle$.
In pulsed optomechanics, the timescale of the interaction of the optical field with the mechanical oscillator is shorter than the mechanical period.
With the additional condition of an optical field on resonance with the cavity and neglecting vacuum fluctuations, the Hamiltonian \eqref{eqn:hamiltonian_rotated} reduces to
\begin{equation}
	H_\mathrm{int}^\mathrm{opt} \simeq \hbar g_0 \alpha^2 Q_\mathrm{M} - \sqrt{2} \hbar g_0 \alpha Q_\mathrm{L} Q_\mathrm{M}~,
\end{equation}
where we chose $\alpha$ to be real and where $Q_\mathrm{L}$ is the optical amplitude quadrature.
Therefore, the unitary operator representing the interaction between the optical field and the mechanical oscillator can be written as
\begin{equation}
	\mathcal{U} = e^{- i\Omega Q_\mathrm{M}} e^{i \varpi Q_\mathrm{L} Q_\mathrm{M}}~,
\end{equation}
where $\varpi = \sqrt{2} g_0 \alpha \tau$, $\Omega = g_0 \alpha^2 \tau$, $\tau \sim 1/\kappa$ is the effective interaction time and $\kappa$ is the bandwidth of the cavity, supposed to be large.
We can interpret the first term in the operator $\mathcal{U}$ as a changing in the momentum of the mechanical oscillator deriving from radiation pressure, while its position, for the second term, becomes correlated with the optical phase quadrature
\begin{equation}
	P_\mathrm{L} = \mathcal{U}^\dagger P_\mathrm{L} \mathcal{U} = e^{-i \varpi Q_\mathrm{L} Q_\mathrm{M}} \frac{i}{\sqrt{2}} (a^\dagger - a) e^{i \varpi Q_\mathrm{L} Q_\mathrm{M}} = P_\mathrm{L} + \varpi Q_\mathrm{M}~,
\end{equation}
where we considered that the mechanical terms commute with the optical ones.
This quadrature can then be measured through balanced homodyne detection.
In this scheme, the signal to be analyzed and a strong local oscillator coherent field, described by the annihilation operator $a_\mathrm{LO}$, are mixed on a symmetric beam splitter.
The signals at the two output ports are thus a combination of the two input signals
\begin{align}
	a_1 = & \frac{1}{\sqrt{2}} (a + i a_\mathrm{LO}) & a_2 = & \frac{1}{\sqrt{2}} (a_\mathrm{LO} + i a)~.
\end{align}
The output fields are then detected with photodetectors and the two signals are subtracted
\begin{equation}
	\langle n_1 \rangle - \langle n_2 \rangle = i A \langle a^\dagger e^{i\theta} - a e^{-i\theta} \rangle~, 
\end{equation}
where $\langle a_\mathrm{LO} \rangle = A e^{i\theta}$.
We then have information on the optical phase, and therefore, on the position of the harmonic oscillator, when the phase of the local oscillator is $\theta=0$.
It is important to notice that so far we considered standard commutation relations, with no GUP corrections.

In what follows, we will consider the interaction in the Hamiltonian \eqref{eqn:hamiltonian_rotated} as
\begin{equation}
	\mathcal{U} = e^{i \Lambda N_\mathrm{L} Q_\mathrm{M}}~, \label{eqn:optomechanical_interaction}
\end{equation}
\emph{i.e.} without separating the effects on the optical field and on the mechanical oscillator, where $N_\mathrm{L} = a^\dagger a$ is the number operator for the optical field and $\Lambda = g_0 \tau$ is the interaction strength.

\section{Pikovski \emph{et al.} Proposed Experiment}

Pikovski \emph{et al.} in \cite{Pikovski2012_1} proposed an experiment to observe possible deformations of HUP using cavity optomechanics.
In this Section, we will review the main points of their proposal, setting the stage for modifications that will allow for enhancements of the Planck scale signal, as we will see in the next Sections.
The main point of this proposal is to let short optical pulses interact with a cavity, of which one of the two mirrors oscillates harmonically.
The request of a short pulse allowes us to neglect the evolution of the mechanical oscillator.
In this case, the intracavity dynamics is governed by the approximated operator $\mathcal{U} = e^{i \Lambda N_\mathrm{L} Q_\mathrm{M}}$, where $\Lambda \simeq g_0/\kappa = 4 \mathcal{F} \sqrt{\frac{\hbar}{2 \pi m c \lambda_\mathrm{L}}}$ is the effective interaction strength, $\mathcal{F}$ is the cavity finesse, and $\lambda_\mathrm{L}$ is the optical wavelength \cite{Pikovski2012_1}.

Furthermore, since we want to map the mechanical momentum, we need to achieve also a direct optomechanical coupling of the same optical pulse to the mechanical momentum.
Following the procedure utilized in \cite{Vanner2011_1} considering the standard theory, such a coupling is possible, letting the harmonic oscillator evolve between pulse round-trips and letting the mechanical position and the momentum interchange after a quarter of the oscillator period.
Indeed, the position quadrature operator $Q_\mathrm{M}$ evolves according to\footnote{Notice that in \cite{Pikovski2012_1} and in \cite{PikovskiPhD}, a different definition is used
$$Q_\mathrm{M} (t) = U Q_\mathrm{M} U^\dagger~,$$
corresponding to a time-reversed version of the definition in this thesis.
Here we use the definition \eqref{eqn:pos_quad_evolution} because it is adherent to the standard notation in QM.}
\begin{equation}
	Q_\mathrm{M} (t) = U^\dagger Q_\mathrm{M} U~, \label{eqn:pos_quad_evolution}
\end{equation}
where $U = \exp[-i \frac{\omega t}{2} (P_\mathrm{M}^2 + Q_\mathrm{M}^2)]$ and $P_\mathrm{M}$ is the momentum quadrature for the mechanical oscillator.
Since we have
\begin{align}
	[P_\mathrm{M}^2,Q_\mathrm{M}] = & - 2 i P_\mathrm{M} & [Q_\mathrm{M}^2,P_\mathrm{M}] = & 2 i Q_\mathrm{M}~,
\end{align}
it is easy to see that the following relation holds
\begin{align}
	Q_\mathrm{M} (t) = & Q_\mathrm{M} \cos(\omega_\mathrm{M} t) + P_\mathrm{M} \sin(\omega_\mathrm{M} t)~.
\end{align}
Therefore, letting the same pulse re-enter in the cavity after a quarter of mechanical period for a full mechanical cycle, the interaction will involve both position and momentum of the mechanical oscillator.
The total interaction is then described by the four-displacement operator, corresponding to four subsequent optomechanical interactions \cite{Pikovski2012_1}
\begin{equation}
	\xi 
	= \mathcal{U}\left(\frac{3T}{4}\right) \mathcal{U}\left(\frac{T}{2}\right) \mathcal{U}\left(\frac{T}{4}\right) \mathcal{U}\left(0\right)
	= e^{- i\Lambda N_\mathrm{L} P_\mathrm{M}} e^{-i\Lambda N_\mathrm{L} Q_\mathrm{M}} e^{i \Lambda N_\mathrm{L} P_\mathrm{M}} e^{i \Lambda N_\mathrm{L} Q_\mathrm{M}}~, \label{eqn:four-displacement_operator}
\end{equation}
where $T$ is the period of the mechanical oscillator.
Using \eqref{eqn:hadamard}, it is possible to write the displacement operator as
\begin{multline}
	\xi 
	= e^{- i\Lambda N_\mathrm{L} P_\mathrm{M}} \exp \{ i \Lambda N_\mathrm{L} e^{-i\Lambda N_\mathrm{L} Q_\mathrm{M}} P_\mathrm{M} e^{i \Lambda N_\mathrm{L} Q_\mathrm{M}} \} = \\
	= e^{- i\Lambda N_\mathrm{L} P_\mathrm{M}} \exp \{ i \Lambda N_\mathrm{L} ( \PM + \Lambda N_\mathrm{L} ) \} = e^{ i \Lambda^2 N_\mathrm{L}^2 }~. \label{eqn:std_four-displacement}
\end{multline}
This operator corresponds to four subsequent optomechanical interactions \eqref{eqn:optomechanical_interaction} between the optical field and the mechanical oscillator during a full period of the mirror oscillation.
We are then interested in studying the evolution of the optical field, in particular of its phase, due to this interaction.
As for the change in the annihilation operator for the optical field and in its expectation value on a coherent state, using this last relation, \eqref{eqn:hadamard}, and \eqref{eqn:nested_Nk_a}, we find
\begin{align}
	\xi^\dagger a \xi = & e^{- i (2 N_\mathrm{L} + 1) \Lambda^2} a~, \\
	\langle \alpha | \xi^\dagger a \xi | \alpha \rangle = & \alpha e^{- i \Lambda^2} e^{-|\alpha|^2 (1 - e^{- 2 i \Lambda^2})}~, \label{eqn:std_phase_displacement}
\end{align}
where we used the properties of coherent states \cite{Bosso2016_2}
\begin{subequations} \label{eqn:coherent_state_prop}
	\begin{align}
		a |\alpha\rangle &= \alpha | \alpha \rangle~, & e^{i\varphi N_\mathrm{L}}|\alpha\rangle &= |\alpha e^{i\varphi}\rangle~,\\
		\langle\alpha|\beta\rangle & = e^{-(|\alpha|^2 + |\beta|^2)/2} e^{\alpha^*\beta} = e^{-|\alpha-\beta|^2/2}e^{i\Im(\alpha^*\beta)}~, & \langle\alpha|N_\mathrm{L}|\alpha\rangle & = |\alpha|^2.
	\end{align}
\end{subequations}
Therefore, the optomechanical interaction represented by the operator $\xi$ imposes an evolution of the coherent state in phase-space.
In particular, we notice that its phase is displaced by the four interactions.
This phase then can be measured using homodyne detection, as we saw earlier.

We will see that the same results, when Planck scale effects are considered, contain information on possible deformations of the commutation relation between the mechanical position and momentum.
We recall indeed that the above expressions were found considering the standard evolution of the quadrature $Q_\mathrm{M}$ and $P_\mathrm{M}$ and, more in general, the standard commutation relation $[\QM,\PM]=i$.
On the other hand, when the GUP commutator \eqref{eqn:GUP} is considered, \emph{i.e.} in terms of quadrature operators
\begin{equation}
	[Q_\mathrm{M},P_\mathrm{M}] = i \left[ 1 - 2 \delta \tilde{\gamma} P_\mathrm{M} + (3 \epsilon + \delta^2) \tilde{\gamma}^2 P_\mathrm{M}^2 \right]~,\qquad \mbox{where}~\tilde{\gamma} = \gamma \sqrt{\hbar m \omega}~, \label{eqn:quadrature_GUP}
\end{equation}
the evolution of the operators changes, as a consequence of the Hamiltonian for the mechanical oscillator \eqref{eqn:perturbed_Hamiltonian}
\begin{equation}
	H_\mathrm{M} = \frac{\hbar \omega}{2} \left[P_{\mathrm{M},0}^2 + Q_{\mathrm{M},0}^2 - 2 \delta \tilde{\gamma} P_{\mathrm{M},0}^3 + (3 \delta^2 + 2 \epsilon) \tilde{\gamma}^2 P_{\mathrm{M},0}^4 \right]~,
\end{equation}
where $Q_{\mathrm{M},0}$ and $P_{\mathrm{M},0}$ are such that $[Q_{\mathrm{M},0},P_{\mathrm{M},0}] = i$.
Notice that we are not modifying the commutation relation for the optical field.
Therefore
\begin{align}
	[a,a^\dagger] = & 1~, & [Q_\mathrm{L},P_\mathrm{L}] = & i~.
\end{align}
{This is because the relevant GUP parameter for the optical part is smaller then the parameter for the mechanical part, once the values for the proposal in \cite{Pikovski2012_1} are considered.
In fact, comparing the parameters in \eqref{eqn:optic_GUP} and \eqref{eqn:quadrature_GUP}, we find
\begin{align}
	\tilde{\Gamma} = & 2.28 \gamma_0 \times 10^{-36}~, & \tilde{\gamma} = & 1.87 \gamma_0 \times 10^{-34}~.
\end{align}
Furthermore, using the results from Chapter \ref{ch:HO}, we expect that the corrections to the interaction part of the Hamiltonian \eqref{eqn:hamiltonian_rotated} is second order in $\tilde{\Gamma}$.
Therefore, the impact of GUP on the optical sector can be neglected.}
Furthermore, notice that the evolution operator cannot be split in a standard and a GUP terms, because
\begin{multline}
	[P_{\mathrm{M},0}^2 + Q_{\mathrm{M},0}^2, - 2 \delta \tilde{\gamma} P_{\mathrm{M},0}^3 + (3 \delta^2 + 2 \epsilon) \tilde{\gamma}^2 P_{\mathrm{M},0}^4] = \\
	= - 12 \delta \tilde{\gamma} i P_{\mathrm{M},0} Q_{\mathrm{M},0} P_{\mathrm{M},0} + 4 (3 \delta^2 + 2 \epsilon) \tilde{\gamma}^2 i P_{\mathrm{M},0} ( Q_{\mathrm{M},0} P_{\mathrm{M},0} + P_{\mathrm{M},0} Q_{\mathrm{M},0} ) P_{\mathrm{M},0}~.
\end{multline}
Using the results of the previous Chapter, in particular \eqref{eqn:expansion_q_time-ev} and the expansions \eqref{eqns:expansions_q&p}, we can find
\begin{subequations} \label{eqns:ev_position_quadrature}
\begin{align}
	Q_{\mathrm{M}}(0) = & Q_{\mathrm{M}} ~,\\
	Q_{\mathrm{M}} \left(\frac{T}{4}\right) = & \PM
	- (2 \QM^2 - \QM \PM - \PM \QM) \delta \tilde{\gamma}
	+ [ 3 \epsilon \QM \PM \QM + \nonumber \\
	& + \delta^2 (4 \PM \QM \PM - 4 \QM^3 - \QM \PM \QM - 2 \PM^3)] \tilde{\gamma}^2 + 3 \pi \chi \QM \bar{n} ~,\\
	Q_{\mathrm{M}} \left(\frac{T}{2}\right) = & - \QM
	- 2 (\QM \PM + \PM \QM) \delta \tilde{\gamma}
	- 8 \delta^2(\PM \QM \PM - \QM^3) \tilde{\gamma}^2 + 6 \pi \chi \PM \bar{n} ~, \\ 
	Q_{\mathrm{M}} \left(\frac{3T}{4}\right) = & - \PM
	+ (2 \QM^2 + \QM \PM + \PM \QM) \delta \tilde{\gamma}
	- [ 3 \epsilon \QM \PM \QM + \nonumber \\
	& - \delta^2 (4 \PM \QM \PM - 4 \QM^3 + \QM \PM \QM + 2 \PM^3) ] \tilde{\gamma}^2 - 9 \pi \chi \QM \bar{n} ~.
\end{align}
\end{subequations}
Notice that these relations are found substituting the phonon number operator $N_\mathrm{M}$ with the average phonon number $\bar{n}$.
This substitution is justified as long as we consider $\bar{n} \gg 1$, but not large enough to make the product $\chi~ \bar{n}$ of order unity.
Using the four terms above, the four-displacement operator becomes
\begin{equation}
	\xi = e^{i\Lambda N_\mathrm{L} Q_\mathrm{M}(3T/4)} e^{i\Lambda N_\mathrm{L} Q_\mathrm{M}(T/2)} e^{i \Lambda N_\mathrm{L} Q_\mathrm{M}(T/4)} e^{i \Lambda N_\mathrm{L} Q_\mathrm{M}(0)}~. \label{eqn:true_four-displacement_operator}
\end{equation}
We can write the operator $\xi$ using Baker-Campbell-Hausdorff (BCH) formula.
For this purpose, we calculated the necessary commutators in Appendix \ref{apx:BCH}, obtaining
\begin{multline}
	\xi = \exp \left\{ 6 i N_\mathrm{L} \Lambda \chi \pi \bar{n} (\PM - \QM) + i N_\mathrm{L}^2 \Lambda^2 \left[(3 \epsilon - 5 \delta^2) \tilde{\gamma}^2 (\QM^2 + \PM^2) + 6 \chi \pi \bar{n} + 1 \right] + \right. \\
	\left. + i N_\mathrm{L}^3 \Lambda^3 \left[(3 \epsilon - 5 \delta^2) \tilde{\gamma}^2 (\PM - \QM)\right]
    + i N_\mathrm{L}^4 \Lambda^4 (2 \epsilon - 3 \delta^2) \tilde{\gamma}^2 \right\} ~. \label{eqn:four-displacement_GUP}
\end{multline}
Notice that for two of the three models considered in \cite{Pikovski2012_1} we find
\begin{align}
	\xi_\beta = & \exp \left\{ 2 i N_\mathrm{L} \Lambda \tilde{\gamma}^2 \pi \bar{n} (\QM - \PM) + \right. \nonumber \\
	& + i N_\mathrm{L}^2 \Lambda^2 \left[ \tilde{\gamma}^2 (\QM^2 + \PM^2) - 2 \tilde{\gamma}^2 \pi \bar{n} + 1 \right] + \nonumber \\
	& \left. + i N_\mathrm{L}^3 \Lambda^3 \left[ \tilde{\gamma}^2 (\PM - \QM)\right]
    + \frac{2}{3} i N_\mathrm{L}^4 \Lambda^4 \tilde{\gamma}^2 \right\} ~, & \mbox{with} & & \delta=0~, ~~ \epsilon = \frac{1}{3} \\
	\xi_\gamma = & \exp \left\{ i N_\mathrm{L}^2 \Lambda^2 \left[- 2 \tilde{\gamma}^2 (\QM^2 + \PM^2) + 1 \right] + \right. \nonumber \\
	& \left. + i N_\mathrm{L}^3 \Lambda^3 \left[- 2 \tilde{\gamma}^2 (\PM - \QM)\right] 
	- i N_\mathrm{L}^4 \Lambda^4 \tilde{\gamma}^2 \right\} ~. & \mbox{with} & & \delta=1~, ~~ \epsilon = 1~.
\end{align}
This operator, and in particular the evolution of the optical field that it generates, will then provide information on the mechanical $[\QM,\PM]$ commutator.
Therefore, as we have done before for the standard case, we consider the mean optical field given by $\langle a \rangle = \langle \alpha | \xi^\dagger a \xi | \alpha \rangle$.
Applying \eqref{eqn:hadamard} and \eqref{eqn:nested_Nk_a}, we find
\begin{multline}
	\xi^\dagger a \xi = \exp \{ 6 i \Lambda \chi \pi \bar{n} (\PM - \QM) + \\
	- i (2 N_\mathrm{L} + 1) \Lambda^2 \left[(3 \epsilon - 5 \delta^2) \tilde{\gamma}^2 (\QM^2 + \PM^2) + 6 \chi \pi \bar{n} + 1 \right] + \\
	+ i (3 N_\mathrm{L}^2 + 3 N_\mathrm{L} + 1) \Lambda^3 \left[(3 \epsilon - 5 \delta^2) \tilde{\gamma}^2 (\PM - \QM)\right] + \\
	- i (4 N_\mathrm{L}^3 + 6 N_\mathrm{L}^2 + 4 N_\mathrm{L} + 1) \Lambda^4 (2 \epsilon - 3 \delta^2) \tilde{\gamma}^2 \} a~.
\end{multline}
Considering the case of a coherent state with large average photon number $\langle \alpha | N_\mathrm{L} | \alpha \rangle \gg 1$, we can neglect lower powers of the number operator $N_\mathrm{L}$ and approximate this expression as
\begin{equation}
	\xi^\dagger a \xi = \exp \{ - i (2 N_\mathrm{L} + 1) \Lambda^2 (6 \chi \pi \bar{n} + 1) \}\exp \{ - 4 i N_\mathrm{L}^3 \Lambda^4 (2 \epsilon - 3 \delta^2) \tilde{\gamma}^2 \} a~. \label{eqn:evolved_annihilation}
\end{equation}
Finally, since we can write
\begin{equation}
	N_\mathrm{L}^3 = (a^\dagger a)^3 = a^\dagger{}^3 a^3 + 3 a^\dagger {}^2 a^2 + a^\dagger a~,
\end{equation}
applying the operator in \eqref{eqn:evolved_annihilation} on a coherent state we have
\begin{multline}
	\langle \alpha| \xi^\dagger a \xi | \alpha \rangle 
	= \alpha e^{- i \Lambda^2 (6 \chi \pi \bar{n} + 1)} \langle \alpha | e^{- 4 i N_\mathrm{L}^3 \Lambda^4 (2 \epsilon - 3 \delta^2) \tilde{\gamma}^2} | \alpha e^{- 2 i \Lambda^2 (6 \chi \pi \bar{n} + 1)} \rangle = \\
	\simeq \alpha e^{- i \Lambda^2 (6 \chi \pi \bar{n} + 1)} e^{-|\alpha|^2(1 - e^{- 2 i \Lambda^2 (6 \chi \pi \bar{n} + 1)})} e^{- 4 i |\alpha|^6 e^{- 6 i \Lambda^2 (6 \chi \pi \bar{n} + 1)} \Lambda^4 (2 \epsilon - 3 \delta^2) \tilde{\gamma}^2} = \\
	= \langle \xi^\dagger a \xi \rangle_0 e^{\Theta}~, \label{eqn:exp_evolved_annihilation}
\end{multline}
where $\langle a \rangle_0$ is the expectation value of the optical field that we have found in the standard case \eqref{eqn:std_phase_displacement} and
\begin{equation}
	\Theta = -|\alpha|^2 e^{-2i\Lambda^2} (1 - e^{- 12 i \Lambda^2 \chi \pi \bar{n}}) 
	 - i \Lambda^2 6 \chi \pi \bar{n}
	 - 4 i |\alpha|^6 e^{- 6 i \Lambda^2 (6 \chi \pi \bar{n} + 1)} \Lambda^4 (2 \epsilon - 3 \delta^2) \tilde{\gamma}^2
\end{equation}
is an extra rotation in phase-space.
We notice a number of features.
First, notice that the extra rotation depends on the third power of the average photon number $|\alpha|^2$, on the average phonon number, and on the difference $(2\epsilon - 3 \delta)^2$.
These two last aspects are particularly interesting and do not appear in the original work \cite{Pikovski2012_1}.
While for low phonon numbers the corresponding term is negligible, for large numbers it suppresses the extra phase shift due to the optomechanical interaction.
In addition, this term vanishes when $\chi=0$, as in \cite{Ali2011_1}.
On the other hand, the difference $(2\epsilon - 3 \delta^2)$, when different from zero, governs the direction of the phase-space rotation.
In particular we see that for the model in \cite{Kempf1995_1} this appears with a minus sign, while for the model in \cite{Ali2011_1} the same term has the opposite sign.
Therefore, not only a possible extra phase-space displacement can be interpreted as an evidence of a modified $[\QM,\PM]$ commutator, but also the direction of the displacement can be used to distinguish between models.

\subsection{Scaling with the Number of Degrees of Freedom} \label{ssec:number_degrees}

In most of the proposals to probe GUP effects, a conceptual problem due to the description in terms of macroscopic coordinates is present.
Indeed, given the extra terms in the commutation relation, when it is applied to the constituents of a composite system, translating the commutation relation in terms of macroscopic coordinates, \emph{e.g.} the \mbox{center-of-mass} coordinates, reduces Planck scale effects with powers of the inverse number of constituents.
For the particular case of the experimental setup in \cite{Pikovski2012_1}, both in the supplementary information of the same paper and in \cite{AmelinoCamelia2013_1}, it is shown that, describing the problem in terms of position and momentum of the center-of-mass of the movable mirror, one could expect a suppression of the Planck scale effects due to the number of system's components.
This can be easily shown as follows.

Consider, for example, the model in \eqref{eqn:GUP} and assume it to be applied to some elementary degrees of freedom $\{q_1,\ldots,q_N,p_1,\ldots,p_N\}$, where $N$ is the total number of constituents of the system under exam \cite{AmelinoCamelia2013_1}.
For simplicity, we will consider a one-dimensional system.
The results can be easily generalized to higher dimensions.
Considering the standard relation between positions and momenta of the constituents and the homologous quantities for the center-of-mass $q_\mathrm{cm}$ and $p_\mathrm{cm}$, respectively, \cite{AmelinoCamelia2013_1}
\begin{align}
	q_\mathrm{cm} = & \frac{1}{N} \sum_{n=1}^N q_n & p_\mathrm{cm} = & \sum_{n=1}^N p_n~, \label{def:center-of-mass}
\end{align}
we find the following relation for the commutation relation of the \mbox{center-of-mass}
\begin{multline}
	[q_\mathrm{cm},p_\mathrm{cm}] 
	= \frac{1}{N} \sum_{n=1}^N \sum_{m=1}^N [q_n,p_m] 
	= \frac{1}{N} \sum_{n=1}^N i \hbar [1 - \delta \gamma p_n + (3 \epsilon + \delta^2) \gamma^2 p_n^2] = \\
	= i \hbar \left[1 - \delta \frac{\gamma}{N} p_\mathrm{cm} + (3 \epsilon + \delta^2)\frac{\gamma^2}{N^2} p_\mathrm{cm}^2 + (3 \epsilon + \delta^2)\frac{\gamma^2}{N} \sum_{n=0}^N \left( p_n^2 - \frac{p_\mathrm{cm}^2}{N^2}\right)\right]~.
\end{multline}
It is interesting to notice that $p_\mathrm{cm}/N$ represents the average value of the constituents' momenta.
Therefore, the last term simply corresponds to the variance of the components' momenta distribution.
We can then write
\begin{equation}
		[q_\mathrm{cm},p_\mathrm{cm}] = i \hbar \left\{1 - \delta \frac{\gamma}{N} p_\mathrm{cm} + (3 \epsilon + \delta^2) \gamma^2 \left[ \frac{p_\mathrm{cm}^2}{N^2} + (\Delta p)^2 \right] \right\}~. \label{eqn:cm_commutation_relation}
\end{equation}
This result is similar to \eqref{eqn:ang_mom_com_rel_same_ang_mom_q}.
On the other hand, since we are considering a translating rigid body ($p_n = p_m = p,~\forall n,~m$), we have $\Delta p = 0$ and \eqref{eqn:cm_commutation_relation} becomes
\begin{equation}
	[q_\mathrm{cm},p_\mathrm{cm}] = i \hbar \left[1 - \delta \frac{\gamma}{N} p_\mathrm{cm} + (3 \epsilon + \delta^2)\frac{\gamma^2}{N^2} p_\mathrm{cm}^2\right]~.
\end{equation}
This result, although important, since it shows that Planck scale effects on macroscopic bodies may scale with the number of their constituents, rests on two assumptions:
the degrees of freedom to which GUP is applied is (arbitrarily) chosen to be a (not well specified) fundamental one, that we referred to as ``the constituents''; the momentum composition law is given by \eqref{def:center-of-mass}.
While the first can sound reasonable, although there is no theoretical aspect that allows for a particular choice of the fundamental degrees of freedom and therefore, the validity will have to come from experiments, the second assumption depends on the particular QG theory in consideration.
For example, as we saw in Sec.~\ref{ssec:QGP&DSR}, DSR and its ``soccer ball'' problem actually predict a different composition law, in which the total momentum contains terms proportional to powers of the number $N$ \cite{Hossenfelder2014_1}.

Assuming that the scaling with the number of constituents is present, it is one of the main theoretical/conceptual problems against the ability to test a modified commutator with macroscopic bodies, as in the current case.
In what follows we will propose how to avoid such suppression, enhancing, at least in principle, the effect of GUP.

\section{Increasing the Number of Mechanical Periods}

Let us consider preliminarily the standard commutation rule, $[\QM,\PM]=i$.
The evolution of the position of a standard HO is periodic
\begin{equation}
	\QM(T) = \QM~,
\end{equation}
where $T$ is the period of the mechanical oscillator.
This implies that, letting the light field circulate in the proposed experimental apparatus for more than one mechanical period, in principle, we can account for the complete optomechanical interaction by considering a power of the four-displacement operator $\xi^N$, where $N$ in this case is the number of mechanical periods that are probed.
Considering $N$ subsequent applications of the same \mbox{four-displacement} operator \eqref{eqn:std_four-displacement} results in a factor $N$ at the exponent
\begin{equation}
	\xi^N = e^{iN\Lambda^2 N_\mathrm{L}^2}~. \label{eqn:canonical_xi}
\end{equation}

To find how this interaction acts on the intracavity field, we need to study the quantity $\left(\xi_{\mathrm{H}}^N\right)^\dagger a \xi_{\mathrm{H}}^N$.
Carrying a similar analysis as the one in the previous Section, we find
\begin{equation}
	\langle \alpha | (\xi^\dagger)^N a \xi^N | \alpha \rangle = \alpha e^{-iN\Lambda^2} e^{-|\alpha|^2(1 - e^{-2iN\Lambda^2})}~.
\end{equation}
We then see that increasing the number of mechanical periods increases the phase rotation of the optical field.
On the other hand, considering a modified commutation relation, we can see that also the extra phase rotation increases, possibly enhancing Planck scale effects \cite{Bosso2016_2}.

Consider now a modified commutator.
As for the evolution of the position operator, the only difference from the expressions in \eqref{eqns:ev_position_quadrature} is in the term multiplying the average phonon number.
Indeed, we easily see that, using \eqref{eqn:expansion_q_time-ev} again, after a period we have
\begin{equation}
	\QM(T) = \QM - 12 \pi \chi \PM \bar{n}~.
\end{equation}
The second term in the RHS is solely responsible for the terms involving $\bar{n}$ and, being second order in $\gamma$, it follows the standard evolution.
After $N$ periods, the position operator is
\begin{equation}
	\QM(NT) = \QM - 12 N \pi \chi \PM \bar{n}~.
\end{equation}
Notice that this expression is an approximation since the second term actually comes from an exponential that we previously expanded in Taylor series up to order $\mathcal{O}(\tilde{\gamma}^2)$.
Since this term involves the product $\omega t$, it grows linearly with the time $t$; therefore, for large number of periods, this approximation is no longer justified.
On the other hand, as we will see below, this term does not give rise to any divergence in the quantities we are concerned about since any increase of this term in any of the operators in \eqref{eqns:ev_position_quadrature} after one period is compensated by similar increases in another operator in \eqref{eqns:ev_position_quadrature}.
We also see that the other terms in \eqref{eqns:ev_position_quadrature} do not pose any problem in this sense; therefore, their presence in the $\xi$ operator does not change when subsequent periods are considered.
For this reason, let us write \eqref{eqns:ev_position_quadrature} as
\begin{subequations}
\begin{align}
	Q_{\mathrm{M}}(N T) = & Q_{\mathrm{M}} + \Xi(0) - 12 N \pi \chi \PM \bar{n} \\
	Q_{\mathrm{M}} \left(NT + \frac{T}{4}\right) = & \PM + \Xi\left(\frac{T}{4}\right) + (12 N + 3) \pi \chi \QM \bar{n} \displaybreak \\
	Q_{\mathrm{M}} \left(NT + \frac{T}{2}\right) = & - \QM + \Xi\left(\frac{T}{2}\right) + (12 N + 6) \pi \chi \PM \bar{n} \\
	Q_{\mathrm{M}} \left(NT + \frac{3T}{4}\right) = & - \PM + \Xi\left(\frac{3T}{4}\right) - (12 N + 9) \pi \chi \QM \bar{n} ~,
\end{align}
\end{subequations}
where $\Xi(t)$ represents the other terms at a time $t$.
We will now apply BCH formula to obtain the displacement operator $\xi_N$ at the $N$-th period.
First notice that
\begin{subequations}
\begin{align}
	\left[Q_{\mathrm{M}} \left(NT + \frac{T}{4}\right),Q_{\mathrm{M}}(N T)\right] = & \left[\PM + \Xi\left(\frac{T}{4}\right),Q_{\mathrm{M}} + \Xi(0)\right]~, \\
	\left[Q_{\mathrm{M}} \left(NT + \frac{3T}{4}\right),Q_{\mathrm{M}} \left(NT + \frac{T}{2}\right)\right] = & \left[-\PM + \Xi\left(\frac{3T}{4}\right),-Q_{\mathrm{M}} + \Xi\left(\frac{T}{4}\right)\right]~.
\end{align}
\end{subequations}
Therefore, since these terms do not depend on $N$, the only difference with respect the operators in \eqref{eqns:2-displacement} is
\begin{align}
	\log(\xi_{N,l}) - \log(\xi_l) = & 12 N \pi \chi [- \QM + \PM] \bar{n}~, \\
	\log(\xi_{N,r}) - \log(\xi_r) = & 12 N \pi \chi [\QM - \PM] \bar{n}~.
\end{align}
Using BCH formula once more, it is easy to see that there is no difference between the displacement operators during the first period and during the $N$-th period.
In fact, the only difference could come from the following terms
\begin{multline}
	\log(\xi_N) - \log(\xi) = \\
	= - 12 N \pi \chi [\QM + \PM,- \QM + \PM] \bar{n} + 12 N \pi \chi [\QM - \PM,\QM + \PM] \bar{n} = 0~.
\end{multline}
Therefore, we see that, even in GUP, the evolution of $\QM(t)$ does not pose any problem, and subsequent interactions for $N$ periods can be accounted for by simply considering $\xi^N$.
It is straightforward to see that this feature leads to enhancing the effects of a modified $[\QM,\PM]$ commutator.

Similarly to what we did for the standard case, we can then look for the results in the case of the displacement operator in \eqref{eqn:four-displacement_GUP} \cite{Bosso2016_2}.
As we showed, considering $4N$ interactions during $N$ mechanical periods results in the following operator
\begin{multline}
	\xi^N = \exp \left\{ 6 i N N_\mathrm{L} \Lambda \chi \pi \bar{n} (\PM - \QM) + \right. \\
	+ i N N_\mathrm{L}^2 \Lambda^2 \left[(3 \epsilon - 5 \delta^2) \tilde{\gamma}^2 (\QM^2 + \PM^2) + 6 \chi \pi \bar{n} + 1 \right] + \\
	\left. + i N N_\mathrm{L}^3 \Lambda^3 \left[(3 \epsilon - 5 \delta^2) \tilde{\gamma}^2 (\PM - \QM)\right]
    + i N N_\mathrm{L}^4 \Lambda^4 (2 \epsilon - 3 \delta^2) \tilde{\gamma}^2 \right\} ~.
\end{multline}
Proceeding as in the previous Section, we find, with the assumption of a large average photon number,
\begin{equation}
	(\xi^\dagger)^N a \xi^N \simeq \exp \left\{- i N (2 N_\mathrm{L} + 1) \Lambda^2 ( 6 \chi \pi \bar{n} + 1 ) - 4 i N N_\mathrm{L}^3 \Lambda^4 (2 \epsilon - 3 \delta^2) \tilde{\gamma}^2 \right\} a ~.
\end{equation}
The expectation value of this operator then gives
\begin{multline}
	\langle \alpha | (\xi^\dagger)^N a \xi^N | \alpha \rangle 
	= \alpha e^{- i N \Lambda^2 ( 6 \chi \pi \bar{n} + 1 )} e^{- 4 i N |\alpha|^6 e^{- 6 i N \Lambda^2 ( 6 \chi \pi \bar{n} + 1 )} \Lambda^4 (2 \epsilon - 3 \delta^2) \tilde{\gamma}^2} \times \\
	\times e^{- |\alpha|^2 [ 1 - e^{- i 2 N \Lambda^2 ( 6 \chi \pi \bar{n} + 1 )}]}~.
\end{multline}
We then see that increasing the number of probed mechanical periods directly increases the effects of GUP since, comparing the last equation with the result in \eqref{eqn:exp_evolved_annihilation}, the GUP parameter $\tilde{\gamma}$ is multiplied by $N$.
Furthermore, we see that the observations relative to \eqref{eqn:exp_evolved_annihilation} still hold in this case.

This enhancement can, in principle, contrast the scaling shown in Sec.~\ref{ssec:number_degrees}.
Neglecting possible deleterious effects, one can theoretically enhance Planck scale effects to any desirable level, since in this case the extra rotation is given by
\begin{multline}
	\Theta_N = -|\alpha|^2 e^{-2 i N \Lambda^2} (1 - e^{- 12 i \Lambda^2 N \chi \pi \bar{n}}) 
	 - i \Lambda^2 6 N \chi \pi \bar{n} + \\
	- 4 i N |\alpha|^6 e^{- 6 i N \Lambda^2 (6 \chi \pi \bar{n} + 1)} \Lambda^4 (2 \epsilon - 3 \delta^2) \tilde{\gamma}^2~,
\end{multline}
or, for small values of $\bar{n}$,
\begin{equation}
	\Theta_N = - 4 i N |\alpha|^6 e^{- 6 i N \Lambda^2} \Lambda^4 (2 \epsilon - 3 \delta^2) \tilde{\gamma}^2~. \label{eqn:extra:rotation}
\end{equation}

\section{Deleterious effects and the signal-to-noise ratio}\label{sec:noise}

\label{sec:deleterious_effects}

\begin{figure}
\centering
\includegraphics[width=0.5\textwidth]{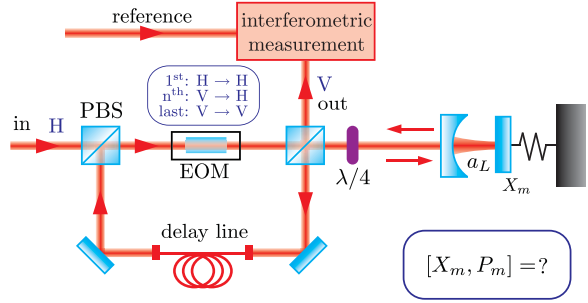}
\caption[Scheme to probe modified commutators.]{Experimental scheme to probe the mechanical position-momentum commutator for a mechanical oscillator \cite{Pikovski2012_1}.
A light pulse is injected into the cavity, one of whose mirrors is let oscillate harmonically.
This optomechanical interaction is repeated another three times in one mechanical period to probe both quadratures of the mechanical oscillator.
The delay line is used to let the mechanical oscillator evolve for a quarter of period before the light pulse could reenter in the cavity.
The Polarizing BeamSplitters (PBS) and the Electro-Optic Modulator (EOM) are used to select and rotate the correct (linear) polarization of the light, while the quarter-wave plate is used to convert the light pulse from linear to circular polarization and \emph{vice versa}.
After a full mechanical period, the light pulse leaves the cycle to measure its phase with balanced homodyne detection.}
\end{figure}

The proposed experiment consists of a coherent light pulse entering in a cavity and interacting with the movable mirror, as described in the previous Sections.
After this first interaction, the light field travels through a delay line, to let the mechanical oscillator evolve for a quarter of period.
After such time, the light pulse reenters in the cavity, interacting a second time with the mechanical system.
In the original proposal, this cycle is repeated other two times before the phase of the optical field is measured.
In our work, we suggest to continue the cycles for $N$ mechanical periods, for a total of $4N$ interactions.

There are many sources of noise and imperfections, both of fundamental and technical origins, that can limit the ability to detect these potential Planck scale deformations.
In this section, we analyze some of them \cite{Bosso2016_2}.
Notable unwanted effects include the intrinsic optical quantum noise, additional (classical) optical phase noise, optical loss, mechanical anharmonicity~\cite{Latmiral2016}, and mechanical decoherence. The signal $\Theta$ then has to be compared to the sensitivity of measuring the optical phase shift $\phi$. If we call $\delta \phi$ the uncertainty of measuring $\phi$, then $\Theta  > \delta \phi$ is desirable.

\subsection{Optical phase noise}

In an ideal experiment with no loss or classical phase noise with a coherent state of light, the phase uncertainty is given by the Heisenberg limit $\delta \phi = 1/(2\sqrt{N_\textrm{r} N_\textrm{p}})$.
Here, $N_\textrm{p}$ is the mean photon number and $N_\textrm{r}$ is the number of independent runs of the experiment.
It is important to compare this uncertainty to the scaling of the signal, which we have shown to increase with the number of mechanical cycles.
Thus, the signal to noise ratio (SNR), in ideal conditions, becomes
\begin{equation}
\frac{\Theta}{\delta \phi} \propto N \sqrt{N_\textrm{r} N_\textrm{p}}~ .
\end{equation}
It is therefore advantageous to use more cycles $N$ over more individual runs $N_\textrm{r}$ of a single cycle experiment.
This is because $N$ amplifies the signal, while $N_\textrm{r}$ reduces the uncertainty in estimating the phase by averaging over many runs.

Using $N$ cycles is advantageous if no deleterious effects are present.
However, using several cycles can also increase the noise in the setup. Thus, for some noise sources, the advantage may be negated, leading to an optimal $N$. We anticipate that the primary noise source will be classical phase noise, which will reduce our ability to estimate the optical phase shift. We then have $\delta \phi \sqrt{N_\textrm{r}} = 1/(2\sqrt{N_\textrm{p}}) + \delta \phi_\textrm{c}$, where $\delta \phi_\textrm{c}$ is the classical phase noise. In any practical application, the classical noise $\delta \phi_\textrm{c}$ will increase with photon number and propagation time.
At some sufficiently large $N$ or $N_\textrm{p}$, the noise is dominated by $\delta \phi_\textrm{c}$ and the signal-to-noise ratio will reach a maximum.
Such classical noise is not fundamental, however, and can be avoided in principle, but it requires significant effort at large photon numbers.

\begin{figure}
\centering
\includegraphics[width=0.50\textwidth]{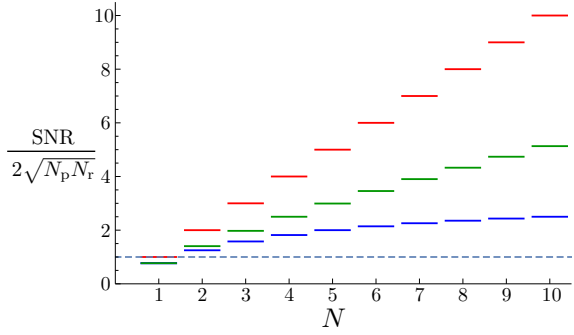}
\caption[Plot of signal-to-noise ratio as a function of the cycle number $N$.]{Plot of signal-to-noise ratio (SNR), \eqref{Eq:SNR}, as a function of the cycle number $N$. The upper trace represents the ideal case with no classical noise ($\zeta_\textrm{c} = \zeta_\textrm{u} = 0$, shown in red). The two other curves show the effect of uncorrelated noise ($\zeta_\textrm{c} = 0,~\zeta_\textrm{u} = 0.3$, shown in green), and correlated noise ($\zeta_\textrm{c} = 0.3,~\zeta_\textrm{u} = 0$, shown in blue). A dashed line at unity is given to aid comparison.}
\label{Fig-SNR}
\end{figure}

Between each interaction, there will be fluctuations that give a random phase shift to the optical field.
We anticipate high-frequency noise, which is uncorrelated between loops, and lower frequency noise, which is correlated between loops. Random, uncorrelated noise, will cause the classical noise to scale as $\delta \phi_\textrm{c} \propto \sqrt{N}$, whereas the scaling of correlated noise may approach $\delta \phi_\textrm{c} \propto N$. The signal-to-noise ratio then becomes
\begin{equation}
\frac{\Theta}{\delta \phi} \propto 2\sqrt{N_\textrm{p} N_\textrm{r}} \, \frac{N}{\zeta_\textrm{c}N + \zeta_\textrm{u}\sqrt{N} + 1} ,
\label{Eq:SNR}
\end{equation}
where $\zeta_\textrm{c}$ and $\zeta_\textrm{u}$ parametrize the strength of the correlated and uncorrelated classical noise, respectively, in units of the optical quantum noise. The size of these terms will be strongly dependent on the experimental realization and in practice both terms will be present to some extent. Examples of how the signal-to-noise ratio can scale are provided in Fig.~\ref{Fig-SNR}.

\subsection{Optical loss}

\begin{figure}
\centering
\includegraphics[width=0.70\textwidth]{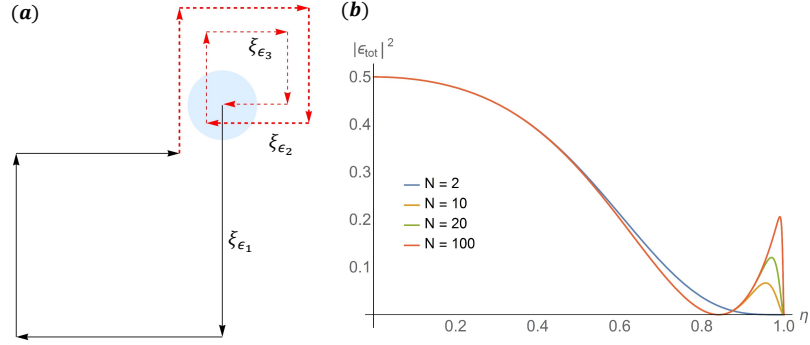}
\caption[(a)  Displacement of the mechanical state using 3 cycles.\protect\\ (b) The parameter $|\epsilon_\textrm{tot}|^2$ as a function of optical loss $\eta$.]{(a) Displacement of the mechanical state (blue circle) using 3 cycles as described by \eqref{Ndisplace}. Due to losses (here $\eta = 0.89$), the first loop (black arrows) does not close and thus causes degradation of the signal as given by Eq.~\eqref{eqn:mean_noise}. Two additional loops (red dashed arrows) bring the mechanical state back to its original position, thus giving $\epsilon_\textrm{tot}=0$ for this particular value of optical loss. (b) The parameter $|\epsilon_\textrm{tot}|^2$ as a function of optical loss $\eta$, as given in Eq.~\eqref{Ndisplace2}. A finite value of $|\epsilon_\textrm{tot}|^2$ reduces the signal.}
\label{Fig-Arrows}
\end{figure}

Using an $N$-cycle scheme has the additional advantage that one can now control the pulses within a sequence.
Two example types of control include displacing the pulse to modify its magnitude and phase~\cite{Khosla2013} and changing the timing of the pulses to interact with a different mechanical quadrature.
This can be used to better close the loop made in mechanical phase-space in the presence of optical loss and thus reduce any unwanted residual correlations between the light and mechanics, which reduces the signal-to-noise ratio (see below and Fig. \ref{Fig-Arrows}).
In the following, we describe a protocol to minimize the effects of optical loss (and related deleterious effects such as pulse shape distortion) without full control of the timing of the pulses as that would require an efficient, phase coherent, tunable optical delay line, which would introduce unwanted experimental complexity. In this model we neglect the added optical noise due to the effective optical self-Kerr effect, and the small change in the optical noise due to optical loss, as these effects will be small compared to the classical phase noise discussed above and the effects loss has on the mean of the optical field discussed below. Then, optical loss can be described by a change of the interaction parameter $\Lambda$ after each pulsed interaction \cite{Pikovski2012_1}. For a single cycle, the four-pulse operator for regular QM becomes
\begin{equation} \label{eqn:four-displacement_operator_noise}
{\xi}_{\varepsilon}
= e^{-i\Lambda_4 {N}_\textsc{L} \PM} e^{-i \Lambda_3 {N}_\textsc{L} \QM} e^{i \Lambda_2 {N}_\textsc{L} \PM} e^{i \Lambda_1 {N}_\textsc{L} \QM} 
= e^{(\epsilon b^{\dagger} - \epsilon^\star b) N_\textsc{l}} e^{i \Lambda_{\epsilon}^2 N_\textsc{l}^2}~,
\end{equation}
where
\begin{align}
	\epsilon = & \frac{1}{\sqrt{2}} \left[ \Lambda_4 - \Lambda_2 + i (\Lambda_1 - \Lambda_3)\right]~, &
	\Lambda_{\epsilon}^2 = & \Lambda_2 \Lambda_3 + \frac{1}{2} \left(\Lambda_1 - \Lambda_3 \right) \left(\Lambda_2 - \Lambda_4 \right)~.\label{eqn:epsilon}
\end{align}
A varying interaction strength thus not only changes the strength of the signal but also introduces an unwanted correlation between the light and the mechanical state. For an initial thermal state $\rho_\textsc{m}^{(th)}$ of the mechanics with the mechanical thermal occupation number $\bar{n}$, the mean phase becomes
\begin{align}\label{eqn:mean_noise}
	\langle a \rangle_{\epsilon} = & \textrm{Tr}[a \xi_{\epsilon} | \alpha \rangle \langle \alpha | \otimes \rho_\textsc{m}^{(th)}  {\xi}_{\epsilon}^{\dagger} ] = \nonumber \\
	= & \textrm{Tr}[ e^{-(\epsilon b^{\dagger} - \epsilon^* b) N_\textsc{l}} a \, e^{(\epsilon b^{\dagger} - \epsilon^* b) N_\textsc{L}} e^{i \Lambda_{\epsilon}^2 N_\textsc{L}^2} | \alpha \rangle \langle \alpha | \otimes \rho_\textsc{m}^{(th)} e^{- i \Lambda_{\epsilon}^2 N_\textsc{L}^2} ] = \nonumber \\
	= & \textrm{Tr}[ e^{- i \Lambda_{\epsilon}^2 N_\textsc{L}^2} a \,  e^{i \Lambda_{\epsilon}^2 N_\textsc{L}^2} | \alpha \rangle \langle \alpha | \otimes \rho_\textsc{m}^{(th)} e^{\epsilon b^{\dagger} - \epsilon^* b} ] = \nonumber \\
	= & \textrm{Tr}_{\textsc{l}}[  e^{- i \Lambda_{\epsilon}^2 (2 N_\textsc{L}+1)} a | \alpha \rangle \langle \alpha |] ~ \textrm{Tr}_{\textsc{m}}[\rho_\textsc{m}^{(th)} e^{\epsilon b^{\dagger} - \epsilon^* b}] 
	= \langle a \rangle_{0, \epsilon } \, e^{-\frac{|\epsilon|^2}{2}\left(1 +2 \bar{n} \right)} ~ ,
\end{align}
where we used the cyclic property of the trace and where $\langle a \rangle_{0, \epsilon} = \alpha e^{- i \Lambda_{\epsilon}^2 - |\alpha|^2 \left(1 - e^{- i 2\Lambda_{\epsilon}^2}\right)}$.
For $|\epsilon| \not = 0$, it can thus be beneficial to cool the mechanical mode of interest. For  the case of optical loss by a constant fraction in between each pulsed interaction, characterized by the intensity efficiency $\eta$, where $\eta = 1$ is no loss, the strengths of each of the four displacements are given by $\Lambda_1 = \Lambda,~\Lambda_2=\eta \Lambda,~\Lambda_3 = \eta^2 \Lambda,~\Lambda_4 = \eta^3 \Lambda$, and thus $\epsilon = \Lambda (1 - \eta^2)(i - \eta)/\sqrt{2}$ and $\Lambda_{\epsilon}^2 = \eta \Lambda^2 (1 + \eta^4)/2$.
However, using a cycle of $N$ loops can reduce this deleterious effect, if one changes the direction of phase-space displacements after the first cycle. In the second  cycle (and all subsequent ones) we can apply the following modified four-displacement operator:
\begin{equation} \label{eqn:four-displacement_operator_noise2}
	{\xi}_{\epsilon_2} 
	= e^{i \Lambda_4^{(2)} {N}_\textsc{L} \PM} e^{i \Lambda_3^{(2)} {N}_\textsc{L} \QM} e^{ - i \Lambda_2^{(2)} {N}_\textsc{L} \PM} e^{- i \Lambda_1^{(2)} {N}_\textsc{L} \QM} 
	= e^{-(\epsilon_2 b^{\dagger} - \epsilon^\star_2 b) N_\textsc{L}} e^{i \Lambda_{\epsilon_2}^{2} N_\textsc{L}^2}~.
\end{equation}
The sign of the correlating term is reversed, while the main phase has the same sign. This remains true for any commutator deformation that is of even power in the mechanical momentum.
Thus, if the above sequence is performed after the sequence given by \eqref{eqn:four-displacement_operator_noise}, the second displacement is subtracted from the first displacement while the signal adds up as in the ideal case:
\begin{equation}
	{\xi}_{\epsilon_2} {\xi}_{\epsilon_1} 
	=  e^{[(\epsilon_1-\epsilon_2) b^{\dagger} - (\epsilon_1-\epsilon_2)^\star b] N_\textsc{L}} e^{i (\Lambda_{\epsilon_1}^{ 2} + \Lambda_{\epsilon_2}^{ 2} ) N_\textsc{L}^2}~.
\end{equation}
Here we assumed loss by a constant fraction as above, such that $\epsilon_2 = \eta^4 \epsilon_1$. In this case the additional phase factor vanishes, which is otherwise $e^{i N_\textsc{l} \Im[\epsilon_2 \epsilon_1^\star]}$.

We can repeat the same interaction multiple times. After $N$ cycles in total, we have \mbox{$\epsilon_N = \eta^{4(N-1)} \epsilon$} and $\Lambda_{\epsilon_N}^{ 2} = \eta^{8(N-1)} \Lambda_{\epsilon}^{2}$. Thus, for $N-1$ further cycles with the sequence given in \eqref{eqn:four-displacement_operator_noise2}, we obtain (without any commutator deformation):
\begin{equation}
	\xi_\textrm{tot} 
	= {\xi}_{\epsilon_N} \ldots {\xi}_{\epsilon_1} 
	= e^{(\epsilon_\textrm{tot} b^{\dagger} - \epsilon_\textrm{tot}^\star b) N_\textsc{L}} e^{i \Lambda_\textrm{tot}^{ 2} N_\textsc{L}^2}~, \label{Ndisplace}
\end{equation}
where
\begin{subequations} \label{Ndisplace2}
\begin{align}
	\epsilon_\textrm{tot} = & \epsilon \frac{1 - 2\eta^4 + \eta^{4N}}{1 - \eta^4} = \Lambda (i - \eta) \frac{1 - 2\eta^4 + \eta^{4N}}{\sqrt{2}(1 + \eta^2)}~, \\
	\Lambda_\textrm{tot} = & \Lambda_{\epsilon}^{ 2} \sum_{j=0}^{N-1} \eta^{8j} = \eta \Lambda^2 \frac{1 - \eta^{8 N}}{2(1-\eta^4)}~.
\end{align}
\end{subequations}
If $\eta \geq 2^{-1/4} \approx 0.84$, the unwanted correlating term can, therefore, be made to completely vanish, see Fig. \ref{Fig-Arrows}.
Depending on the value of $\eta$, the additional loops given by \eqref{eqn:four-displacement_operator_noise2} may overcompensate for the losses, in which case some of the loops could be performed in the original direction as in \eqref{eqn:four-displacement_operator_noise}.
For $\eta < 0.84$, the effect of loss can still be reduced with the above scheme.
In this case the optical phase is still affected as in \eqref{eqn:mean_noise}, but with a new effective parameter $|\epsilon|^2 \rightarrow |\epsilon_\textrm{tot}|^2$.

\subsection{Mechanical decoherence}

The mechanical resonator undergoes dissipation and decoherence during the protocol, as the mechanical mode is coupled to a thermal bath.
This degrades the overall signal, which was estimated in \cite{Pikovski2012_1} for a single cycle.
For $N$ cycles, the mechanical decoherence will be increased as the system spends a longer time coupled to the bath between initialization of the protocol and final readout.
Building on this earlier work, we can estimate the increase in decoherence to be proportional to $N$, as the overall coupling time to the bath is $2 \pi N/ \omega_\mathrm{M}$.
This reduces the mean of the optical field to approximately $\langle \alpha \rangle \rightarrow \langle \alpha \rangle [1 - \Lambda^2 N k_\mathrm{B} T / ( \hbar \omega_\mathrm{M} Q)]$, where $Q$ is the mechanical quality factor and $T$ the temperature.
This unwanted effect can be minimized by performing the experiment at cryogenic temperatures and using mechanical resonators with a high $\omega_\mathrm{M} Q$ product.
{For example, for the parameters in \cite{Pikovski2012_1} for the linear and quadratic model and for $T<100$ mK, $Q>10^6$, and $N \sim 10^5$ we find
\begin{equation}
	\frac{\Lambda^2 N k_\mathrm{B} T}{\hbar \omega_\mathrm{M} Q} \lesssim 10^{-10}~.
\end{equation}}
In contrast to other proposals that aim to directly study mechanical non-classicality, this protocol is more robust to mechanical thermal occupation and we do not anticipate that this effect will be a primary source of degradation to the signal-to-noise ratio.

\vspace{1EM}
Finally, as we saw in this Chapter, considering the proposal in \cite{Pikovski2012_1} with multiple mechanical periods and the expedients that we analyzed here, it is possible in principle to detect deformations of the commutation relation for the movable mirror of the cavity.
\chapter{Conclusions}

\begin{quote}
	\begin{verse}
		Lo duca e io per quel cammino ascoso \\
		intrammo a ritornar nel chiaro mondo; \\
		e sanza cura aver d'alcun riposo,
		
		salimmo s\`u, el primo e io secondo, \\
		tanto ch'i' vidi de le cose belle \\
		che porta 'l ciel, per un pertugio tondo. 
		
		E quindi uscimmo a riveder le stelle.  
	\end{verse}
	
	\begin{flushright}
		Divina Commedia, Inferno, Canto XXXIV -- Dante Alighieri
	\end{flushright}
\end{quote}

The current theoretical framework of Physics, governed by QM and GR, has many bright as well as dark sides.
Although both theories can successfully describe a large number of observations, they are obviously not complete.
On one side, for example, QM cannot account for a dynamically evolving spacetime background.
On the other side, GR breaks down when it deals with black holes, one of its own predictions.
Therefore, a quantum theory of gravity is not only desirable from the aesthetic point of view of a unique fundamental theory, but also for the necessary search for a fully consistent theory that could solve the problems of QM and GR.
Further motivations come from black hole physics, with the revolutionary discovery of thermodynamic aspects of these objects.
In particular, starting from a classical description, an analogy between black hole physics and thermodynamics was achieved.
Such an analogy was then deepened by semiclassical aspects, with the theoretical discovery of black holes thermal radiation.
In this way, the definition of a temperature and of an entropy for black holes is not only the consequence of an analogy but has a more direct meaning.

In the past decades, several approaches to QG have been pursued.
In this thesis, we presented two of them, namely ST and LQG, and described DSR, a proposal for modified kinematical aspects in Special Relativity.
Each of them has its own target and motivation.
For example, while ST pursues, with the introduction of a new fundamental element, a unification of all four known fundamental interactions, besides the quantization of gravity, LQG focuses on this last aspect trying to build a background-independent theory of QG.
Finally, DSR questions Lorentz symmetry in a theory of QG, partially solving it by introducing a new invariant scale, usually identified with the Planck scale.
However, these approaches are still at a developing stage; therefore, none of them can be chosen as a definite model or description of QG.

An important aspect worth pointing out is that none of these theories have experimental corroborations, nor did any experiment or observation show any deviation from standard GR or QM.
Thus, these approaches to QG are tentative approaches, in the sense that, unless internal inconsistencies are discovered or they are disproved by future experiments, none of these theories can be discarded.
Therefore, one of the guides for the attempt to quantize gravity is high mathematical rigor.
It is also necessary to treat new approaches, like QFT on manifolds in LQG, in which the experience developed by physicists in other areas, like QFT on metric spaces, could be useless or misleading.
One is therefore forced to appeal to new or not widely known mathematical tools, like spin network theory, spectral-representation theory, infinite-dimensional measure theory and differential geometry over infinite-dimensional spaces.
Further requirements concern the low-energy limit of these theories, that has to give back the classical theory of GR.
Moreover, since the thermodynamic theory of black holes is based on very solid grounds, one would also expect that a definitive theory of QG can recover what is found in the semi-classical approach.

Unfortunately, we are still far from a complete and consistent theory.
Every model has its own problems, and more work is required to solve them and to obtain a new and solid theoretical framework.

Nonetheless, as we saw in Chapter \ref{ch:GUP}, further guidance can come from experiments and phenomenology.
In particular, in this thesis, we focused on one aspect of QGP, namely GUP.
A large number of theories predicts a modification of the Heisenberg Uncertainty Principle \cite{Garay1995_1,Gross1988_1,Amati1989_1,Maggiore1993_1,Maggiore1993_2,Scardigli1999_1,AmelinoCamelia2002_1}, where the commutation relation between canonical coordinates and momenta is modified by momentum-dependent terms.
Motivated by the possibility of testing this modification and Planck scale effects in low-energy quantum systems, we have considered in detail the effects of GUP on several theoretical as well as experimental systems.
Starting with the most general GUP model {up to and including quadratic terms}, we showed that in all the examined contexts, GUP gives rise to potentially observable effects.
We will briefly review them here, focusing on results and future steps.
Finally, we will present possible new ideas and applications of GUP that may resolve themselves into new possible testing grounds for a future theory of gravity.

\section{GUP and Angular Momentum}

In Chapter \ref{ch:angular-momentum} we examined how GUP affects the theory of angular momentum \cite{Bosso2017}.
We found the modified spectrum of the angular momentum operators $L_z$ and $L^2$.
We then saw that these modifications lead to corrected energy levels of the hydrogen atom, and its behavior in an external magnetic field.
When applied to magnetic field interactions, it leads to different values for the Larmor frequency and for the splitting in the Stern--Gerlach experiment.
We finally showed how the modified algebra of the total angular momentum of a multi-particle system depends on the number of components and presented interesting Planck scale modifications to the CG coefficients.
It is worth noting that all the modifications derived here are potentially observable, allowing for new tests on QGP.

It is worthwhile to also note that we computed GUP corrections of the angular momentum spectrum and applied it to the appropriate Schr\"odinger equation governing non-relativistic systems, such as the hydrogen atom and the Stern--Gerlach experiment.
While the corrections to the spectrum will equally apply to relativistic systems (such as relativistic hydrogen atom), to compute GUP corrections to energy eigenvalues and eigenvectors, the emission spectra etc., one would have to use the Dirac equation.

Comparing the results of Chapter \ref{ch:angular-momentum} with experimental data, we can obtain upper bounds for the GUP parameter $\gamma_0$.
For instance, let us assume that the energy levels of the relativistic hydrogen atom are modified by terms proportional to $\gamma_0^2 \langle p_0^2 \rangle / M_\mathrm{Pl}^2 c^2$ (similar to (\ref{eqn:energy_levels_GUP}) for the non-relativistic case).
For an estimation of the frequency of the 2$S$ - 1$S$ transition, using \eqref{eqn:h-spectrum_GUP} and \eqref{eqn:exp_p_r2}, we find an expected deviation from the standard frequency of the order
\begin{equation}
	|\nu_{2S-1S} - \nu_{2S-1S}| \simeq R_\infty \frac{\gamma_0^2}{M_\mathrm{Pl}^2 c} \frac{\hbar^2}{a_0^2 4} \sim \gamma_0^2 10^{-34} \mathrm{Hz}~.
\end{equation}
Comparing this with the result in \cite{Matveev2013}, we see that $\gamma_0 \lesssim 10^{17}$.
This further motivates a study of GUP for the relativistic hydrogen atom, 
in which the relativistic corrections should impose a much tighter bound on $\gamma_0$.
Similarly, for the Stern--Gerlach experiment, one can obtain an estimation of the parameter $\gamma_0$ by directly comparing the relative error in the splitting $\delta z$ of an experiment with 
\eqref{eqn:cor_Stern--Gerlach}.
For example, for a relative error of 10\%, one obtains $\gamma_0 \lesssim 10^{14}$.
Notice that more accurate experiments would result in more stringent bounds.

Some issues remain to be better understood, \emph{e.g.} the dependence of the angular momentum algebra on linear momentum and some ambiguity in the CG coefficients, as we saw at the end of Chapter \ref{ch:angular-momentum}.
Furthermore, we replaced the operator representing the GUP modification in some formulae with its expectation value.
While this suffices to estimate Planck scale effects, and while we showed in the Subsec. \ref{ssec:spectrum_w/p} that this is a reasonable assumption based on diagonality of $L^2$ and $L_z$ operators on respective eigenstates, in the future we would like to study this further, to see if additional corrections result by retaining the operator form.
The results presented here can be applied to look for QG signatures, \emph{e.g.} in spectroscopic as well as astrophysical observations.
Furthermore, assuming that also the spin algebra obeys similar modifications, they can be applied to a number of quantum systems interacting with magnetic fields, in atomic and nuclear physics.

\clearpage

\subsection{GUP-modified Hydrogen Atom}

In this Subsection, we would like to analyze the magnitude of the error made in the approximation used in Sec.~\ref{sec:h-atom}, where we considered the extra terms in the Hamiltonian \eqref{eqn:hamiltonian} as expectation values following the approximation
\begin{equation}
	\mathcal{C} |\Psi\rangle \rightarrow \langle \Psi | \mathcal{C} | \Psi \rangle | \Psi \rangle~. \label{eqn:approx}
\end{equation}
Although this approximation was functional to our scope, that is, looking for possible deviations on the hydrogen atom spectrum induced by GUP, this aspect requires more dedicated work.
We expect that, in applying this approximation, we have an error in our results of the same order of the uncertainty on $1 - \mathcal{C}$.
In fact, while this uncertainty would vanish for the approximation in \eqref{eqn:approx}, when it is not considered we would have the following uncertainty for $(1 - \mathcal{C})$ on a hydrogen atom eigenstate 
\begin{equation}
	\Delta (1 - \mathcal{C}) = \sqrt{\langle n l m | (1-\mathcal{C})^2 | n l m \rangle - (\langle n l m |1-\mathcal{C} | n l m \rangle)^2}~.
\end{equation}
For a quadratic model ($\delta=0$), this expression, up to second order in $\gamma$, reduces to
\begin{equation}
	\Delta (1 - \mathcal{C}) = \gamma^2 \epsilon \frac{\alpha^2 m^2 c^2}{n^2} \sqrt{\left[ \frac{2 n (4n^2 + 1) (2l + 1) - 1}{n^2 (2l + 1)^2} - 4 \right]}~.
\end{equation}
Notice that the term under squared root is always positive.
Therefore, we can propagate this uncertainty in the GUP-modified energy spectrum \eqref{eqn:energy_levels_GUP}, finding
\begin{equation}
	\Delta E = \left| \frac{\partial E}{\partial (1 - \langle \CE \rangle) } \right| \Delta (1 - \CE) 
	= \alpha^2 \frac{m c^2}{n^2 (1 - \langle \CE \rangle)} \Delta (1 - \CE) \simeq \alpha^2 \frac{m c^2}{n^2} \Delta (1 - \CE)~,
\end{equation}
and for the relative error
\begin{equation}
	\frac{\Delta E}{E} \simeq \Delta (1 - \mathcal{C})~.
\end{equation}
This estimation then suggests that the error originating from the approximation \eqref{eqn:approx} decreases with $n$.
In particular, in the limit $n \rightarrow \infty$, the error vanishes, $\Delta (1 - \CE) \rightarrow 0$ $\forall l<n$.
On the other hand, for small $n$ we have the larger error.
This can be explained recalling that for $n \rightarrow \infty$, the electron of the hydrogen atom becomes free and $\langle p_0 \rangle \rightarrow 0$, $\Delta p_0 \rightarrow 0$.
Moreover, this condition is interesting since it shows that GUP effects are more relevant when the hydrogen atom is in its ground state.

{It is important to stress here that comparisons between the results of this thesis concerning the hydrogen atom energy levels and experiments cannot be performed at this stage.
Since our results are besed on the non-relativistic theory, any deviation due to GUP is to be considered as a deviation from the non-relativistic hydrogen atom.
However, experimental results, as in \cite{Matveev2013}, are different from the theoretical values in the non-relativistic theory for 1 part over 10000, while the sensitivity achieved in recent experiments is of the order of $10^{-15}$.
Notice that a description of the hydrogen atom in QFT agrees with experiments, the difference with experimental results being of the same order of magnitude of their uncertainties.
Given this agreement between experiments and the relativistic theory for the hydrogen atom, the latter is then necessary when we want to compare any deviation from the standard theory with observations.
Therefore, an analysis of the hydrogen atom in QFT with GUP is required, so to have a better estimation of the energy levels and, in particular, of energy levels with small $n$, where GUP effects are larger.
Finally, as we saw earlier, to achieve a better resolution of GUP effects on low-energy states, future research has to consider a description of the hydrogen atom without the approximation \eqref{eqn:approx}.}

\section{Harmonic Oscillator with Minimal Length}

In Chapter \ref{ch:HO} we considered the implication of the existence of a minimal length through the implementation of GUP (\ref{eqn:perturbed_Hamiltonian}) by looking at perturbations from the $\gamma=0$ case \cite{Bosso2017a}.
We were thus able to find normalized eigenstates (new number states) {\eqref{eqn:perturbed_state}}, eigenvalues {\eqref{eqn:energ_eigenstate}}, and a new set of ladder operators {\eqref{def:new_ops}}.
We then focused our attention on coherent and squeezed states.

One of our most notable results is the spacing of the energy ladder as a function of the   linear and quadratic terms $\delta$ and $\epsilon$ in the GUP.
There are three distinct cases.
When $\delta^2 < \epsilon$, the spacing of the energy ladder increases with the number $n$.
For the case $\delta^2 = \epsilon$, the spacing does not depend on $n$, obtaining a regular ladder, as in the standard theory. Finally, for $\delta^2 > \epsilon$ the correction is negative.
Therefore, the spacing of the energy ladder decreases with the number $n$ up to a maximum number $n_{\mathrm{max}}$, corresponding to the maximum value of the energy.
This simply indicates that GUP breaks down for large enough energies, where Planck scale effects become relevant and a full theory of QG is necessary since higher order terms cannot be neglected.

For coherent states, defined as eigenstates of the new annihilation operator, we find that {they are still} minimal uncertainty states.
Squeezed states introduce additional interesting features.
The position uncertainty has a lower bound  -- not present in the standard theory -- that depends on the particular GUP model under consideration.  
As for the momentum uncertainty, its maximal squeezing depends on the model and on the choice of the coefficients $\delta$ and $\epsilon$.
That is, for some choices of the coefficients, a lower bound for the squeeze parameter exists, since beyond this limit higher order Planck scale effects cannot be neglected.

Our results   are potentially testable.
For example, direct application to mechanical oscillators can be tested, \emph{e.g.} as in \cite{Pikovski2012_1,Marin2013_1,Bawaj2014_1}.
In fact, mainly motivated by the results in \cite{Kempf1995_1}, many groups already proposed experiments to test them.
So far this class of experiments focused only on the energy spectrum of the HO.
On the other hand, experiments on resonant mass detectors could in principle also test the results concerning coherent and squeezed states \cite{Hollenhorst1979}.

A different class of tests could be performed in the astrophysical scenario.
In fact, as we showed in Sec.~\ref{sec:HO_GUP} and in particular in (\ref{eqn:detection}), massive systems constitute ideal observational tools to probe GUP effects.
However, this is somewhat controversial.
If GUP applies only to fundamental constituents of matter (see for example \cite{AmelinoCamelia2013_1}), then Planck scale effects on mesoscopic and macroscopic objects should be small.
However, if the GUP can be applied to the center-of-mass of an optomechanical oscillator, and to other systems, (\emph{e.g.} \cite{Pikovski2012_1,Marin2013_1,Bawaj2014_1,Bosso2016_2}), then Planck scale effects may be observable in mesoscopic and macroscopic  systems. Should such be the case, then controlling other variables to isolate GUP effects will be a major challenge.

It is interesting to comment on the choice of mass scale in \eqref{Mpscale}.
It is generally expected that this is the Planck mass, but on empirical grounds, we could replace $M_{\mathrm{Pl}}$ in \eqref{Mpscale} with any mass scale that has yet to be probed by experiments. 
A conservative value would be just above the LHC scale, \emph{i.e.} $ E_{\mathrm{LHC}} \sim 10$ TeV.
This might come from some hitherto undiscovered intermediate scales, or one deriving from a large 5-th dimension with order TeV Planck energy in five dimensions.
Our estimated corrections  would then be amplified by a factor of $M_{\mathrm{Pl}} c^2 / E_{\mathrm{LHC}} \sim 10^{15}$, or equivalently by a different value of $\gamma_0$ in \eqref{Mpscale}.
Since this ratio appears squared in \eqref{eqn:unc_period}, the most interesting scenario would be the optomechanical oscillator experiments, for which the periods would vary from seconds to hours, potentially rendering the calculated effects observable. 
Alternatively, a scale \mbox{$E\sim 10^{11}$} GeV would yield oscillation times of seconds for the LIGO detector, again observable at
least in principle.  Depending on the temporal resolution of the detectors, these arguments can set lower bounds to the relevant energy scale for the GUP (see Table \ref{tbl:unc_osc_period}).

\section{Observing Planck scale effects with Quantum Optomechanics}

In Chapter \ref{ch:QO}, we presented a possible experiment to test quantum gravitational deformations of the HUP and developed a strategy to enhance its sensitivity \cite{Bosso2016_2}.
Using an extended sequence of optomechanical interactions, we could accumulate a greater signal even for weak optomechanical couplings.
Cavity optomechanics is well-suited to probe GUP for the \mbox{center-of-mass} motion of a macroscopic mechanical oscillator.
An analysis of classical optical phase-noise for this approach shows how the advantage provided by this scheme diminishes with increasing amounts of phase noise.
Additionally, for a quadratic GUP model, it is possible to dramatically reduce the effects of optical loss by reversing the direction of phase-space displacements after the first cycle.
Making the practical assumption that optical loss is constant between each pulsed interaction, this technical problem can be mitigated.

The enhancement of the signal-to-noise ratio provided by the proposed scheme can also help to constrain the different possible interpretations of the generalized uncertainty principle.
As we showed in Sec.~\ref{ssec:number_degrees} and as was previously noted in \cite{Pikovski2012_1} and \cite{AmelinoCamelia2013_1}, it is an interesting open question which degrees of freedom are affected by the GUP and whether expected signals will differ depending on where the model is applied.
If the GUP commutators are applied to the ``fundamental degrees of freedom'' (which have to be specified independently of the model, \emph{e.g.} atoms or elementary particles), then the resulting GUP effects are reduced in comparison to application of the GUP deformations to the centre-of-mass motion.
For example, applying GUP to a composite system with $M$ fundamental constituents will result in a reduction in the signal by $M^k$, where $k$ lies between 1 and 2 depending on the correlations of the individual particles [22].
Thus, any experimental improvements will not only help to determine the magnitude of any GUP correction to standard QM, but can also shed light on the question of what level such deformations may apply.
Indeed, for an $N$-cycle experimental scheme, we would obtain $\Theta \simeq \frac{8 N}{3 M^k} (2 \epsilon - 3 \delta^2) \tilde{\gamma}^2 |\alpha|^6 \lambda^4 e^{- i 6 N \lambda^2}$, and thus, this reduction can be partially compensated by using multiple cycles.

\section{Future Research Topics}

In analyzing the results of this thesis' previous Chapters, we saw that much more can still be done on these particular topics, both on the theoretical/conceptual side, as well as on the experimental side.
On the other hand, many questions on GUP and on the existence of a minimal length are still open and need to be addressed.
In the following, we briefly review some speculative aspects for which research is needed.
Given their nature, we consider it worthwhile to review them here.

\subsection{Foundations of Quantum Optics with GUP} \label{ssec:fund_QO}

Applications to Quantum Optics of the results shown in Chapter \ref{ch:HO} can lead to important and potentially observable features \cite{Bosso}.
Indeed, starting from a description of the electromagnetic field in terms of amplitude and phase quadratures, it is possible to include the GUP commutation relation to obtain a perturbed Hamiltonian.
Following similar steps and definitions of Chapter \ref{ch:HO}, one can then construct a GUP-modified theory of Quantum Optics.
This modification can have important implications in many optical experiments, as for example the one presented in Chapter \ref{ch:QO}, where both the optical and the mechanical parts should be treated following the results found in Chapter \ref{ch:HO}.
Furthermore, the implementations of these results to Quantum Optics will also affect experiments involving squeezed states, with particular reference to future evolutions of gravitational wave interferometers.
This is especially relevant as the use of squeezed states has been planned for LIGO interferometers \cite{Dwyer2013_1}.
Therefore, while on one hand further study is still necessary to understand some of the features highlighted in Chapter \ref{ch:HO}, on the other hand, we may be one step closer to the experimental evidence of Planck scale Physics.

To see how the results of Chapter \ref{ch:HO} can imply a modified quantum description of electromagnetic fields, let us consider the standard steps for developing a quantum theory of optics.
Following \cite{berestetskii2012quantum}, we define generalized position and momentum for an electromagnetic (classical) field, namely $q_k$ and $p_k$, respectively, as given by
\begin{align}
	q_k = & \sqrt{\frac{\hbar}{2 \omega_k}} ( a_k^\star + a_k )~, & p_k = & i \sqrt{\frac{\hbar \omega_k}{2}} ( a_k^\star - a_k)~,
\end{align}
where $a_k$ and $a_k^\star$ are the complex amplitudes for the mode $k$.
In this way, the vector potential is written as
\begin{equation}
	\vec{A}
	= \sum_{\vec{k}} \sqrt{\frac{\hbar}{2 \epsilon_0 V \omega_k}} \vec{\epsilon}_k \left[ a_k e^{-i\Theta} + a_k^\star e^{i\Theta} \right]
	= \sum_{\vec{k}} \sqrt{\frac{1}{\epsilon_0 V}} \vec{\epsilon}_k \left[ q_k \cos \Theta + \frac{1}{\omega_k} p_k \sin \Theta \right]~,
\end{equation}
where $\Theta = \omega_k t - \vec{k} \cdot \vec{r}$ and $V$ is the volume of the cavity with periodic boundary conditions inside which the radiation is defined.
As for the transverse electric and magnetic fields, we have
\begin{subequations}
\begin{align}
	\vec{E}_\mathrm{T} = & - \frac{\partial \vec{A}}{\partial t} = \sum_{\vec{k}} \sqrt{\frac{1}{\epsilon_0 V}} \vec{\epsilon}_k \left[ \omega_k q_k \sin \Theta - p_k \cos \Theta \right]~, \\
	\vec{B} = & \vec{\nabla} \times \vec{A} = \sum_{\vec{k}} \sqrt{\frac{\mu_0}{V}} \hat{k} \times \vec{\epsilon}_k \left[ \omega_k q_k \sin \Theta - p_k \cos \Theta \right]~.
\end{align}
\end{subequations}
The Hamiltonian for the mode $k$ of the radiation in the cavity is given by
\begin{equation}
	H_k = \frac{1}{2} \int_\mathrm{cavity} \left( \epsilon_0 \overline{E_{\mathrm{T}k}^2} + \mu_0^{-1} \overline{B_k^2} \right) dV~,
\end{equation}
where the bars denote a cycle average
\begin{align}
	\overline{E_{\mathrm{T}k}^2} = & \frac{1}{2 \epsilon_0 V} \left( \omega_k^2 q_k^2 + p_k^2 \right)~, & \overline{B_k^2} = & \frac{\mu_0}{2 V} \left( \omega_k^2 q_k^2 + p_k^2 \right)~.
\end{align}
Therefore, the Hamiltonian reads
\begin{equation}
	H_k 
	= \frac{1}{2} \left( p_k^2 + \omega_k q_k^2\right)~.
\end{equation}
Notice that the dimension of $p_k$ is that of momentum divided by (mass)$^{1/2}$, as well as $q_k$ has the dimension of position multiplied by (mass)$^{1/2}$.
Therefore, in promoting the quantities defined here to operators and in imposing a GUP-inspired commutation relation, we consider the following \emph{Ansatz}
\begin{equation}
	[q_k,p_{k'}] = i \hbar \delta_{k,k'} \left(1 - 2 \delta \Gamma p_k + (3 \epsilon + \delta^2) \Gamma^2 p_k^2 \right)~,
\end{equation}
where
\begin{equation}
	\Gamma = \frac{\gamma_0}{\sqrt{M_\mathrm{Pl}} c}~.
\end{equation}
Notice that this \emph{Ansatz} is based on the formal equivalence between the generalized positions and momenta for an electromagnetic field and the position and momentum of an HO.
Indeed, we can in principle apply the GUP assumption of a modified commutator
\begin{equation}
	[q_i,p_j] = i \hbar \delta_{ij} \rightarrow [q_i,p_j] = i \hbar f(\mathbf{p})
\end{equation}
to every couple of position-like and momentum-like quantities.
On the other hand, for the quadrature operators, defined as
\begin{align}
	Q_k = & \sqrt{\frac{\omega_k}{\hbar}} q_k~, & P_k = & \sqrt{\frac{1}{\hbar \omega_k}} p_k~,
\end{align}
we find
\begin{equation}
	[Q_k,P_{k'}] = i \delta_{k,k'} \left(1 - 2 \delta \tilde{\Gamma} P_k + (3 \epsilon + \delta^2) \tilde{\Gamma}^2 P_k^2 \right)~, \label{eqn:optic_GUP}
\end{equation}
where
\begin{equation}
	\tilde{\Gamma} = \gamma_0 \sqrt{\omega_k t_{\mathrm{Pl}}}~,
\end{equation}
$t_{\mathrm{Pl}}$ being the Planck time.
Therefore, we have a theory that is completely equivalent to the one considered in Chapter \ref{ch:HO}, and thus we are allowed to use the same results.
Finally, we notice that in this case Planck scale effects are achieved for high frequencies/short period oscillations, as well as large values of $P_k$.

\subsection{Noise in Interferometers} \label{ssec:LIGO}

Using this GUP-modified version of Quantum Optics, we can, for example, investigate the sources of noise in a Michelson-Morley interferometer and how they are changed because of Planck scale effects.
There are two main sources of noise in interferometry, namely radiation pressure noise and shot noise \cite{Dwyer2013_1}.
The first is due to fluctuations in the radiation pressure of the optical field on the mirrors of the interferometer.
It is proportional to the fluctuations entering at the unused port.
On the other hand, shot noise can be interpreted as noise due to random arrival of photons on the mirrors.
It is inversely proportional to the laser amplitude.
Shot noise can therefore be reduced, increasing the radiation pressure noise.
The increase of the latter can be cured by letting a suitable squeezed field enter at the dark port.
Therefore, any change of the two quadratures' behavior will eventually produce relevant observable effects on the radiation pressure noise, limiting the efficiency of the use of squeezed states.

Notice that, in terms of $\sim$-operators defined in Sec.~\ref{sec:new_operators}, we have
{
\begin{subequations} \label{eqn:operators_beam_splitter}
\begin{align}
	\frac{1}{\sqrt{2}}(Q + i P) = & \tilde{a} + i \frac{\delta}{2^{3/2}} \tilde{\Gamma} \left[3 \tilde{a}^2 + 3 \left(2 \tilde{N} + 1\right) - \tilde{a}^\dagger{}^2\right] + \nonumber \\
	& - \frac{\tilde{\Gamma}^2}{4} \left[ \delta^2 (6 \tilde{a}^3 + 4 \tilde{a} \tilde{N}_{a} + 6 \tilde{N}_{a} \tilde{a}^\dagger+ 3 \tilde{a}^\dagger {}^3) + \epsilon (\tilde{a}^3 - 6 \tilde{N}_{a} \tilde{a}^\dagger + \tilde{a}^\dagger {}^3) \right] \\
	\frac{1}{\sqrt{2}}(Q - i P) = & \tilde{a}^\dagger - i \frac{\delta}{2^{3/2}} \tilde{\Gamma} \left[3 \tilde{a}^\dagger {}^2 + 3 \left(2 \tilde{N} + 1\right) - \tilde{a}^2\right] + \nonumber \\
	& - \frac{\tilde{\Gamma}^2}{4} \left[ \delta^2 (6 \tilde{a}^\dagger {}^3 + 4 \tilde{N}_{a} \tilde{a}^\dagger + 6 \tilde{a} \tilde{N}_{a} + 3 \tilde{a} {}^3) + \epsilon (\tilde{a}^\dagger {}^3 - 6 \tilde{a} \tilde{N}_{a} + \tilde{a} {}^3) \right]~,
\end{align}
\end{subequations}
}
where we have dropped the subscript $k$ since we will consider a single mode.
We can then follow the same steps leading to eq. (2.14) in \cite{Caves1981_1} obtaining instead
\begin{subequations} \label{eqn:beam_splitter}
\begin{align}
	Q_{2,1} + i P_{2,1} = & \frac{e^{i\Delta}}{\sqrt{2}} \left[ ( Q_{1,1} + i P_{1,1}) + e^{i\mu} ( Q_{1,2} + i P_{1,2}) \right] ~,\\
	Q_{2,2} + i P_{2,2} = & \frac{e^{i\Delta}}{\sqrt{2}} \left[ ( Q_{1,2} + i P_{1,2}) - e^{-i\mu} ( Q_{1,1} + i P_{1,1}) \right] ~,
\end{align}
\end{subequations}
where the first index distinguishes between input and output fields for the beam splitter of the interferometer, while the second distinguishes the two channels.
In the previous equations, $\Delta$ represents the absolute phase shift, while $\mu$ is the relative phase difference of the beam splitter.
Although these quantities depend on the particular beam splitter in use, we will consider the case with
\begin{align}
	\Delta = & 0~, & \mu = & \frac{\pi}{2}~,
\end{align}
corresponding to the case of a symmetrical beam splitter.
These relations will allow us to find the first order and second order terms in GUP parameter for the output fields in terms of the input fields.

Let us call $\tilde{a}_i$ and $\tilde{b}_i$ the annihilation operators for the input and output channels, respectively. {with $i=1,2$}.
Furthermore, let us assume that $\tilde{b}_i$ can be written in terms of $\tilde{a}_i$ as
\begin{equation}
	\tilde{b}_i = A_i + \delta \tilde{\Gamma} B_i + \tilde{\Gamma}^2 (\delta^2 C_i + \epsilon D_i)~,
\end{equation}
where $A_i$, $B_i$, $C_i$ and $D_i$ are functions of $\tilde{a}_j$ and $\tilde{a}^\dagger_j$, with $j=1,2$.
Using eqs. (\ref{eqn:operators_beam_splitter}) and (\ref{eqn:beam_splitter}), we can then write the following equations
\begin{subequations}
\begin{equation}
	A_i = \frac{1}{\sqrt{2}} \left[ \tilde{a}_i + i \tilde{a}_j \right] ~,
\end{equation}
\vspace{-3em}
{
\begin{multline}
	B_i + \frac{i}{2^{3/2}} (3 A_i^2 + 3( 2 A_i^\dagger A_i + 1) - A_i^\dagger {}^2) = \\
	= \frac{i}{4} \left[ \left(3 \tilde{a}_i^2 + 3 \left(2 \tilde{N}_i + 1\right) - \tilde{a}_i^\dagger{}^2 \right) + i \left(3 \tilde{a}_j^2 + 3 \left(2 \tilde{N}_j + 1\right) - \tilde{a}_j^\dagger{}^2 \right) \right] ~,
\end{multline}
\vspace{-4em}
\begin{multline}
	C_i + \frac{i}{2^{3/2}} (3 A_i B_i + 3 B_i A_i + 3 ( 2 A_i^\dagger B_i + 2 B_i^\dagger A_i ) - A_i^\dagger B_i^\dagger - B_i^\dagger A_i^\dagger) + \\
	- \frac{1}{4} \left(4 A_i^3 + 3 A_i A_i^\dagger A_i  + 6 A_i^\dagger A_i A_i^\dagger + 3 A_i^\dagger A_i^\dagger A_i^\dagger \right) = \\
	= - \frac{1}{2^{5/2}} \left[ \left(6 \tilde{a}_i^3 + 4 \tilde{a}_i \tilde{N}_{a_i} + 6 \tilde{N}_{a_i} \tilde{a}_i^\dagger + 3 \tilde{a}_i^\dagger {}^3\right) + i \left(6 \tilde{a}_j^3 + 4 \tilde{a}_j \tilde{N}_{a_j} + 6 \tilde{N}_{a_j} \tilde{a}_j^\dagger + 3 \tilde{a}_j^\dagger {}^3 \right) \right]~.
\end{multline}
\vspace{-2em}
\begin{multline}
	D_i - \frac{1}{4} \left( A_i^3 - 6 A_i^\dagger A_i A_i^\dagger + A_i^\dagger {}^3 \right) = \\
	= - \frac{1}{2^{5/2}} \left[ \left( \tilde{a}_i^3 - 6 \tilde{N}_{a_i} \tilde{a}_i^\dagger + \tilde{a}_i^\dagger {}^3 \right) + i \left( \tilde{a}_j^3 - 6 \tilde{N}_{a_j} \tilde{a}_j^\dagger + \tilde{a}_j^\dagger {}^3 \right) \right]~,
\end{multline}
}
\end{subequations}
where $i,j=1,2$, and $i\not=j$.
We used $\tilde{N}_{a_i}$ to indicate the number operator associated with $\tilde{a}_i$.
The first equations correspond to the results of the standard theory, and we can use them to find higher order terms in $\tilde{b}_i$.
In particular, we have
{
\begin{subequations}
\begin{align}
		B_i = & \frac{i}{2^{5/2}} \left\{3 \left[\tilde{a}_i^2 (\sqrt{2} - 1) + i \tilde{a}_j^2 (\sqrt{2} - i) \right] + \right. \nonumber \\
		& \qquad + 3 \left[ (2 \tilde{N}_{a_i} + 1) (\sqrt{2} - 1) + i (2 \tilde{N}_{a_2} + 1) (\sqrt{2} + i) \right] + \nonumber \\
	& \left. - \left[ \tilde{a}_i^\dagger {}^2 (\sqrt{2} - 1) + i \tilde{a}_j^\dagger {}^2 (\sqrt{2} - i) \right] - 2 i (3 \tilde{a}_i \tilde{a}_j + 3 \tilde{a}_i^\dagger \tilde{a}_j - 3 \tilde{a}_i \tilde{a}_j^\dagger - \tilde{a}_i^\dagger \tilde{a}_j^\dagger) \right\} ~,\\
	C_i = & \frac{1}{2^{7/2}} \left\{ \left[6 (2^{3/2} - 3) \tilde{a}_i^3
		+ 4 (2^{3/2} - 3) \tilde{a}_i \tilde{N}_{a_i} 
		+ 6 (2^{3/2} - 3) \tilde{N}_{a_i} \tilde{a}_i^\dagger 
		- 3 \tilde{a}_i^\dagger {}^3 \right] + \right. \nonumber \\
	& - \left[ 6 (i+\sqrt{2}) \tilde{a}_j^3
		+ 2 (6 i + 13 \sqrt{2}) \tilde{a}_j \tilde{N}_{a_j}
		+ 6 (i - \sqrt{2}) \tilde{N}_{a_j} \tilde{a}_j^\dagger
		+ 3 (i + 2^{3/2}) \tilde{a}_j^\dagger {}^3 \right] + \nonumber \\
	& -  \left[ 6 i \tilde{a}_i \left[ (2^{3/2} - 3) \tilde{a}_i + (3 i - \sqrt{2}) \tilde{a}_j \right] \tilde{a}_j
		+ 4 i (2 \tilde{N}_{a_i} + 1) \tilde{a}_j + \right. \nonumber \\
	&	\qquad - 2 (2 + 5 \sqrt{2} i) \tilde{a}_i^\dagger \tilde{a}_j^2
		+ 4 (2 \sqrt{2} - 1) i \tilde{a}_i^2 \tilde{a}_j^\dagger 
		+ 2 (2 - 9 \sqrt{2} i) \tilde{a}_i (2 \tilde{N}_{a_j} + 1)+ \nonumber \\
	& \qquad + 6 \left[  (2\sqrt{2} + 1) i \tilde{a}_i^\dagger {}^2 \tilde{a}_j 
		+ (2 \sqrt{2} - 1) i (2 \tilde{N}_{a_i} + 1) \tilde{a}_j^\dagger 
		- (\sqrt{2} i - 1) \tilde{a}_i^\dagger (2 \tilde{N}_{a_j} + 1) + \right. \nonumber \\
	& \left. \left. \left. \qquad \qquad - (1 + \sqrt{2} i) \tilde{a}_i \tilde{a}_j^\dagger {}^2 \right]
		+ 3 \tilde{a}_i^\dagger (3 \tilde{a}_i^\dagger + (2^{3/2} i + 3) \tilde{a}_j^\dagger) \tilde{a}_j^\dagger \right] \right\}~,\\
	D_i = & - \frac{1}{2^{7/2}} \left\{ \left[3 (2 \tilde{a}_i^3 - 6 \tilde{N}_{a_i} \tilde{a}_i^\dagger + \tilde{a}_i^\dagger {}^3 )
		+ i ( 2 \tilde{a}_j^3
		- 6 \tilde{N}_{a_j} \tilde{a}_j^\dagger 
		+ 3 \tilde{a}_j^\dagger {}^3) \right] + \right. \nonumber \\
	& \qquad + 6 i \tilde{a}_i (\tilde{a}_i + i \tilde{a}_j) \tilde{a}_j
		+ 6 \left[\tilde{a}_i \tilde{a}_j^\dagger {}^2 - i \tilde{a}_i^\dagger {}^2 \tilde{a}_j + i (2 \tilde{N}_{a_i} + 1) \tilde{a}_j^\dagger - \tilde{a}_i^\dagger (2 \tilde{N}_{a_j} + 1)\right] + \nonumber \\
	& \left. \qquad -  3 i \tilde{a}_i^\dagger ( \tilde{a}_i^\dagger - i \tilde{a}_j^\dagger) \tilde{a}_j^\dagger \right\}~.
\end{align}
\end{subequations}
}

The differential momentum transferred on the end mirrors of the two arms of the interferometer is proportional to the difference in number of the photons in each arm
\begin{equation}
	\tilde{b}_2^\dagger \tilde{b}_2 - \tilde{b}_1^\dagger \tilde{b}_1 = W + \tilde{\Gamma} \delta X + \tilde{\Gamma}^2 ( \delta^2 Y + \epsilon Z )
\end{equation}
where
{
\begin{subequations}
\begin{align}
	W = & i \tilde{a}_1 \tilde{a}_2^\dagger + \mathrm{h.c.} ~, \\
	X = & - \frac{1}{8} \left[(1 - i) \tilde{a}_1^3
		+ 3 (1 + i) \tilde{a}_i \tilde{N}_{a_1}
		- 3 (1 - 2^{3/2} + i) (2 \tilde{N}_{a_1} + 1) \tilde{a}_2 + \right. \nonumber \\
	& \qquad \left. + 3 (1 + 2^{3/2} + i) \tilde{a}_1^2 \tilde{a}_2^\dagger
		+ (3 - 2^{3/2} - 3 i) \tilde{a}_1^2 \tilde{a}_2 - (\mbox{terms with } 1 \leftrightarrow 2) \right] + \mathrm{h.c.} ~, \\
	Y = & \frac{1}{16} \left[- 3 \sqrt{2} (1 + 2 i) \tilde{a}_1^4
		- 6 \sqrt{2} (2 + i) \tilde{a}_1^3 \tilde{a}_2
		- 6 (- 1 + \sqrt{2} + \sqrt{2} i) \tilde{a}_1 (2 \tilde{N}_{a_1} + 1) \tilde{a}_1 + \right. \nonumber \\
	& \qquad - 6 (2^{3/2} - 4 i + 5 \sqrt{2} i) \tilde{a}_1 \tilde{N}_{a_1} \tilde{a}_2
		- 6 (- 2^{3/2} + 4 i + \sqrt{2} i) \tilde{a}_1^3 \tilde{a}_2^\dagger + \nonumber \\
	& \qquad + 6 (\sqrt{2} i - 4 i + \sqrt{2} - 3) (2 \tilde{N}_{a_1} + 1) \tilde{a}_2^2	
		+ 15 \sqrt{2} \tilde{N}_{a_1}(\tilde{N}_{a_1} + 1) + \nonumber \\
	& \qquad - 2 (- 6 \sqrt{2} - 22 i + 15 \sqrt{2} i) \tilde{a}_1 \tilde{N}_{a_1} \tilde{a}_2^\dagger
		+ 2 (3 \sqrt{2} + 4) i \tilde{a}_1^2 \tilde{a}_2^\dagger {}^2 + \nonumber \\
	& \qquad \left. - (\mbox{terms with } 1 \leftrightarrow 2) \right] + \mathrm{h.c.} ~, \\
	Z = & \frac{1}{4} \left[- \tilde{a}_1 (2 \tilde{N}_{a_1} + 1) \tilde{a}_1
		- 3 i \tilde{a}_1 (2 \tilde{N}_{a_1} - \tilde{a}_1^2) \tilde{a}_2
		- 2 i \tilde{a}_1^3 \tilde{a}_2^\dagger
		+ 3 (2 \tilde{N}_{a_1} + 1) \tilde{a}_2^2 + \right. \nonumber \\
	& \qquad \left. - (\mbox{terms with } 1 \leftrightarrow 2) \right] + \mathrm{h.c.} ~.
\end{align}
\end{subequations}
}
Following \cite{Caves1981_1}, we will consider the expectation value of this operator on a coherent state for the input channel 1 and a squeezed state for the input channel 2
\begin{equation}
	|\Psi \rangle = S_2 (\xi) D_1 (\alpha) |0\rangle~, \label{eqn:state_ligo}
\end{equation}
obtaining for the difference between the momentum transferred on the two mirrors at the end of each arm
\begin{equation}
	\langle \mathcal{P} \rangle \propto \tilde{\Gamma} \delta \langle X \rangle + \tilde{\Gamma}^2 (\delta^2 \langle Y \rangle + \epsilon \langle Z \rangle)~,
\end{equation}
where
{
\begin{subequations} \label{eqns:exp_P}
\begin{align}
	\langle X \rangle = & \frac{\alpha}{8} \left\{(i-1) (3 \alpha^\star {}^2 + \alpha^2)
		- 6 \left[(i + 1) \cosh^2 r - \sqrt{2} \cosh (2 r)\right] + \right. \nonumber \\
	& \left. \qquad + \sinh (2 r) \left[2^{3/2} (i \sin \theta - e^{- i \theta}) - 3 (1 - i) \cos \theta\right]\right\} + \mathrm{h.c.} ~, \\
	\langle Y \rangle = & \frac{1}{32} \left\{
		- 6 \sqrt{2} (1 + 2 i) \alpha^4 
		+ 24 \left[1 - \sqrt{2} (1 + i)\right] \alpha^2 |\alpha|^2 + \right. \nonumber \\
	& \qquad + 24 \left\{3 - 2 \left[\sqrt{2} (i + 1) - i\right]\right\} \alpha^2
		+ 24 \left[3 - \sqrt{2} (1 + i) + 4 i\right] \alpha^2 \sinh^2 r + \nonumber \\
	& \qquad - 4 (4 + 3 \sqrt{2}) i \alpha^2 \sinh (2 r) e^{- i \theta}
		+ 12 \left[3 - 4 i - \sqrt{2} (1 - i)\right] |\alpha|^2 \sinh (2 r) e^{- i \theta} + \nonumber \\
	& \qquad + \frac{9 (1 - 2 i) \sinh^2 (2 r) e^{- 2 i \theta}}{\sqrt{2}}
		+ 12 \left\{3 - 2 \left[\sqrt{2} (1 - i) + i\right]\right\} \sinh (2 r) e^{- i \theta} + \nonumber \\
	& \qquad \left. + 36 \left[\sqrt{2} (i - 1) + 1\right] \sinh^2 r \sinh (2 r) e^{- i \theta} + \mathrm{h.c.} \right\} + \nonumber \\
	& - \frac{15 \sqrt{2} \left[3 \sinh^2 (2 r) - 4 |\alpha|^2 (2 + |\alpha|^2)\right]}{32} ~, \\
	\langle Z \rangle = & - \frac{1}{2} \alpha^2 (|\alpha|^2 + 3 \cosh^2 r) + \mathrm{h.c.} 
		- \frac{3}{2} (|\alpha|^2 + \cosh^2 r) \sinh (2 r) \cos \theta ~.
\end{align}
\end{subequations}
}
These expressions show a number of interesting and potentially observable features.
First, we notice that the zeroth order in $\tilde{\Gamma}$ vanishes; therefore, we can recover the standard result in the limit $\tilde{\Gamma} \rightarrow 0$.
Furthermore, recalling that $\theta$ and $\alpha$ are functions of time (they represent the orientation of the squeezed state and the position of the coherent state in phase-space, respectively, as given by Fig. \ref{fig:squeezed_state}), we notice that only one term is associated with a constant quantity, corresponding to a constant difference between the momenta transferred to the two mirrors
{
\begin{equation}
	\langle \bar{\mathcal{P}} \rangle = - \frac{b \hbar \omega}{c} \tilde{\Gamma}^2 \delta^2 \frac{15 \sqrt{2} \left[3 \sinh^2 (2 r) - 4 |\alpha|^2 (2 + |\alpha|^2)\right]}{16}~,
\end{equation}
}
where $\beta$ is the number of reflections in one arm.
We then see that the linear term in GUP leads to a violation of isotropy of space.
Notice that this feature is second order in $\tilde{\Gamma}$ and is present only in GUP models with a term linear in the momentum.
Furthermore, it can be interpreted as a displaced equilibrium of the two mirrors: since one arm contains more photons, the corresponding mirror will feel an averagely higher pressure.
Consequently, the length of the arm is changed, now being a function of the squeeze parameter $r$.
Therefore, we would expect a shift in the interference fringes of the interferometer when a squeezed state is injected.
This effect is thus in principle observable.

Finally, notice that some of the terms in (\ref{eqns:exp_P}) can give rise to beats when the frequencies associated with the squeezed and coherent states are close but not equal.
Therefore, even though one would expect that the oscillations caused by these terms are well outside the range of frequencies to which the interferometer is sensitive, for some values of these frequencies the terms in (\ref{eqns:exp_P}) can give rise to beats in the detectable range of frequencies.
For instance, consider $\langle Z \rangle$.
We can write
{
\begin{multline}
	\langle Z \rangle = \left\{ \frac{3}{2} \cosh^2 r \sinh (2 r) - |\alpha|^2 \left[ |\alpha|^2 + 3 \cosh^2 r - \frac{3}{2} \sinh (2 r)\right]\right\} \cos (2 \omega_c t) + \\ 
		- 3 (|\alpha|^2 + \cosh^2 r) \sinh (2 r) \left[\cos \left(\frac{2 \omega_c + \omega_s}{2} t\right) \cos \left(\frac{2 \omega_c - \omega_s}{2} t \right) \right]~, \label{eqn:beat}
\end{multline}
}
where $\omega_c$ and $\omega_s$ are the frequencies of the coherent and squeezed states, respectively.
We can then see that when $2\omega_c \sim \omega_s$ an interferometer sensitive to low frequencies, like LIGO, can potentially observe a quadratic dependence on the momentum of the uncertainty principle.

As for the variance $(\Delta \mathcal{P})^2${, in the case $\alpha \in \mathbb{R}$ and $\theta=0$ \cite{Caves1981_1}, we have
\begin{equation}
	(\Delta \mathcal{P})^2 \propto  \Lambda_0 + \tilde{\Gamma} \delta \Lambda_1 + \tilde{\Gamma}^2 [ \delta^2 \Lambda_{2\delta} + \epsilon \Lambda_{2\epsilon} ]~,
\end{equation}
where
\begin{subequations}
\begin{align}
	\Lambda_0 = & \alpha^2 e^{2r} + \sinh^2 r~, \\
	\Lambda_1 = & \frac{3 \alpha}{4} \left[2 - (3 + 4 \alpha^2) e^{2 r} - e^{4 r}\right]~, \displaybreak\\
	\Lambda_{2\delta} = & \frac{\alpha^4}{4} (14 \cosh(2r) + 25 e^{-2r} + 9) + \nonumber \\
		& - \frac{\alpha^2}{8} \left[- 3 (2^{7/2} - 25) + 42 (\sqrt{2} - 1) e^{-2r} + 9 (2^{5/2} - 7) e^{4r} + 6 (2^{3/2} - 23) \cosh (2r) + \right. \nonumber \\
		& \qquad \left. + 8 (3 \sqrt{2} - 8) \cosh (4 r)\right] + \nonumber \\
		& + \frac{1}{32} \left[- 67 + 24 \sqrt{2} + 12 (5 - 2^{5/2}) e^{-2r} - 36 (\sqrt{2} - 1) e^{4r} + 29 \cosh (2r)+ \right. \nonumber \\
		& \qquad \left.  + 12 \sqrt{2} \sinh (2r) - 9 (2^{5/2} - 15) \cosh (4r) + 15 \cosh (6 r)\right]~, \\
	\Lambda_{2\epsilon} = & \frac{\alpha^4}{2} e^{2 r}
		- \frac{3}{8} \alpha^2 \left[1 - 8 e^{2r} + 7 e^{4r} - 2 \sinh (2r)\right] + \nonumber \\
		& - \frac{3}{16} \left[- 3 + 3 e^{4 r} - 8 \sinh (2r) + \sinh (4r)\right] ~.
\end{align}
\end{subequations}}
Also in this case, we can identify a number of features.
We observe that, while the first term corresponding to the limit $\tilde{\Gamma} \rightarrow 0$ is positive definite, the other terms are not.
In particular, we see that, for $\delta>0$, the linear term can be negative.
We also notice that the second order terms of this expansion contain terms proportional to $e^{-4r}$ and $e^{4r}$ and are therefore more sensitive on $r$ than the leading term.
We would then expect these terms to become dominant for large $|r|$, with larger deviations from the standard results.
On the other hand, we also notice that some of these terms can lead to negative values of the uncertainty $(\Delta \mathcal{P})^2$, signaling the breakdown of this model close to the Planck scale.

This variance is associated with the radiation noise.
To characterize the shot noise, one has to consider the number of photons at the output ports of the interferometer (not of the beam splitter, as previously done for the radiation noise).
Once the shot noise is computed for a similar state as \eqref{eqn:state_ligo}, one can find the set of parameters that minimize the total noise in the interferometer.
Eventually, as we can guess from the results for the radiation pressure noise, the minimal noise regime will depend on $\tilde{\Gamma}$ and $r$.
Therefore, we expect that in an interferometric experiment it is possible to observe Planck scale effects.

\subsection{GUP Breaks the Equivalence Principle?}

It is interesting to note that a quadratic GUP model resembles the relativistic correction to the classical dispersion relation
\begin{equation}
	\frac{p^2}{2m} = \frac{p_0^2}{2m} \left( 1 + 2 \frac{\gamma_0^2}{M_{\mathrm{Pl}}^2 c^2} \epsilon p_0^2 \right)~, \qquad \sqrt{p_0^2 c^2 + m^2 c^4} \simeq m c^2 + \frac{p_0^2}{2 m} \left(1  - \frac{p_0^2}{4 m^2 c^2}\right)~,
\end{equation}
where we explicitly wrote the dependence on the Planck scale for the GUP correction.
Therefore, starting from a standard, non-relativistic theory, we would expect both corrections to be present, that is
\begin{equation}
	E 
	= \frac{p_0^2}{2m} \left[ 1 + \frac{p_0^2}{c^2} \left( 2 \frac{\gamma_0^2 \epsilon}{M_{\mathrm{Pl}}^2} - \frac{1}{4 m^2} \right) \right] 
	= \frac{p_0^2}{2m} \left[ 1 + \frac{p_0^2}{c^2} \left( \frac{8 \gamma_0^2 \epsilon m^2 - M_{\mathrm{Pl}}^ 2 }{4 M_{\mathrm{Pl}}^2 m^2} \right) \right]~.
\end{equation}
We can then define a quantity with unit of mass,
\begin{equation}
	\mu = \frac{M_{\mathrm{Pl}} m}{ \sqrt{M_{\mathrm{Pl}}^2 - 8 \gamma_0^2 \epsilon m^2}}~, \label{def:mu_vs_m}
\end{equation}
such that
\begin{equation}
	E 
	= \frac{p_0^2}{2m} \left( 1 - \frac{p_0^2}{4 \mu^2 c^2} \right)~.
\end{equation}
This equation can be obtained as the limit of small momenta and small masses of the following dispersion relation
\begin{equation}
	E^2 
	= \mu^2 c^4 + p_0^2 c^2 + 2 \gamma^2 \epsilon p_0^4 c^2 
	= \mu^2 c^4 + p^2 c^2~, \label{eqn:disp_rel_GUP+SR}
\end{equation}
where we used the expansion $p^2 = p_0^2 (1 + 2 \gamma^2 \epsilon p_0^2)$.
We then see that including GUP in a relativistic theory is equivalent to changing the definition of mass.
Evidently, from \eqref{def:mu_vs_m} we see our observations are valid only for
\begin{equation}
	m < \frac{ M_{\mathrm{Pl}}}{2\sqrt{2 \gamma_0^2 \epsilon}} \equiv m_\mathrm{max}~, \label{eqn:condition_max_mass}
\end{equation}
beyond which the ``new mass'' $\mu$ becomes complex.
Furthermore, while for $m/M_\mathrm{Pl} \ll 1 $ we have $\mu \simeq m$, for large masses (compared to particle masses), \emph{i.e.} $m \simeq m_\mathrm{max}$, $\mu$ diverges.
looking at \eqref{eqn:disp_rel_GUP+SR}, we also observe that $\mu$ can be interpreted as a GUP-corrected inertial mass, as long as the condition in \eqref{eqn:condition_max_mass} is fulfilled.
We could then wonder what $m$ is.
When \eqref{def:mu_vs_m} is reversed we find
\begin{equation}
	m = \frac{M_{\mathrm{Pl}} \mu}{ \sqrt{M_{\mathrm{Pl}}^2 + 8 \gamma_0^2 \epsilon \mu^2}}~.
\end{equation}
We then see that $m$ grows slower than $\mu$ and that for $\mu \rightarrow \infty$, $m$ will tend to $M_\mathrm{Pl}/8 \gamma_0 \epsilon$.
Therefore, if we consider processes, like collisions between two particles of masses $\mu_1$ and $\mu_2$, while the final state will have a final mass $\mu = \mu_1 + \mu_2$ (we are assuming a classical collision process, with conservation of mass), we will have $m \not = m_1 + m_2$, where $m_1$ and $m_2$ are the $m$-type masses associated with the two bodies in collision.
The mass $m$ will actually be a function of $m_1$, $m_2$, and $M_\mathrm{Pl}$.
This last term, and in particular the fact that it contains $G$, may be a hint that $m$ is associated with the gravitational mass.
Thus, assuming this association, we have that GUP breaks the Equivalence Principle, imposing a non-linear relation between the inertial and gravitational mass.
Work on this topic is currently in progress.

Condition \eqref{eqn:condition_max_mass} also demonstrates another issue.
Everyday life objects, like a soccer ball, have a mass $m \sim 10^8 M_\mathrm{Pl}$ and no experiment so far has shown any deviation from the Equivalence Principle \cite{Will2014}.
Therefore, a consistent theory that includes GUP and the effects that we have shown in this Subsection has to take also these last issues into consideration.

\subsection{Relativistic QM / QFT with GUP}

As we pointed out several times throughout this thesis, future questions concern how relativistic QM, and ultimately QFT, is changed by GUP.
This and similar topics have been already explored in the past, from different perspectives and with different methodologies (\emph{e.g.}, see \cite{Das2010_1,Kober2010,Pramanik2013_1}), but to our knowledge, no systematic analysis of QFT with minimal length has been conducted.
The necessity of this study comes from the expected relevance of QG effects in high-energetic (relativistic) processes or, reversing this argument, by the expectation that relativistic physics is an easier ground on which QGP can be performed (see, for instance, the comments concerning experimental bounds for the GUP parameter using hydrogen atom spectroscopy).

A possible way to approach this problem, similar to the technique of \cite{Kober2010}, is to consider similar QFT Lagrangians, substituting momentum terms as in \eqref{eqn:expansion_momentum}, \emph{i.e.}
\begin{equation}
	\partial_\mu \rightarrow \tilde{\partial}_\mu \equiv \partial_\mu \left(1 + \Gamma^2 \partial_\nu \partial^\nu \right)~, \label{eqn:sub_derivative}
\end{equation}
where $\Gamma$, containing information on the Planck scale, is the GUP parameter in natural units and where we considered a quadratic model for simplicity.
Furthermore, notice that we extended the prescription in \eqref{eqn:expansion_momentum} in a covariant way.
In particular, using the correspondence principle we can write the following correction for the energy of a system
\begin{equation}
	E \rightarrow E = E_0(1 + \Gamma^2 m^2)~,
\end{equation}
where $m^2 = p^\mu p_\mu$ is the inertial mass of the system.

With this prescription, for example, we have for the case of Dirac fields
\begin{equation}
	\mathcal{L} = \bar{\Phi} (i \dslash \tilde{\partial} - m) \Phi~.
\end{equation}
For the Euler--Lagrange equations, using $\bar{\Phi}$ as independent fields, we find
\begin{equation}
	\frac{\partial \mathcal{L}}{\partial \bar{\Phi}} - \partial_\mu \frac{\partial \mathcal{L}}{\partial (\tilde{\partial}_\mu \bar{\Phi})} = \frac{\partial \mathcal{L}}{\partial \bar{\Phi}} = (i \dslash \tilde{\partial} - m) \Phi = 0~,
\end{equation}
that corresponds to the GUP-modified Dirac equation.
Equivalently we can write
\begin{equation}
	\left\{- i \alpha^i \tilde{\partial}_i + \beta m\right\} \Phi = i \tilde{\partial}_0 \Phi~.
\end{equation}
It is worth noticing that the conjugate momentum of the field $\Phi$ remains the standard one
\begin{equation}
	\Pi = \frac{\partial \mathcal{L}}{\partial ( \tilde{\partial}_0 \Phi)} = i \Phi^\dagger~.
\end{equation}
Similarly, one can use the same substitution \eqref{eqn:sub_derivative} in the Klein--Gordon and Proca equation, being able at this point of studying systems of scalar, vector, and tensor fields, rewriting Feynman diagram rules, studying loops and, ultimately, looking for effects of a minimal length in relativistic systems, as the relativistic hydrogen atom.

A different procedure follows Sec.~\ref{ssec:fund_QO}.
As in \cite[Ch. XXI]{Messiah}, we can start with a classical field and a given equation of motion.
Once the fields are expressed in terms of their normal coordinates, one can quantize them imposing a modified commutation (anticommutation) relation for each of the normal coordinates and the respective momenta.
In this way, each field is described as an infinite linear combination of harmonic oscillators satisfying similar conditions of Chapter \ref{ch:HO}.
An immediate consequence is that creating two pairs of particles will cost a quantity of energy different from simply the double of the energy required for the creation of a single pair.
This is due to the different energy ladder of the modified HO \eqref{eqn:energ_eigenstate}.

\section{Closing Remarks}

The discovery of the Uncertainty Principle is one of the revolutionary events that signaled the end of Classical Physics and the rise of Quantum Physics.
Its implications are in strong contrast with those of Newtonian Mechanics, where uncertainties are simply consequences of our incomplete knowledge of a physical system and of unavoidable experimental errors.

In the present times, we may be at the dawn of a new revolution, of similar entity as the previous one:
the fact that the Uncertainty Principle is no longer valid seems to be a solid characteristic, one that we expect to be present in a future theory of QG.
The final form of this new Uncertainty Principle and, ultimately, the final form of the quantum theory of gravity are yet to be discovered, and a new great effort, similar to the effort that marked the beginning of the twentieth century, is underway.
There is, though, a fundamental differences between the development of the current physical theories and of QG.
When QM and GR were forged, clear evidence of something new, something that could not be accounted for in the well-established theories of Newtonian Mechanics and of Maxwell's Electromagnetism, led to radical changes in the description of Nature.
On the contrary, since the experimental tools at our disposal are not always adequate to help us in the task of developing a quantum theory of gravity, we are at the present time forced to proceed blindly, without the light of any experimental support.
We know that Nature cannot be completely described by our current theories, that something new is still out there, but we do not have any experimental evidence for it, nor any experimental evidence against QM or GR.

On the other hand, if we knew where to look at, we may be very close to finding this evidence, directly or indirectly.
The role of QGP is therefore decisive for the future development of Physics.
Through different approaches and studying different (expected) QG effects, QGP could point us towards where and how to look.
In this search, no stone should be left unturned, since the prize is the next leap in the millennial journey of Physics.

\bibliography{thesis} 

\appendix

\chapter{Clebsch-Gordan Coefficients}
\chaptermark{apx-CG}
\label{apx:CG}
In this Appendix, we will calculate several CG coefficients for different values of the total azimuthal and magnetic quantum numbers, L and M, referring these values to the maximum values $L_{\mathrm{max}} = l_1+l_2$ and $M_{\mathrm{max}} = L_{\mathrm{max}}$, where $l_1$ and $l_2$ are the azimuthal quantum numbers of the individual systems.
In what follows, for a more compact notation, the states with respect which the expectation values are considered are indicated by subscripts, \emph{i.e.}
\begin{equation}
	\langle \CE \rangle_{l,m} = \langle lm | \CE | lm \rangle~.
\end{equation}

\section{$L=L_{\mathrm{max}}$, $M=L_{\mathrm{max}}$}
For this case, the state represented by the total angular momentum can be related to just one combination of the states concerning the individual angular momenta, that is
\begin{equation}
	|L_\mathrm{max},L_\mathrm{max}\rangle = |l_1;l_2\rangle~.
\end{equation}
This means that the CG coefficient for this case is simply
\begin{equation}
	\langle l_1;l_2|L_\mathrm{max},L_\mathrm{max}\rangle = 1~. \label{eqn:CG_max_max}
\end{equation}

\section{$M=L_{\mathrm{max}}-1$}
Two states are possible
\begin{align}
		|l_1-1;l_2\rangle~, & & |l_1;l_2-1\rangle~.
\end{align}

\subsection{$L = L_{\mathrm{max}}$}
Applying the lowering operator $L_- = l_{1,-} + l_{2,-}$ we find
\begin{multline}
	|L_\mathrm{max},L_\mathrm{max}-1\rangle \propto  L_-|L_\mathrm{max},L_\mathrm{max}\rangle = (l_{1,-} + l_{2,-})|l_1;l_2\rangle = \\
	= \hbar[(1- \langle \mathcal{C} \rangle_{l_1, l_1})\sqrt{2 l_1}|l_1-1;l_2\rangle + (1- \langle \mathcal{C} \rangle_{l_2, l_2} ) \sqrt{2 l_2}|l_1;l_2-1\rangle]~,
\end{multline}
where we used the result in (\ref{eqn:CG_max_max}) and the relation (\ref{eqn:L+-}).
Since both these coefficients are positive (we are assuming that $\langle \mathcal{C} \rangle$ is smaller than 1), the Condon--Shortley phase convention is already fulfilled, we need just to normalize this combination since
\begin{equation}
	||L_-|L_\mathrm{max},L_\mathrm{max} -1 \rangle||^2 = 2\hbar^2 [(1- \langle \mathcal{C} \rangle_{l_1, l_1} )^2 l_1 + (1- \langle \mathcal{C} \rangle_{l_2, l_2} )^2 l_2]~.
\end{equation}
Thus, the two CG coefficients for this case are
\begin{subequations} \label{eqn:CG_0-1}
\begin{align}
	\langle l_1-1;l_2|L_{\mathrm{max}},L_{\mathrm{max}}-1\rangle & 
	= \frac{(1- \langle \mathcal{C} \rangle_{l_1, l_1} )\sqrt{l_1}}{\sqrt{(1- \langle \mathcal{C} \rangle_{l_1, l_1} )^2 l_1 + (1- \langle \mathcal{C} \rangle_{l_2, l_2} )^2 l_2}}~, \\
	\langle l_1;l_2-1|L_{\mathrm{max}},L_{\mathrm{max}}-1\rangle &
	= \frac{(1- \langle \mathcal{C} \rangle_{l_2, l_2} )\sqrt{l_2}}{\sqrt{(1- \langle \mathcal{C} \rangle_{l_1, l_1} )^2 l_1 + (1- \langle \mathcal{C} \rangle_{l_2, l_2} )^2 l_2}}~.
	\end{align}
\end{subequations}

\subsection{$L = L_{\mathrm{max}}-1$}
In this case, we will find the two CG coefficients for $|L_{\mathrm{max}}-1,L_{\mathrm{max}}-1\rangle$ applying the orthonormality condition between this state and $|L_{\mathrm{max}},L_{\mathrm{max}}-1\rangle$.
The state in this case can be written as a linear combination of $|l_1-1;l_2\rangle$ and $| l_1;l_2-1\rangle$
\begin{equation}
	|L_\mathrm{max}-1,L_\mathrm{max}-1\rangle = G_{10}|l_1-1;l_2\rangle + G_{01}| l_1;l_2-1\rangle~.
\end{equation}
From the orthogonality condition we find
\begin{multline}
	\langle L_\mathrm{max},L_\mathrm{max} -1 |L_\mathrm{max}-1,L_\mathrm{max}-1\rangle = \\
	= \frac{(1- \langle \mathcal{C} \rangle_{l_1, l_1} )\sqrt{l_1}}{\sqrt{(1- \langle \mathcal{C} \rangle_{l_1, l_1} )^2 l_1 + (1- \langle \mathcal{C} \rangle_{l_2, l_2} )^2 l_2}} G_{10} + \frac{(1- \langle \mathcal{C} \rangle_{l_2, l_2} )\sqrt{l_2}}{\sqrt{(1- \langle \mathcal{C} \rangle_{l_1, l_1} )^2 l_1 + (1- \langle \mathcal{C} \rangle_{l_2, l_2} )^2 l_2}} G_{01} = \\
	= 0~,
\end{multline}
obtaining
\begin{equation}
	G_{01} = - \frac{(1- \langle \mathcal{C} \rangle_{l_1, l_1} )\sqrt{l_1}}{(1- \langle \mathcal{C} \rangle_{l_2, l_2} )\sqrt{l_2}}G_{10}~.
\end{equation}
Normalizing the state
\begin{equation}
	|G_{10}|^2 \left[1 + \frac{(1- \langle \mathcal{C} \rangle_{l_1, l_1} )^2 l_1}{(1- \langle \mathcal{C} \rangle_{l_2, l_2} )^2 l_2} \right] = |G_{10}|^2 \frac{(1- \langle \mathcal{C} \rangle_{l_1, l_1} )^2 l_1 + (1- \langle \mathcal{C} \rangle_{l_2, l_2} )^2 l_2}{(1- \langle \mathcal{C} \rangle_{l_2, l_2} )^2 l_2} = 1~,
\end{equation}
and imposing the Condon--Shortley phase convention
\begin{equation}
	\langle l_1; L_\mathrm{max}-1-l_1| L_\mathrm{max}-1,L_\mathrm{max}-1\rangle = \langle l_1; l_2-1| L_\mathrm{max}-1,L_\mathrm{max}-1\rangle \geq 0~.
\end{equation}
we have
\begin{subequations}
	\begin{align}
		\langle l_1-1;l_2|L_{\mathrm{max}}-1,L_{\mathrm{max}}-1\rangle &
		= - \frac{(1- \langle \mathcal{C} \rangle_{l_2, l_2} ) \sqrt{l_2}}{\sqrt{(1- \langle \mathcal{C} \rangle_{l_1, l_1} )^2 l_1 + (1- \langle \mathcal{C} \rangle_{l_2, l_2} )^2 l_2}}~, \\
		\langle l_1;l_2-1|L_{\mathrm{max}}-1,L_{\mathrm{max}}-1\rangle &
		= \frac{(1- \langle \mathcal{C} \rangle_{l_1, l_1} ) \sqrt{l_1}}{\sqrt{(1- \langle \mathcal{C} \rangle_{l_1, l_1} )^2 l_1 + (1- \langle \mathcal{C} \rangle_{l_2, l_2})^2 l_2}}.
	\end{align}
\end{subequations}

\section{$M=L_{\mathrm{max}}-2$}
In this case, the three possible states are
\begin{align}
		|l_1-2;l_2\rangle~,&& |l_1-1;l_2-1\rangle~, && |l_1;l_2-2\rangle~.
\end{align}

\subsection{$L=L_{\mathrm{max}}$}
The CG coefficients for this case are found acting one time with $L_-$ on the state \newline\mbox{$|L_{\mathrm{max}},L_{\mathrm{max}}-1\rangle$}
\begin{align}
	L_- |L_{\mathrm{max}},L_{\mathrm{max}}-2\rangle \propto & (l_{1,-} + l_{2,-})[\langle l_1 - 1;l_2|L_{\mathrm{max}},L_{\mathrm{max}}-1\rangle |l_1 - 1;l_2\rangle + \nonumber \\
	& + \langle l_1;l_2 - 1|L_{\mathrm{max}},L_{\mathrm{max}}-1\rangle |l_1;l_2-1\rangle] = \nonumber \\
	= & \hbar \{\langle l_1 - 1;l_2|L_{\mathrm{max}},L_{\mathrm{max}}-1\rangle[(1- \langle \mathcal{C} \rangle_{l_1, l_1-1} ) \sqrt{4l_1 - 2}|l_1 - 2;l_2\rangle + \nonumber \\
	& + (1 - \langle \mathcal{C} \rangle_{l_2, l_2} )\sqrt{2l_2}|l_1-1;l_2-1\rangle] + \nonumber \\
	& + \langle l_1;l_2 - 1|L_{\mathrm{max}},L_{\mathrm{max}}-1\rangle[(1- \langle \mathcal{C} \rangle_{l_1, l_1} )\sqrt{2l_1}|l_1 - 1;l_2-1\rangle + \nonumber \\
	& + (1- \langle \mathcal{C} \rangle_{l_2, l_2-1} )\sqrt{4l_2 - 2}|l_1;l_2-2\rangle]\} = \nonumber \\
	= & \hbar\left\{\frac{\sqrt{2}(1- \langle \mathcal{C} \rangle_{l_1, l_1} ) (1- \langle \mathcal{C} \rangle_{l_1, l_1-1} ) \sqrt{l_1(2l_1-1)}}{\sqrt{(1- \langle \mathcal{C} \rangle_{l_1, l_1}  )^2l_1 + (1- \langle \mathcal{C} \rangle_{l_2, l_2} )^2l_2}}|l_1 - 2;l_2\rangle + \right. \nonumber \\
	& \frac{2\sqrt{2}(1- \langle \mathcal{C} \rangle_{l_1, l_1} )(1- \langle \mathcal{C} \rangle_{l_2, l_2} )\sqrt{l_1l_2}}{\sqrt{(1-  \langle \mathcal{C} \rangle_{l_1, l_1} )^2l_1 + (1- \langle \mathcal{C} \rangle_{l_2, l_2} )^2l_2}}|l_1-1;l_2-1\rangle + \nonumber \\
	& + \left. \frac{\sqrt{2}(1- \langle \mathcal{C} \rangle_{l_2, l_2} )(1- \langle \mathcal{C} \rangle_{l_2, l_2-1} ) \sqrt{l_2(2l_2-1)}}{\sqrt{(1- \langle \mathcal{C} \rangle_{l_1, l_1} )^2l_1 + (1- \langle \mathcal{C} \rangle_{l_2, l_2} )^2l_2}}|l_1;l_2-2\rangle \right\}~,
\end{align}
where we used the relation (\ref{eqn:L+-}) and the coefficients in (\ref{eqn:CG_0-1}).
Normalizing this last result we find
\begin{subequations}
\begin{align}
	\langle l_1 - 2;l_2| L_\mathrm{max}, L_\mathrm{max}-2\rangle &
	= \frac{(1- \langle \mathcal{C} \rangle_{l_1, l_1} ) (1- \langle \mathcal{C} \rangle_{l_1, l_1-1} ) \sqrt{l_1(2l_1-1)}}{\Omega_0}~, \\
	\langle l_1 - 1;l_2 - 1| L_\mathrm{max}, L_\mathrm{max}-2\rangle &
	= \frac{2(1- \langle \mathcal{C} \rangle_{l_1, l_1} )(1- \langle \mathcal{C} \rangle_{l_2, l_2} )\sqrt{l_1l_2}}{\Omega_0}~, \\
	\langle l_1;l_2 - 2| L_\mathrm{max}, L_\mathrm{max}-2\rangle &
	= \frac{(1- \langle \mathcal{C} \rangle_{l_2, l_2} )(1- \langle \mathcal{C} \rangle_{l_2, l_2-1} )\sqrt{l_2(2l_2-1)}}{\Omega_0}~.
	\end{align}
\end{subequations}
where
\begin{multline}
	\Omega_0 = \left[(1- \langle \mathcal{C} \rangle_{l_1, l_1} )^2 (1- \langle \mathcal{C} \rangle_{l_1, l_1-1} )^2 l_1(2l_1-1) + 4 (1- \langle \mathcal{C} \rangle_{l_1, l_1} )^2 (1- \langle \mathcal{C} \rangle_{l_2, l_2} )^2 l_1l_2 + \right.\\
	+ \left. (1- \langle \mathcal{C} \rangle_{l_2, l_2} )^2 (1- \langle \mathcal{C} \rangle_{l_2, l_2-1} )^2 l_2(2l_2-1) \right]^{1/2}~.
\end{multline}
Since these coefficients are all positive, the phase convention is already fulfilled.

\subsection{$L=L_{\mathrm{max}} - 1$}

To find the CG coefficients for this case we apply $L_-$ on the state $|L_{\mathrm{max}} - 1, L_{\mathrm{max}} - 1\rangle$
\begin{align}
	L_-|L_{\mathrm{max}} - 1, L_{\mathrm{max}} - 1\rangle = & (l_{1,-} + l_{2,-})[\langle l_1 - 1;l_2|L_{\mathrm{max}}-1,L_{\mathrm{max}}-1\rangle |l_1 - 1;l_2\rangle + \nonumber \\ 
	& \langle l_1;l_2 - 1|L_{\mathrm{max}}-1,L_{\mathrm{max}}-1\rangle |l_1;l_2-1\rangle] = \nonumber \\
	= & \hbar \langle l_1 - 1;l_2|L_{\mathrm{max}}-1,L_{\mathrm{max}}-1\rangle \times \nonumber \\
	& \qquad \times [(1- \langle \mathcal{C} \rangle_{l_1, l_1-1} ) \sqrt{4l_1 -2}|l_1 - 2;l_2\rangle + \nonumber \\
	& \qquad \quad + (1- \langle \mathcal{C} \rangle_{l_2, l_2} ) \sqrt{2l_2}|l_1 - 1;l_2-1\rangle] + \nonumber \\
	& + \hbar \langle l_1;l_2-1|L_{\mathrm{max}}-1,L_{\mathrm{max}}-1\rangle \times \nonumber \\
	& \qquad \times [(1- \langle \mathcal{C} \rangle_{l_1, l_1} ) \sqrt{2l_1}|l_1 - 1;l_2-1\rangle + \nonumber \\
	& \qquad \quad + (1- \langle \mathcal{C} \rangle_{l_2, l_2-1} ) \sqrt{4l_2 - 2}|l_1;l_2-2\rangle] \propto \nonumber \\
	\propto & - \hbar (1- \langle \mathcal{C} \rangle_{l_1, l_1-1} ) (1- \langle \mathcal{C} \rangle_{l_2, l_2} ) \sqrt{4l_1 -2}\sqrt{l_2}|l_1 - 2;l_2\rangle + \nonumber \\
	& + \hbar [(1- \langle \mathcal{C} \rangle_{l_1, l_1} )^2 \sqrt{2} l_1 - (1- \langle \mathcal{C} \rangle_{l_2, l_2} )^2 \sqrt{2} l_2]|l_1 - 1;l_2-1\rangle + \nonumber \\
	& + \hbar (1- \langle \mathcal{C} \rangle_{l_1, l_1} ) (1- \langle \mathcal{C} \rangle_{l_2, l_2-1} ) \sqrt{l_1}\sqrt{4l_2 - 2}|l_1;l_2-2\rangle~.
\end{align}
Normalizing we thus find
\begin{subequations}
\begin{align}
	\langle l_1 - 2;l_2| L_\mathrm{max}-1, L_\mathrm{max}-2\rangle &
	= \frac{(1- \langle \mathcal{C} \rangle_{l_1, l_1-1} ) (1- \langle \mathcal{C} \rangle_{l_2, l_2} ) \sqrt{4l_1 -2}\sqrt{l_2}}{\Omega_1}~, \\
	\langle l_1 - 1;l_2 - 1| L_\mathrm{max} -1, L_\mathrm{max}-2\rangle & 
	= - \frac{(1- \langle \mathcal{C} \rangle_{l_1, l_1} )^2 \sqrt{2} l_1 - (1- \langle \mathcal{C} \rangle_{l_2, l_2} )^2 \sqrt{2} l_2}{\Omega_1}~, \\
	\langle l_1;l_2 - 2| L_\mathrm{max} - 1, L_\mathrm{max}-2\rangle &
	= - \frac{(1- \langle \mathcal{C} \rangle_{l_1, l_1} ) (1- \langle \mathcal{C} \rangle_{l_2, l_2-1} ) \sqrt{l_1}\sqrt{4l_2 - 2}}{\Omega_1}~,
	\end{align}
\end{subequations}
with
\begin{multline}
	\Omega_1 =\left[ (1- \langle \mathcal{C} \rangle_{l_1, l_1-1} )^2 (1- \langle \mathcal{C} \rangle_{l_2, l_2} )^2 (4l_1-2) l_2 + (1- \langle \mathcal{C} \rangle_{l_1, l_1} )^4 2 l_1^2 + \right. \\
	+ (1- \langle \mathcal{C} \rangle_{l_2, l_2} )^4 2 l_2^2 - (1- \langle \mathcal{C} \rangle_{l_1, l_1} )^2 (1- \langle \mathcal{C} \rangle_{l_2, l_2} )^2 4 l_1 l_2 + \\
	\left. + (1- \langle \mathcal{C} \rangle_{l_1, l_1} )^2 (1- \langle \mathcal{C} \rangle_{l_2, l_2-1} )^2 l_1 (4l_2 - 2) \right]^{1/2}~.
\end{multline}

\subsection{$L=L_{\mathrm{max}} - 2$}
As first step, let us define the CG coefficients for this case in the following way
\begin{equation}
	|L_{\mathrm{max}} - 2, L_{\mathrm{max}} - 2\rangle = G_{20}|l_1-2,l_2\rangle + G_{11}|l_1-1,l_2-1\rangle + G_{02}|l_1,l_2-2\rangle~.
\end{equation}
Consider the orthogonality between the state $|L_{\mathrm{max}} - 2, L_{\mathrm{max}} - 2\rangle$ and \mbox{$|L_{\mathrm{max}} - 1,L_{\mathrm{max}} - 2\rangle$}
\begin{multline}
	-G_{20} (1- \langle \mathcal{C} \rangle_{l_1, l_1-1} ) (1- \langle \mathcal{C} \rangle_{l_2, l_2} ) \sqrt{4l_1 -2}\sqrt{l_2} + G_{11}[(1- \langle \mathcal{C} \rangle_{l_1, l_1} )^2 \sqrt{2} l_1 + \\ 
	- (1- \langle \mathcal{C} \rangle_{l_2, l_2} )^2 \sqrt{2} l_2] + G_{02} (1- \langle \mathcal{C} \rangle_{l_1, l_1} ) (1- \langle \mathcal{C} \rangle_{l_2, l_2-1} ) \sqrt{l_1}\sqrt{4l_2 - 2} = 0 \label{eqn:-2-2_step1}
\end{multline}
and the orthogonality between the first state and $|L_{\mathrm{max}}, L_{\mathrm{max}} - 2\rangle$
\begin{multline}
	G_{20} (1- \langle \mathcal{C} \rangle_{l_1, l_1-1} ) (1- \langle \mathcal{C} \rangle_{l_1, l_1} ) \sqrt{l_1(2l_1-1)}  + 2G_{11}(1- \langle \mathcal{C} \rangle_{l_1, l_1} ) (1- \langle \mathcal{C} \rangle_{l_2, l_2} ) \sqrt{l_1l_2} \\
	+ G_{02} (1- \langle \mathcal{C} \rangle_{l_2, l_2} ) (1- \langle \mathcal{C} \rangle_{l_2, l_2-1} ) \sqrt{l_2(2l_2-1)} = 0~.
\end{multline}
From the latter we have
\begin{equation}
	G_{11} = - G_{20} \frac{1- \langle \mathcal{C} \rangle_{l_1, l_1-1}}{1- \langle \mathcal{C} \rangle_{l_2, l_2}} \frac{\sqrt{2l_1-1}}{2\sqrt{l_2}} 
	- G_{02} \frac{1- \langle \mathcal{C} \rangle_{l_2, l_2-1}}{1- \langle \mathcal{C} \rangle_{l_1, l_1}} \frac{\sqrt{2l_2-1}}{2\sqrt{l_1}}~. \label{eqn:-2-2_step2}
\end{equation}
Inserting this last result in (\ref{eqn:-2-2_step1}) we find
\begin{multline}
	- G_{20} \frac{1- \langle \mathcal{C} \rangle_{l_1, l_1-1}}{1- \langle \mathcal{C} \rangle_{l_2, l_2}} \frac{\sqrt{2l_1-1}}{\sqrt{2l_2}} \left[ (1- \langle \mathcal{C} \rangle_{l_2, l_2})^2 l_2 + (1- \langle \mathcal{C} \rangle_{l_1, l_1} )^2 l_1 \right] + \\
	+ G_{02} \frac{1- \langle \mathcal{C} \rangle_{l_2, l_2-1}}{1- \langle \mathcal{C} \rangle_{l_1, l_1}} \frac{\sqrt{2l_2-1}}{\sqrt{2l_1}} \left[ (1- \langle \mathcal{C} \rangle_{l_2, l_2})^2 l_2 + (1- \langle \mathcal{C} \rangle_{l_1, l_1} )^2 l_1 \right] = 0 \Rightarrow\\
	\Rightarrow G_{20} = G_{02} \frac{(1- \langle \mathcal{C} \rangle_{l_2, l_2}) (1- \langle \mathcal{C} \rangle_{l_2, l_2-1})}{(1- \langle \mathcal{C} \rangle_{l_1, l_1}) (1- \langle \mathcal{C} \rangle_{l_1, l_1-1})} \frac{\sqrt{l_2(2l_2-1)}}{\sqrt{l_1(2l_1-1)}}~,
\end{multline}
whence, using this result in (\ref{eqn:-2-2_step2}) we obtain
\begin{equation}
	G_{11} = - G_{02} \frac{1- \langle \mathcal{C} \rangle_{l_2, l_2-1}}{1- \langle \mathcal{C} \rangle_{l_1, l_1}} \frac{\sqrt{2l_2-1}}{\sqrt{l_1}}~.
\end{equation}
Imposing the normalization condition
\begin{multline}
	\frac{|G_{02}|^2}{(1- \langle \mathcal{C} \rangle_{l_1, l_1})^2 (1- \langle \mathcal{C} \rangle_{l_1, l_1-1})^2} \left[ (1- \langle \mathcal{C} \rangle_{l_2, l_2})^2 (1- \langle \mathcal{C} \rangle_{l_2, l_2-1})^2 + \right. \\
	(1- \langle \mathcal{C} \rangle_{l_1, l_1-1})^2 (1- \langle \mathcal{C} \rangle_{l_2, l_2-1})^2 (2l_1-1) (2l_2-1) + \\
	\left. (1- \langle \mathcal{C} \rangle_{l_1, l_1})^2 (1- \langle \mathcal{C} \rangle_{l_1, l_1-1})^2 l_1(2l_1-1) \right] \frac{l_2(2l_2-1)}{l_1(2l_1-1)} = 1
\end{multline}
and the Condon--Shortley phase convention we have
\begin{subequations}
\begin{align}
	\langle l_1-2;l_2| L_\mathrm{max} - 2, L_\mathrm{max}-2\rangle = &
	\frac{(1- \langle \mathcal{C} \rangle_{l_2, l_2}) (1- \langle \mathcal{C} \rangle_{l_2, l_2-1})}{\Omega_2} ~, \\
	\langle l_1-1 ;l_2-1| L_\mathrm{max} - 2, L_\mathrm{max}-2\rangle = & 
	- \frac{(1- \langle \mathcal{C} \rangle_{l_2, l_2-1}) (1- \langle \mathcal{C} \rangle_{l_1, l_1-1})}{\Omega_2} \sqrt{\frac{2l_1-1}{l_2}} ~, \displaybreak \\
	\langle l_1;l_2 - 2| L_\mathrm{max} - 2, L_\mathrm{max}-2\rangle = & 
	\frac{(1- \langle \mathcal{C} \rangle_{l_1, l_1}) (1- \langle \mathcal{C} \rangle_{l_1, l_1-1})}{\Omega_2} \sqrt{\frac{l_1(2l_1-1)}{l_2(2l_2-1)}}~,
	\end{align}
\end{subequations}
where
\begin{multline}
	\Omega_2 = \left[(1- \langle \mathcal{C} \rangle_{l_2, l_2})^2 (1- \langle \mathcal{C} \rangle_{l_2, l_2-1})^2 + \right. \\
	+ (1- \langle \mathcal{C} \rangle_{l_1, l_1-1})^2 (1- \langle \mathcal{C} \rangle_{l_2, l_2-1})^2 (2l_1-1) (2l_2-1) + \\
	\left. + (1- \langle \mathcal{C} \rangle_{l_1, l_1})^2 (1- \langle \mathcal{C} \rangle_{l_1, l_1-1})^2 l_1(2l_1-1) \right]^{1/2}~.
\end{multline}

\section{$M = L_{\mathrm{max}} -3$}

In this last case, four states are possible
\begin{align}
	|l_1-3;l_2\rangle~,&& |l_1-2;l_2-1\rangle~, && |l_1-1;l_2-2\rangle~, && |l_1;l_2-3\rangle~.
\end{align}

\subsection{$L = L_\mathrm{max}$}

Let us apply $L_-$ on the state $|L_\mathrm{max}, L_\mathrm{max} - 2\rangle$
\begin{align}
	L_- |L_\mathrm{max}, L_\mathrm{max} - 2\rangle \propto 
	& (l_{1,-} + l_{2-}) [\langle l_1 - 2; l_2|L_\mathrm{max}, L_\mathrm{max} - 2\rangle |l_1 - 2; l_2\rangle + \nonumber \\
	& \langle l_1 - 1; l_2 - 1|L_\mathrm{max}, L_\mathrm{max} - 2\rangle |l_1 - 1; l_2 - 1\rangle + \nonumber \\
	& \langle l_1; l_2 - 2|L_\mathrm{max}, L_\mathrm{max} - 2\rangle |l_1; l_2 - 2\rangle ] = \nonumber
\end{align}
\vspace{-3em}
\begin{align}
	\propto & (1- \langle \mathcal{C} \rangle_{l_1, l_1} ) (1- \langle \mathcal{C} \rangle_{l_1, l_1-1} ) (1 - \langle \CE \rangle_{l_1,l_1-2}) \sqrt{l_1(2l_1-1)(3l_1 - 3)} |l_1-3;l_2\rangle + \nonumber \\
	& + 3 (1- \langle \mathcal{C} \rangle_{l_1, l_1} ) (1- \langle \mathcal{C} \rangle_{l_1, l_1-1} ) (1- \langle \mathcal{C} \rangle_{l_2, l_2} ) \sqrt{l_1(2l_1-1)l_2} |l_1-2; l_2-1\rangle + \nonumber \\
	& + 3 (1- \langle \mathcal{C} \rangle_{l_1, l_1} )(1- \langle \mathcal{C} \rangle_{l_2, l_2} ) (1 - \langle \CE \rangle_{l_2,l_2-1}) \sqrt{l_1l_2(2l_2-1)} |l_1-1; l_2-2\rangle + \nonumber \\
	& + (1- \langle \mathcal{C} \rangle_{l_2, l_2}) (1- \langle \mathcal{C} \rangle_{l_2, l_2-1}) (1 - \langle \CE \rangle_{l_2,l_2-2}) \sqrt{l_2(2l_2-1)(3l_2 - 3)} |l_1; l_2-3\rangle~.
\end{align}
After normalizing, we have
\begin{subequations}
\begin{align}
	\langle l_1 - 3; l_2|L_\mathrm{max}, L_\mathrm{max} - 3\rangle & = \frac{(1- \langle \mathcal{C} \rangle_{l_1, l_1} ) (1- \langle \mathcal{C} \rangle_{l_1, l_1-1} ) (1 - \langle \CE \rangle_{l_1,l_1-2})}{\Omega_3} \times \nonumber \\
	& \qquad \times  \sqrt{l_1(2l_1-1)(3l_1 - 3)}~,\\
	\langle l_1 - 2; l_2 - 1|L_\mathrm{max}, L_\mathrm{max} - 3\rangle & = \frac{3 (1- \langle \mathcal{C} \rangle_{l_1, l_1} ) (1- \langle \mathcal{C} \rangle_{l_1, l_1-1} ) (1- \langle \mathcal{C} \rangle_{l_2, l_2} )}{\Omega_3} \times \nonumber \\
	& \qquad \times  \sqrt{l_1(2l_1-1)l_2} ~,\\
	\langle l_1 - 1; l_2 - 2|L_\mathrm{max}, L_\mathrm{max} - 3\rangle & = \frac{3 (1- \langle \mathcal{C} \rangle_{l_1, l_1} )(1- \langle \mathcal{C} \rangle_{l_2, l_2} ) (1 - \langle \CE \rangle_{l_2,l_2-1})}{\Omega_3} \times \nonumber \\
	& \qquad \times  \sqrt{l_1l_2(2l_2-1)}~,\\
	\langle l_1; l_2 - 3|L_\mathrm{max}, L_\mathrm{max} - 3\rangle & = \frac{(1- \langle \mathcal{C} \rangle_{l_2, l_2}) (1- \langle \mathcal{C} \rangle_{l_2, l_2-1}) (1 - \langle \CE \rangle_{l_2,l_2-2})}{\Omega_3} \times \nonumber \\
	& \qquad \times \sqrt{l_2(2l_2-1)(3l_2 - 3)}~,
\end{align}
\end{subequations}
where
\begin{multline}
	\Omega_3 = [ (1- \langle \mathcal{C} \rangle_{l_1, l_1} )^2 (1- \langle \mathcal{C} \rangle_{l_1, l_1-1} )^2 (1 - \langle \CE \rangle_{l_1,l_1-2})^2 l_1(2l_1-1)(3l_1 - 3) + \\
	+ 9 (1- \langle \mathcal{C} \rangle_{l_1, l_1} )^2 (1- \langle \mathcal{C} \rangle_{l_1, l_1-1} )^2 (1- \langle \mathcal{C} \rangle_{l_2, l_2} )^2 l_1(2l_1-1)l_2 + \\
	+ 9 (1- \langle \mathcal{C} \rangle_{l_1, l_1} )^2 (1- \langle \mathcal{C} \rangle_{l_2, l_2} )^2 (1 - \langle \CE \rangle_{l_2,l_2-1})^2 l_1l_2(2l_2-1) + \\
	+ (1- \langle \mathcal{C} \rangle_{l_2, l_2})^2 (1- \langle \mathcal{C} \rangle_{l_2, l_2-1})^2 (1 - \langle \CE \rangle_{l_2,l_2-2})^2 l_2(2l_2-1)(3l_2 - 3) ]^{1/2}
\end{multline}
\chapter{Useful Relations} \label{apx:exps}

\newtheorem{result}{Result}

In this Appendix, we will prove a few useful relations.
For a more compact notation, we define
\begin{equation}
	[^n A, B] = \underbrace{[A,[A\ldots,[A,[A,}_{n~\mathrm{times}} B\underbrace{]]\ldots]]}_{n~\mathrm{times}}~.
\end{equation}
The first result consists in a simple relation for expanded nested commutators.

\begin{result} \label{res:com_binom}
Given two operators, $A$ and $B$, the following relation for the nested commutator holds
\begin{equation}
	[^n A, B] = \sum_{i=0}^n (-1)^i \binom{n}{i} A^{n-i} B A^i~. \label{eqn:com_binom}
\end{equation}
\end{result}

\begin{proof}
We will proceed by induction.
Notice that the result is trivial for $n=0$ and $n=1$.
For $n=2$ we have
\begin{equation}
	[^2 A,B] = [A,AB - BA] = A^2 B - 2 ABA + BA^2~.
\end{equation}
Let us assume that \eqref{eqn:com_binom} is valid for a generic natural number $n$.
For $n+1$ we find
\begin{multline}
	[^{n+1} A,B] = \left[A, \sum_{i=0}^n (-1)^i \binom{n}{i} A^{n-i} B A^i\right] = \sum_{i=0}^n (-1)^i \binom{n}{i} A^{n-i} [A,B] A^i = \\
	= A^{n+1} B + \sum_{i=1}^{n} (-1)^i \left\{\binom{n}{i} + \binom{n}{i-1}\right\} A^{n-i+1} B A^i + (-1)^{n+1} B A^{n+1}~.
\end{multline}
Using the following property of binomial coefficients
\begin{equation}
	\binom{n}{i} + \binom{n}{i-1} = n! \frac{n+1}{i!(n-i+1)!} = \binom{n+1}{i}~, \label{eqn:sum_binom}
\end{equation}
we finally find
\begin{multline}
	[^{n+1} A,B] = A^{n+1} B + \sum_{i=1}^{n} (-1)^i \binom{n+1}{i} A^{n-i+1} B A^i + (-1)^{n+1} B A^{n+1} = \\
	= \sum_{i=0}^{n+1} (-1)^i \binom{n+1}{i} A^{n+1-i} B A^i~. \qedhere
\end{multline}
\end{proof}

We can then prove a notorious result.
\begin{result}
Given two operators, $A$ and $B$, the following relation holds
\begin{equation}
    e^A B e^{-A} = B + [A,B] + \ldots + \frac{1}{n!} [^n A,B] + \ldots~. \label{eqn:hadamard}
\end{equation}
\end{result}

\begin{proof}
Let us write the exponential of an operator as sums of operators
\begin{align}
    e^{A} = & \sum_{i=0}^\infty \frac{A^i}{i!}~, & e^{-A} =& \sum_{i=0}^\infty (-1)^i \frac{A^i}{i!}~.
\end{align}
In this way, we can write
\begin{equation}
    e^{A} B e^{-A} = \left[ \sum_{i=0}^\infty \frac{A^i}{i!} \right] B \left[\sum_{j=0}^\infty (-1)^j \frac{A^j}{j!} \right] = \sum_{n=0}^\infty \frac{1}{n!} \sum_{i=0}^n (-1)^i \binom{n}{i}A^{n-i} B A^i~,
\end{equation}
where in the last equality we expressed the sum collecting terms with the same power of $A$ and noticing that
\begin{equation}
	\frac{1}{(n-i)!} \frac{1}{i!} = \binom{n}{i} \frac{1}{n!}~.
\end{equation}
Using the Result \ref{res:com_binom}, we find
\begin{equation}
    e^{A} B e^{-A} = \sum_{n=0}^\infty \frac{1}{n!} [^n A,B]~.
\end{equation}
\end{proof}

Another interesting result is the following.
\begin{result}
Given three operators, $A$, $B$, and $C$, the following relation holds
\begin{equation}
	[^n A, BC] = \sum_{i=0}^n \binom{n}{i} [^{n-i} A,B] [^i A,C]~. \label{eqn:nest_com_2}
\end{equation}
\end{result}

\begin{proof}
Let us prove it by induction.
For $n=0$ and $n=1$ it is trivial.
For $n=2$ we have
\begin{equation}
	[^2A,BC] = [A,B[A,C] + [A,B]C] = [^2 A,B]C + 2 [A,B][A,C] + B[^2 A,C]~.
\end{equation}
Let us then assume that \eqref{eqn:nest_com_2} is valid for an arbitrary natural number $n$.
We than find
\begin{multline}
	[^{n+1} A,BC] = \sum_{i=0}^n \binom{n}{i} \left( [^{n-i+1} A,B] [^i A,C] + [^{n-i} A, B] [^{i+1} A,B] \right) = \\
	= [^{n+1} A,B] C + \sum_{i=1}^n \left[ \binom{n}{i} + \binom{n}{i-1} \right] [^{n-i+1} A,B] [^i A,C] + B [^{n+1} A,C] = \\
	= \sum_{i=0}^{n+1} \binom{n+1}{i} [^{n+1-i} A,B] [^i A,C]~,
\end{multline}
where we used the property \eqref{eqn:sum_binom}.\qedhere
\end{proof}

\section{Commutators Involving Annihilation and Creation Operators}

In this section we want to find compact relations concerning commutators of annihilation and creation operators.
We start from the standard commutators
\begin{align}
	[a,a^\dagger] = & 1~, & [a,N] = & a~, & [a^\dagger,N] = & -a^\dagger~. \label{eqn:std_com}
\end{align}
We want to prove the following relations.
\begin{result}
The following commutation relations hold
\begin{align}
	[a,N^k] = & \left[ (N + 1)^k - N^k \right] a~, & [a^\dagger,N^k] = & - a^\dagger \left[ (N + 1)^k - N^k \right], & & \mbox{with } k \geq 1~.
\end{align}
\end{result}

\begin{proof}
	We will proceed by induction.
	The case $k=1$ is trivially satisfied by \eqref{eqn:std_com}.
	For $k=2$ we have
	\begin{equation}
		[a,N^2] = N [a,N] + [a,N] N = (2 N + 1)a~.
	\end{equation}
	Assuming the relation in the hypothesis to be true for a generic $k$, we want to prove it for $k+1$.
	In this case we have
	\begin{multline}
		[a,N^{k+1}] 
		= N [a,N^k] + [a,N]N^k 
		= \left[ \sum_{i=0}^{k-1} \binom{k}{i} N^{i+1} \right] a + a N^k = \\
		= \left[ \sum_{i=0}^{k-1} \binom{k}{i} N^{i+1} \right] a + \left[ \sum_{i=0}^k \binom{k}{i} N^i \right] a
		= a + \left\{ \sum_{i=1}^k \left[\binom{k}{i-1} + \binom{k}{i} \right] N^i \right\} a =\\
		= a + \left[ \sum_{i=1}^k \binom{k+1}{i} N^i \right] a
		= \left[ (N + 1)^k - N^k \right] a~.
	\end{multline}
	The second relation is found noticing that
	\begin{equation}
		[a^\dagger,N^k] = - \left( [a,N^k] \right)^\dagger = - a^\dagger \left[ (N + 1)^k - N^k \right]~.
	\end{equation}
\end{proof}

We now want to prove the following relation for nested commutators.
\begin{result}
	The following relation holds
	\begin{equation}
		[^m N^k,a] = (-1)^m \left[ (N + 1)^k - N^k \right]^m a \label{eqn:nested_Nk_a}
	\end{equation}
\end{result}

\begin{proof}
	First, notice that
	\begin{align}
		[^1 N^k,a] = & - \left[ (N + 1)^k - N^k \right] a \\
		[^2 N^k,a] = & - \left[ N^k, \left[ (N + 1)^k - N^k \right] a \right] = \left[ (N + 1)^k - N^k \right]^2 a
	\end{align}
	Let us assume the hypothesis true for a generic $m$, we then find
	\begin{equation}
		[^{m+1} N^k,a] = (-1)^m \left[ N^k , \left[ (N + 1)^k - N^k \right]^m a \right] = (-1)^{m+1} \left[ (N + 1)^k - N^k \right]^{m+1} a~.
	\end{equation}
\end{proof}
Similarly, we can find
\begin{equation}
	[^m N^k,a^\dagger] = a^\dagger \left[ (N + 1)^k - N^k \right]^m~.
\end{equation}
\chapter{Explicit Expression of $\xi$ Operator} \label{apx:BCH}

In this Appendix, we will compute the necessary commutators to apply BCH formula for the operator in \eqref{eqn:true_four-displacement_operator}
\begin{equation}
	\xi = e^{i\Lambda N_\mathrm{L} Q_\mathrm{M}(3T/4)} e^{i\Lambda N_\mathrm{L} Q_\mathrm{M}(T/2)} e^{i \Lambda N_\mathrm{L} Q_\mathrm{M}(T/4)} e^{i \Lambda N_\mathrm{L} Q_\mathrm{M}(0)}~.
\end{equation}
Let us first recall the involved operators
\begin{subequations}
\begin{align}
	Q_{\mathrm{M}}(0) = & Q_{\mathrm{M}}\\
	Q_{\mathrm{M}} \left(\frac{T}{4}\right) = & \PM
	- (2 \QM^2 - \QM \PM - \PM \QM) \delta \tilde{\gamma}
	+ [ 3 \epsilon \QM \PM \QM + \nonumber \\
	& + \delta^2 (4 \PM \QM \PM - 4 \QM^3 - \QM \PM \QM - 2 \PM^3)] \tilde{\gamma}^2 + 3 \pi \chi \QM \bar{n} \\
	Q_{\mathrm{M}} \left(\frac{T}{2}\right) = & - \QM
	- 2 (\QM \PM + \PM \QM) \delta \tilde{\gamma}
	- 8 \delta^2(\PM \QM \PM - \QM^3) \tilde{\gamma}^2 + 6 \pi \chi \PM  \bar{n} \\
	Q_{\mathrm{M}} \left(\frac{3T}{4}\right) = & - \PM
	+ (2 \QM^2 + \QM \PM + \PM \QM) \delta \tilde{\gamma}
	- [ 3 \epsilon \QM \PM \QM + \nonumber \\
	& - \delta^2 (4 \PM \QM \PM - 4 \QM^3 + \QM \PM \QM + 2 \PM^3) ] \tilde{\gamma}^2 - 9 \pi \chi \QM \bar{n} ~.
\end{align}
\end{subequations}
Furthermore, remind that the first terms for the BCH formula are given by
\begin{equation}
	e^A e^B \simeq \exp\left\{A + B + \frac{1}{2} [A,B] + \frac{1}{12} ([A,[A,B]] + [B,[B,A]]) - \frac{1}{24} [B,[A,[A,B]]]\right\} \ldots ~.
\end{equation}
To second order in $\tilde{\gamma}$, this expansion terminates.
In what follows, we will initially split the operator $\xi$ in the left and right couples.
Finally, we will apply BCH formula once more to find a unique expression for the four-displacement operator.

\paragraph{1st Commutators}

For the first commutators we find
\begin{subequations}
\begin{align}
	\left[\QM\left(\frac{T}{4}\right),\QM(0)\right] 
	= & - i
	- 2 i \delta \tilde{\gamma} (\QM - \PM)
	+ [\delta^2 (2 + 5 i \PM^2 - 4 i \PM \QM + i \QM^2) + \nonumber \\
	& - 3 i \epsilon (\PM^2 + \QM^2)] \tilde{\gamma}^2 ~, \displaybreak\\
	\left[\QM\left(\frac{3T}{4}\right),\QM\left(\frac{T}{2}\right)\right]
	= & - i
	+ 2 i \delta \tilde{\gamma} (\QM - \PM) + \nonumber \\
	& + [\delta^2 (-2 + 5 i \PM^2 + 4 i \PM \QM + 9 i \QM^2) - 3 i \epsilon (\PM^2 + \QM^2)] \tilde{\gamma}^2 ~,
\end{align}
\end{subequations}

\paragraph{2nd Commutators}

For the second commutators we find
\begin{subequations}
\begin{align}
	\left[ \QM \left(\frac{T}{4}\right),\left[\QM\left(\frac{T}{4}\right),\QM(0)\right] \right]
	= & 2 [ \delta^2 (3 \QM-2 \PM) - 3 \epsilon \QM] \tilde{\gamma}^2 - 2 \delta \tilde{\gamma} ~,\\
	\left[ \QM \left(0\right),\left[\QM\left(0\right),\QM\left(\frac{T}{4}\right)\right] \right]
	= & - 2 [\delta^2 (2 \QM - 3 \PM) + 3 \epsilon \PM] \tilde{\gamma}^2 + 2 \delta \tilde{\gamma} ~,\\
	\left[ \QM \left(\frac{3T}{4}\right),\left[\QM\left(\frac{3T}{4}\right),\QM\left(\frac{T}{2}\right)\right] \right]
	= & - 2 [\delta^2 (3 Q - 2 \PM) - 3 \epsilon Q] \tilde{\gamma}^2 - 2 \delta \tilde{\gamma} ~,\\
	\left[ \QM \left(\frac{T}{2}\right),\left[\QM\left(\frac{T}{2}\right),\QM\left(\frac{3T}{4}\right)\right] \right]
	= & 2 [ \delta^2 (2 \QM - 3 \PM) + 3 \epsilon \PM ] \tilde{\gamma}^2 + 2 \delta \tilde{\gamma} ~.
\end{align}
\end{subequations}

\paragraph{3rd Commutators}

For the third commutators we find
\begin{subequations}
\begin{align}
	\left [ \QM(0) , \left[ \QM \left(\frac{T}{4}\right),\left[\QM\left(\frac{T}{4}\right),\QM(0)\right] \right] \right]
	= & - 4 i \delta^2 \tilde{\gamma}^2 ~,\\
	\left [ \QM\left(\frac{T}{2}\right) , \left[ \QM \left(\frac{3T}{4}\right),\left[\QM\left(\frac{3T}{4}\right),\QM\left(\frac{T}{2}\right)\right] \right] \right]
	= & - 4 i \delta^2 \tilde{\gamma}^2 ~.
\end{align}
\end{subequations}

Using the commutators above, we can then find
\begin{subequations} \label{eqns:2-displacement}
\begin{align}
	\xi_\mathrm{l} = & e^{i\Lambda N_\mathrm{L} Q_\mathrm{M}(3T/4)} e^{i\Lambda N_\mathrm{L} Q_\mathrm{M}(T/2)} = \\
	= & \exp\{
	i N_\mathrm{L} \Lambda [
	- (\PM + \QM)
    - \delta \tilde{\gamma} (\PM \QM + \QM \PM - 2 \QM^2)
    - 3 \epsilon \tilde{\gamma}^2 \QM \PM \QM + \nonumber \\
	& + \delta^2 \tilde{\gamma}^2 (2 \PM^3 + 4 \QM^3 - 4 \PM \QM \PM + \QM \PM \QM)
	+ 3 \pi \chi (2 \PM - 3 \QM) (1 + \bar{n})] + \nonumber \\
	& + \frac{i}{2} N_\mathrm{L}^2 \Lambda^2 [1
        + 2 \delta \tilde{\gamma} (\PM - \QM)
        + \delta^2 \tilde{\gamma}^2 (2 - 5 \PM^2 - 9 \QM^2 - 4 \PM QM) + \nonumber \\
	& \qquad - 3 \epsilon \tilde{\gamma}^2 (\PM^2 + \QM^2)] + \nonumber \\
	& - \frac{i}{6} (3 \epsilon - \delta^2) N_\mathrm{L}^3 \tilde{\gamma}^2 \Lambda^3 (\PM + \QM)
    + \frac{i}{6} \delta^2 N_\mathrm{L}^4 \tilde{\gamma}^2 \Lambda^4\} \\
	\xi_\mathrm{r} = & e^{i \Lambda N_\mathrm{L} Q_\mathrm{M}(T/4)} e^{i \Lambda N_\mathrm{L} Q_\mathrm{M}(0)} = \\
	= & \exp \{- i N_\mathrm{L} \Lambda [- (\PM + \QM) 
	- \delta \tilde{\gamma} (\PM \QM + \QM \PM - 2 \QM^2))
	- 3 \epsilon \tilde{\gamma}^2 \QM \PM \QM + \nonumber \\
	& + \delta^2 \tilde{\gamma}^2 (2 \PM^3 + 4 \QM^3 - 4 \PM \QM \PM + \QM \PM \QM)
	- 3 \pi \chi \QM (1 + \bar{n})] + \nonumber \\
	& + \frac{i}{2} N_\mathrm{L}^2 \Lambda^2[
        1
        - 2 \delta \tilde{\gamma} (\PM - \QM)
        - \delta^2 \tilde{\gamma}^2 (2 + 5 \PM^2 + \QM^2 - 4 \PM \QM) + \nonumber \\
	& \qquad + 3 \epsilon \tilde{\gamma}^2 (\PM^2 + \QM^2) ] + \nonumber \\
	& + \frac{i}{6} (3 \epsilon - \delta^2) N_\mathrm{L}^3 \tilde{\gamma}^2 \Lambda^3 (\PM + \QM)
	+ \frac{i}{6} \delta^2 N_\mathrm{L}^4 \tilde{\gamma}^2 \Lambda^4\} ~.
\end{align}
\end{subequations}
Computing the commutators between $A = \log(\xi_\mathrm{r})$ and $B = \log(\xi_\mathrm{l})$, we obtain
\begin{subequations}
\begin{align}
	[A,B] = & 2 i N_\mathrm{L}^3 \tilde{\gamma}^2 \Lambda^3 \left[ 5 (\QM - \PM) \delta^2 - 3 (\QM - \PM) \epsilon \right] + 12 i \pi \chi N_\mathrm{L}^2 \Lambda^2 \\
	[A,[A,B]] = & 4 i N_\mathrm{L}^4 \tilde{\gamma}^2 \Lambda^4 (3 \epsilon - 5 \delta^2) \\
	[B,[B,A]] = & 4 i N_\mathrm{L}^4 \tilde{\gamma}^2 \Lambda^4 (3 \epsilon - 5 \delta^2) \\
	[B,[A,[A,B]]] = & 0~.
\end{align}
\end{subequations}
Using these expressions, we finally find
\begin{multline}
	\xi = \exp \left\{ 6 i N_\mathrm{L} \Lambda \chi \pi \bar{n} (\PM - \QM) 
	+ i N_\mathrm{L}^2 \Lambda^2 \left[(3 \epsilon - 5 \delta^2) \tilde{\gamma}^2 (\QM^2 + \PM^2) + 6 \chi \pi \bar{n} + 1 \right] + \right. \\
	\left. + i N_\mathrm{L}^3 \Lambda^3 \left[(3 \epsilon - 5 \delta^2) \tilde{\gamma}^2 (\PM - \QM)\right]
    + i N_\mathrm{L}^4 \Lambda^4 (2 \epsilon - 3 \delta^2) \tilde{\gamma}^2 \right\}~.
\end{multline}

\end{document}